\documentclass[11pt]{report}

\usepackage{amsmath,amstext,amsfonts}
\usepackage{amssymb,latexsym}
\usepackage{cite}
\usepackage[figuresright]{rotating}
\usepackage{newcent}
\usepackage{afterpage}

\def\spacebetweentables{\vspace{1.5cm}}

\let\oldtabular\tabular
\let\oldendtabular\endtabular
\renewenvironment{tabular}{\small\oldtabular}{\oldendtabular}

\newcommand{\captionfonts}{\small}
\makeatletter  % Allow the use of @ in command names
\long\def\@makecaption#1#2{%
  \vskip\abovecaptionskip
  \sbox\@tempboxa{{\captionfonts #1: #2}}%
  \ifdim \wd\@tempboxa >\hsize
    {\captionfonts #1: #2\par}
  \else
    \hbox to\hsize{\hfil\box\@tempboxa\hfil}%
  \fi
  \vskip\belowcaptionskip}
\makeatother   % Cancel the effect of \makeatletter
\begin{document}

\sloppy

\unitlength=1mm

% greek letters
\def\a{{\alpha}}
\def\b{{\beta}}
\def\D{{\Delta}}
\def\d{{\delta}}
\def\e{{\epsilon}}
\def\g{{\gamma}}
\def\G{{\Gamma}}
\def\k{{\kappa}}
\def\l{{\lambda}}
\def\L{{\Lambda}}
\def\m{{\mu}}
\def\n{{\nu}}
\def\o{{\omega}}
\def\O{{\Omega}}
\def\S{{\Sigma}}
\def\s{{\sigma}}
\def\th{{\theta}}
\def\w{{\omega}}

% altdeutsch
\def\cC{{\mathcal C}}
\def\cB{{\mathcal B}}
\def\cT{{\mathcal T}}
\def\cQ{{\mathcal Q}}
\def\cL{{\mathcal L}}
\def\cO{{\mathcal O}}
\def\cA{{\mathcal A}}
\def\cH{{\mathcal H}}
\def\cR{{\mathcal R}}
\def\cJ{{\mathcal J}}
\def\cK{{\mathcal K}}
\def\cI{{\mathcal I}}
\def\cZ{{\mathcal Z}}
\def\cD{{\mathcal D}}

% others
\def\ol#1{{\overline{#1}}}
\def\Dtilde{{\tilde{\Delta}}}
\def\Dslash{D\hskip-0.65em /}
\def\qslash{D\hskip-0.65em /}
\def\vslash{v\hskip-0.50em /}
\def\Bslash{B\hskip-0.65em /}
\def\diag{\text{diag}}
\def\tr{\text{tr}}
\def\str{\text{str}}
\def\det{\text{det}}
\def\goesto{{\mathop{\longrightarrow}}}
\def\CPT{{$\chi$PT}}
\def\QCPT{{Q$\chi$PT}} 
\def\PQCPT{{PQ$\chi$PT}}
\def\HMCPT{{HM$\chi$PT}}
\def\order{{\mathcal O}}
\def\Lag{{\mathcal L}}
\def\LQCD{{\Lambda_{\text{QCD}}}}
\def\Lchi{{\Lambda_\chi}}

\def\eqref#1{{(\ref{#1})}}

\def\JFV{{J_{\text{FV}}}}
\def\KFV{{K_{\text{FV}}}} 
\def\IFV{{I_{\text{FV}}}} 
\def\HFV{{H_{\text{FV}}}} 

\def\erf{{\text{erf}}}
%%%%%%%%%%%%%%%%%%%%%%%%%%%%%%%%%%%%

\def\la{\langle}
\def\ra{\rangle}
\def\mev{\text{ MeV}}
\def\gev{\mbox{ GeV}}

%\prelimpages
 
\title{\Huge Chiral Perturbation Theory on the Lattice and its Applications}
\author{{\LARGE Daniel Arndt}}
\date{June 2004}

\maketitle

\begin{abstract}
Chiral perturbation theory (\CPT),
the low-energy effective theory of QCD,
can be used to
describe QCD observables in the low-energy region
in a model-independent way.
At any given order in the chiral expansion, 
\CPT\ introduces a finite
number of parameters that encode the short-distance
physics and that must be determined from experiment
or numerical lattice QCD simulations.
In this thesis, we calculate a number of hadronic observables
in the quenched and partially quenched versions
of \CPT:

Chiral
corrections to $B^{(*)}\rightarrow D^{(*)}$
at zero recoil are investigated in quenched \CPT.
We study in detail
the 
charge radii of the meson and baryon octets,
electromagnetic properties of the baryon decuplet,
and the baryon decuplet to octet electromagnetic
transitions in both, quenched and partially quenched
\CPT.
We further show how effects due to the finite size of the lattice
can be accounted for in heavy meson \CPT\ and calculate,
as explicit examples,
neutral $B$ meson mixing 
and the heavy-light meson decay constants. 
We also demonstrate how one can account for
effects due to finite lattice spacing in the low-energy theories,
considering as an example electromagnetic 
meson and baryon
properties.

The results of our calculations 
are crucial to extrapolate quenched
and partially quenched lattice data from the heavier
light quark masses used on the lattice to the physical values.
\end{abstract}

\pagenumbering{roman}

\tableofcontents
%\listoffigures
%\listoftables

\chapter*{Glossary}
%\addcontentsline{toc}{chapter}{Glossary}
\thispagestyle{plain}

\begin{tabular}{l l}
  QCD:\ \ Quantum Chromodynamics. & EFT:\ \ Effective Field Theory.\\ \\
  LQCD:\ \ Lattice QCD. & HQET:\ \ Heavy Quark Effective Theory.\\ \\
  QQCD: \ \ Quenched QCD. & CKM:\ \ Cabibbo-Kobayashi-Maskawa.\\ \\
  PQQCD: \ \ Partially QQCD. & LEC:\ \ low-energy constant.\\ \\
  \CPT: \ \ Chiral Perturbation Theory. & LO:\ \ leading order.\\ \\
  \QCPT:\ \ Quenched \CPT. & NLO:\ \ next-to-LO.\\ \\
  \PQCPT:\ \ Partially \QCPT. & NNLO:\ \ next-to-NLO.\\ \\
  \HMCPT:\ \ Heavy Meson \CPT. & \\ \\
  \phantom{xxxxxxxxxxxxxxxxxxxxxxxxxxxxxxxxxxxxx} & 
\end{tabular}

\chapter*{Acknowledgments}
%\addcontentsline{toc}{chapter}{Acknowlegements}
\thispagestyle{plain}

First of all, I would like to thank my research advisor,
Martin Savage, under whose guidance I have worked in the 
past five years.
Martin taught me not only 
what I know about low-energy QCD,
but also how research should be done.
His enthusiasm, dedication, and way of thinking physics
are truly inspiring
and
without his generous support this thesis
would not have been possible.
In short---I couldn't have asked for a better advisor.
I also want to thank Martin for
his patience with me
when,
time and again,
the weather was 
irresistible and the mountains were calling,
causing me to 
leave my windowless office and
stray 
from the true path of physics.
Many thanks also to
the other members of my graduate committee,
Eric Adelberger,
Steve Ellis, 
Wick Haxton, 
David Kaplan, 
and
Sarah Keller.

Most projects of the past years  
were done with my collaborators
Silas Beane, 
Paddy Fox, 
David Lin,
Martin Savage, 
and
Brian Tiburzi.
Two of these collaborations were especially memorable:
It was a great pleasure to work with Paddy on a
very interesting early project
involving nuclear and particle theory as well as astrophysics.
And I want to thank Brian for an intense and enjoyable collaboration
on a series of papers this past summer.
The research contained in this thesis
was supported in part by the
U.S.\ Department of Energy
under grant
DE-FG-03-97ER41014.

I found the atmosphere in the fourth floor very enjoyable
and conductive to learning.
Thanks to the
theory students and postdocs,
past and present,
with whom I have shared so many years and
who were always happy to chat physics and beyond: 
Matthias B{\"u}chler, 
Chen-Shan Chin, 
Jason Cooke, 
Will Detmold, 
Rob Fardon,
Kyung Kim, 
Pavel Kovtun, 
Andre Kryjevski, 
Tom Luu,
Antony Miceli, 
Gautam Rupak, 
Noam Shoresh, 
Mithat {\"U}nsal,
Ruth Van de Water, 
Andr{\'e} Walker-Loud,
and 
Daisuke Yamada.
All these people provided an environment 
that I looked forward to every day.
Thanks also to my fellow countrymen and regular skiing buddies
Milan Diebel
and
Andreas Zoch.

Life in Seattle would have been only half the fun
had I not taken advantage of the city's unique location
so close to the mountain ranges of the 
Pacific Northwest.
It is a great pleasure to
thank 
my frequent hiking, climbing, and skiing companions
Lisa Good\-enough, 
Jule Gust,
Jim Prager, 
Dustin Shigeno, 
Markus Wagner,
and
Gary Yngve 
for numerous fine outings in the Cascades. 
(Special thanks to Dustin for a very memorable climb of
Mount Rainier!)
Many of the trips I did in the last years where 
done through the 
Seattle Mountaineers' climbing program.
The Mountaineers not only enabled me to learn
how to savely travel the wilderness,
I also got to know many people 
beyond the physics department 
which enlarged my horizon and certainly helped keep me sane
throughout graduate school.
There are too many people 
in the Mountaineers whom
I'd like to thank.
Let me just name a few:
Rob Brown, 
Jim Farris,
and
Priscilla Moore.

Last but not least I want to thank 
my parents in Germany
and 
my sister in South Africa
for supporting me during the last five years and for enduring all these
telephone conversations at very odd hours.
Danke!

%\textpages

\pagenumbering{arabic}

\chapter{Introduction}
\label{chapter:introduction}

Quantum chromodynamics (QCD), 
a part of the very successful
standard model of particle physics,
was formulated over 30~years
ago. 
It is the theory that describes
the interaction of quarks and gluons, which are the building blocks
of hadrons.
In principle, it not only enables the calculation of properties
of protons and neutrons, that make up the nuclei of atoms,
but of the nuclei themselves.
In short, QCD describes all hadronic properties of matter.

Unfortunately, 
even though QCD is simple enough to be written down in the form of
a few partial differential equations,
solving it 
to calculate even basic hadronic properties,
such as the mass or the magnetic moment of the proton,
is very complicated and still poses a challenge.
In a similar theory, quantum electrodynamics,
the fundamental coupling constant is small 
at low energies
and
observables
can be arranged in the form of a series, 
called a perturbative expansion.
Therefore, in calculating a property,
such as the anomalous magnetic moment of the electron,
one only needs to consider the
first few terms in this series that dominate%
---usually a cumbersome but easy task---%
whereas subsequent terms can be neglected
because they are small.

\begin{figure}[tb]
  \centering
  \includegraphics[width=0.75\textwidth]%
                     {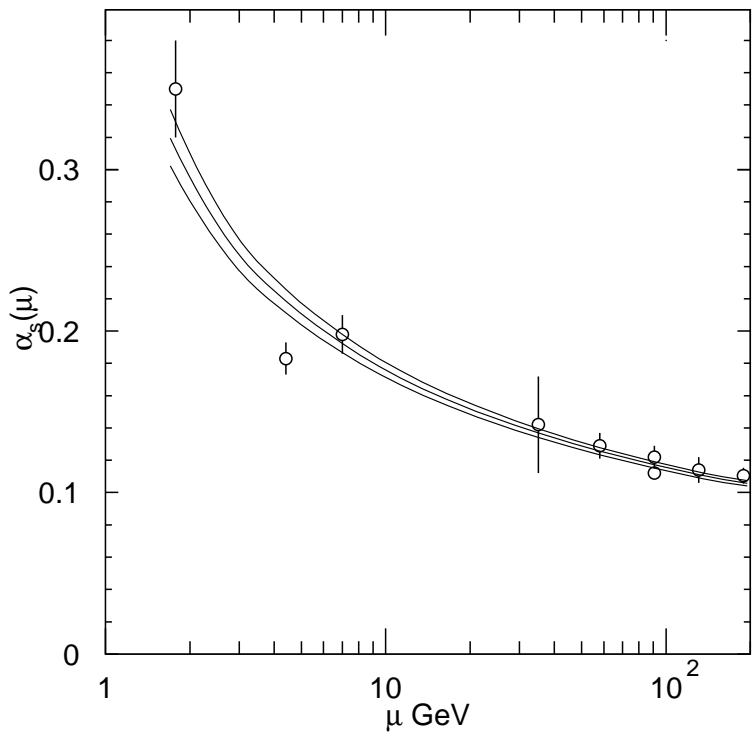}
  \caption[Measured $\a_S$ plotted against the momentum $\mu$
    at which the measurement was made]{
    Measured $\a_S$ plotted against the momentum $\mu$
    at which the measurement was made.
    The three lines show the central value and the $\pm 1 \sigma$
    of the Particle Data Group's average.
    The data points are from several experimental
    measurements.
    Figure taken from Ref.~\cite{Hagiwara:2002fs}.
  }
  \label{introduction-F:runningalpha}
\end{figure}

In QCD the picture is different.
The fundamental coupling constant in QCD, $\a_S$,
depends upon the energy
exchanged in the process under consideration
(see Fig.~\ref{introduction-F:runningalpha})
in a different way:
$\a_S$ is small at large energies, such
as occur during a particle collision in a large
particle accelerator or in a quark-gluon plasma.
Here, perturbative techniques are applicable.
At energies $\lesssim 1$~GeV, however,
the coupling constant becomes large, $\a_S\sim 1$, 
and a perturbative expansion in powers of $\a_S$ fails.
The series does not converge,
instead of becoming smaller,
subsequent terms get bigger and bigger.
This is what makes solving QCD so complicated in 
the low energy region
which is relevant  
for hadronic properties
because that is
where quarks and gluons bind together into composite states.

\begin{figure}[tb]
  \centering
  \includegraphics[width=0.75\textwidth]%
                     {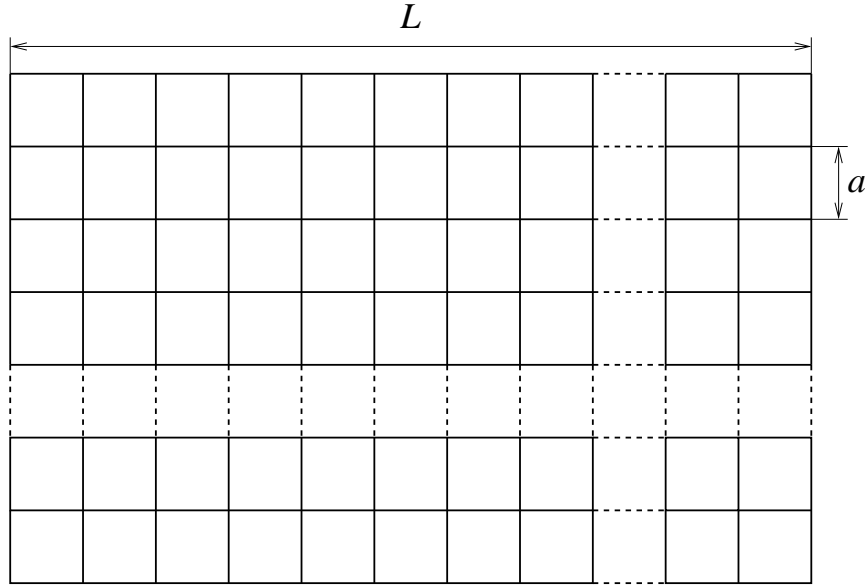}
  \caption[Lattice size $L$ and lattice spacing $a$]{
    Lattice size $L$ and lattice spacing $a$.
    In general, these can be different for each of
    the four dimensions.  However, often only the
    time direction has different $L$ and $a$.
  }
  \label{introduction-F:lattice}
\end{figure}

A way to solve this problem is to use lattice QCD (LQCD).
Here,
one simulates QCD 
with the help of computers
on a finite-sized 4-dimensional 
grid (or lattice) 
that represents points in discretized space-time
(see Fig.~\ref{introduction-F:lattice}).
Since even for today's most powerful computers 
this task is very time-consuming, 
theorists use
additional approximations
to simplify the calculation:
among others,
they neglect (or partially neglect) 
contributions of quark-antiquark pairs that constantly pop 
out of and disappear into
the vacuum 
(the so-called ``quenched'' and ``partially quenched'' approximations)
and that are very costly to calculate; 
and 
they simulate with light quarks 
(of the up and down flavors) that are several times
more massive than in nature.

Because of such approximations, lattice theorists need
to know how to connect their results to
QCD of the real world.
In particular, 
for each property they measure on the lattice
they need to know how to extrapolate 
from the heavier quarks they use down to the quark masses of
nature. 
A model-independent way to do this extrapolation is to use
a low-energy effective theory 
that exploits the symmetries of QCD
and is formulated in terms of 
the relevant degrees of freedom in the low-energy region, 
mesons and baryons, rather than
quarks and gluons:
chiral perturbation theory (\CPT).
Since
the quark mass, $m_q$, dependence is
ex\-plic\-it in \CPT\ (the low-energy
constants are independent of $m_q$)
it is the only rigorous tool for extrapolating
LQCD results down to physical quark masses.
Since most simulations today use either the
quenched or the partially quenched approximations
of QCD, one has to use the quenched or
partially quenched versions of \CPT\ to do the appropriate extrapolations.

The research contained in this thesis involves 
the calculation of a number of
hadronic properties
using \CPT\ as well as 
quenched \CPT\ (\QCPT) 
and partially quenched \CPT\ (\PQCPT).
Our results for \QCPT\ are necessary to
extrapolate existing quenched lattice data 
of these properties
to the physical regime.%
\footnote{Note that our results
in the present form 
cannot be used to
extrapolate staggered lattice simulations.}
Moreover,   
because of the conceptual ad\-vances 
in lattice computing algorithms
made in the past few years
and because of the availability of faster computers,
many of these properties
will be simulated with improved precision in partially quenched QCD,
although it will be a long time
before real simulations with light physical quarks become feasible.
Therefore these lattice results
need to be extrapolated to real-world QCD
using our results for PQ$\chi$PT.

Besides (partial) quenching and simulating at heavier light quarks,
there is a number of further 
artifacts that come about by using LQCD
and that must be taken into account and included
in the (P)Q$\chi$PT treatment.
As steps in this directions,
we have included $\order(a)$ effects 
(due to the non-zero lattice spacing, $a$,
the ``graininess'' of the discrete lattice)
in the calculation of baryon properties 
and
finite $L$ effects 
(due to the finite size of the lattice box, $L$)
in
the calculation of properties of heavy quark systems.

\begin{center} \rule{0.5\textwidth}{0.4pt}\end{center}

This thesis contains work carried out over the past two years
and it is laid out as follows.
In Chapter~\ref{chapter:CPT} we 
give a brief introduction into 
QCD, \CPT, LQCD, \QCPT, and \PQCPT\
that is needed for all subsequent chapters.
We elaborate on the 
implications of the quenched and partially quenched
theories for the extrapolation
of 
lattice QCD simulations
carried out at an unphysical regime
to the physical regime.
The subsequent chapters deal with the calculation of a number
of hadronic properties.
Chapters~\ref{chapter:B2D} and \ref{chapter:fv}
involve heavy mesons.
In Chapter~\ref{chapter:B2D}, we
study 
the semileptonic $B^{(*)}\to D^{(*)}$ decays
in the heavy quark limit and
calculate the lowest order chiral 
corrections from the breaking
of heavy quark symmetry 
at the zero recoil point in \QCPT~\cite{Arndt:2002ed}.
In Chapter~\ref{chapter:fv}, we incorporate
finite volume effects in the calculation of properties
of heavy quark systems.
In particular, we investigate how the scale $\D$, which
comes from the breaking of heavy quark symmetry,
influences finite volume effects.
This work was carried out in collaboration
with David Lin~\cite{Arndt:2004bg}.
In Chapters~\ref{chapter:baryon_ff}--\ref{chapter:trans},
we calculate a number of hadronic properties
in the baryon sector in \QCPT\ and \PQCPT. 
Whereas Chapter~\ref{chapter:baryon_ff} involves the
baryon octet, Chapter~\ref{chapter:decuplet} and 
Chapter~\ref{chapter:trans} deal with the baryon decuplet and
the baryonic octet-decuplet transition, respectively.
In Chapter~\ref{chapter:fa}, we extend the calculation of the
subsequent three chapters by incorporating finite $a$ effects.
These chapters are work done in collaboration with
Brian Tiburzi%
~\cite{Arndt:2003ww,Arndt:2003we,Arndt:2003vd,Arndt:2004we}.
Finally, in Chapter~\ref{chapter:conclusions} we summarize and conclude.
Several appendices contain supplemental material that
has been taken out of the main text in order to improve readability.

Most of the work contained in this thesis has been published
previously:
\begin{itemize}
\item
Daniel Arndt,
{\it Chiral $1/M_Q^2$ Corrections to
$B^{(*)}\rightarrow D^{(*)}$ at Zero Recoil in 
Quenched Chiral Perturbation Theory},
Phys.\ Rev.\ D {\bf 67}, 074501 (2003).

\item
Daniel Arndt and Brian C.\ Tiburzi,
{\it Charge Radii of the Meson and Baryon Octets in 
Quenched and Partially Quenched 
Chiral Perturbation Theory},
Phys.\ Rev.\ D {\bf 68}, 094501 (2003).

\item
Daniel Arndt and Brian C.\ Tiburzi,
{\it Electromagnetic Properties of the Baryon Decuplet
in Quenched and Partially Quenched Chiral Perturbation Theory},
Phys.\ Rev.\ D {\bf 68}, 114503 (2003),
Erratum-ibid.\ D {\bf 69}, 059904 (2004).

\item
Daniel Arndt and Brian C.\ Tiburzi,
{\it Baryon Decuplet to Octet 
Electromagnetic Transitions 
in Quenched and Partially Quenched 
Chiral Perturbation Theory},
Phys.\ Rev.\ D {\bf 69}, 014501 (2004).

\item
Daniel Arndt and Brian C. Tiburzi,
{\it  Hadronic Electromagnetic Properties at 
Finite Lattice Spacing},
Phys.\ Rev.\ D {\bf 69}, 114503 (2004).

\item
Daniel Arndt and C.J.\ David Lin,
{\it  Heavy Meson Chiral Perturbation Theory 
in Finite Volume},
Phys.\ Rev.\ D in press,
{[hep-lat/0403012]}.
\end{itemize}

\chapter{QCD, Chiral Perturbation Theory, and the Lattice}
\label{chapter:CPT}

In this chapter 
we introduce the part of the standard model that
describes the interactions of quarks and gluons, QCD.
Since QCD can only be solved perturbativly at energies
well above the energy scale relevant for hadronic properties,
we describe \CPT, which is QCD's 
low-energy effective theory.
\CPT---%
as an effective field theory (EFT)---% 
introduces a number of unknown parameters
that encode the underlying short-distance physics and that
have to be fixed by comparison to either 
experimental measurements
or to results from numerical lattice QCD (LQCD) 
simulations.
We briefly describe LQCD and
the quenched and partially quenched approximations
that are used frequently
and
introduce the low-energy chiral effective theories
that can be used to extrapolate results from lattice
simulations that employ these approximations:
\QCPT\ and \PQCPT.
Lastly, we comment on how reliable it is 
to predict QCD properties 
from lattice simulations that use
the 
quenched and partially quenched approximations.

\section{QCD and Chiral Symmetries}

The Lagrangian of QCD is given by 
\begin{equation}
  \cL_{\text{QCD}}\label{CPT-eq:LQCD}
  =
  -\frac{1}{4}G_{\mu\nu}^AG^{A\,\mu\nu}
  +
  \sum_{a,b=u,d,s}\bar{q}_a(i\Dslash-m_q)_{ab} q_b
\end{equation}
where the eight gauge bosons $A_\mu^A$ are contained in
the gluon field strength 
tensor $G^A_{\mu\nu}$ that is given by 
\begin{equation}
  G^A_{\mu\nu}
  =
  \partial_{\mu} A_{\nu}^A-\partial_{\nu} A_{\mu}^A-gf^{ABC}A_\mu^B A_\nu^C
.\end{equation}
The structure constants $f^{ABC}$ are defined by
\begin{equation}
  [T^A,T^B]=if^{ABC}T^C
\end{equation}
and the $T_A$ are the eight generators of color $SU(3)$.
The Lagrangian also contains the
triplet of quarks
$q=(u,d,s)$
of the up, down, and strange flavors%
\footnote{
Although the standard model has
six flavors of quarks that, in principle, should all be included,
only the three lightest flavors
are relevant for the calculation of hadronic properties
at energies  $\lesssim 1$~GeV.
The quarks of the charm, top, and bottom flavors
have masses that are typically much larger than
$1$~GeV.
}
with mass matrix 
\begin{equation}
  m_q=\diag(m_u,m_d,m_s)
.\end{equation}
The quarks
are minimally coupled to the gluon fields via
\begin{equation}
  D_\mu=\partial_\mu+igA_\mu^AT^A
.\end{equation}

Assuming that one is in a regime where perturbation theory is applicable,
one can
calculate how the strong coupling constant 
\begin{equation}
  \a_S(\mu)=\frac{g^2(\mu)}{4\pi}
\end{equation}
depends on the renormalization scale $\mu$.
From the QCD $\b$ function calculated to
$\order(g^3)$
one finds
\begin{equation} \label{CPT-eq:aS}
  \a_S(\mu)=\frac{12\pi}{(33-2N_f)\log(\mu^2/\LQCD^2)}
.\end{equation}
This means that, as long as the number of quark flavors $N_f$ is
smaller than 16, $\a_S$ becomes larger with
decreasing $\mu$,
a behavior known as {\it asymptotic freedom}. 
Moreover, if $\mu\to\LQCD$ then $\a_S$ blows up.
Of course, in that case the theory is not perturbative in
the first place. 
However, $\LQCD$,
which 
can be determined from fitting Eq.~\eqref{CPT-eq:aS}
to experimental measurements
to be about 200~MeV
(see Fig.~\ref{introduction-F:runningalpha}),
can still be viewed as the scale
where QCD becomes strongly coupled.

In the limit of vanishing quark masses ($m_q\to 0$)
the quark part of Eq.~\eqref{CPT-eq:LQCD}
becomes simply
\begin{equation}
   \cL
   =
   \bar{q}i\Dslash q
   =
  \bar{q}_Li\Dslash q_L + \bar{q}_Ri\Dslash q_R
\end{equation}
which exhibits an exact global $SU(3)_L\times SU(3)_R$ symmetry.
This means that 
the left- and right-handed quark fields,
\begin{equation}
  q_L=\frac{1-\g^5}{2}q
\quad
\text{and}
\quad
  q_R=\frac{1+\g^5}{2}q
,\end{equation}
transform under independent $SU(3)$ flavor space rotations,
\begin{equation}\label{CPT-eq:LRtrafos}
  q_L\rightarrow Lq_L,\quad q_R\rightarrow Rq_R
,\end{equation}
with $L\in SU(3)_L$ and $R\in SU(3)_R$.

In nature, the masses of the light quarks are not zero.
If they are turned on
then the term
\begin{equation}
  \bar{q}m_qq=\bar{q}_Lm_qq_R+\bar{q}_Rm_qq_L
\end{equation}
appears in
the Lagrangian 
which
is only invariant 
if $L=R$.
In that case the symmetry
is broken down to its diagonal subgroup:
$SU(3)_L\times SU(3)_R\to SU(3)_V$.
Although the masses of the light quarks are not zero,
they are nevertheless small compared to the scale $\LQCD$.
One would therefore expect nature to exhibit at least an
approximate $SU(3)_L\times SU(3)_R$ symmetry.
However, such a symmetry is not seen.  What one does see 
is evidence for just a single
$SU(3)$.
One therefore assumes that the symmetry 
$SU(3)_L\times SU(3)_R$ 
is 
spontaneously
broken down to the observed $SU(3)$.
This symmetry breaking is accomplished by
the formation of scalar quark bilinears  $\bar{q}q$
that have a non-zero vacuum expectation value:
\begin{equation}
  \langle\bar{q}_R^i q_L^j\rangle=\l\delta^{ij}
.\end{equation}
Under an $SU(3)_L\times SU(3)_R$
transformation this vacuum expectation value becomes
\begin{equation}
  \langle\bar{q}_R^i q_L^j\rangle\to \l(LR^\dagger)^{ij}
,\end{equation}
which means that for $\l\neq0$
the vacuum expectation value is only 
unchanged if $L=R$ and
the chiral symmetry
$SU(3)_L\times SU(3)_R$
is spontaneously broken down to its diagonal subgroup
$SU(3)_V$.
The eight broken subgroups cause the appearance
of eight massless Goldstone bosons
that are the fluctuations along the directions where
the potential is constant.
These eight Goldstone bosons are believed to be realized in nature
as the pseudoscalar meson octet.  
The fact that the pseudoscalar mesons are light but not massless
reflects the fact that the $SU(3)_L\times SU(3)_R$
is only an approximate symmetry of the Lagrangian.
The Goldstone bosons
can be represented
by a $3\times3$ matrix $\S$
that transforms as
\begin{equation} \label{CPT-eq:Sigmatrans}
  \S\to L\S R^\dagger
\end{equation}
and can be written as
\begin{equation} \label{CPT-eq:sigma}
  \S=\exp\left(\frac{2i\Phi}{f}\right)
,\end{equation}
where $\Phi$ is a traceless hermitian matrix given by
\begin{equation}
  \Phi=
    \left(
      \begin{array}{ccc}
        \frac{1}{\sqrt{2}}\pi^0+\frac{1}{\sqrt{6}}\eta & \pi^+ & K^+ \\ 
        \pi^- & -\frac{1}{\sqrt{2}}\pi^0+\frac{1}{\sqrt{6}}\eta & K^0 \\
        K^- & \bar{K}^0 & -\frac{2}{\sqrt{6}}\eta
      \end{array}
    \right)
\end{equation}
and $f$ is a constant with dimensions of mass,
known as the pion decay constant.

At energies below $\L_\chi$,%
\footnote{Actually, energies should be below the
mass of the $\rho$, $m_\rho=770$~MeV,
since the $\rho$ is not included in the EFT.
This is accomplished by treating $m_\rho\sim\Lambda_\chi$.}
these Goldstone bosons are the only degrees of
freedom 
and one
can write down an effective Lagrangian that
describes their interactions.
In principle, any term with the correct dimensions that
obeys all the symmetries of the QCD Lagrangian in 
Eq.~\eqref{CPT-eq:LQCD} can be included in such
an effective Lagrangian.
However, since the number of such terms is infinite,
one has to use a truncation scheme that limits the
number of terms to be included:
Because higher order terms contain more and more derivatives,
they are suppressed by the scale $\L_\chi$.
For the theory with massless quarks, the lowest order term is
\begin{equation}
  \frac{f^2}{8}\tr\left(\partial^\mu\S^\dagger\partial_\mu\S\right)
.\end{equation}
If $m_q$ is non-zero, then the term $\bar{q}m_q q$ 
in Eq.~\eqref{CPT-eq:LQCD}
is not 
invariant under $SU(3)_L\times SU(3)_R$.
This can be fixed by treating $m_q$ as an 
independent field, a so-called {\it spurion},
that is assumed to transform as
\begin{equation}
  m_q\to Lm_qR^\dagger  
.\end{equation} 
Under simultaneous chiral
transformations of the quark and spurion fields 
the
mass term in Eq.~\eqref{CPT-eq:LQCD} is invariant.
Using the spurion technique one
can then include terms into the EFT 
Lagrangian that involve $m_q$.  Doing so yields
the lowest order Lagrangian 
of \CPT\
that is of order $p^2/\L_\chi^2$~\cite{Gasser:1985gg,Gasser:1985ux}
\begin{equation} \label{CPT-eq:Lcpt}
  \cL
  =
  \frac{f^2}{8}\tr\left(\partial^\mu\S^\dagger\partial_\mu\S\right)
  +
  \l\,\tr\left(m_q\S+m_q^\dagger\S^\dagger\right)
  +
  \dots
.\end{equation}
This Lagrangian includes all possible terms
up to order $p^2/\L_\chi^2$;
terms that are of higher order in the $p/\L_\chi$
chiral expansion (more derivatives, more powers
of $m_q$) have been neglected.

The Lagrangian in Eq.~\eqref{CPT-eq:Lcpt}
is valid as long as $p \ll \Lchi$
and  $p/\Lchi$
can be used as a small expansion parameters.
It can be systematically expanded to include
terms that are of higher order in $p/\Lchi$.
Each term is accompanied by a constant 
($f$ and $\l$ in the above lowest order Lagrangian)
that is {\it a priori} unknown.
Observables receive contributions from both
long-range and short range physics;
the long-range contribution arises from
the  
(non-analytic) structure of 
pion loop contributions, while the short-range
contribution is encoded in these low-energy constants that
appear in the chiral Lagrangian and are unconstrained in
\CPT. 
These constants must
be determined from experiment or lattice simulations.

By expanding Eq.~\eqref{CPT-eq:Lcpt} to lowest order in 
the meson fields $\Phi$
one can calculate the masses of the 
Goldstone bosons in terms of the quark masses, $m_q$,
and the constants 
$f$ and $\l$.  
For the masses of off-diagonal meson,
that are made up of the (anti-)quarks $q$ and $ q'$,
one finds
\begin{equation}\label{CPT-eq:hanswurst}
  m_{qq'}^2
  =
  \frac{4\l}{f^2}\left(m_q+m_q'\right)
.\end{equation}
For example, this gives 
\begin{equation}
  m_{\pi^\pm} = \frac{4\l}{f^2}\left(m_u+m_d\right),\quad
  m_{K^\pm} = \frac{4\l}{f^2}\left(m_u+m_s\right),
\end{equation}
and
\begin{equation}
  m_{K^0} = m_{\bar{K}^0} = \frac{4\l}{f^2}\left(m_d+m_s\right)
.\end{equation}
Clearly, in the isospin limit, where $m_u=m_d$, the
charged and neutral kaons have equal mass.
Similarly one finds for the masses of the mesons on the diagonal
\begin{equation}
  m_{\pi^0} = \frac{4\l}{f^2}\left(m_u+m_d\right)
  \quad\text{and}\quad
  m_{\eta} = \frac{4\l}{3f^2}\left(m_u+m_d+4m_s\right)
,\end{equation}
so that $m_{\pi^0}=m_{\pi^\pm}$ in the isospin limit.
In nature, the kaons are much heavier than the pions, which
reflects the fact that
$m_s\gg m_u,m_d$.

\section{Lattice QCD}
If the coupling $\a_S$ is small then one can use perturbative
methods to calculate the vacuum expectation value of an
operator $O$ from the path integral%
\footnote{
LQCD is formulated in Euclidean space, 
which can be accomplished
by a Wick rotation from Minkowski space 
$t\to-it_E$.
We will use Euclidean space in this subsection only.
}
\begin{equation} \label{CPT-eq:expval}
  \langle O \rangle
  =
  \frac{1}{\cZ}
  \int\cD A\,\cD\bar{q}\,\cD q\,
  O
  \exp
  \left(-\int d^4x\,\cL\right)
\end{equation}
with the generating functional $\cZ$ defined as
\begin{equation}
  \cZ
  =
  \int\cD A\,\cD\bar{q}\,\cD q
  \exp
  \left(-\int d^4x\,\cL\right) 
\end{equation}
by expanding the exponential in powers of the 
interacting part of the Lagrangian and solving the 
functional integral analytically for 
the first
few terms 
in the series.
This approach fails in 
the strong coupling region (for energies smaller $\lesssim\LQCD$)
because the expansion parameter
$\a_S$ becomes large.
A way to solve QCD in the strong coupling region has been 
proposed by Wilson~\cite{Wilson:1974sk}
and it involves putting QCD on a 4-dimensional 
discrete space-time lattice and solving it 
numerically using computers.
This method, known as lattice QCD, basically involves two steps:
\begin{enumerate}
  \item
  The infinite dimensional 
  functional integral in Eq.~\eqref{CPT-eq:expval}
  needs to be discretized so that it can be
  calculated in a finite number of steps.
  This is accomplished by discretizing space-time and
  putting QCD in a 4-dimensional space-time lattice.
  In Wilsons formulation of LQCD,
  the fermionic fields (quarks) live on the lattice sites
  whereas the gauge fields (gluons) are defined on the links
  which are
  the lines that connect neighboring lattice sites.
  \item
  Even after this 
  discretization solving the functional integral
  means summing over an enormous number of paths in
  configuration space.  However, 
  for most of these paths the exponential
  is tiny; the integral is dominated only by a small number
  of paths. 
  In lattice simulations one tries to exploit this by
  sampling only a small number of gauge configurations
  that minimize the action using Monte Carlo methods.
  Then one can approximate $O(A)$ as the average over this
  finite ensemble of gauge configurations
  \begin{equation}
    \langle O(A)\rangle
    \approx
    \frac{1}{N}
    \sum_c O(A_c)
  \end{equation}
  where $N$ is the number of configurations in the ensemble.
\end{enumerate} 
 
\begin{figure}[tb]
  \centering
  \includegraphics[width=0.5\textwidth]%
               {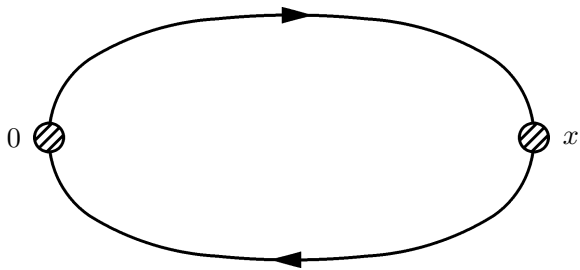}
  \caption[Quark line diagram representing the 
  pion correlation function]%
  {Quark line diagram representing the 
  pion correlation function.
  The hatched blobs represent sources for the pion at the
  indicated positions. The quarks
  are represented by solid lines.
  While only the quarks connected to the sources are shown,
  this diagram must be evaluated with all quark and gluon
  contributions.}
  \label{CPT-F:pioncorrelator}
\end{figure}

As an example, consider a pion 
that, being a pseudoscalar, can be represented
by 
$\pi(x)=\bar{d}(x)\g^5 u(x)$ (see Fig.~\ref{CPT-F:pioncorrelator}).
The correlation function for this pion
can be written as
\begin{eqnarray} \label{CPT-eq:correlation}
  \langle 0|\bar{\pi}(x)\pi(0) | 0\rangle
  &=&
  \langle 0| \bar{u}(x)\g^5 d(x)\bar{d}(0)\g^5 u(0)| 0\rangle 
                \nonumber        \\
  &=&
  \frac{1}{\cZ}
  \int\cD A\,\cD \bar{q}\,\cD q\,
  \bar{u}(x)\g^5 d(x)\bar{d}(0)\g^5 u(0)
  \exp\left(-\int d^4x\,\cL\right)
                          \nonumber \\
  &=&
  \frac{1}{\cZ}
  \int\cD A\,
  \tr\left[G_d(x,0)\g^5G_u(0,x)\g^5\right]
  \exp\left(-S_G[A]\right)\det[\Dslash+m_q]
               \nonumber \\
  &=&
  \left\langle
    \tr\left[G_d(x,0)\g^5G_u(0,x)\g^5\right]
  \right\rangle_{\text{sampled over gauge field configurations}} 
                          \label{CPT-eq:expectval2}
\end{eqnarray} 
where $S_G[A]$ is the pure Yang-Mills part of the
gauge field action
and 
$G_q^{-1}=\Dslash+m_q$ 
is the inverse propagator for a quark of flavor $q$
[the discretized version of which appears in the 
last line of Eq.~(\ref{CPT-eq:expectval2})].
In an LQCD simulation one approximates the expectation value 
by the average
over the weighted samples.

The computing power available today
puts severe restrictions on what can be simulated:
Typically, the size of the lattice, $L$, is limited to
a few fermi
($\sim$2--4~fm);
obviously it should be at least as big as 
the Compton wave length of the lightest particle
one wants to simulate.
Moreover, the lattice spacing, $a$, should be as small 
as possible 
so that discretization artifacts are kept to a minimum;
typically $a\sim L/5$, so that a typical box
size would be $5\times5\times5\times5$.

But even with these constraints it turns out that lattice simulations 
with realistic quark masses 
($m_{u,d}\sim 5$~MeV, $m_s\sim 100$~MeV) 
are not feasible with the
computational power that is available today.

\section{Quenching and Partial Quenching}
\label{section:CPT-quenching}
The fermion determinant in 
Eq.~\eqref{CPT-eq:correlation}
is very expensive to compute
since it typically scales $\sim m_q^{-2.5}$.
In contrast, 
the propagators are
much less costly to calculate as they scale $\sim m_q^{-1}$.

The mass that appears in the fermion determinant is
the mass for quarks that are generated in the gauge field
background, {\it i.e.}, it is only assigned
to quarks that are generated {\it dynamically} 
from vacuum polarization in the gluonic background.
These so-called ``sea'' quarks 
are not connected to the sources of the correlator.
The quarks that are connected to the sources, and that
have their mass appearing in the propagators, are called 
``valence'' quarks.
Since in a lattice simulation
the calculation of the fermion determinant 
(that involves only sea quarks)
is independent of the calculation of the propagators 
(involving solely valence quarks)
one has the freedom to
vary the masses of the sea and valence
quarks independently.

As an extreme way to save computing time
one can omit calculating the
fermion determinant completely.
This is called the quenched approximation of QCD (QQCD).
Effectively, this is a theory without sea quarks
as they are treated as being infinitely heavy.
Although simulating QQCD is much less
costly than simulating full QCD
(by a factor $\sim 1000$)
it turns out that there exists, as will be explained shortly,
no known connection between QQCD and QCD.
Although there are hints 
that quenching might not make much difference
for certain observables,
it does introduces uncontrolled systematic errors.

A less severe approximation is
partially quenched QCD (PQQCD).
Unlike in QQCD, where the sea quark masses
are set to infinity,
they are kept finite in PQQCD.
Sea quarks are thereby retained as dynamical degrees of freedom
and the fermion determinant is no longer
equal to one. 
However, 
by efficaciously giving the sea quarks larger masses, 
the fermion determinant becomes much less costly to calculate than
in full QCD.
The main advantage of PQQCD, compared to QQCD, is that 
there does exist a known analytic connection to QCD:
By setting the sea quark masses equal to the valence
quark masses one recovers QCD.

How would one in practice 
do a perturbative
QQCD or PQQCD calculation?
The obvious approach is to write down
all QCD Feynman diagrams that contribute to a certain order
in perturbation theory.  
Then, for QQCD, one simply disregards all diagrams 
that contain virtual quark loops
since these consist of sea quarks.
For PQQCD, one
assigns the sea quark mass to the quarks that appear in
virtual loops.
This method has been used in, for example, 
Refs.~\cite{Labrenz:1996jy,Leinweber:2002qb}.
Although dropping or modifying individual diagrams
is very illustrative,
this methods is somewhat artificial.

A more systematic way, that does not require 
modification of individual diagrams,
is to include ghost quarks (that have bosonic statistics) 
in the quenched theory
and to introduce
ghost and sea quarks in the partially quenched theory.
In the next two subsections we will introduce the field theoretical
formulation of QQCD and PQQCD.
We will also explain
their effective low energy theories,
\QCPT\ and \PQCPT,
that are needed to properly extrapolate lattice data
from the heavier light quark masses used on the lattice
to realistic masses.

Note that, although in general the number of valence and
sea quark flavors need not be identical,
we use the case of flavor $SU(3)$ 
and work with three valence and three sea quark flavors
throughout most of this thesis.
The case of flavor $SU(2)$,
with two valence and two sea quark flavors, 
is very similar and will be
explained when appropriate.

\subsection{QQCD and \QCPT}
\begin{figure}[tb]
  \centering
  \includegraphics[width=0.66\textwidth]%
               {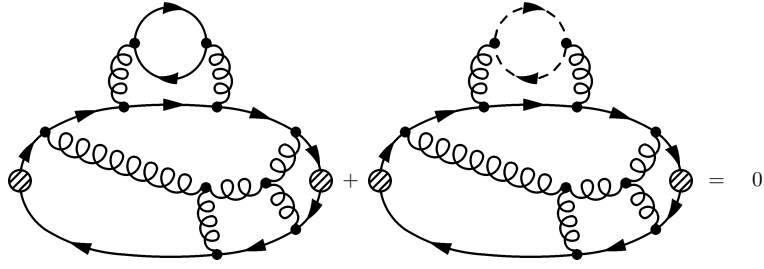}
  \caption[Cancellation of valence and ghost quark loops
  in QQCD]%
  {Cancellation of valence and ghost quark loops
  in QQCD.  Dashed lines represent ghost quarks.
  Since loops containing valence and ghost quarks of the
  same flavor have the opposite sign,
  the first two diagrams cancel completely, 
  effectively removing any diagram with closed quark loops.}
  \label{CPT-F:pion-quenched}
\end{figure}

In QQCD the quark part of the Lagrangian is written as%
~\cite{Savage:2001dy}
\begin{equation}\label{CPT-eqn:LQQCD}
  {\cal L}
  =
  \sum_{a,b=u,d,s}\bar{q}_a(i\Dslash-m_q)_{ab} q_b
  + \sum_{\tilde{a},\tilde{b}=\tilde{u},\tilde{d},\tilde{s}}
      \bar{\tilde{q}}_{\tilde{a}}
      (i\Dslash-m_{\tilde{q}})_{\tilde{a}\tilde{b}} 
      \tilde{q}_{\tilde{b}}
  =
  \sum_{j,k=u,d,s,\tilde{u},\tilde{d},\tilde{s}}
  \bar{Q}_j(i\Dslash-m_Q)_{jk} Q_k
.\end{equation}
Here, in addition to the fermionic light valence quarks $u$, $d$, and $s$ 
their bosonic counterparts $\tilde{u}$, $\tilde{d}$, and $\tilde{s}$
have been added.
These six quarks are in the fundamental representation of
the graded group $SU(3|3)$%
~\cite{BahaBalantekin:1981kt,BahaBalantekin:1981qy,BahaBalantekin:1982bk}
and have been 
accommodated in the six-component vector
\begin{equation}
  Q=(u,d,s,\tilde{u},\tilde{d},\tilde{s})
\end{equation}
that obeys the graded equal-time commutation relation
\begin{equation} \label{CPT-eqn:commutation}
  Q^\a_i({\bf x}){Q^\b_j}^\dagger({\bf y})
  -(-1)^{\eta_i \eta_j}{Q^\b_j}^\dagger({\bf y})Q^\a_i({\bf x})
  =
  \d^{\a\b}\d_{ij}\d^3({\bf x}-{\bf y})
,\end{equation}
where $\a$ and $\b$ are spin and $i$ and $j$ are flavor indices.
The graded equal-time commutation relations for two $Q$'s and two
$Q^\dagger$'s can be written analogously.
The grading factor 
\begin{equation}
   \eta_k
   = \left\{ 
       \begin{array}{cl}
         1 & \text{for } k=1,2,3 \\
         0 & \text{for } k=4,5,6
       \end{array}
     \right.
\end{equation}
takes into account the different statistics for
fermionic and bosonic quarks.
The quark mass and charge matrices are given by 
\begin{equation}
  m_Q=\text{diag}(m_u,m_d,m_s,m_u,m_d,m_s)
\end{equation}
and
\begin{equation}
  \cQ=\text{diag}\left(\frac{2}{3},-\frac{1}{3},-\frac{1}{3},
                       \frac{2}{3},-\frac{1}{3},-\frac{1}{3}
                 \right)
,\end{equation}
respectively, so that diagrams with closed ghost quark loops cancel 
those with valence quarks
as illustrated in
Fig.~\ref{CPT-F:pion-quenched}.

For massless quarks,
the Lagrangian in Eq.~(\ref{CPT-eqn:LQQCD}) exhibits a graded symmetry
$SU(3|3)_L \otimes SU(3|3)_R \otimes U(1)_V$ that is assumed 
to be spontaneously broken down to $SU(3|3)_V \otimes U(1)_V$. 
The low-energy effective theory of QQCD that emerges by 
expanding about the physical vacuum state is \QCPT.
The dynamics of the emerging 36~pseudo-Goldstone mesons 
can be described at lowest 
order in the chiral expansion by the $\order(E^2)$ Lagrangian%
\footnote{
Here, $E\sim p$, $m_\pi$ where $p$ is an external momentum.
}~\cite{Morel:1987xk,Sharpe:1992ft,Bernard:1992mk,%
Bernard:1992ep,Golterman:1994mk}
\begin{equation}\label{CPT-eqn:Lchi}
  {\cal L} =
  \frac{f^2}{8}
    \str\left(D^\mu\Sigma^\dagger D_\mu\Sigma\right)
    + \l\,\str\left(m_Q\Sigma+m_Q^\dagger\Sigma^\dagger\right)
    + \a\partial^\mu\Phi_0\partial_\mu\Phi_0
    - \mu_0^2\Phi_0^2
\end{equation}
where
$\S$ is defined in Eq.~\eqref{CPT-eq:sigma}
and
\begin{equation}
  \Phi=
    \left(
      \begin{array}{cc}
        \pi & \chi^{\dagger} \\ 
        \chi & \tilde{\pi}
      \end{array}
    \right)
.\end{equation}
Here the $\pi$, $\tilde{\pi}$, and $\chi$ are $3\times3$ matrices
of pseudo Goldstone bosons with quantum numbers of $\bar{q}q$ pairs,
pseudo Goldstone bosons with quantum numbers of 
$\bar{\tilde{q}}\tilde{q}$ pairs, and 
pseudo Goldstone fermions with quantum numbers 
of $\bar{\tilde{q}}q$ pairs, respectively:
\begin{equation}
  \pi=
    \left(
      \begin{array}{ccc}
        \eta_u & \pi^+ & K^+ \\ 
        \pi^- & \eta_d & K^0 \\
        K^- & \bar{K^0} & \eta_s
      \end{array}
    \right),\quad
  \tilde{\pi}=
    \left(
      \begin{array}{ccc}
        \tilde{\eta}_u & \tilde{\pi}^+ & \tilde{K}^+ \\ 
        \tilde{\pi}^- & \tilde{\eta}_d & \tilde{K}^0 \\
        \tilde{K}^- & \bar{\tilde{K^0}} & \tilde{\eta}_s
      \end{array}
    \right),
  \quad\text{and}\,\,
  \chi=
    \left(
      \begin{array}{ccc}
        \chi_{\eta_u} & \chi_{\pi^+} & \chi_{K^+} \\ 
        \chi_{\pi^-} & \chi_{\eta_d} & \chi_{K^0} \\
        \chi_{K^-} & \chi_{\bar{K^0}} & \chi_{\eta_s}
      \end{array}
    \right)
.\end{equation}
The pion decay constant is
$f=132$~MeV,
and we have defined the gauge-covariant derivative
$D_\mu\S=\partial_\mu\S+ie\cA_\mu[\cQ,\S]$.
The str() denotes a supertrace over flavor indices
defined as
\begin{equation}
  \str(X)=\sum_{i=1}^6(-1)^{\eta_i}X_{ii}
.\end{equation}
Upon expanding the Lagrangian in \eqref{CPT-eqn:Lchi} one finds that
to lowest order
the mesons with quark content $Q\bar{Q'}$
are canonically normalized when
their masses are given by
\begin{equation}\label{CPT-eqn:mqq}
  m_{QQ'}^2=\frac{4\lambda}{f^2}(m_Q+m_{Q'})
.\end{equation}
One also finds that the propagator for off-diagonal (flavored)
Goldstone mesons composed of (ghost-) quarks $Q$ and $Q'$
is given by
\begin{equation}\label{CPT-eqn:Gqq}
  G_{QQ'}(p)=\frac{i}{p^2-m_{QQ'}^2+i\e}
.\end{equation}

The flavor-singlet field $\Phi_0$ is defined as  
\begin{equation}
  \Phi_0
  =
  \frac{1}{\sqrt{6}}\str(\Phi)
  =
  \frac{1}{\sqrt{2}}(\eta'-\tilde{\eta}')
.\end{equation}
$\Phi_0$ is
invariant under $SU(3|3)_L \otimes SU(3|3)_R \otimes U(1)_V$
and thus arbitrary functions of it can be included in the
Lagrangian.
To lowest order in the chiral expansion only the two operators included
in Eq.~(\ref{CPT-eqn:Lchi}) with parameters
$\a$ and $\mu_0$ remain and are understood to be inserted 
perturbativly~\cite{Bernard:1992mk}.
Notice that this singlet field $\Phi_0$ is not heavy as in \CPT\
and therefore cannot be 
integrated out. It introduces a new vertex,
the so-called hairpin
with the propagator
\begin{equation}
  G_{\eta_a\eta_a}
  =
  \frac{i}{p^2-m_{\eta_a}^2+i\e}
  +
  \frac{i(\mu_0^2-\a p^2)}{\left(p^2-m_{\eta_a}^2+i\e\right)^2}
\end{equation}
that exhibits a double pole which causes 
quenching artifacts
and is ultimately responsible for the sick behavior 
of the quenched theory.

\subsection{PQQCD and \PQCPT}
\begin{figure}[tb]
  \centering
  \includegraphics[width=\textwidth]%
               {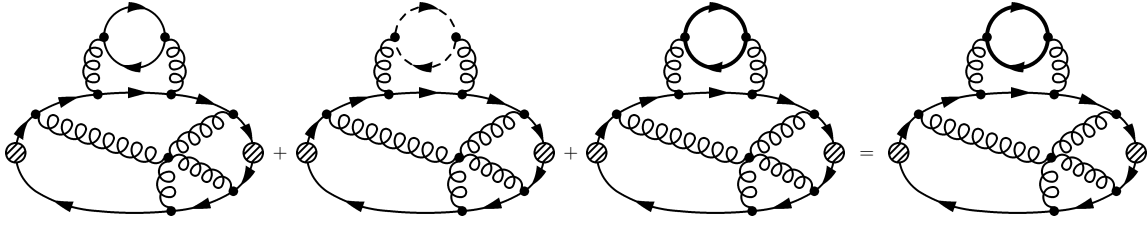}
  \caption[Cancellation between valence, ghost, and sea quark loops
  in PQQCD]%
  {Cancellation between valence and ghost quark loops
  in PQQCD.  As in Fig.~\ref{CPT-F:pion-quenched},
  dashed lines represent ghost quarks whereas fat solid lines
  represent the (heavier) sea quarks.
  Like in QQCD, valence quark loops are canceled by their
  ghostly counterparts.  The inclusion of the (fermionic) 
  sea quarks
  effectively replaces the valence quark masses in loops by sea quark
  masses.  Obviously, in the QCD limit, where valence and sea
  quarks have equal mass, one recovers QCD.}
  \label{CPT-F:pion-partiallyquenched}
\end{figure}

The physics for flavor off-diagonal mesons in PQQCD
is very similar to the QQCD case.
The quark part of the Lagrangian is 
extended once again by including 
three light fermionic sea quarks $j$, $l$, and $r$
and can be
written as%
~\cite{Sharpe:2000bn,Sharpe:2001fh,Sharpe:2000bc,Sharpe:1999kj,
Golterman:1998st,Sharpe:1997by,Bernard:1994sv,Shoresh:2001ha}
\begin{eqnarray}\label{CPT-eqn:LPQQCD}
  {\cal L}
  &=&
  \sum_{a,b=u,d,s}\bar{q}_a(i\Dslash-m_q)_{ab} q_b
  + \sum_{\tilde{a},\tilde{b}=\tilde{u},\tilde{d},\tilde{s}}
      \bar{\tilde{q}}_{\tilde{a}}
      (i\Dslash-m_{\tilde{q}})_{\tilde{a}\tilde{b}} 
      \tilde{q}_{\tilde{b}}
  +
  \sum_{a,b=j,l,r}
  \bar{q}_{\text{sea},a} (i\Dslash-m_{\text{sea}})_{ab} q_{\text{sea},b}
               \nonumber \\
  &&=
  \sum_{j,k=u,d,s,\tilde{u},\tilde{d},\tilde{s},j,l,r}
  \bar{Q}_j(i\Dslash-m_Q)_{jk} Q_k
.\end{eqnarray}
These nine quarks are in the fundamental representation of
the graded group $SU(6|3)$%
~\cite{BahaBalantekin:1981kt,BahaBalantekin:1981qy,BahaBalantekin:1982bk}
and have been 
accommodated in the nine-component vector
\begin{equation}
  Q=(u,d,s,j,l,r,\tilde{u},\tilde{d},\tilde{s})
\end{equation}
that obeys the graded equal-time commutation relation
in Eq.~(\ref{CPT-eqn:commutation}).
Now, however, the grading factor is
\begin{equation}
   \eta_k
   = \left\{ 
       \begin{array}{cl}
         1 & \text{for } k=1,2,3,4,5,6 \\
         0 & \text{for } k=7,8,9
       \end{array}
     \right.
.\end{equation}
The quark mass matrix is given by 
\begin{equation} \label{CPT-eq:mQPQ}
  m_Q=\text{diag}(m_u,m_d,m_s,m_j,m_l,m_r,m_u,m_d,m_s)
\end{equation}
so that,
in a perturbative expansion,
diagrams with closed ghost quark loops cancel 
those with valence quarks
just like in QQCD.
Effects of virtual quark loops are,
however, present due to the contribution of the finite-mass 
sea quarks
(see Fig.~\ref{CPT-F:pion-partiallyquenched}).

It has been recently realized~\cite{Golterman:2001yv} that 
the light quark electric charge matrix $\cQ$ is not uniquely
defined in PQQCD.  
The only constraint one imposes is for
the charge matrix $\cQ$ to have vanishing
supertrace. Thus, as in QCD, no new operators
involving the singlet component are subsequently introduced.
Following~\cite{Chen:2001yi} we use
\begin{equation} \label{CPT-eqn:chargematrix}
  \cQ
  =
  \diag
  \left(
    \frac{2}{3},-\frac{1}{3},-\frac{1}{3},q_j,q_l,q_r,q_j,q_l,q_r
  \right)
\end{equation}
so that QCD is recovered in the limit 
$m_j\to m_u$, $m_l\to m_d$, and $m_r\to m_s$
independently of the $q$'s. 

For massless quarks,
the Lagrangian in Eq.~(\ref{CPT-eqn:LPQQCD}) exhibits a graded symmetry
$SU(6|3)_L \otimes SU(6|3)_R \otimes U(1)_V$ that is assumed 
to be spontaneously broken down to $SU(6|3)_V \otimes U(1)_V$. 
The low-energy effective theory of PQQCD that emerges by 
expanding about the physical vacuum state is \PQCPT.
The dynamics of the emerging 80~pseudo-Goldstone mesons 
can be described at lowest 
order in the chiral expansion by the Lagrangian
given in Eq.~\eqref{CPT-eqn:Lchi}
with $\Sigma$ as defined in Eq.~\eqref{CPT-eq:sigma}
but $\Phi$ now being extended to include mesons that
contain sea quarks
\begin{equation} \label{CPT-eq:Phi}
  \Phi=
    \left(
      \begin{array}{cc}
        M & \chi^{\dagger} \\ 
        \chi & \tilde{M}
      \end{array}
    \right)
.\end{equation}
The $M$, $\tilde{M}$, and $\chi$ are matrices
of pseudo-Goldstone bosons with quantum numbers of $q\ol{q}$ pairs,
pseudo-Goldstone bosons with quantum numbers of 
$\tilde{q}\ol{\tilde{q}}$ pairs, 
and pseudo-Goldstone fermions with quantum numbers of $\tilde{q}\ol{q}$ pairs,
respectively.
Explicitly they are given by
\begin{equation} \label{CPT-eq:MMtilde}
  M=
    \left(
      \begin{array}{cccccc}
        \eta_u & \pi^+ & K^+ & J^0 & L^+ & R^+\\ 
        \pi^- & \eta_d & K^0 & J^- & L^0 & R^0\\
        K^- & \bar{K^0} & \eta_s & J^-_s & L^0_s & R^0_s\\
        \bar{J}^0 & J^+ & J^+_s & \eta_j & Y^+_{jl} & Y^+_{jr}\\
        L^- & \bar{L}^0 & \bar{L}^0_s & Y^-_{jl} & \eta_l & Y^0_{lr}\\
        R^- & \bar{R}^0 & \bar{R}^0_s & Y^-_{jr} & \bar{Y}^0_{lr} & \eta_r
      \end{array}
    \right),\quad
  \tilde{M}=
    \left(
      \begin{array}{ccc}
        \tilde{\eta}_u & \tilde{\pi}^+ & \tilde{K}^+\\ 
        \tilde{\pi}^- & \tilde{\eta}_d & \tilde{K}^0\\
        \tilde{K}^- & \tilde{\bar{K}}^0 & \tilde{\eta}_s
      \end{array}
    \right)
,\end{equation}
and
\begin{equation} \label{CPT-eq:chi}
  \chi=
    \left(
      \begin{array}{ccc}
        \chi_{\eta_u} & \chi_{\pi^+} & \chi_{K^+} \\ 
        \chi_{\pi^-} & \chi_{\eta_d} & \chi_{K^0} \\
        \chi_{K^-} & \chi_{\bar{K^0}} & \chi_{\eta_s}
      \end{array}
    \right)
.\end{equation}
Meson masses and non-singlet propagators are similar to the
quenched case as given in
Eqs.~\eqref{CPT-eqn:mqq} and \eqref{CPT-eqn:Gqq}.

The flavor singlet field given by $\Phi_0=\str(\Phi)/\sqrt{6}$
is, in contrast to the \QCPT\ case, rendered heavy by the $U(1)_A$
anomaly
and can therefore be integrated out in \CPT.
Analogously, its mass $\mu_0$ can be taken to be 
on the order of the chiral symmetry breaking scale, 
$\mu_0\to\Lambda_\chi$.  
In this limit the 
flavor singlet propagator becomes independent of the
coupling $\a$ and 
deviates from a simple pole form~\cite{Sharpe:2000bn,Sharpe:2001fh}:
\begin{equation}
  G_{\eta_a\eta_b}
  =
  \frac{i\delta^{ab}}{q^2-m_{\eta_a}^2+i\e}
  -
  \frac{i}{3}
  \frac{(q^2-m_{jj}^2)(q^2-m_{rr}^2)}
    {\left(q^2-m_{\eta_a}^2+i\e\right)
     \left(q^2-m_{\eta_b}^2+i\e\right)
     \left(q^2-m_X^2+i\e\right)}
.\end{equation}
This can be more compactly written in a form
that only contains single poles:
\begin{eqnarray}
  G_{\eta_a\eta_b}
  &=&
  \delta^{ab}P\left(m_{\eta_a}\right)
  -\frac{1}{3}
   \frac{(m_{jj}^2-m_{\eta_a}^2)(m_{rr}^2-m_{\eta_a}^2)}
        {(m_{\eta_a}^2-m_{\eta_b}^2)(m_{\eta_a}^2-m_X^2)}
   P\left(m_{\eta_a}\right)   \nonumber \\
  &&+\frac{1}{3}
   \frac{(m_{jj}^2-m_{\eta_b}^2)(m_{rr}^2-m_{\eta_b}^2)}
        {(m_{\eta_a}^2-m_{\eta_b}^2)(m_{\eta_b}^2-m_X^2)}
   P\left(m_{\eta_b}\right)
  -\frac{1}{3}
   \frac{(m_X^2-m_{jj}^2)(m_X^2-m_{rr}^2)}
        {(m_X^2-m_{\eta_a}^2)(m_X^2-m_{\eta_b}^2)}
   P\left(m_X\right)     \nonumber \\
\end{eqnarray}
where
\begin{equation}
  P(m)=\frac{i}{q^2-m^2+i\e}
\end{equation}
and
$m_X$ is given by
$m_X^2=(m_{jj}^2+2m_{rr}^2)/3$.

\subsection{Inclusion of the Baryon Octet and Decuplet in \QCPT\ and \PQCPT}
Just as there are mesons in QQCD [PQQCD]%
\footnote{
Here, we explain 
the inclusion of baryons for the quenched case;
the partially quenched case is very similar and included 
in square brackets.}
with quark content
$\ol{Q}_iQ_j$ that contain
valence~(v) and ghost~(g)
[v, g, and sea(s)] 
quarks,
there are baryons 
with quark compositions $Q_iQ_jQ_k$ that
contain these two [three] types of quarks.
Restrictions on the baryon fields ${\mathcal B}_{ijk}$
come from the fact that these fields
must reproduce the familiar octet and decuplet
baryons when $i$, $j$, $k=1$-$3$%
~\cite{Labrenz:1996jy,Chen:2001yi,Savage:2002fm}.
To this end, one decomposes the irreducible representations
of $SU(3|3)_V$ [$SU(6|3)_V$] into 
irreducible representations of 
$SU(3)_{\text{v}} \otimes SU(3)_{\text{g}}
 \otimes U(1)$
[$SU(3)_{\text{v}} \otimes SU(3)_{\text{s}} \otimes SU(3)_{\text{g}}
 \otimes U(1)$].

\subsubsection{Baryon Octet}
The method to construct the octet baryons is to use the
interpolating field
\begin{equation}
  \cB_{ijk}^\g
  \sim
  \left(Q_i^{\a,a}Q_j^{\b,b}Q_k^{\g,c}-Q_i^{\a,a}Q_j^{\g,c}Q_k^{\b,b}\right)
  \e_{abc}(C\g_5)_{\a\b}
,\end{equation}
which when restricted to $i$, $j$, $k=1$-$3$ 
has non-zero overlap with the octet baryons.
Under $SU(3|3)_V$ [$SU(6|3)_V$],
where $Q_i\longrightarrow U_{ij}Q_j$ and 
$\overline{Q_i}\longrightarrow \overline{Q_j}U_{ji}^\dagger$, 
$\cB_{ijk}$ transforms as
\begin{equation}
  \cB_{ijk}
  \longrightarrow
  (-)^{\eta_m(\eta_j+\eta_n)+(\eta_m+\eta_n)(\eta_k+\eta_l)}
  U_{im}U_{jn}U_{kl}\cB_{mnl}
.\end{equation}
Using the commutation relations in Eq.~(\ref{CPT-eqn:commutation})
one sees that $\cB_{ijk}$ satisfies the symmetries
\begin{eqnarray}
  \cB_{ijk}&=&(-)^{1+\eta_j\eta_k}\cB_{ikj}, \nonumber \\
  0&=&\cB_{ijk}+(-)^{1+\eta_i\eta_j}\cB_{jik}
        +(-)^{1+\eta_i\eta_j+\eta_j\eta_k+\eta_k\eta_i}\cB_{kji}
                   \label{CPT-eqn:Bsymmetries}
.\end{eqnarray}  

The spin-1/2 baryon octet $B_{ijk}=\cB_{ijk}$,
where the
indices $i$, $j$, and $k$ are restricted to $1$-$3$,
is contained as an 
$(\bf 8,\bf 1)$ [$(\bf 8,\bf 1,\bf 1)$] 
of
$SU(3)_{\text{v}} \otimes SU(3)_{\text{g}}$
[$SU(3)_{\text{v}} \otimes SU(3)_{\text{s}} \otimes SU(3)_{\text{g}}$]
in the 
$\bf 70$ [$\bf 240$]
representation.
The octet baryons, written in the familiar two-index notation
\begin{equation}
  B=
    \left(
      \begin{array}{ccc}
        \frac{1}{\sqrt{6}}\L+\frac{1}{\sqrt{2}}\S^0 & \S^+ & p \\ 
        \S^- & \frac{1}{\sqrt{6}}\L-\frac{1}{\sqrt{2}}\S^0 & n \\
        \Xi^- & \Xi^0 & -\frac{2}{\sqrt{6}}\L
      \end{array}
    \right)
,\end{equation}
are embedded in $B_{ijk}$ as~\cite{Labrenz:1996jy}
\begin{equation}
  B_{ijk}
  =
  \frac{1}{\sqrt{6}}
  \left(
    \e_{ijl}B_{kl}+\e_{ikl}B_{jl}
  \right)
.\end{equation}
As explained in Ref.~\cite{Labrenz:1996jy}, 
it is convenient to switch to the three-index ``quark flow''
notation $B_{ijk}$ as opposed to the familiar
two-index notation of the octet baryons.
The reason that the two-index notation is possible
at all 
is due to the
fact that a $3\times3$ matrix contains
8~elements plus an overall constant.

Besides the conventional octet baryons that contain valence quarks,
$qqq$,
there are also baryon fields with other types of quarks
contained in the $\bf 70$ ($\bf 240$).
Since we are only interested 
in calculating one-loop diagrams that have octet baryons
in the external states,
we will need only the $\cB_{ijk}$ 
that contain at least two valence quarks.
We use the explicit construction in \cite{Savage:2001dy,Chen:2001yi}.
For example, 
baryons that consist of two valence and one ghost quark 
are denoted by the tensors
$_{\tilde a}{\tilde s}_{bc}$ and
$_{\tilde a}{\tilde t}_{bc}$ that
transform as a
${\bf 27}=({\bf 6},{\bf 3})+(\bf\bar 3,\bf 3)$
of 
$SU(3)_{\text{v}} \otimes SU(3)_{\text{g}}$
[${\bf 27}=({\bf 6},{\bf 1},{\bf 3})+(\bf\bar 3,{\bf 1},{\bf 3})$
of 
$SU(3)_{\text{v}} \otimes SU(3)_{\text{s}} \otimes SU(3)_{\text{g}}$].
For completeness,
we list the
transformations for octet baryons containing any combination
of quarks in
Table~\ref{CPT-table:SU3Q}
\begin{table}[tb]
\centering
\caption[Embedding of the baryon octet and decuplet into $SU(3|3)_V$]{
        Embedding of the baryon octet and decuplet into $SU(3|3)_V$ for QQCD.}
\label{CPT-table:SU3Q}
\begin{tabular}{c | c   c | c   c}\hline\hline
  & \multicolumn{2}{c|}{Octet} & \multicolumn{2}{c}{Decuplet} \\
  & $SU(3)_{\text{v}}\otimes SU(3)_{\text{g}}$ & dim & $SU(3)_{\text{v}}\otimes SU(3)_{\text{g}}$ & dim \\ \hline
  $qqq$ & $(\bf8,\bf1)$ & $\bf8$ & $(\bf10,\bf1)$ & $\bf10$ \\
  $qq\tilde{q}$ & $(\bf6,\bf3)\oplus(\bf\bar3,\bf3)$ & $\bf27$ & $(\bf6,\bf3)$ & $\bf18$\\
  $q\tilde{q}\tilde{q}$ & $(\bf3,\bf6)\oplus(\bf3,\bf\bar3)$  & $\bf27$ & $(\bf3,\bf\bar3)$  & $\bf9$\\
  $\tilde{q}\tilde{q}\tilde{q}$ & $(\bf1,\bf8)$ & $\bf8$ & $(\bf1,\bf1)$ & $\bf1$\\
  \hline
  & & $\bf70$ & & $\bf38$\\
  \hline\hline
\end{tabular}
\end{table}
for QQCD and in
Table~\ref{CPT-table:SU3PQ} for PQQCD.
\begin{table}[tb]
\centering
\caption[Embedding of the baryon octet and decuplet for $SU(6|3)_V$]{
        Embedding of the baryon octet and decuplet for $SU(6|3)_V$ for PQQCD.}
\label{CPT-table:SU3PQ}
\begin{tabular}{c | c   c | c   c}\hline\hline
  & \multicolumn{2}{c|}{Octet} & \multicolumn{2}{c}{Decuplet} \\
  & $SU(3)_{\text{v}}\otimes SU(3)_{\text{s}}\otimes SU(3)_{\text{g}}$ & dim & $SU(3)_{\text{v}}\otimes SU(3)_{\text{s}}\otimes SU(3)_{\text{g}}$ & dim \\ \hline
  $qqq$ & $(\bf8,\bf1,\bf1)$ & $\bf8$ & $(\bf10,\bf1,\bf1)$ & $\bf10$\\
  $qqq_{\text{s}}$ & $(\bf6,\bf3,\bf1)\oplus(\bf\bar3,\bf3,\bf1)$ & $\bf27$ & $(\bf6,\bf3,\bf1)$ & $\bf18$ \\
  $qq_{\text{s}}q_{\text{s}}$ & $(\bf3,\bf6,\bf1)\oplus(\bf3,\bf\bar3,\bf1)$ & $\bf27$ & $(\bf3,\bf6,\bf1)$ & $\bf18$\\
  $q_{\text{s}}q_{\text{s}}q_{\text{s}}$ & $(\bf1,\bf8,\bf1)$ & $\bf8$ & $(\bf1,\bf10,\bf1)$ & $\bf10$\\
  $qq\tilde{q}$ & $(\bf6,\bf1,\bf3)\oplus(\bf\bar3,\bf1,\bf3)$ & $\bf27$ & $(\bf6,\bf1,\bf3)$ & $\bf18$\\
  $qq_{\text{s}}\tilde{q}$ & $(\bf3,\bf3,\bf3)\oplus(\bf3,\bf3,\bf3)$ & $\bf54$ & $(\bf3,\bf3,\bf3)$ & $\bf27$\\
  $q_{\text{s}}q_{\text{s}}\tilde{q}$ & $(\bf1,\bf3,\bf6)\oplus(\bf1,\bf3,\bf\bar3)$ & $\bf27$ & $(\bf1,\bf6,\bf3)$ & $\bf18$\\
  $q\tilde{q}\tilde{q}$ & $(\bf3,\bf1,\bf6)\oplus(\bf3,\bf1,\bf\bar3)$ & $\bf27$ & $(\bf3,\bf1,\bf\bar3)$ & $\bf9$\\
  $q_{\text{s}}\tilde{q}\tilde{q}$ & $(\bf1,\bf3,\bf6)\oplus(\bf1,\bf3,\bf\bar3)$ & $\bf27$ & $(\bf1,\bf3,\bf\bar3)$ & $\bf9$\\
  $\tilde{q}\tilde{q}\tilde{q}$ & $(\bf1,\bf1,\bf8)$ & $\bf8$ & $(\bf1,\bf1,\bf1)$ & $\bf1$\\
  \hline
  & & $\bf240$ & & $\bf138$\\
  \hline\hline
\end{tabular}
\end{table}
In Appendix~\ref{CPT-SU2} we list the transformations of the
doublet and quartet baryons for the two flavor case $SU(2)$
(Tables~\ref{CPT-table:SU2Q} and \ref{CPT-table:SU2PQ}).

\subsubsection{Baryon Decuplet}
Similarly, 
the familiar spin-3/2 decuplet baryons are embedded
in the $\bf 38$ [$\bf 138$].  
Here,
one uses the interpolating field
\begin{equation}
  \cT_{ijk}^{\a,\mu}
  \sim
  \left(
    Q_i^{\a,a}Q_j^{\b,b}Q_k^{\g,c}
    +Q_i^{\b,b}Q_j^{\g,c}Q_k^{\a,a}
    +Q_i^{\g,c}Q_j^{\a,a}Q_k^{\b,b}
  \right)
  \e_{abc}
  \left(C\g^\mu\right)_{\b\g}
\end{equation}
that describes the $\bf 38$ [$\bf 138$] 
dimensional representation of $SU(3|3)_V$ [$SU(6|3)_V$]
and has non-zero overlap with the decuplet baryons when 
the indices are restricted
to $i$, $j$, $k=1$-$3$.
Due to the commutation relations in Eq.~\eqref{CPT-eqn:commutation},
$\cT_{ijk}$ satisfies the symmetries
\begin{equation} \label{CPT-eqn:Tsymmetries}
  \cT_{ijk}=(-)^{1+\eta_i\eta_j}\cT_{jik}=(-)^{1+\eta_j\eta_k}\cT_{ikj}
.\end{equation}

The decuplet baryons 
are then readily embedded in $\cT$ by construction:
$T_{ijk}=\cT_{ijk}$, where
the indices $i$, $j$, $k$ are restricted to $1$-$3$.
They transform as a $(\bf 10, \bf1)$ 
[$(\bf 10, \bf 1, \bf1)$]
under
$SU(3)_{\text{v}} \otimes SU(3)_{\text{g}}$
[$SU(3)_{\text{v}} \otimes SU(3)_{\text{s}} \otimes SU(3)_{\text{g}}$].
Because of Eq.~(\ref{CPT-eqn:Tsymmetries}), $T_{ijk}$ is
a totally symmetric tensor.  
Our normalization convention is such that $T_{111}=\D^{++}$.
For the spin-3/2 baryons that contain 
two valence quarks---%
the only ones relevant for our purpose---% 
we use the states constructed in%
~\cite{Savage:2001dy,Chen:2001yi}.
For example,
spin-3/2 baryons consisting of two valence and one ghost quark
transform as $(\bf 6,\bf3)$ [$(\bf 6,\bf 1,\bf3)$]
under 
$SU(3)_{\text{v}} \otimes SU(3)_{\text{g}}$
[$SU(3)_{\text{v}} \otimes SU(3)_{\text{s}} \otimes SU(3)_{\text{g}}$].
For completeness, we list the transformations for 
the remaining decuplet baryons in 
Table~\ref{CPT-table:SU3Q} (QQCD) and
Table~\ref{CPT-table:SU3PQ} (PQQCD);
the transformations for the two flavor case are given
in Appendix~\ref{CPT-SU2}.

\subsubsection{Free Lagrangian for Baryons}
At leading order in the heavy baryon expansion, the 
free $SU(3)$ Lagrangian for the $\cB_{ijk}$ and 
$\cT_{ijk}$ is given by~\cite{Labrenz:1996jy}
\begin{eqnarray} \label{CPT-eqn:L}
  {\mathcal L}
  &=&
  i\left(\ol\cB v\cdot{\mathcal D}\cB\right)
  +2\a_M\left(\ol\cB \cB{\mathcal M}_+\right)
  +2\b_M\left(\ol\cB {\mathcal M}_+\cB\right)
  +2\sigma_M\left(\ol\cB\cB\right)\str\left({\mathcal M}_+\right)
                              \nonumber \\
  &&-i\left(\ol\cT^\mu v\cdot{\mathcal D}\cT_\mu\right)
  +\D\left(\ol\cT^\mu\cT_\mu\right)
  +2\g_M\left(\ol\cT^\mu {\mathcal M}_+\cT_\mu\right)
  -2\ol\sigma_M\left(\ol\cT^\mu\cT_\mu\right)\str\left({\mathcal M}_+\right)
,\nonumber \\
\end{eqnarray}
where 
${\mathcal M}_+
  =\frac{1}{2}\left(\xi^\dagger m_Q \xi^\dagger+\xi m_Q \xi\right)$
with $\xi^2=\Sigma$.
The covariant derivatives of $\cB_{ijk}$ and $\cT_{ijk}$ 
both have the form
\begin{equation} \label{CPT-eqn:coderiv}
  ({\mathcal D}^\mu \cB)_{ijk}
  =
  \partial^\mu \cB_{ijk}
  +(V^\mu)_{il}\cB_{ljk}
  +(-)^{\eta_i(\eta_j+\eta_m)}(V^\mu)_{jm}\cB_{imk}
  +(-)^{(\eta_i+\eta_j)(\eta_k+\eta_n)}(V^\mu)_{kn}\cB_{ijn}
.\end{equation}

The brackets in (\ref{CPT-eqn:L}) are shorthands for field
bilinear invariants originally employed in~\cite{Labrenz:1996jy}
\begin{align} \label{CPT-eq:bracketnotation}
  \left(\ol\cB\G\cB\right)&=\ol\cB^\a_{kji}\G^\b_\a\cB_{ijk,\b}, \\
  \left(\ol\cB\G Y\cB\right)&=\ol\cB^\a_{kji}\G^\b_\a Y_{il}\cB_{ljk,\b},
  \quad
  \left(\ol\cB \G\cB Y\right)=(-)^{(\eta_i+\eta_j)(\eta_k+\eta_n)}
                             \ol\cB^\a_{kji}\G^\b_\a Y_{kn}\cB_{ijn,\b}, \\
  \left(\ol\cT^\mu\G\cT_\mu\right)&=
                           \ol\cT^{\mu,\a}_{kji}\G^\b_\a\cT_{ijk,\b\mu}, \\
  \left(\ol\cT^\mu\G Y\cT_\mu\right)&=\ol\cT^{\mu\a}_{kji}\G^\b_\a Y_{il}\cT_{ljk,\b\mu},
   \quad \text{and} \quad
  \left(\ol\cB\G Y^\mu\cT_\mu\right)=\ol\cB^\a_{kji}\G^\b_\a \left(Y^\mu\right)_{il}\cT_{ljk,\b\mu}
,\end{align}
which ensure that the contraction of flavor indices maintains proper 
transformations under chiral rotations.
To lowest order in the chiral expansion, Eq.~\eqref{CPT-eqn:L} gives the 
propagators
\begin{equation}
  \frac{i}{v\cdot k},\quad
  \frac{iP^{\mu\nu}}{v\cdot k-\D}
\end{equation}
for the spin-1/2 and spin-3/2 baryons, respectively.
Here, $v$ is the velocity and $k$ the residual momentum of the
heavy baryon which are related to the momentum $p$ by
$p=M_B v+k$.
$M_B$ denotes the (degenerate) mass of the octet baryons
and $\D$ the decuplet--baryon mass splitting.
The polarization tensor
\begin{equation}
  P^{\mu\nu}
  =
  \left(v^\mu v^\nu-g^{\mu\nu}\right)-\frac{4}{3}S^\mu S^\nu
\end{equation}
reflects the fact that the Rarita-Schwinger field 
$\cT^\mu_{ijk}$ contains both spin-1/2 and spin-3/2 pieces;
only the latter remain as propagating
degrees of freedom (see \cite{Jenkins:1991ne}, for example).

\section{Extrapolation of Lattice Data}
\label{section:CPT-extrapol}
If unquenched lattice simulations with light enough
quarks
were possible today
then one could simply use \CPT\ to extrapolate 
to the physical quark masses.
Unfortunately, now and in the foreseeable future this is not the case
and one is bound to simulate using the quenched or partially quenched
approximations
and to extrapolate 
to the physical quark masses
using the appropriate low-energy effective theories, 
\QCPT\ and \PQCPT.
The next question then is:
What statements about QCD can be made from
extrapolated QQCD or PQQCD lattice data?

\begin{figure}[tb]
  \centering
  \includegraphics[width=0.5\textwidth]%
               {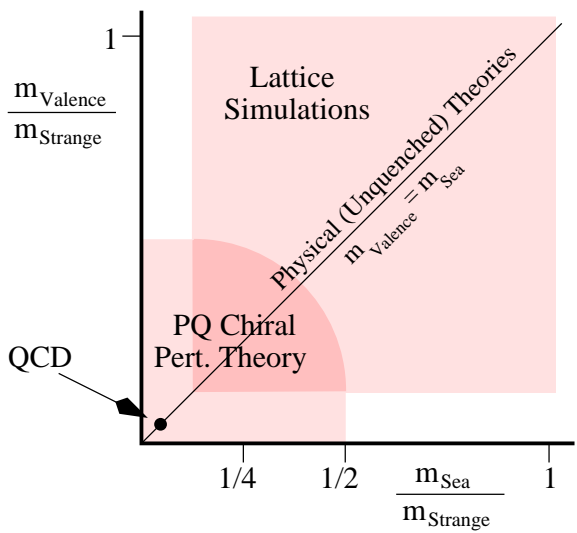}
  \caption[Domains for QCD and PQQCD in quark mass space]%
  {Domains for QCD and PQQCD in quark mass space.
   Full QCD is represented by the point on the diagonal.
   \CPT\ ``lives'' on the diagonal.
   Partial quenching opens up the parameter space
   to the 2-dimensional gray region that
   has a large overlap with the region where \PQCPT\
   is applicable.
   The low-energy parameters for \PQCPT\ (which are the
   same as for \CPT) can be determined
   from this overlap with improved precision.
   Figure from Ref.~\cite{Sharpe:2003vy} courtesy Ruth Van de Water.}
  \label{CPT-F:ruthplot}
\end{figure}

Since PQQCD retains a $U(1)_A$ anomaly,
the equivalent to the singlet field in QCD is heavy (on the order
of
the chiral symmetry breaking scale $\L_\chi$) and can be integrated out%
~\cite{Sharpe:2000bn,Sharpe:2001fh}%
---just like in QCD.
Therefore, the low-energy constants appearing in \PQCPT\ are the same
as those appearing in \CPT.
By fitting \PQCPT\ to partially quenched
lattice data one can determine these constants  
and make physical predictions for QCD.
The advantage of PQQCD is that,
since one can vary the sea quark masses
independently from the valence quark masses,
one has an enlarged parameter space (more adjustable ``knobs'')
and can hope to determine the low-energy constants
with greater accuracy
by fitting to a larger number of
partially quenched lattice results
(see Fig.~\ref{CPT-F:ruthplot}).

For example, since the valence and ghost quarks
have equal masses, 
the contribution of valence quarks in
disconnected quark loop diagrams is
eliminated by the ghost quarks.
The effects of disconnected loop diagrams
are solely due to sea quarks
and the physics of the sea sector
can be explored by varying the sea quark masses.
Furthermore, in the limit where the masses
of the sea quarks become equal to those of the valence 
and ghost quarks, one recovers QCD.

In processes that involve electroweak gauge fields,
the
``theory space'' of PQQCD is enlarged once more
since one 
can chose arbitrary values for the charges of
the ghost and sea quarks, $q_j$, $q_l$, and $q_r$ in
Eq.~(\ref{CPT-eqn:chargematrix}).
For example, if one choses $q_j=q_l=q_r=0$
then photons can only couple to valence quarks.
In the case $q_j=2/3$, $q_l=-1/3$, and $q_r=-1/3$
contributions of the valence and ghost sectors cancel
and photons can only couple to sea quarks~\cite{Savage:2002fm}.

In QQCD the answer to the question raised above is different.  
The problem with the quenched approximation is
that the Goldstone boson singlet is no longer affected by
the $U(1)_A$ anomaly as in QCD.
In other words, the QQCD equivalent of the $\eta'$
that is heavy in QCD remains light and must be included in the
\QCPT\ Lagrangian.
This requires the addition of new operators and hence new
low-energy constants.
In general, the low-energy constants appearing in the \QCPT\
Lagrangian are unrelated to those in \CPT\ and
extrapolated quenched lattice data is unrelated to QCD.
Although there is some empirical evidence
that the difference between QQCD and QCD
for some observables 
is small at large quark masses
both theories deviate considerable in the low quark mass region.
In fact, several examples show that
the behavior of meson loops near the chiral limit is
frequently
misrepresented in \QCPT%
~\cite{Booth:1994rr,Savage:2001dy,Arndt:2002ed,Arndt:2003we}.
In the following chapters,
we find this is additionally true for a number of
meson and baryon hadronic properties.

\chapter{Chiral $1/M^2$ 
         corrections to $B^{(*)}\rightarrow D^{(*)}$
         at Zero Recoil in \QCPT}
\label{chapter:B2D}

In this chapter, we study 
the semileptonic $B^{(*)}\to D^{(*)}$ decays
in the limit that the heavy quark masses are infinite. 
We calculate the lowest order chiral 
corrections, which are of ${\mathcal O}(1/M^2)$, from the breaking
of heavy quark symmetry 
at the zero recoil point in \QCPT.
These results will aid in the extrapolation of 
quenched lattice calculations 
from the light quark masses used on the lattice
down to the physical ones.

\section{Introduction}
The Cabibbo-Kobayashi-Maskawa (CKM) matrix describes
the flavor mixing among the quarks; 
its elements are
fundamental input parameters for the standard model.
Their precise knowledge is not only  
crucial to determine the standard model
but also to shed light on the origin of $CP$ violation.
The matrix element that parametrizes the amount of mixing 
between the $b$ and $c$ quarks,
$V_{cb}$,
can be extracted from the
exclusive semileptonic $B$ meson decays 
$B\to Dl\nu$ and $B\to D^*l\nu$,
where $l=e,\mu$.
Heavy quark effective theory (HQET)
(for a recent review, see~\cite{Manohar:2000dt}),
which is exact in the limit of infinite masses $M$ for the heavy quarks,
predicts the width
of the process $B\to D^*l\nu$ as
\begin{equation}
  \frac{d\Gamma}{d\o}
  (B\to D^*)
  =
  \frac{G_F^2|V_{cb}|^2}{48\pi^3}
  {\mathcal K}(\o)
  {\mathcal F}_{B\to D^*}(\o)^2
,\end{equation}
where $\o=v\cdot v'$ is the scalar product of the 4-velocities 
$v$ and $v'$ of the
$B$ and $D^*$ mesons, respectively.
${\mathcal K}(\o)$ is a known kinematical factor and
${\mathcal F}(\o)$ is a form factor whose value at the kinematical
point $\o=1$ is ${\mathcal F}(1)=1$ in the $M\to\infty$ limit.
There are, however, perturbative and nonperturbative corrections 
to ${\mathcal F}(1)$,
\begin{equation}\label{B2D-eqn:F1}
  {\mathcal F}_{B\to D^*}(1)
  =
  \eta_A+\d_{1/M^2}+\dots
,\end{equation}
where the parameter $\eta_A\approx 0.96$ is a QCD radiative correction
known to two-loop order~\cite{Czarnecki:1996gu}
and $\d_{1/M^2}$ are non-perturbative corrections 
of ${\mathcal O}(1/M^2)$ to the
infinite mass limit of HQET.  
Note that, according to Luke's theorem~%
\cite{Luke:1990eg}, there are no ${\mathcal O}(1/M)$ corrections
at zero-recoil.
One chooses the 
zero-recoil point 
because, for $\o=1$, ${\mathcal F}_{B\to D^*}$ can be expressed in
terms of a single form factor $h_{A_1}$ given by
\begin{equation}
  \frac{\langle D^*(v,\e')|\bar{c}\gamma^\mu\gamma_5b|B(v)\rangle}
       {\sqrt{m_Bm_{D^*}}}
  =
  -2ih_{A_1}(1)\e'^{*\mu}
.\end{equation}
This is in contrast with the general case $\o>1$ for which
${\mathcal F}_{B\to D^*}(\o)$ is a linear combination of several
different form factors of $B\to D^*l\nu$ mediated by vector and
axial vector currents.

Several experiments,
most recently by CLEO~\cite{Briere:2002ew}, have determined the product
$[{\mathcal F}_{B\to D^*}(1)|V_{cb}|]^2$ by 
measuring $d\Gamma_{B\to D^*}/d\o$ and extrapolating it to
the zero-recoil point.
The mixing parameter $|V_{cb}|$ can then be extracted once the value
${\mathcal F}_{B\to D^*}(1)$ that encodes the strong interaction physics
has been evaluated.
The uncertainty in $|V_{cb}|$ is therefore determined by the 
experimental errors and by theoretical uncertainties in the
determination of ${\mathcal F}_{B\to D^*}(1)$.
Presently, the theoretical uncertainties dominate.%
\footnote{
Similarly, one can use the decay $B\to Dl\nu$ to extract
$[{\mathcal F}_{B\to D}(1)|V_{cb}|]^2$ from the measured
$d\Gamma_{B\to D}/d\o$. However, $d\Gamma_{B\to D}/d\o$ is
more heavily suppressed by phase space near $\o=1$ than 
$d\Gamma_{B\to D^*}/d\o$.
In addition, the $B\to D$ channel is experimentally more challenging.
Thus the extraction of $|V_{cb}|$ from this channel is less precise
but serves as a consistency check.
} 

A model-independent way
of calculating ${\mathcal F}(1)$
is provided by numerical lattice QCD simulations.
Recently, such calculations have been performed%
~\cite{Simone:1999nv,El-Khadra:2001rv,Hashimoto:2001nb,Kronfeld:2002cc}
for the decays $B\to D^{(*)}l\nu$
using QQCD.
Several systematic uncertainties, 
such as from statistics and lattice space
dependence, contribute to the error of these calculations.
Another contribution to the uncertainties comes
from the chiral extrapolation
of the light quark mass.
This extrapolation can be done by matching QQCD 
to \QCPT\
and calculating the non-analytic corrections $\d_{1/M^2}$
in Eq.~(\ref{B2D-eqn:F1}) in \QCPT.
The formally dominant contributions to these corrections
come from the hyperfine mass splitting between the heavy pseudoscalar
and vector mesons that stems from the inclusion of
heavy quark symmetry breaking operators of
${\mathcal O}(1/M)$
in the Lagrangian.

In QCD, the corrections due to $D$~meson 
hyperfine splitting 
have been calculated in \CPT\
by Randall and Wise%
~\cite{Randall:1993qg}.  
A more complete treatment, 
involving additional corrections due to
$B$~meson hyperfine splitting,
${\mathcal O}(1/M)$ axial vector coupling corrections, 
and ${\mathcal O}(1/M)$ corrections to the current,
has been given in~\cite{Boyd:1995pq}.
Recently, the $D$ meson 
hyperfine splitting corrections have also been
determined in \PQCPT~\cite{Savage:2001jw} 
for PQQCD.

In this chapter, we calculate the ${\mathcal O}(1/M^2)$
corrections in Eq.~(\ref{B2D-eqn:F1})
due to $D$ and $B$~meson hyperfine splitting 
in \QCPT.
These corrections are---%
upon expanding in powers of the hyperfine splitting $\D$---%
of order
$\LQCD^{3n/2}/(M^nm_q^{n/2})$ for $n\ge 2$
and formally larger than 
those coming from the inclusion of ${\mathcal O}(1/M)$
heavy quark symmetry breaking operators in the Lagrangian and
current which 
are suppressed by powers of 
$\LQCD/M$.
This argument is similar to the one that 
applies to \CPT~\cite{Manohar:2000dt}.
Our \QCPT\ calculation can be used to extrapolate lattice results%
~\cite{Hashimoto:2001nb} that
use the quenched approximation down to the physical light quark masses.
So far, this extrapolation has been based upon the \CPT\ calculation%
~\cite{Randall:1993qg}.
Using \QCPT\ 
should therefore give a better estimate of the uncertainties related 
to the chiral extrapolation.

A central role in the lattice calculation
of $B\to D^*$%
~\cite{Hashimoto:2001nb,Kronfeld:2002cc}
is played by the double ratios of matrix elements
\begin{equation}\label{B2D-eqn:Rplus}
  {\mathcal R}_+
  =
  \frac{\langle D|\bar{c}\g^0b|B\rangle\langle B|\bar{b}\g^0c|D\rangle}
       {\langle D|\bar{c}\g^0c|D\rangle\langle B|\bar{b}\g^0b|B\rangle}
,\end{equation}
\begin{equation}\label{B2D-eqn:R1}
  {\mathcal R}_1
  =
  \frac{\langle D^*|\bar{c}\g^0b|B^*\rangle\langle B^*|\bar{b}\g^0c|D^*\rangle}
       {\langle D^*|\bar{c}\g^0c|D^*\rangle\langle B^*|\bar{b}\g^0b|B^*\rangle}
,\end{equation} 
and
\begin{equation}\label{B2D-eqn:RA1}
  {\mathcal R}_{A_1}
  =
  \frac{\langle D^*|\bar{c}\g^j\g_5b|B\rangle
        \langle B^*|\bar{b}\g^j\g_5c|D\rangle}
       {\langle D^*|\bar{c}\g^j\g_5c|D\rangle
        \langle B^*|\bar{b}\g^j\g_5b|B\rangle}
.\end{equation}
In these ratios,
statistical fluctuations are highly correlated and cancel
to a large degree.
The
${\mathcal O}(1/M^2)$ correction to the double ratios 
can therefore be calculated fairly
accurately and used to derive the 
${\mathcal O}(1/M^2)$ correction to the matrix elements
themselves.
For this reason, 
we also calculate ${\mathcal O}(1/M^2)$ corrections to the
decay $B^*\rightarrow D^*$
in addition to 
the  
experimentally accessible decays
$B\rightarrow D$ and $B\rightarrow D^*$,
and thus the corrections to 
${\mathcal R}_+$, ${\mathcal R}_1$, and ${\mathcal R}_{A_1}$.

\section{Quenched Heavy Meson Chiral Perturbation Theory}
\label{B2D-sec:QHMCPT}
The $D$ mesons with quantum numbers of $c\bar{Q}$ can be written as 
a six-component vector
\begin{equation}
  D=(D_u,D_d,D_s,D_{\tilde{u}},D_{\tilde{d}},D_{\tilde{s}})
.\end{equation} 
Heavy quark symmetry is provided by combining creation and annihilation
operators for the pseudoscalar and vector mesons, 
$D$ and $D^*$, respectively,
together into the field $H^D$:
\begin{eqnarray}
  H^D&=&\frac{1+\vslash}{2}(\Dslash^*+i\gamma_5D), \\
  \bar{H}^D=\gamma^0H^{D\dagger}\gamma^0
       &=&({\Dslash^*}^\dagger+i\gamma_5D^\dagger)\frac{1+\vslash}{2}
,\end{eqnarray}
where $v$ denotes the velocity of a heavy meson.
In HQET the momentum of a heavy quark is only changed
by a small residual momentum of 
${\mathcal O}(\Lambda_{\text{QCD}})$.
Hence, $v$ is not changed and $H$ is usually denoted by an index
$v$ which we have dropped here to unclutter the formalism.
In the heavy quark limit, 
the dynamics of the heavy mesons are described by the Lagrangian%
~\cite{Booth:1995hx,Sharpe:1996qp}
\begin{equation}\label{B2D-eqn:LD}
  {\cal L}_D
  =
  -i\,\tr[\bar{H}_a^Dv_\mu(\partial^\mu\d_{ab}+i V_{ba}^\mu)H_b^D]
  +g\,\tr(\bar{H}_a^DH_b^D\gamma_\nu\gamma_5 A_{ba}^\nu) 
  +\g\,\tr(\bar{H}_a^DH_a^D\gamma_\mu\gamma_5)\,\str A^\mu
\end{equation}
where the traces tr() are over Dirac indices and supertraces str()
over the flavor indices are implicit.
The additional coupling term involving $\Phi_0\sim\str A^\mu$ is a 
feature of \QCPT\ and not present in \CPT.
The light-meson fields are
\begin{equation}
  A_{\mu}
  =
  \frac{i}{2}(\xi^\dagger\partial_\mu\xi-\xi\partial_\mu\xi^\dagger)
  =
  -\frac{1}{f}\partial_\mu\Phi+{\mathcal O}(\Phi^3)
\end{equation}
and
\begin{equation}
  V_{\mu}
  =
  \frac{i}{2}(\xi^\dagger\partial_\mu\xi+\xi\partial_\mu\xi^\dagger)
  =
  \frac{i}{2f^2}[\Phi,\partial_\mu\Phi]+{\mathcal O}(\Phi^4)
.\end{equation}
Expanding the Lagrangian ${\cal L}_D$  to lowest order in the
meson fields leads to the (derivative) 
couplings $DD^*\partial\phi$ and $D^*D^*\partial\phi$
whose coupling constants are equal as a consequence of
heavy quark spin symmetry.
At leading order in the $1/M$ expansion,
the $DD\partial\phi$ coupling vanishes by parity.

An analogous formalism applies to the fields $B$ and $B^*$ 
which are combined into $H^B$.
Note that the 
axial coupling $g$ is the same for $H^D$ and $H^B$ mesons
at this order in the $1/M$ expansion 
as dictated by heavy quark flavor symmetry.

We do not include terms of order $m_q\sim\sqrt{m_\pi}$ 
in the Lagrangian as
explicit chiral symmetry breaking effects are suppressed compared to
the leading corrections.  The presence of these terms is implied by
the nonzero meson masses $m_{qq}$.

\section{Matrix Elements of 
         \protect$\bar{B}^{(*)}\rightarrow D^{(*)}l\bar{\nu}$}
The non-zero hadronic matrix elements for 
$B^{(*)}\rightarrow D^{(*)}$ can be defined 
in terms of the 16 independent form factors 
$h_\pm$, $h_V$, $h_{A_{1,2,3}}$, and $h_{1\dots 10}$ 
as%
~\cite{Falk:1993wt,Manohar:2000dt} 
\begin{equation}\label{B2D-eqn:matrixelementfirst}
  \frac{\langle D(v')|\bar{c}\gamma^\mu b|B(v)\rangle}
       {\sqrt{m_Bm_D}}
  =
  h_+(\o)(v+v')^\mu+h_-(\o)(v-v')^\mu 
,\end{equation}
\begin{equation}
  \frac{\langle D^*(v',\e')|\bar{c}\gamma^\mu b|B(v)\rangle}
       {\sqrt{m_Bm_{D^*}}}
  =
  -h_V(\o)\varepsilon^{\mu\nu\a\b}\e'^*_\nu v'_\a v_\b
,\end{equation}
\begin{equation}
  \frac{\langle D^*(v',\e')|\bar{c}\gamma^\mu\gamma_5b|B(v)\rangle}
       {\sqrt{m_Bm_{D^*}}}
  =
  -ih_{A_1}(\o)(\o+1)\e'^{*\mu}
      +ih_{A_2}(\o)(v\cdot\e'^*)v^\mu
      +ih_{A_3}(\o)(v\cdot\e'^*)v'^\mu 
,\end{equation}
\begin{eqnarray}
  \frac{\langle D^*(v',\e')|\bar{c}\gamma^\mu b|B^*(v,\e)\rangle}
       {\sqrt{m_{B^*}m_{D^*}}}
  &=&
  -(\e'^*\cdot \e)[h_1(\o)(v+v')^\mu+h_2(\o)(v-v')^\mu]
  +h_3(\o)(\e'^*\cdot v)\e^\mu
               \nonumber \\
  &&
  +h_4(\o)(\e\cdot v')\e'^{*\mu}
  -(\e\cdot v')(\e'^*\cdot v)[h_5(\o)v^\mu+h_6(\o)v'^\mu]
, \nonumber \\
\end{eqnarray}
and
\begin{eqnarray}\label{B2D-eqn:matrixelementlast}
  \frac{\langle D^*(v',\e')|\bar{c}\gamma^\mu\gamma_5b|B^*(v,\e)\rangle}
       {\sqrt{m_{B^*}m_{D^*}}}
  &=&
  i\varepsilon^{\mu\a\k\d}
  \left\{
    \e^*_\k\e_\d\left[h_7(\o)(v+v')^\mu+h_8(\o)(v-v')^\mu\right]
    \right. \nonumber \\
    &&\left. \phantom{dadasd}
    +v'^\a v^\b
     \left[h_9(\o)(\e'^*\cdot v)\e^\mu+h_{10}(\o)(\e\cdot v')\e'^{*\mu}\right]
  \right\}
. \nonumber \\
\end{eqnarray}
Here, $\o=v\cdot v'$ and
$v$ ($\e$) and $v'$ ($\e'$)
are the velocities (polarization vectors)
of the initial state $B^{(*)}$ meson
and final state $D^{(*)}$ meson, respectively.
Note that we will not explicitly calculate matrix elements
of $B^*\to D$ as these can be easily related to the
$B\to D^*$ calculation by a Hermitian conjugation of the matrix
elements and an interchange of the $c$ and $b$ quarks,
i.e., $B^{(*)}\leftrightarrow D^{(*)}$.

In the heavy quark limit the matrix elements
in Eqs.\ (\ref{B2D-eqn:matrixelementfirst})--(\ref{B2D-eqn:matrixelementlast})
are reproduced by the operator
\begin{equation}
  \bar{c}\gamma^\mu(1-\gamma_5)b
  \rightarrow 
  -\xi(\omega)\tr[\bar{H}^D_{v'}\gamma^\mu(1-\gamma_5)H^B_v]
.\end{equation}
Here, $\xi(\o)$
is the universal Isgur-Wise function~\cite{Isgur:1989vq,Isgur:1990ed}
with the normalization $\xi(1)=1$.
To lowest order in the heavy quark expansion one finds
\begin{equation}
  h_+(\o)
  =h_V(\o)
  =h_{A_1}(\o)=h_{A_3}(\o)
  =h_1(\o)=h_3(\o)=h_4(\o)=h_7(\o)=\xi(\o)
\end{equation}
and the remaining 8 form factors vanish.

The discussion of the $\bar{B}^{(*)}\rightarrow D^{(*)}l\bar{\nu}$
matrix elements is similar for different flavors of the light quark $q$
content of the $B^{(*)}$ and $D^{(*)}$ mesons;
it applies equally to $q=u$, $d$, or~$s$
as the theory splits into three similar copies
of a one-flavor theory.
In the limit of light quark $SU(3)_V$ flavor symmetry the matrix elements
(and in particular the Isgur-Wise function) are therefore
independent of the
light quark flavor.
However, in nature the masses of the $u$, $d$, and $s$ quarks are
different and $SU(3)_V$ is not an exact symmetry.
Therefore our results will include terms that 
depend upon $m_q$ via the meson masses
$m_{qq}$ defined in Eq.~(\ref{CPT-eqn:mqq}).

\section{\protect$1/M^2$ Corrections}
The lowest order heavy quark symmetry violating operator 
that can be in\-clud\-ed in the Lagrangian ${\mathcal L}_D$ in
Eq.~(\ref{B2D-eqn:LD}) is the dimension-three operator 
$(\l_{D_2}/M_D)
  \tr\left(\bar{H_a}^D\sigma^{\mu\nu}H_a^D\sigma_{\mu\nu}\right)$.
It violates heavy-quark spin and flavor symmetries and
comes from the QCD magnetic moment operator
$\bar{c}\sigma^{\mu\nu}G_{\mu\nu}^AT^Ac$,
where $G_{\mu\nu}^A$ is the gluon field strength tensor and
$T^A$ with $A=1,\dots,8$ are the eight color $SU(3)$ generators.
This operator gives rise to a
mass difference between the $D$ and $D^*$
mesons of
\begin{equation}
  \D_D=m_{D^*}-m_D=-8\frac{\l_{D_2}}{\bar{M}_D}
.\end{equation}
This effect can be taken into account by modifying the $D$ and $D^*$
propagators which become
\begin{equation}
  \frac{i\d_{ab}}{2(v\cdot k+3\D_D/4+i\e)}
  \quad\text{and}\quad
  \frac{-i\d_{ab}(g_{\mu\nu}-v_\mu v_\nu)}{2(v\cdot k-\D_D/4+i\e)}
,\end{equation}
respectively,
so that in the rest frame, where $v=(1,0,0,0)$, an on-shell $D$ has
residual energy of $-3\D_D/4$ and an on-shell $D^*$ has residual
energy of $\D_D/4$.
A similar effect due to the inclusion of
a QCD magnetic moment operator for the $b$ quark
applies to the $B^{(*)}$ mesons.

There are no corrections to the matrix elements for the semileptonic decays
$B^{(*)}\rightarrow D^{(*)}e\nu$ of $\order(1/M)$ at zero-recoil 
according to Luke's theorem~\cite{Luke:1990eg}.  
The leading corrections enter at $\order(1/M^2)$.
In addition to tree-level contributions from the insertion of 
$\order(1/M^2)$ suppressed operators 
into the heavy quark Lagrangian or the current 
there 
are one-loop contributions from wave function renormalization
and vertex correction.
These one-loop diagrams have a non-analytic dependence on the
meson mass $m_{qq}$ and depend on the subtraction point $\mu$.
This dependence on $\mu$ is canceled by the tree-level contribution
of the $\order(1/M^2)$ operators.

Because of the absence of disconnected
quark loops in QQCD, which manifests itself
as a cancellation between intermediate pseudo Goldstone bosons and 
pseudo Goldstone fermions in loops in \QCPT, the only loop diagrams
that survive are those that contain a hairpin interaction or 
a $\g$ coupling [see Eq.~\eqref{B2D-eqn:LD}].

\begin{figure}[tb]
%\begin{fullpage}
    \includegraphics[width=\textwidth]{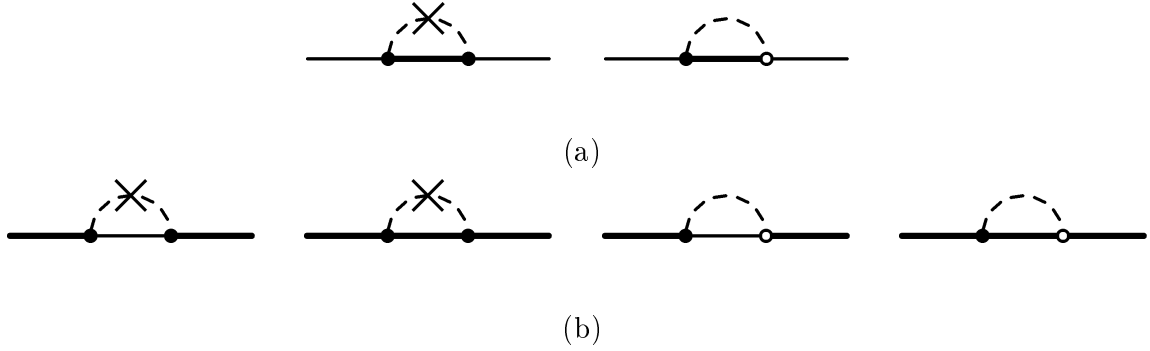}%
    \caption%
      [Graphs contributing to wavefunction renormalization for heavy
      pseudoscalar and vector mesons]%
      {Graphs contributing to wavefunction renormalization for heavy
      (a) pseudoscalar and (b) vector mesons.
      A thin (thick) line denotes a heavy pseudoscalar (vector) meson,
      a dashed line denotes the $\Phi_0$, while a dashed-crossed
      line denotes the insertion of a hairpin.
      A full (empty) vertex denotes a $g$ ($\g$)
      coupling.
  }
  \label{B2D-F:wf-renorm}
%\end{fullpage}
\end{figure}

The wave function renormalization contributions for
the pseudoscalar and vector meson, $Z_{D/B}$ and $Z^*_{D/B}$, 
respectively, 
come from the one-loop diagrams shown in 
Fig.~\ref{B2D-F:wf-renorm} and
have been calculated in%
~\cite{Booth:1994rr,Booth:1995hx,Sharpe:1996qp}.
%\begin{fullpage}
%\end{fullpage}
%\afterpage{\clearpage}
Including the $\a$ coupling we find, for these diagrams,
\begin{eqnarray}
  Z
  &=&
    1
    +\frac{ig^2\mu_0^2}{f^2}H_1(\D)
    -\frac{ig^2\a}{f^2}H_2(\D)
    +\frac{6i\g g}{f^2}F_1(\D)
\end{eqnarray}
and
\begin{eqnarray}
  Z^*
  &=&
    1
    +\frac{ig^2\mu_0^2}{3f^2}H_1(-\D)
    -\frac{ig^2\a}{3f^2}H_2(-\D)
    +\frac{2i\g g}{f^2}F_1(-\D)
   \nonumber \\
  &&   
    +\frac{2ig^2\mu_0^2}{3f^2}H_1(0)
    -\frac{2ig^2\a}{3f^2}H_2(0)
    +\frac{4i\g g}{f^2}F_1(0)
.\end{eqnarray}
The functions $H_1$, $H_2$, and $F_1$ come from loop integrals and
are given in Appendix~\ref{appendix-B2D}.
Note that in the heavy quark limit where $\D=0$
one recovers
$Z=Z^*$, as required by heavy quark symmetry.

\begin{figure}[tb]
%\begin{fullpage}
    \includegraphics[width=\textwidth]{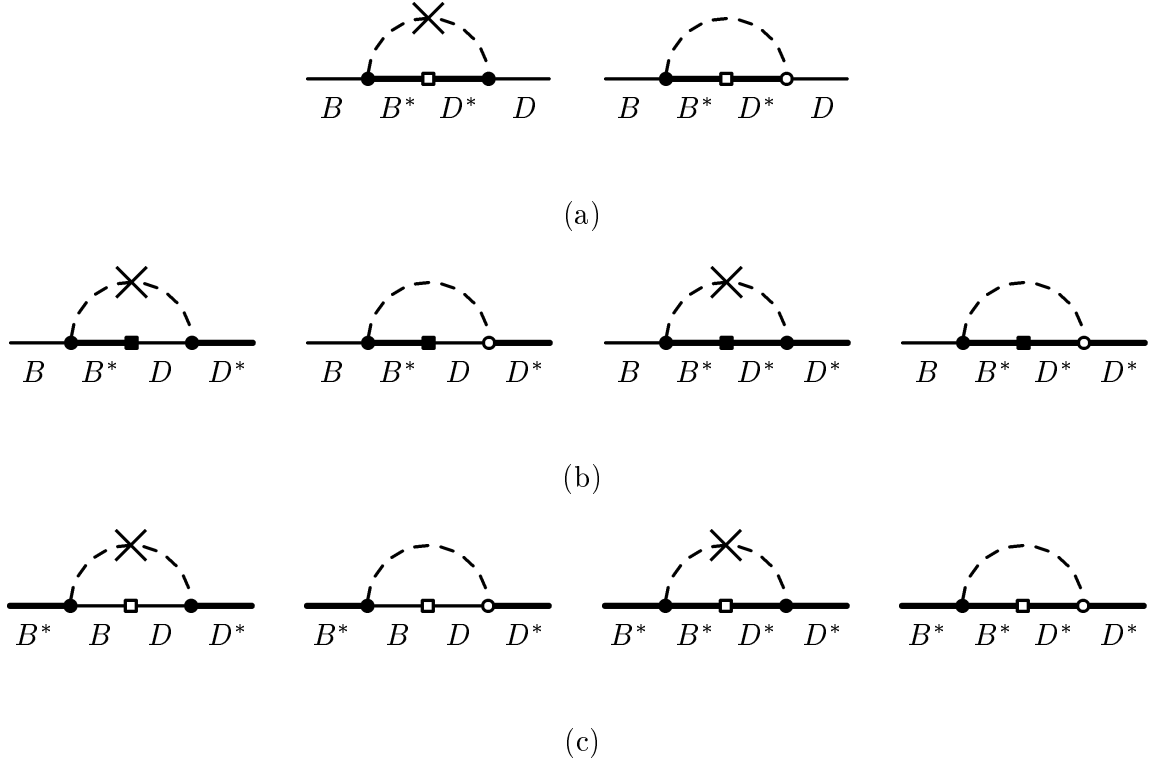}%
    \caption%
      [\QCPT\ graphs which contribute to the vertex correction 
      of the form factors 
      $h_+(1)$, $h_{A_1}(1)$, and $h_1(1)$]%
      {\QCPT\ graphs which contribute to the vertex correction 
      of the form factors 
      (a) $h_+(1)$, (b) $h_{A_1}(1)$, and (c) $h_1(1)$.
      A full (empty) square denotes the insertion of the
      operator $\bar{c}\g^\mu\g_5b$ ($\bar{c}\g^\mu b$). 
  }
  \label{B2D-F:vertex-corr}
%\end{fullpage}
\end{figure}

The vertex corrections come from one-loop diagrams.
The non-vanishing contributions
are shown in Fig.~\ref{B2D-F:vertex-corr}.
Combining the wave function renormalization and vertex corrections 
and including a local counterterm to cancel the dependence
on the renormalization scale $\mu$, we 
find the following corrections for the form factors:
\begin{eqnarray}\label{B2D-eqn:hplus}
  \d h_+(1)
  &=&
  X_+(\mu)
  +\frac{Z_B-1}{2}+\frac{Z_D-1}{2}
               \nonumber \\
              &&
    -\frac{ig^2}{f^2}
    \left[
      \mu_0^2\,H_5(\D_B,\D_D)-\a\,H_8(\D_B,\D_D)
    \right]
    -\frac{6ig\g}{f^2}G_5(\D_B,\D_D)
  \nonumber \\
  &\rightarrow&
  X_+(\mu)
  +\frac{1}{(4\pi f)^2}
  \left(
    \frac{g^2\mu_0^2}{3m^2}
   -\left[
      \frac{g^2\a}{3}-2g\g
    \right]
    \log\frac{m^2}{\mu^2}
  \right)
  (\D_B-\D_D)^2
               \nonumber \\
              &&
  +{\mathcal O}(\{\D_B,\D_D\}^3)
,\end{eqnarray}
\begin{eqnarray}
  \d h_{A_1}(1)
  &=&
  X_{A_1}(\mu)
  +\frac{Z_B-1}{2}+\frac{Z_D^*-1}{2}
               \nonumber \\
              &&
    -\frac{ig^2}{3f^2}
    \left[
      \mu_0^2\,H_5(\D_B,-\D_D)-\a\,H_8(\D_B,-\D_D)
    \right]
    -\frac{2ig\g}{f^2}G_5(\D_B,-\D_D)
               \nonumber \\
              &&
    -\frac{2ig^2}{3f^2}
    \left[
      \mu_0^2\,H_5(\D_B,0)-\a\,H_8(\D_B,0)
    \right]
    -\frac{4ig\g}{f^2}G_5(\D_B,0)
  \nonumber
\end{eqnarray}
\begin{eqnarray}\label{B2D-eqn:hA1}
  \phantom{\d h_{A_1}(1)}
  &\rightarrow&
  X_{A_1}(\mu)
   +\frac{1}{(4\pi f)^2}
   \left(
     \frac{g^2\mu_0^2}{9m^2}
    -\left[
       \frac{g^2\a}{9}-\frac{2g\g}{3}
     \right]
     \log\frac{m^2}{\mu^2}
   \right)
   \left(3\D_B^2+\D_D^2+2\D_B\D_D\right)
               \nonumber \\
              &&
  +{\mathcal O}(\{\D_B,\D_D\}^3)
,\end{eqnarray}
and
\begin{eqnarray}\label{B2D-eqn:h1}
  \d h_1(1)
  &=&
  X_1(\mu)
  +\frac{Z_B^*-1}{2}+\frac{Z_D^*-1}{2}
               \nonumber \\
              &&
    -\frac{ig^2}{3f^2}
    \left[
      \mu_0^2\,H_5(-\D_B,-\D_D)-\a\,H_8(-\D_B,-\D_D)
    \right]
    -\frac{2ig\g}{f^2}G_5(-\D_B,-\D_D)
               \nonumber \\
              &&
    -\frac{2ig^2}{3f^2}
    \left[
      \mu_0^2\,H_5(0,0)-\a\,H_8(0,0)
    \right]
    -\frac{4ig\g}{f^2}G_5(0,0)
  \nonumber \\
  &\rightarrow&
  X_1(\mu)
   +\frac{1}{(4\pi f)^2}
   \left(
     \frac{g^2\mu_0^2}{9m^2}
    -\left[
       \frac{g^2\a}{9}-\frac{2g\g}{3}
     \right]
     \log\frac{m^2}{\mu^2}
   \right)
   \left(\D_B-\D_D\right)^2
               \nonumber \\
              &&
  +{\mathcal O}(\{\D_B,\D_D\}^3)
,\end{eqnarray}
which are defined by $h_+(1)=1+\d h_+(1)$
and analog expressions for $\d h_{A_1}(1)$ and $\d h_1(1)$.
The functions $H_5$, $H_8$, and $G_5$ come from loop-integrals 
that are 
listed in Appendix~\ref{appendix-B2D} and we have defined $m=m_{qq}$.
The insertions of tree-level $\order(1/M^2)$ operators
are represented by 
the functions $X_+(\mu)$, $X_{A_1}(\mu)$, and $X_1(\mu)$
which are independent of $m$  and exactly cancel the
$\mu$ dependence of the logarithm.
These functions can be extracted from lattice simulations by measuring
the zero-recoil form factors for a varying mass of the light quark.

Experimentally, 
$\D_D\approx 142\,\text{MeV}$ and 
$\D_B\approx 46\,\text{MeV}$ so that the ratios
$\D_D/m$ and $\D_B/m$,
which enter the form factor corrections through
the function
$R(\D/m)$ (defined in Appendix~\ref{appendix-B2D}), 
are $\order(1)$.
On the lattice, however, one can vary all 
quark masses.
Expanding first in powers of $\D$ and then taking the chiral limit
$m\to 0$
one finds the formal limits given in 
Eqs.\ (\ref{B2D-eqn:hplus})--(\ref{B2D-eqn:h1})
where we have only kept the pieces non-analytic in $m$.
This demonstrates 
that the terms linear in $\D_D$ and $\D_B$, although
present in wave function renormalization and vertex corrections,
cancel as required by Luke's theorem~\cite{Luke:1990eg}.
The leading order corrections are 
${\mathcal O}(\{\D_B,\D_D\}^2)$.

As a consistency check one can 
restore heavy quark flavor symmetry by taking $\D_B=\D_D$.
Since the ${\mathcal O}(1/M^2)$ corrections to $h_+(1)$ and $h_1(1)$
are proportional to $(\D_B-\D_D)^2$ they disappear as they should
since the charge associated with the operators
$\bar{c}\g_\mu c$ and $\bar{b}\g_\mu b$
is conserved.
This argument does not apply for the $B\to D^*$ transition matrix 
element in the limit $\D_B=\D_D$ since there is no conserved axial charge
associated with the operators 
$\bar{c}\g_\mu\g_5c$ and $\bar{b}\g_\mu\g_5b$.

In the chiral limit, 
the term proportional
to $\mu_0^2$
has a $1/m^2$ singularity and
dominates over the terms proportional
to $\a$ and $\g$ that are only logarithmically divergent.  
This is analogous to a term of the form
$(m_{qq}^2-m_{jj}^2)/m_{qq}^2$ found by 
Savage~\cite{Savage:2001jw}
for \PQCPT\ (here, $m_{qq}$ and $m_{jj}$ are valence and sea quark 
masses, respectively).
In the limit $m_{jj}\to m_{qq}$ this term, however, vanishes as 
\PQCPT\ goes to \CPT\ where the dominant term is $\log m_{qq}$. 
In \QCPT, on the other hand, the
$1/m_{qq}^2$ pole
persists, revealing the sickness
of QQCD where the hairpin interactions give a completely
different chiral behavior than in QCD.

The size of $\mu_0$ can be estimated from the 
$\eta$-$\eta'$ mass splitting~\cite{Sharpe:1992ft}, 
large $N_C$ arguments~\cite{Witten:1979vv,Veneziano:1979ec}
($N_C$ being the number of colors),
or lattice calculations.
These estimates imply 
$\mu_0\approx 500-900\,\text{MeV}$;
for the purpose of dimensional analysis we use
$\mu_0\sim \order(\LQCD)$.
Taking $g\sim \order(1)$
we therefore find that 
$\d h_+$, $\d h_{A_1}$, and $\d h_1$ are of order
$\D^n/m^n\sim \LQCD^{3n/2}/(M^nm_q^{n/2})$ for $n\ge 2$
and thus larger than tree-level heavy quark symmetry breaking operators
that are suppressed by $\LQCD/M$.

To show the dependence of the zero-recoil form factors 
on the 
mass of the light spectator quark it is necessary to know the 
numerical values of the parameters $\mu_0$, $g$, $\a$, and $\g$.
In determining reasonable values for these couplings we 
follow the discussion by Sharpe and Zhang~\cite{Sharpe:1996qp}.
Assuming that $g$ is similar to the \CPT\ value we use
$g^2=0.4$.
The hairpin coupling $\a$ is proportional to $1/N_C$, 
and thus assumed to be small; we use two values, $\a=0$ and $\a=0.7$.
The coupling $\g$ is known to be suppressed by
$1/N_C$ compared to $g$, the sign is undetermined. 
We take $-g\le\g\le g$
(see~\cite{Sharpe:1996qp} and references therein).

\begin{figure}[tb]
  \centering
    \includegraphics[width=0.75\textwidth]{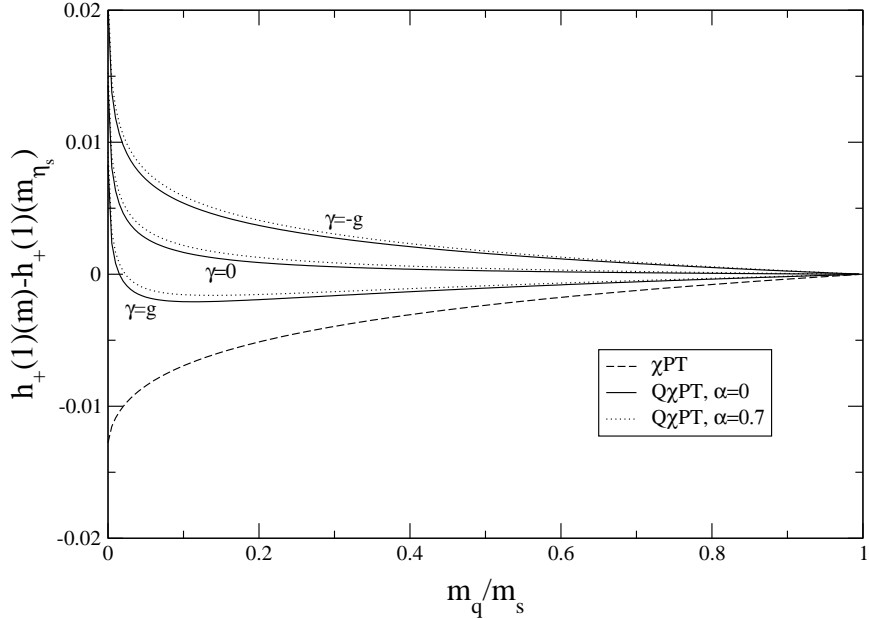}%
    \caption%
      [Light spectator quark dependence of $h_+(1)$ in \QCPT]%
      {Dependence of $h_+(1)$ on the mass $m_q$ of the light spectator quark 
      in \QCPT.  For comparison, the \CPT\ result from%
      ~\cite{Randall:1993qg} is also shown (dashed line).
      The result has been normalized to unity for $m_q=m_s$.
      We have chosen $\mu_0=700\,\text{MeV}$ and
      $g^2=0.4$.}
  \label{B2D-F:plus-plot}
\end{figure}

\begin{figure}[tb]
  \centering
    \includegraphics[width=0.75\textwidth]{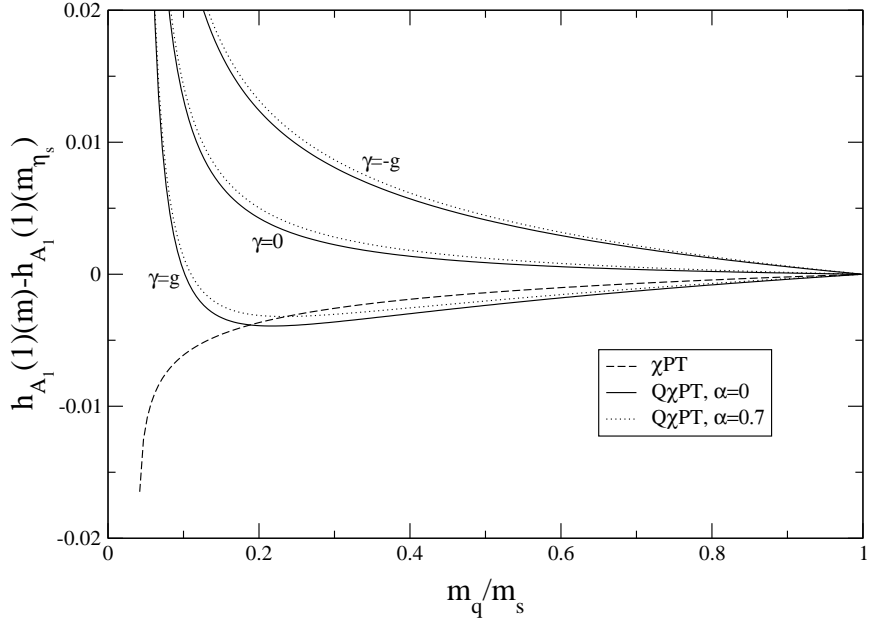}%
    \caption[Light spectator quark dependence of $h_{A_1}(1)$ in \QCPT]%
      {Dependence of $h_{A_1}(1)$ on the mass of the light spectator quark
      in \QCPT.  The dashed line denotes the \CPT\ result%
      ~\cite{Randall:1993qg}.
      The numerical values for the parameters are those used in
      Fig.~\ref{B2D-F:plus-plot}.}
  \label{B2D-F:A1-plot}
\end{figure}

With these parameters, 
the dependence of $h_+(1)$ and $h_{A_1}(1)$ on the 
mass of the light spectator quark $m_q$
is show in 
Figs.~\ref{B2D-F:plus-plot} and~\ref{B2D-F:A1-plot},
respectively.
The graphs are plotted against $m_q$ 
in units of the 
strange quark mass $m_s$ with $m_q/m_s=m^2/m_{\eta_s}^2$ where 
$m_{\eta_s}^2=2m_K^2$.
The behavior of $h_+(1)$ in \QCPT\
is dominated at small $m$ by the $1/m^2$ pole 
that is non-existent in \CPT.
Lattice calculations of $h_+(1)$%
~\cite{El-Khadra:2001rv} 
show
a small downward trend for 
decreasing $m_q$ down to the chiral limit that is 
similar to the downward trend 
seen from the \CPT\ calculation (dashed line).
The same behavior (down to $m_q\approx0.1 m_s$)
can also be seen for \QCPT\ for a certain choice of
parameters (e.g., $\g$ positive).
The case of $h_{A_1}(1)$ is different as there is a pole at $m=\D_D$ 
which is close to the physical pion mass.  Here, 
both $D^*$ and $\pi$ can be on-shell and 
the decay $B\to D^*\pi$ becomes kinematically allowed.
Lattice calculations of $h_{A_1}(1)$%
~\cite{Hashimoto:2001nb} for $m_q=(0.6\dots1) m_s$
show
a small downward trend for  
decreasing $m_q$ similar to the downward trend 
seen from the \CPT\ calculation 
(dashed line in Fig.~\ref{B2D-F:A1-plot}).
A similar trend down to $m_q\approx0.2 m_s$
can also be seen in the \QCPT\ calculation 
for a relatively large positive value 
of $\g$.

Although the downward trend in the lattice data for the two cases 
seems significant as 
the statistical errors are highly correlated, the 
uncertainty is still relatively high (typically $\pm0.01$)
and the existing lattice data can be accommodated by a wide
range of values for the parameters in the \QCPT\ Lagrangian.

As can be seen in the figures, the variation of the quenched result
is primarily due to the parameter $\g$ as
the sensitivity of the result 
to the value of $\a$ is very small. 
We have also checked how the result depends on the parameter
$g$ in the reasonable range $0.1<g^2<0.5$
and found that
the change from the value $g^2=0.4$ 
is at most 25\% for $g^2=0.5$, still well within the statistical errors
of the lattice data.

Finally, we calculate the double ratios defined in
Eqs. (\ref{B2D-eqn:Rplus})--(\ref{B2D-eqn:RA1}) using the 
results in Eqs.\ (\ref{B2D-eqn:hplus})--(\ref{B2D-eqn:h1}).
We find 
\begin{equation}
  {\mathcal R}_+
  =
  1+2\d h_+(1)
,\end{equation}
\begin{equation}
  {\mathcal R}_1
  =
  1+2\d h_1(1)
,\end{equation}
and
\begin{eqnarray}
  {\mathcal R}_{A_1}
  &=&
  1
  +
  \tilde{X}_{A_1}(\mu)
               \nonumber \\
               &&
  -\frac{ig^2}{3f^2}
   \left\{
     \mu_0^2
     \left[
       H_5(\D_B,-\D_D)+H_5(\D_D,-\D_B)-H_5(\D_D,-\D_D)-H_5(\D_B,-\D_B)
     \right]
               \right.\nonumber \\
               && \left.
     \phantom{ggggg}
     -\a
     \left[
       H_8(\D_B,-\D_D)+H_8(\D_D,-\D_B)-H_8(\D_D,-\D_D)-H_8(\D_B,-\D_B)
     \right]     
   \right\}
               \nonumber \\
               &&
   -\frac{2ig\g}{f^2}
     \left[
       G_5(\D_B,-\D_D)+G_5(\D_D,-\D_B)-G_5(\D_D,-\D_D)-G_5(\D_B,-\D_B)
     \right]    
  \nonumber \\
  &\rightarrow&
  1
  +
  \tilde{X}_{A_1}(\mu)
%               \nonumber \\
%               &&  
   -\frac{1}{(4\pi f)^2}
   \left(
     \frac{2g^2\mu_0^2}{9m^2}
    -\left[
       \frac{2g^2\a}{9}-\frac{4g\g}{3}
     \right]
     \log\frac{m^2}{\mu^2}
   \right)
   \left(\D_B-\D_D\right)^2
               \nonumber \\
              &&
  +{\mathcal O}(\{\D_B,\D_D\}^3)
,\end{eqnarray}
where 
$\tilde{X}_{A_1}(\mu)$
is the counterterm associated with ${\mathcal R}_{A_1}$.

\section{Conclusions}
Knowledge of the $B^{(*)}\to D^{(*)}$ form factors
at the zero-recoil point
is crucial to extract the value of $V_{cb}$ from experiment.
In the limit that the heavy quarks are infinitely heavy,
HQET predicts that the form factors 
$h_+$, $h_{A_1}$, and $h_1$ are equal, $h_+(1)=h_{A_1}(1)=h_1(1)=\xi(1)$.
The formally dominant 
correction due to breaking of heavy quark symmetry
comes from the inclusion of a $\order(1/M)$ dimension-three
operator in the Lagrangian that leads 
to
hyperfine-splitting between the heavy pseudoscalar and vector mesons. 
These leading order corrections are ${\mathcal O}(\{\D_B,\D_D\}^2$
as required by Luke's theorem.

Recent lattice simulations using the quenched approximation of
QCD have made a big step forward in determining these zero-recoil
form factors.  Presently, however, the simulations use
light quark masses that are much heavier than the physical ones
and therefore rely on a chiral extrapolation down
to the physical quark masses. 
In this 
chapter
we have calculated the dominant corrections
to the form factors $h_+$, $h_{A_1}$, and $h_1$
in \QCPT\ and determined the non-analytic dependence 
on the light quark masses via the light meson masses $m_{qq}$. 
Using these results, instead of the \CPT\ calculation, 
to extrapolate the
QQCD lattice measurements of these form factors
down to the physical pion mass should give a more reliable
estimate of the errors associated with the chiral extrapolation.
We have also calculated the corrections to certain double ratios
that are used in lattice QCD calculations of the decay
$B\to D^*$.

\chapter{\HMCPT\ in a Finite Volume}
\label{chapter:fv}

In this chapter, 
we study finite volume effects in heavy quark systems 
in the framework of heavy meson chiral perturbation theory 
(\HMCPT)
for
QCD, QQCD, and PQQCD.
A novel feature of this investigation is the role played by the 
scales $\D$ and ${\delta}_{s}$, where  
$\D$ is the mass difference
between the heavy-light vector and pseudoscalar mesons of the
same quark content, and ${\delta}_{s}$ is the difference 
of the masses of the $u$ and $d$, and the mass of the $s$ quark
that is
due to light flavour $SU(3)$ breaking.  The 
primary conclusion of this chapter is that finite
volume effects arising from the propagation of Goldstone mesons in 
the effective theory
can be altered by the presence of these scales.
Since $\D$ varies significantly 
with the heavy quark mass, these volume effects can be amplified
in both heavy and light quark mass extrapolations.

As an explicit example, we present 
results for $B$ parameters of neutral $B$ meson mixing matrix elements
and heavy-light decay constants to
one-loop order in finite volume \HMCPT\ for
full, quenched, and $N_{f}=2+1$ partially quenched QCD.
Our calculation shows that for 
high-precision
determinations of the phenomenologically interesting $SU(3)$ breaking 
ratios, finite volume effects are significant in 
QQCD and not negligible in PQQCD, 
although they are generally small in QCD.

\section{Introduction}
\label{fv-sec:intro}
Numerical calculations of hadronic properties using LQCD have provided
significant inputs to particle physics phenomenology.  In particular, the
joint effort between experiment and theory to investigate the unitarity
triangle in the CKM matrix from $B$ meson
decays and mixing has made
impressive progress~\cite{Battaglia:2003in}, in which LQCD has played 
an important role.  Nevertheless, current lattice calculations are 
still subject to various systematic errors.  In this chapter, we address 
finite volume effects which arise 
in lattice calculations for heavy-light meson systems
from the light degrees of freedom.  Our framework is
\HMCPT\ with first 
order $1/M$ and chiral corrections.
We assume the mass hierarchy
\begin{equation}\label{fv-eq:mass_hierarchy}
  m_{QQ'} \ll \Lambda_{\chi} \ll M ,
\end{equation}
where $m_{QQ'}$ is the mass of any Goldstone meson 
given in Eqs.~\eqref{CPT-eq:hanswurst} 
and \eqref{CPT-eqn:mqq}
and
$M$ is the mass of the heavy-light meson.
Under this
assumption, we discard corrections of the size
\begin{equation} \label{fv-eq:mgp_over_mp}
 \frac{m_{QQ'}}{M} .
\end{equation}
Concerning the finite volume, we work with the
condition that
\begin{equation}\label{fv-eq:MpiL_big}
 m_{QQ'} L \gg 1,
\end{equation}
where $L$ is the spatial extent of the cubic box.  Therefore,
given that $f_{\pi} L/\sqrt{2}$ ($f_{\pi} \approx 132 \mev$) will be close to 
one in lattice simulations in the near future, 
one can still neglect the chiral symmetry restoration effects resulting from
the Goldstone zero momentum modes~\cite{Gasser:1987ah,Leutwyler:1992yt} 
when Eq.~(\ref{fv-eq:MpiL_big}) is satisfied.

The main task of this work is to study the volume effects due to the presence
of the scales 
\begin{equation}\label{fv-eq:Delta_def}
 \D = M_{P^{\ast}} - M_{P} ,
\end{equation}
and
\begin{equation}\label{fv-eq:delta_s_def}
 \delta_{s} = M_{P_{s}} - M_{P} ,
\end{equation}
where $P^{\ast}$ and $P$ are the heavy-light vector and pseudoscalar mesons 
containing a $u$ or $d$ anti-quark,%
\footnote{We work in the isospin limit.}%
and $P_{s}$ is the 
heavy-light pseudoscalar meson with an $s$ anti-quark. The
scale $\D$ appears due to the breaking of heavy quark spin symmetry
that is of $\cO(1/M)$ and $\delta_{s}$ comes from light flavour $SU(3)$ 
breaking in the heavy-light meson masses.  Under the assumption
of Eq.~(\ref{fv-eq:mass_hierarchy}), $\D$ is independent of the
light quark mass, and $\delta_{s}$ does not contain any $1/M$
corrections, at the order
we are working. 

In the real world, both $\D$ and $\delta_{s}$ are not very 
different from the pion mass.  In fact~\cite{Hagiwara:2002fs},
$M_{B_{s}} - M_{B} =  91 \mev$,
$M_{D_{s}} - M_{D} =  104 \mev$,
$M_{B^{\ast}} - M_{B} = 46 \mev$,
and
$M_{D^{\ast}} - M_{D} = 142 \mev$.
In current lattice simulations, these mass splittings 
vary between 0 and $\sim 150$
MeV.  Therefore it is important to include them
in the investigation of finite volume effects.
Equation~(\ref{fv-eq:MpiL_big}) implies that the Compton
wavelength of the Goldstone meson is small compared
to the size of the box.  Therefore
finite volume effects mainly result from the propagation of 
the Goldstone mesons
to the boundary.
However, as shown in Section~\ref{fv-sec:FV}, 
$\D$ and $\delta_{s}$ can, 
in a non-trivial way, alter these 
effects.  In particular, since $\D$ varies 
with the heavy quark mass,
finite volume effects can be significantly amplified 
in heavy quark mass extrapolations.

This chapter is organized as follows.  
In Section~\ref{fv-sec:HMChPT} we summarize
the ingredients of HM$\chi$PT relevant to this work.  
We mainly expand the treatment in Section~\ref{B2D-sec:QHMCPT}
to the partially quenched case.
Section~\ref{fv-sec:FV} is 
devoted to the discussion of HM$\chi$PT in finite volume,
emphasising the role played by $\delta_{s}$ and
$\D$.  
We then present an explicit
calculation of neutral $B$ meson mixing and heavy-light decay constants in 
Section~\ref{fv-sec:BBbar_mixing} and discuss the phenomenological impact that
finite volume effects can have.  We conclude in 
Section~\ref{fv-sec:conclusion}.   
Some mathematical formulae and results are summarized in 
Appendix~\ref{fv-app}.

\section{Partially Quenched Heavy Meson Chiral Perturbation Theory}
\label{fv-sec:HMChPT}
The chiral Lagrangian for the Goldstone mesons
in the three theories 
QCD, QQCD, and PQQCD has been discussed in detail in
Chapter~\ref{chapter:CPT} and we will not repeat it here. 
The inclusion of the heavy-light mesons into
the quenched version of \HMCPT\ 
has been discussed in Section~\ref{B2D-sec:QHMCPT}.
Here we will expand this discussion and also include
the cases of QCD and PQQCD.

\HMCPT\ was first proposed in
Refs.~\cite{Burdman:1992gh, Wise:1992hn,Yan:1992gz}, with the
generalization to 
quenched and partially quenched theories given in 
Refs.~\cite{Booth:1995hx, Sharpe:1996qp}.  The $1/M$ and
chiral corrections were studied by 
Boyd and Grinstein~\cite{Boyd:1995pa} in QCD and by 
Booth~\cite{Booth:1994rr}
in QQCD.  The spinor field appearing in this effective theory is 
\begin{equation}\label{fv-eq:H_field}
 H^{(Q)}_{a} = \frac{1 + \vslash}{2} 
 \left( 
   P^{\ast (Q)}_{a,\mu} \gamma^{\mu} - P^{(Q)}_{a}\gamma_{5}
 \right)
,\end{equation}
where $P^{(Q)}_{a}$ and $P^{\ast (Q)}_{a,\mu}$ annihilate 
pseudoscalar and vector
mesons containing a heavy quark $Q$ and a light anti-quark of flavour $a$.  

The HM$\chi$PT Lagrangian, to lowest order in the chiral and
$1/M$ expansion, for $D$ mesons containing a heavy quark $c$ and a
light anti-quark of flavour $a$ is then
given in Eq.~\eqref{B2D-eqn:LD}.
The Lagrangian for a heavy $B$ meson containing a $b$ quark 
is analogous.
In QCD, the term proportional to $\g$ is non-existent as
the $\eta'$ is integrated out.
We do not formally distinguish the coupling $g$ in the three theories
with the understanding that the numerical values are different.
The HM$\chi$PT Lagrangian for mesons containing a 
heavy anti-quark $\bar{Q}$ and a light quark of flavour $a$ is obtained
by applying the charge conjugation operation to the Lagrangian
in Eq.~\eqref{B2D-eqn:LD}~\cite{Grinstein:1992qt}.
At this order, 
the propagators for $P^{(Q)}_{a}$ and $P^{\ast (Q)}_{a}$ mesons are
\begin{equation}
 \frac{i}{2 (v\cdot k + i\epsilon)}
 \quad\text{and}\quad
 \frac{-i (g_{\mu\nu} - v_{\mu}v_{\nu})}{2 (v\cdot k + i\epsilon)}
,\end{equation}
respectively.

The effects of chiral and heavy quark symmetry breaking have been 
systematically studied in full~\cite{Boyd:1995pa} and
quenched~\cite{Booth:1994rr} HM$\chi$PT.  
Amongst them, the only 
relevant feature necessary for the purpose of this work, {\it i.e.},
the investigation of finite volume effects, 
are the shifts to the masses of the heavy-light mesons.
These shifts are from the heavy quark spin breaking term
\begin{equation}\label{fv-eq:HQ_spin_breaking_term}
\frac{\lambda_{2}}{M}
{\tr_D} \left(
 \bar{H}^{(Q)}_{a}\sigma_{\mu\nu} H^{(Q)}_{a} \sigma^{\mu\nu}
\right)
,\end{equation}
and the chiral symmetry breaking terms
\begin{equation}\label{fv-eq:su3_violating_terms}
 \lambda_{1}B_{0}\tr_D \left( 
  \bar{H}^{(Q)}_{a} 
  \left[
   \xi m_{Q} \xi + 
   \xi^{\dagger} M_Q \xi^{\dagger}
  \right]_{ab}
  H^{(Q)}_{b}
 \right)
+ \lambda_{1}^{\prime}B_{0}\tr_D \left( 
  \bar{H}^{(Q)}_{a} 
  H^{(Q)}_{a}
 \right)
 \left[
   \xi m_Q \xi + 
   \xi^{\dagger} m_Q \xi^{\dagger}
  \right]_{bb} .
\end{equation}
We choose to work with the effective theory in which the 
heavy-light pseudoscalar
mesons that contain a heavy quark and a $u$ or $d$ valence 
anti-quark are massless.
Notice that the term proportional to $\lambda^{\prime}_{1}$ in 
Eq.~(\ref{fv-eq:su3_violating_terms}) causes a universal shift to
all the heavy-light meson masses.
This means that the
masses appearing in the propagators of heavy vector mesons and 
any meson containing an $s$ anti-quark (valence or ghost) are shifted 
as follows:
\begin{equation}\label{fv-eq:shift_start}
  \frac{-i (g_{\mu\nu} - v_{\mu} v_{\nu})}
   {2 (v\cdot k - \D + i\epsilon)},
  \quad
  \frac{i}{2 (v\cdot k - \delta_{s} + i\epsilon)},
  \quad\text{and}\quad
  \frac{-i (g_{\mu\nu} - v_{\mu} v_{\nu})}
   {2 (v\cdot k - \D - \delta_{s} + i\epsilon)} ,
\end{equation}
for $P^{\ast}$, $P_{s}$, and $P^{\ast}_{s}$, respectively.  The mass
shifts can be written in terms of the couplings in 
Eqs.~(\ref{fv-eq:HQ_spin_breaking_term}) and  
(\ref{fv-eq:su3_violating_terms}),
$\D = -8\lambda_{2}/M$ ,
and
\begin{equation}\label{fv-eq:delta_s_lambda}
 \delta_{s} = 2\lambda_{1} B_{0} (m_s-m_u) .
\end{equation}

In PQQCD, there are two additional mass shifts
because the
sea quarks have different masses from the valence and ghost quarks:
\begin{equation}\label{fv-eq:tilde_delta_s_def}
 \tilde{\delta}_{s} = M_{\tilde{P}_{s}} - M_{\tilde{P}}
      = 2\lambda_{1} B_{0} (m_r-m_j) ,
\end{equation}
and
\begin{equation}\label{fv-eq:delta_sea_def}
 \delta_{\mathrm{sea}} = M_{\tilde{P}} - M_{P}
 = 2\lambda_{1} B_{0} (m_j-m_u) .
\end{equation}
where $\tilde{P}$ ($\tilde{P}_{s}$) is
the heavy-light pseudoscalar meson with a $d$ ($s$) sea 
anti-quark.  The propagators of the heavy mesons containing 
sea anti-quarks are:
\begin{equation}
  \frac{i}
   {2 (v\cdot k - \delta_{\mathrm{sea}} + i\epsilon)},
  \quad
  \frac{-i (g_{\mu\nu} - v_{\mu} v_{\nu})}
   {2 (v\cdot k - \D - \delta_{\mathrm{sea}} 
  + i\epsilon)}, 
\end{equation}
\begin{equation} \label{fv-eq:shift_end}
  \frac{i}{2 (v\cdot k - \delta_{\mathrm{sea}} - \tilde{\delta}_{s} 
   + i\epsilon)},
  \quad\text{and}\quad
  \frac{-i (g_{\mu\nu} - v_{\mu} v_{\nu})}
   {2 (v\cdot k  - \D 
    - \delta_{\mathrm{sea}} - \tilde{\delta}_{s} + i\epsilon)}
\end{equation}
for $\tilde{P}$, $\tilde{P}^*$ (vector meson with a $d$ sea anti-quark), 
$\tilde{P}_s$, 
and $\tilde{P}^*_s$ (vector meson with an $s$ sea anti-quark), respectively.

\section{Finite Volume Effects}\label{fv-sec:FV}
In this section, we discuss generic features of finite volume effects
in HM$\chi$PT.  For clarity, we use the symbol $\Delta$ for one of 
($\D$, $\delta_{s}$, $\tilde{\delta}_{s}$, 
$\delta_{\mathrm{sea}}$) or any
sum amongst them.

In the limit where the heavy quark mass goes to infinity and the light quark
masses are equal, all the heavy mesons in HM$\chi$PT become on-shell 
static sources, and there is a velocity superselection rule when the
momentum transfer involved in the scattering of the heavy meson system
is fixed~\cite{Georgi:1990um}.  For illustration, consider the 
vertex with coupling $g$ in 
the Lagrangian in
Eq.~(\ref{B2D-eqn:LD}).  The heavy-light meson
$P$ can scatter into $P^{\ast}_{(s)}$ by emitting a Goldstone meson with
mass $m_{QQ'}$ through this vertex.  
The momenta of the mesons $P$ and $P^{\ast}_{(s)}$
are
$M_{P}v_{\mu}$,
and
$M_{P^{\ast}_{(s)}}v_{\mu} + k_{\mu} = M_{P}v_{\mu} + k_{\mu}$,
where the velocity 
$v_{\mu}=(1,0,0,0)$ in the rest frame of the heavy mesons, and 
$k_{\mu}$ is the soft momentum carried by the Goldstone meson.
The infinitely heavy $P$ and $P^{\ast}_{(s)}$ mesons
do not propagate in space.
Therefore, when such a system is in
a cubic spatial box, 
finite volume effects result entirely from the
propagation of the Goldstone meson to the boundary with momentum
$k\sim m_{QQ'}$. In this case the volume effects behave like
$\exp{(-m_{QQ'}L)}$ multiplied by a polynomial in
$1/L$.

The breaking of heavy quark spin and $SU(3)$ light flavour symmetries 
in HM$\chi$PT can induce a mass difference
$M_{P^{\ast}_{(s)}} = M_{P} + \Delta$,
which complicates the above picture.  
In this scenario,
the $P^{\ast}_{(s)}$ is still regarded 
as a static source, but it is off-shell with the virtuality $\Delta$.
The period during which the Goldstone meson can propagate to the
boundary is limited by the time uncertainty conjugate to
this virtuality, {\it i.e.},

\begin{equation}\label{fv-eq:virtuality}
\delta t \sim \frac{1}{\Delta} .
\end{equation}
This means that finite volume effects, which arise from the propagation of the 
Goldstone mesons in such a system, will decrease as $\Delta$ increases. 
Eq.~(\ref{fv-eq:virtuality}) also indicates that the suppression of the 
volume effects by a non-zero $\Delta$ is controlled by the parameter
\begin{equation}\label{fv-eq:mass_ratio}
 \frac{m_{QQ'}}{\Delta} .
\end{equation}

To see explicitly how this phenomenon appears in a calculation,
we consider a typical sum in one-loop HM$\chi$PT, with a 
Goldstone propagator and a heavy-light vector meson propagator in 
the loop, in a cubic box with periodic boundary conditions:
\begin{equation}\label{fv-eq:calJ}
 \cJ(m,\Delta) 
 = -i  \frac{1}{L^{3}}
 \sum_{\vec{k}} \int \frac{d k_{0}}{2\pi}
 \frac{1}{(k^{2}-m^2 + i\epsilon )
 (v\cdot k - \Delta + i\epsilon )}  ,
\end{equation}
where the spatial momentum $\vec{k}$ is quantized in finite 
volume as
$\vec{k} = (2\pi/L)\vec{i}$,
with $\vec{i}$ being a three dimensional integer vector.
Using the Poisson summation formula, it is straightforward to show that
$\cJ(m,\Delta) = J(m,\Delta)+\JFV(m,\Delta)$,
where 
\begin{equation}\label{fv-eq:J}
 J(m,\Delta) = -i \int \frac{d^{4}k}{(2\pi)^{4}}
 \frac{1}{(k^{2}-m^{2} +i\epsilon )
 (v\cdot k - \Delta + i\epsilon )}  ,
\end{equation}
is the infinite volume limit of 
$\cJ(m,\Delta)$, and  
\begin{equation}\label{fv-eq:JFV}
 \JFV(m,\Delta) 
 =
 \frac{1}{4\pi^2} 
 \sum_{\vec{n}\not = \vec{0}} \int_{0}^{\infty}
 d |\vec{k}| \frac{|\vec{k}|}{\w(\w + \Delta)}
 \frac{\sin(|\vec{k}| |\vec{n}| L)}{|\vec{n}| L}
,\end{equation}
with
$\w = \sqrt{|\vec{k}|^{2}+m^{2}}$,
is the finite volume correction to $J(m,\Delta)$.  
In the asymptotic limit where $mL\gg 1$ it can be shown that 
(with $n = |\vec{n}|$)
\begin{equation}\label{fv-eq:JFV_asymp}
 \JFV(m,\Delta) 
 = 
\sum_{\vec{n}\not=\vec{0}} \frac{1}{8\pi n L}
\exp(- n m L) \,\cA
,\end{equation}
where
\begin{eqnarray}\label{fv-eq:JFV_asymptotic}
  \cA 
  &=&
  \exp(z^2) [1 - \erf(z)]
  +
  \frac{1}{n m L} 
  \left[
    \frac{1}{\sqrt{\pi}} 
    \left( 
      \frac{z}{4} - \frac{z^{3}}{2}
    \right )
    + 
    \frac{z^{4}}{2}\exp(z^{2})[1 - \erf(z)]
  \right]
               \nonumber\\
  &&-
  \frac{1}{(n m L)^2}
  \left[
    \frac{1}{\sqrt{\pi}}
    \left( 
      \frac{9z}{64}-\frac{5z^3}{32}+\frac{7z^5}{16}+\frac{z^7}{8} 
    \right)
    -
    \left( 
      \frac{z^6}{2}+\frac{z^8}{8}
    \right)
    \exp(z^2)[1 - \erf(z)]
  \right]
                \nonumber\\
  &&+
  \cO\left(\frac{1}{(n m L)^3}\right)
,\end{eqnarray}
with
\begin{equation}
  z=\frac{\Delta}{m}\sqrt{\frac{n m L}{2}}
.\end{equation}
The quantity ${\mathcal{A}}$ is the alteration of finite volume
effects due to the presence of a non-zero $\Delta$.  
It multiplies the factor
$\exp(-n m L)$, which results from the propagation
of the Goldstone meson to the boundary.
It is possible to analytically compute the higher order corrections 
of ${\mathcal{A}}$ in powers of 
$1/(n m L)$.  This way, one can achieve any 
desired numerical precision.  
Here it is clear that this alteration of volume effects is
controlled by the ratio in Eq.~(\ref{fv-eq:mass_ratio}).

\begin{figure}[tb]
  \centering
  \includegraphics[width=0.75\textwidth]{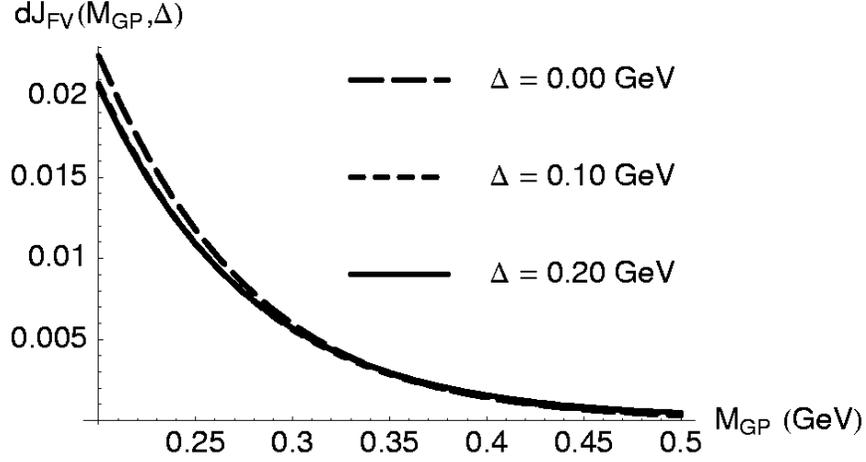}
  \caption[$d\JFV(m,\Delta)$ plotted against $m$]{
    \label{fv-fig:dJFV}$d\JFV(m,\Delta)$ 
    plotted against $m$ for three different $\Delta$.
    This function indicates the deviation
    (in percent) of the asymptotic form of $\JFV$ from the definition
    in Eq.~(\ref{fv-eq:JFV}). 
    In this plot, 
    $L=$2.5~fm and $m = 0.197$~GeV corresponds
    to $m L = 2.5$ whereas $m = 0.32$~GeV corresponds
    to $m L = 4$.  
    The curve for $\Delta=0.1$~GeV is hidden by that for $\Delta=0.2$~GeV.}
\end{figure}

\begin{figure}[tb]
  \centering
  \includegraphics[width=0.75\textwidth]{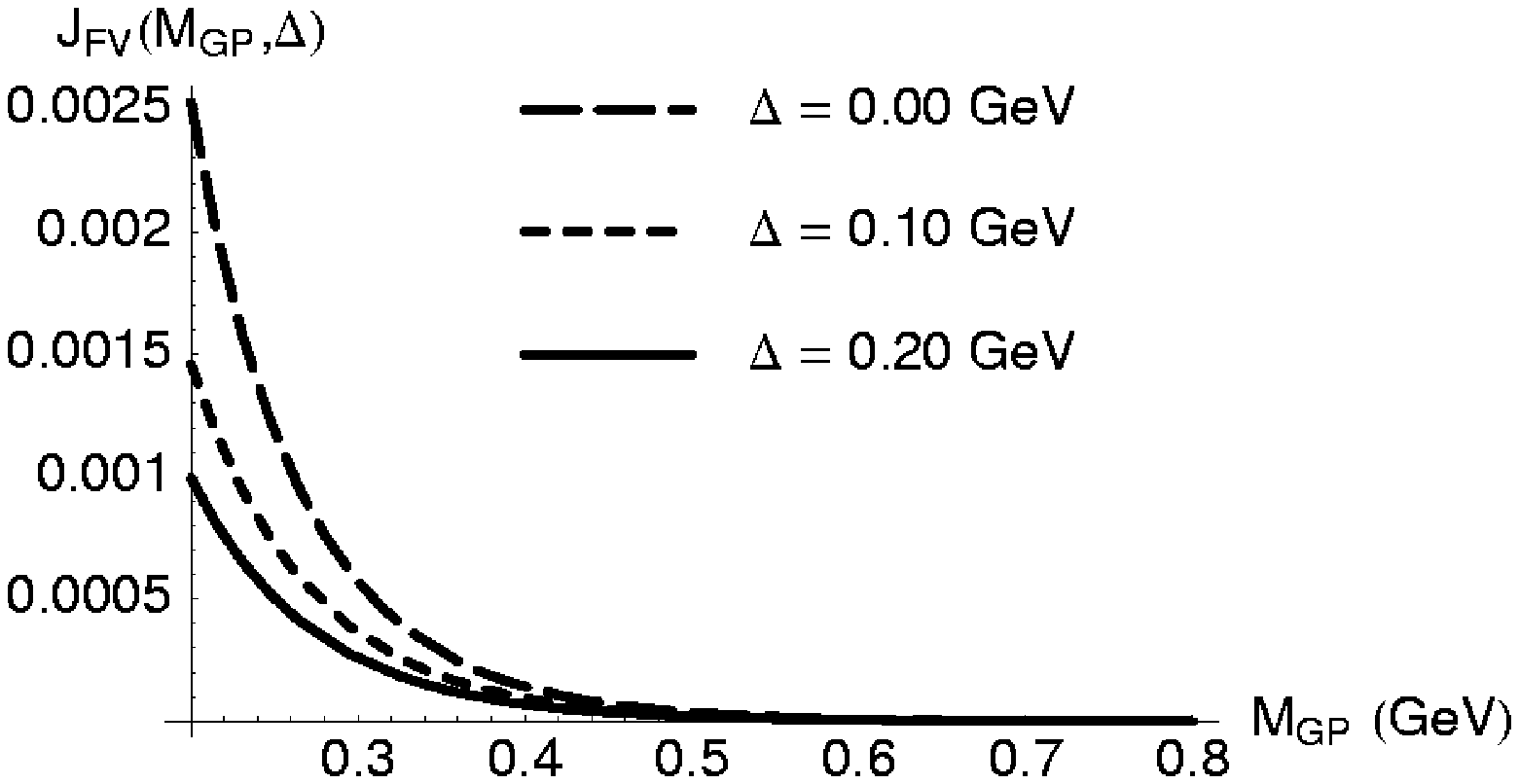}
  \caption[$\JFV(m,\Delta)$ plotted against $m$]{
    \label{fv-fig:JFV}$\JFV(m,\Delta)$ 
    plotted as 
    a function of $m$, for three different $\Delta$.
    Again, 
    $L=$2.5 fm.}
\end{figure}

Next, we consider different limits of ${\mathcal{A}}$ at  
fixed $m$ and $L$.  When
$\Delta=0$, clearly ${\mathcal{A}}=1$.  If
$\Delta$ is very small compared to $m$, such that 
$z\ll 1$, ${\mathcal{A}}$ is dominated by
the $[1/(m L)]^0$ term, {\it i.e.},
${\mathcal{A}} \approx \exp(z^2)[ 1-\erf(z)]$.
Since $\erf(z)$ grows much faster than
exp($z^{2}$) in this regime, ${\mathcal{A}}$ will decrease as $\Delta$
increases.  When $\Delta$ is of $\cO(m)$ or 
larger, $z\gg 1$, and one can perform an asymptotic expansion of
the error function.  It can be shown that in this situation, 
$\cA \sim 1/z$.
That is, ${\mathcal{A}}$ also decreases as $\Delta$ increases.
We have also numerically checked that this is true when $z\approx 1$.
This means that the asymptotic formula in Eq.~(\ref{fv-eq:JFV_asymp})
reproduces the physical picture outlined in the beginning of this section
for any $\Delta$. To demonstrate how fast the asymptotic form
in Eq.~(\ref{fv-eq:JFV_asymptotic}) converges to 
Eq.~(\ref{fv-eq:JFV}), we define 
\begin{equation}\label{fv-eq:dJFV}
 d\JFV(m,\Delta) 
 = 
 \frac{\JFV^{\mathrm{num}}(m,\Delta) - \JFV^{\mathrm{asymp}}(m,\Delta)}
 {\JFV^{\mathrm{num}}(m,\Delta)}
,\end{equation}
where $\JFV^{\mathrm{num}}$ is the function $\JFV$
evaluated numerically [Eq.~(\ref{fv-eq:JFV})], and 
$\JFV^{\mathrm{asymp}}$ is the asymptotic form in 
Eq.~(\ref{fv-eq:JFV_asymptotic}).  
In Fig.~\ref{fv-fig:dJFV},
we plot $d\JFV$ as a function of $m$ with three
choices of $\Delta$.  
It is clear from this plot that 
$\JFV$ is
approximated well (to $\le 3\%$) by the asymptotic form
when $m L \ge 2.5$.  We use the asymptotic forms for
integrals of this type throughout this work.  Also, in this paper 
we only include  the terms with $|\vec{n}|=1$,$\sqrt{2}$, $\sqrt{3}$, 
$\sqrt{4}$, and $\sqrt{5}$ in the Poisson summation formula.  
We have confirmed that truncating the 
sum at $|\vec{n}|=\sqrt{5}$ is a very good approximation (to $\sim 3\%$)
when $m L \ge 2.5$.
The function $\JFV(m,\Delta)$ 
is plotted against $m$
in Fig.~\ref{fv-fig:JFV}, 
with $L=2.5$~fm and three choices 
of $\Delta$.  It is clear from
this plot that $\Delta$ can significantly alter the 
finite volume effects in 
$\cJ(m,\Delta)$.

\begin{figure}[tb]
  \centering
  \includegraphics[width=0.75\textwidth]{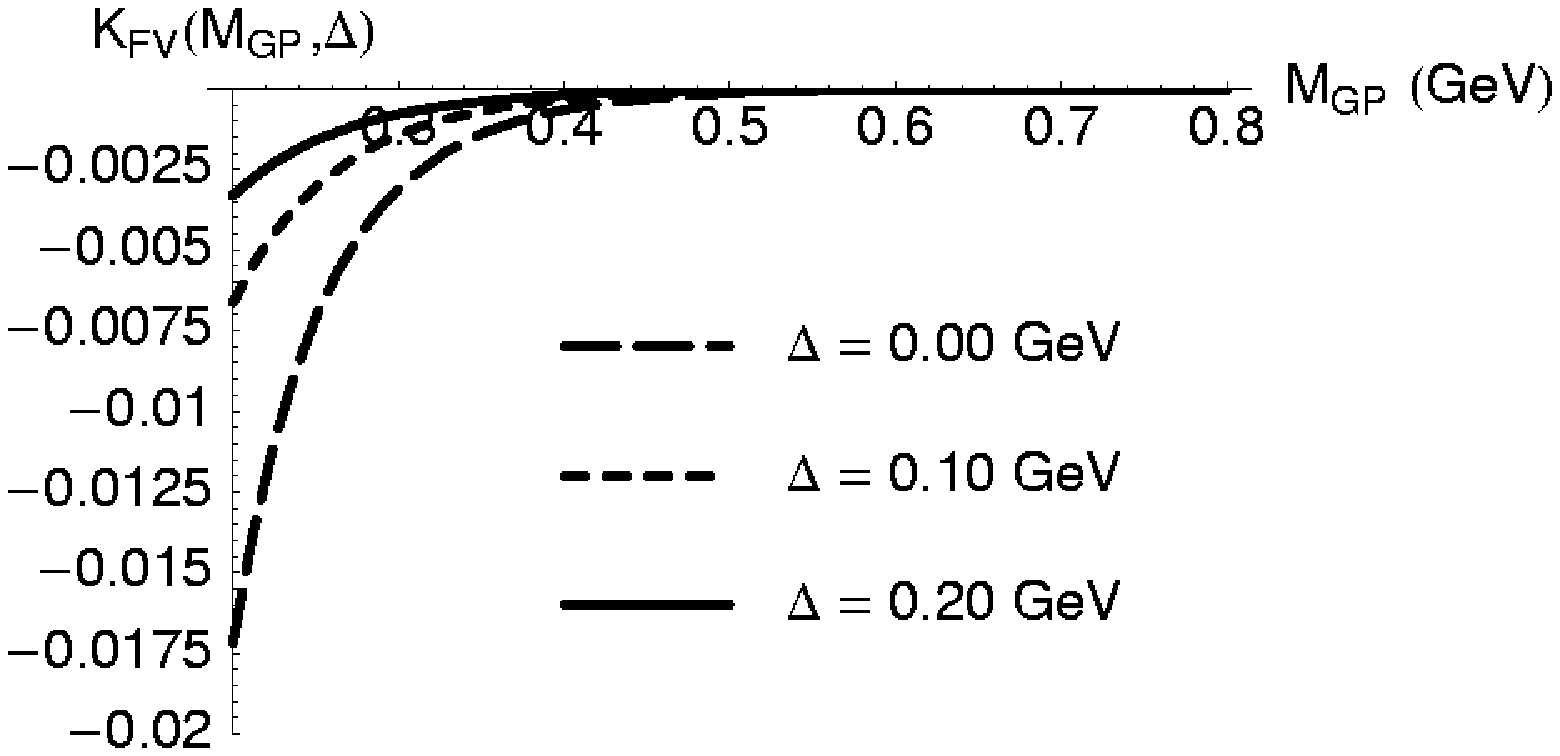}
  \caption[$\KFV(m,\Delta)$ plotted against $m$]{
    \label{fv-fig:KFV}$\KFV(m,\Delta)$ 
    plotted as 
    a function of $m$ for three different 
    $\Delta$.  
    Again,
    $L=2.5$ fm.}
\end{figure}

Another typical sum that appears in one-loop HM$\chi$PT in finite volume is
\begin{equation}\label{fv-eq:calK}
 \cK(m,\Delta) 
 = 
 -\frac{i}{L^3}
 \sum_{\vec{k}} \int \frac{d k_{0}}{2\pi}
 \frac{1}{(k^{2}-m^2 + i\epsilon)(v\cdot k - \Delta + i\epsilon)^2}
.\end{equation}
It is straightforward to show that
$\cK(m,\Delta) = K(m,\Delta) + \KFV(m,\Delta)$,
where
\begin{equation}
 K(m,\Delta) = 
 \frac{\partial}{\partial \Delta}J(m,\Delta)
\end{equation} 
is the infinite volume limit of 
$\cK(m,\Delta)$ and
\begin{equation}\label{fv-eq:KFV_asymptotic}
 \KFV(m,\Delta) = 
 \frac{\partial}{\partial \Delta}\JFV(m,\Delta)
\end{equation}
is the finite volume correction to $K(m,\Delta)$.  
The function 
$\KFV(m,\Delta)$ is 
plotted against $m$
in Fig.~\ref{fv-fig:KFV}, 
with $L=2.5$ fm and three choices of $\Delta$.  
As expected,
$|\KFV(m,\Delta)|$ also decreases when  
$\Delta$ increases
at fixed $m$ and $L$.

\section{Neutral $B$ Mixing and Heavy-Light Decay Constants}
\label{fv-sec:BBbar_mixing}
The study of neutral $B$ meson mixing allows the extraction of the 
magnitude of the CKM matrix element $V_{td}$, 
and hence the determination of one of the sides of the unitarity triangle.
The frequency of the $B_{d}{-}\bar{B}_{d}$ oscillations, which is given
by the mass difference, $\Delta m_{d}$, in this mixing system has been
well measured by various experimental 
collaborations~\cite{Battaglia:2003in}.  
It is also calculable in the standard model via
an operator product expansion in which the top quark and $W$ boson are 
integrated out.  Resumming the next-to-leading order (NLO) short-distance
QCD effects, one obtains
\begin{equation}\label{fv-eq:delta_md}
  \Delta m_{d} 
  = 
  \frac{G_{F}}{8 \pi^{2}} 
   M^{2}_{W}
  |V_{td} V^{\ast}_{tb}|^{2} 
  \eta_{B} S_{0}(x_{t}) C_{B}(\mu)
  \frac{|\la\bar{B}_{d}|\cO^{\Delta B = 2}_{d}(\mu)|B_{d}\ra|}{2 M_{B}}
,\end{equation}
where $\mu$ is the renormalisation scale, $x_{t}=m^{2}_{t}/M^{2}_{W}$, and 
$S_{0}(x_{t})\approx 0.784 x_{t}^{0.76}$ (to better than $1\%$) 
is the relevant Inami-Lim function~\cite{Inami:1981fz}.  The 
coefficients $\eta_{B}=0.55$ and
$C_{B}(\mu)$ are from short-distance QCD 
effects~\cite{Buras:1990fn,Buchalla:1996vs}. The
matrix element of the four-quark operator 
\begin{equation}\label{fv-eq:4q_op_bd}
 \cO^{\Delta B=2}_{d} 
 = 
 [\bar{b}\gamma^{\mu}(1 - \gamma_{5})d]
 [\bar{b}\gamma_{\mu}(1 - \gamma_{5})d]
\end{equation}
between $B_{d}$ and $\bar{B}_{d}$ states contains all the long-distance
QCD effects in Eq.~(\ref{fv-eq:delta_md}), and has to be calculated 
non-perturbatively.   Since $|V_{tb}|=1$ to good
accuracy and $\Delta m_{d}$ has been well measured, a 
high-precision calculation of 
$\la\bar{B}_{d}|\cO^{\Delta B = 2}_{d}(\mu)|B_{d}\ra$ enables 
a clean determination of $|V_{td}|$.

The frequency of the rapid $B_{s}{-}\bar{B}_{s}$ oscillations can be
precisely measured at the Tevatron and LHC~\cite{Battaglia:2003in}. 
Therefore an alternative approach
is to consider the ratio
\begin{equation}\label{fv-eq:delta_ms_over_delta_md}
 \frac{\Delta m_{s}}{\Delta m_{d}} =
 \left | \frac{V_{ts}}{V_{td}} \right |^{2} 
 \frac{M_{B_{d}}}{M_{B_{s}}}
 \left | \frac{\la\bar{B}_{s}|\cO^{\Delta B = 2}_{s}|B_{s}\ra}
  {\la\bar{B}_{d}|\cO^{\Delta B = 2}_{d}|B_{d}\ra}
 \right |
,\end{equation}
in which many theoretical uncertainties cancel.
Here $\Delta m_{s}$ is the mass difference in the $B_{s}{-}\bar{B}_{s}$
system and 
$\cO^{\Delta B=2}_{s} = [\bar{b}\gamma^{\mu}
(1 - \gamma_{5})s][\bar{b}\gamma_{\mu}(1 - \gamma_{5})s]$.  
The unitarity
of the CKM matrix implies $|V_{ts}|\approx |V_{cb}|$ to a few percent, and
$|V_{cb}|$ can be precisely extracted by analysing 
semileptonic $B$ decays~\cite{Battaglia:2003in}.  Therefore a clean
measurement of $\Delta m_{s}/\Delta m_{d}$ will yield an accurate
determination of $|V_{td}|$.

The matrix elements in Eq.~(\ref{fv-eq:delta_ms_over_delta_md}) are 
conventionally parameterized as
\begin{equation}\label{fv-eq:me_to_BfB}
\la\bar{B}_{q}|\cO^{\Delta S = 2}_{q}|B_{q}\ra =
 \frac{8}{3} M^{2}_{B_{q}} f^{2}_{B_{q}} B_{B_{q}}(\mu) ,
\end{equation}
where the parameter $B_{B_{q}}$ measures the deviation from the 
vacuum-saturation approximation of the matrix element, and
$q=d$ or $s$.  The decay constant $f_{B_{q}}$ is defined by
\begin{equation}
 \la 0 |\bar{b}\gamma_{\mu}\gamma_{5}q|B_{q}(\vec{p})\ra =
 i p_{\mu} f_{B_{q}} .
\end{equation}

LQCD provides a reliable tool for calculating these 
non-perturbative QCD quantities from first principles.%
\footnote{Some
selected reviews in the long history of lattice calculations for the $B$ 
mixing system can be found in Refs.~\cite{Flynn:1998ca, 
Sachrajda:2000ci, Flynn:2000hx, Bernard:2000ki, Ryan:2001ej, Yamada:2002wh, 
Lellouch:2002nj, Becirevic:2003hf, Kronfeld:2003sd}.} 
Since $\Delta m_{s}/\Delta m_{d}$ will be measured to very good accuracy,
it is important to have clean theoretical calculations for 
[the $SU(3)$ breaking ratios of] the matrix 
elements, decay constants and $B$ parameters involved. 
Current
lattice calculations have to be combined with effective theories in 
order to obtain these matrix elements at the physical quark masses.  
This procedure can introduce significant systematic errors and 
dominate the uncertainties in the $SU(3)$ breaking ratio%
\cite{Kronfeld:2002ab,Becirevic:2002mh}
\begin{equation}\label{fv-eq:xi_def}
 \xi = \frac{f_{B_{s}} \sqrt{B_{B_{s}}}}{f_{B}\sqrt{B_{B}}} ,
\end{equation}
which is the key theoretical input for future high-precision determination
of $|V_{td}|$ via the study of neutral $B$ mixing.%
\footnote{Notice that
the symbol $\xi$ as defined in Eq.~(\ref{fv-eq:xi_def}) is in 
the traditional notation in $B$ physics, and has nothing to do with 
the Goldstone field $\xi$.}
However, the use of effective theory also offers a framework 
for studying finite volume effects in lattice 
calculations~\cite{Sharpe:1992ft,Bernard:1996ez, Golterman:1997wb, 
Golterman:1998st, Golterman:1998af, Lin:2002nq, Lin:2002aj, Lin:2003tn, 
Becirevic:2003wk, Detmold:2004qn, Beane:2003da, Beane:2003yx, 
Colangelo:2003hf, AliKhan:2003cu, Beane:2004tw}.  
We will
demonstrate in this section that finite volume effects might turn out to
exceed the current quoted systematic errors for quantities such as $\xi$.

\subsection{One-Loop Calculation in a Finite Volume}
\label{fv-sec:one_loop_calculation}
Here, we discuss one-loop calculations for the $B$ 
parameters and heavy-light decay constants mentioned above in finite
volume HM$\chi$PT including the appropriate mass shifts to the first
non-trivial order of the chiral and $1/M$ expansion.  The inclusion
of other first-order corrections in these quantities
is straightforward.  It simply introduces additional
low-energy constants (LECs)
which account for short-distance physics and do not give rise to finite volume
effects at this order, so we will not discuss this issue here.
We have performed the calculation for QCD, QQCD and PQQCD
with the mass shifts given in
Eqs.~(\ref{fv-eq:shift_start})--(\ref{fv-eq:shift_end}).

For the purpose of this work,
the axial current 
$\bar{b}\gamma_{\mu}\gamma_{5}q_{a}$ 
is 
\begin{equation}\label{fv-eq:axial_current}
  A_{\mu} 
 = 
 \frac{\kappa}{2} \tr_{\mathrm{D}}
 \left[ \gamma_{\mu}\gamma_{5} H^{(Q)}_{b}\xi^{\dagger b}_{a}\right]
,\end{equation}
and the four-quark operator $\cO^{\Delta P_{a}=2}$ (when $P_{a}=B_{d,s}$, 
$\cO^{\Delta P_{a}=2}$ becomes $\cO^{\Delta B=2}_{d,s}$) is
\begin{equation}\label{fv-eq:Oaa}
  O^{aa} 
  = 
  4\beta 
  \left[ 
    \left(\xi P^{\ast(Q)\dagger}_{\mu}\right)^{a}
    \left(\xi P^{\ast(\bar{Q})\mu}\right)^{a}
    +
    \left(\xi P^{(Q)\dagger}\right)^{a}
    \left(\xi P^{(\bar{Q})}\right)^{a}
  \right] 
\end{equation}
in HM$\chi$PT~\cite{Grinstein:1992qt}, where $\kappa$ and $\beta$ are 
the low-energy constants which have to be determined
from experiments or lattice calculations.  
Notice that the index
$a$ in Eq.~(\ref{fv-eq:Oaa}) is not summed over.  
Again, the inclusion of the chiral and $1/M_{P}$ corrections in these
operators simply introduces additional LECs and we do not investigate
this aspect here.
We assume that $\kappa$ and $\beta$ are the same in QCD,
QQCD, and PQQCD.  Also, $A_{\mu}$ and $O^{aa}$ can couple
to the $\eta^{\prime}$ in QQCD, but the couplings are $1/N_{c}$
suppressed~\cite{Sharpe:1996qp}, and we neglect them.

\begin{figure}[tb]
  \centering
  \includegraphics[width=0.5\textwidth]{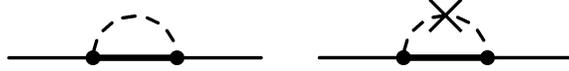}
  \caption[Wave-function renormalization diagrams for heavy-light mesons]{
    \label{fv-fig:diagrams-wf}Wave-function renormalization diagrams 
    for heavy-light mesons}
\end{figure}

\begin{figure}
  \centering
  \includegraphics[width=\textwidth]{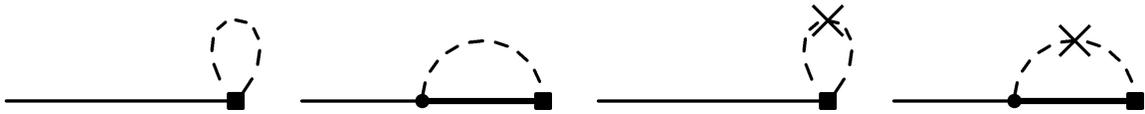}
  \caption[One-loop diagrams contributing to the
    decay constants]{
    \label{fv-fig:fB_diagrams}Diagrams contributing to the one-loop
    calculation of decay constants.  The thin (thick) solid 
    lines are 
    the heavy-light pseudoscalar (vector) mesons.  The dashed 
    lines are Goldstone mesons, and the crosses are the ``double poles''
    which appear in (P)Q$\chi$PT.  The open squares are the 
    operators defined in Eq.~(\ref{fv-eq:axial_current}) and the dots are
    vertices from the HM$\chi$PT Lagrangian.}
\end{figure}

\begin{figure}
  \centering
  \includegraphics[width=\textwidth]{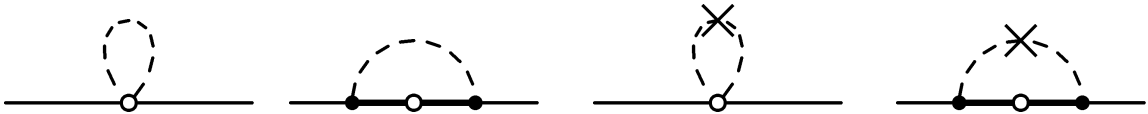}
  \caption[One-loop diagrams contributing to the
    $B$ parameters]{
    \label{fv-fig:BB_diagrams}Diagrams contributing to the one-loop
    calculation of the $B$ parameters.   The open circles are the 
    operators defined in Eq.~(\ref{fv-eq:Oaa}).}
\end{figure}

The diagrams contributing to $f_{P_{(s)}}$ and $B_{P_{(s)}}$ are 
presented in Figs.~\ref{fv-fig:diagrams-wf}--\ref{fv-fig:BB_diagrams}.
Note that 
only diagrams with an intermediate heavy meson depend on the heavy meson
mass shifts.  

Although this is the first one-loop calculation for these decay constants and 
$B$ parameters in finite volume,
some results in the infinite volume limit
already exist in the literature: $f_{P_{(s)}}$ have been calculated at the
lowest order in QCD%
~\cite{Grinstein:1992qt}, QQCD~\cite{Sharpe:1996qp,Booth:1995hx},
and PQQCD~\cite{Sharpe:1996qp}, and up to first-order
corrections in the chiral and
$1/M$ expansions in QCD~\cite{Boyd:1995pa} and 
QQCD~\cite{Booth:1994rr}.  The $B$ parameters have been
calculated only at 
lowest order~\cite{Grinstein:1992qt, Sharpe:1996qp}.  Our results, 
as presented in Appendix~\ref{fv-sec:results}, agree 
with all these previous calculations in the appropriate limits.

\subsection{Phenomenological Impact}
\label{fv-sec:phenomenology}
We have used the one-loop results in Appendix~\ref{fv-sec:results}
to investigate the impact of finite volume effects on $\xi$.  
In this work, we only intend to estimate
the possible size of errors in this quantity, and will
leave the actual comparison with lattice data to a future publication.
Following the usual procedure in lattice calculations for $\xi$, 
we study two $SU(3)$ breaking ratios
\begin{equation}\label{fv-eq:xi_f}
 \xi_{f} = \frac{f_{B_{s}}}{f_{B}} 
\end{equation}
and
\begin{equation}\label{fv-eq:xi_B}
 \xi_{B} = \frac{B_{B_{s}}}{B_{B}}
,\end{equation}
in terms of which,  
$\xi = \xi_{f}\sqrt{\xi_{B}}$.
Furthermore, we define 
$(\xi_{f})_{\mathrm{FV}}$
and
$(\xi_{B})_{\mathrm{FV}}$
to be the contributions from finite volume effects, {\it i.e.}, those from
the volume-dependent part in the one-loop results presented in 
Subsection~\ref{fv-sec:one_loop_calculation}.  To be more precise,
we use the formulae collected in Appendix~\ref{fv-sec:results} to calculate
the volume corrections 
{\em with respect to the lowest-order values} of 
$f_{B_{s}}$ ($B_{B_{s}}$) and $f_{B}$ ($B_{B}$), then take the difference
between the results as an estimate of $(\xi_{f})_{\mathrm{FV}}$
[$(\xi_{B})_{\mathrm{FV}}$].  Since these $SU(3)$ ratios are not very
different from unity (at most $\sim 20\%$), this is a reasonable estimate
of these effects.

Traditionally, many quenched lattice simulations of $B_{B_{(s)}}$ and
$f_{B_{(s)}}$ were performed using $L \sim 1.6$~fm.  Therefore
we present our estimate for finite volume effects 
in QQCD with this box size.  For comparison,
we adopt the same volume for QCD.  
As for PQQCD, we work with $L = 2.5$~fm where most current
high-precision simulations are carried out~\cite{Davies:2003ik}.
Throughout this subsection, we ensure that the condition
$M_{\pi} L \ge 2.5$
holds in all the plots presented here.

We first discuss the procedure in QCD and QQCD.
When studying
the light quark mass dependence of
$(\xi_{f})_{\mathrm{FV}}$ and $(\xi_{B})_{\mathrm{FV}}$, we follow a 
strategy similar to that in Ref.~\cite{Kronfeld:2002ab}.  That is, we
use the Gell-Mann-Okubo formulae to express
$M_{K}$ and $M_{\eta}$ in terms of $M_{\pi}$ and $m_{ss}$: 
\begin{equation}
 M^{2}_{K} = \frac{m^{2}_{ss} + M^{2}_{\pi}}{2}
 \quad\text{and}\quad
 M^{2}_{\eta} = \frac{2 m^{2}_{ss} + M^{2}_{\pi}}{3}
.\end{equation}
We investigate the situation where a lattice calculation is performed at the
physical strange quark mass $(m_{s})_{\mathrm{phys}}$, 
but the up and down quark mass $m_u=m_d$ is varied.
By using $(M_{K})_{\mathrm{phys}}=0.498$~GeV and 
$(M_{\pi})_{\mathrm{phys}}=0.135$~GeV%
~\cite{Hagiwara:2002fs}, we fix
$(m_{ss})_{\mathrm{phys}} = 2 B_{0} (m_{s})_{\mathrm{phys}} = 0.691$~GeV
as an input parameter in our analysis.  
Notice that $(m_{ss})_{\mathrm{phys}}$
is not the mass of a ``physical'' meson, and the subscript just means this
mass is 
estimated by using physical kaon and pion masses.
To the same order, we
can adopt Eq.~(\ref{fv-eq:delta_s_lambda}) to write
$\delta_{s} = \lambda_{1} \left( m^{2}_{ss} - M^{2}_{\pi}\right)$,
and use $(m_{ss})_{\mathrm{phys}}$, $(M_{\pi})_{\mathrm{phys}}$ and physical 
$M_{B_{s}}-M_{B}=0.091$~GeV~\cite{Hagiwara:2002fs} to 
determine
\begin{equation}\label{fv-eq:lambda_value}
  \lambda_{1} = 0.1982 \gev^{-1}
.\end{equation}
This determines how $\delta_{s}$ varies with $M_{\pi}$.
We have also tried to use vanishing pion mass and 
$M_{D_{s}}-M_{D}=0.104$~GeV~\cite{Hagiwara:2002fs} to fix 
$(m_{ss})_{\mathrm{phys}}$ 
and $\lambda_{1}$, and the results
presented in this subsection are not sensitive to this
variation from the values quoted above.

\begin{figure}[tb]
  \centering
  \includegraphics[width=0.49\textwidth]{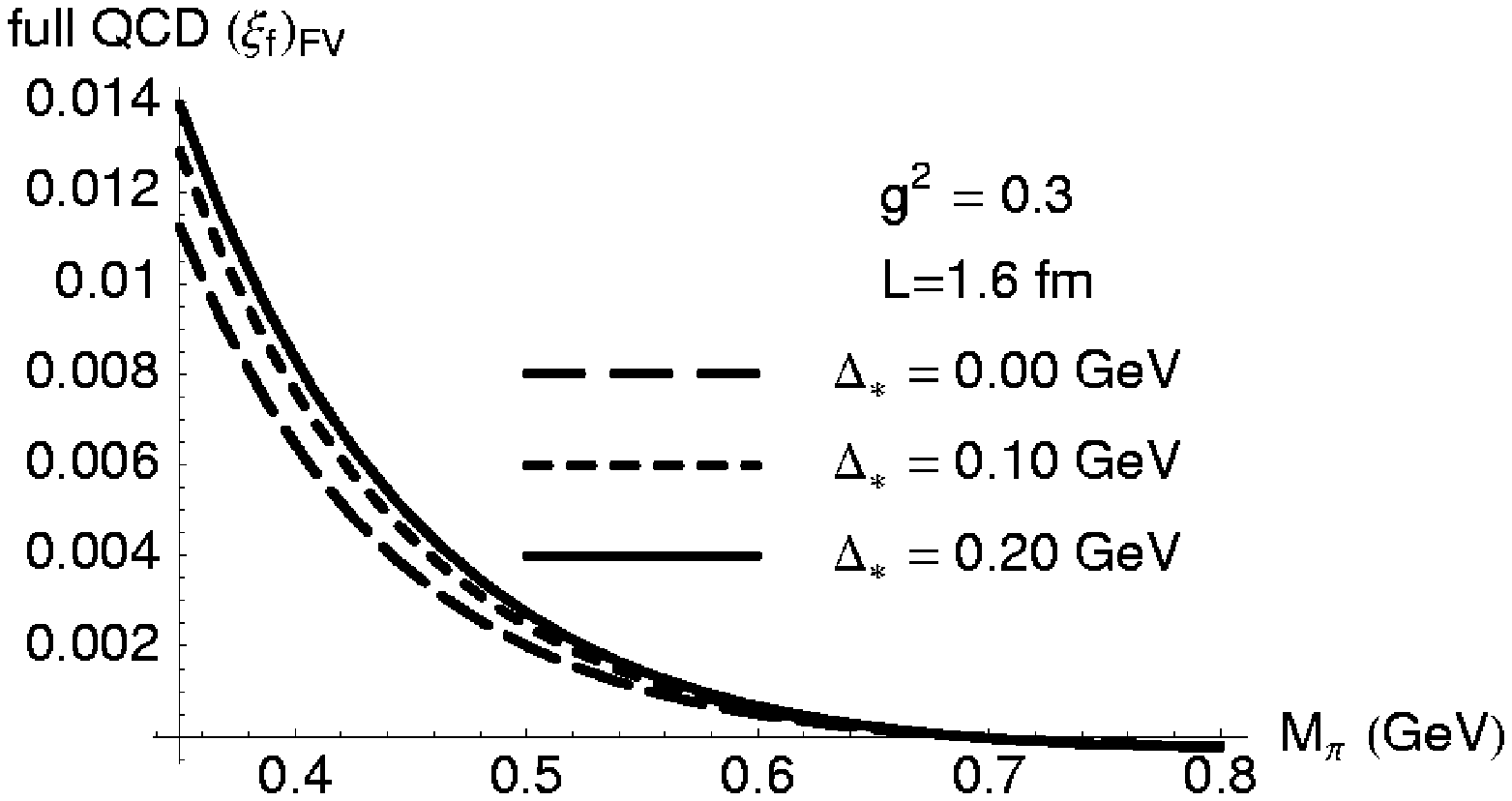}
  \includegraphics[width=0.49\textwidth]{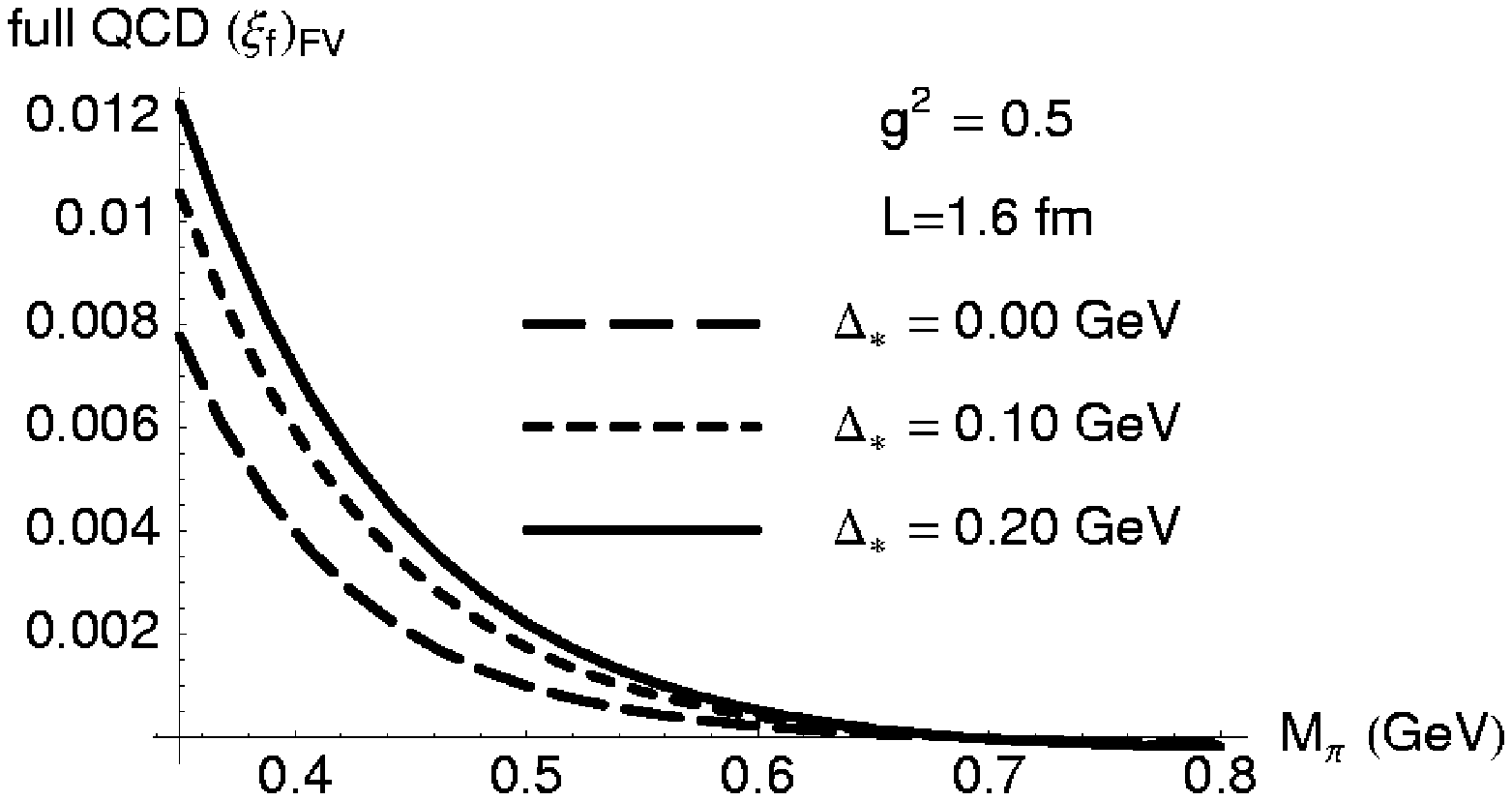}
  \caption[$(\xi_{f})_{\mathrm{FV}}$ in 
    QCD plotted against
    $M_{\pi}$ with $L=1.6$~fm]
    {\label{fv-fig:fBs_over_fB}$(\xi_{f})_{\mathrm{FV}}$ in 
    full QCD plotted against
    $M_{\pi}$ with $L=1.6$~fm. 
    The pion mass $M_{\pi} = 0.35$~GeV corresponds
    to $M_{\pi} L = 2.8$, and $M_{\pi} = 0.5$~GeV corresponds
    to $M_{\pi} L = 4$.}
\end{figure}

\begin{figure}[tb]
  \centering
  \includegraphics[width=0.49\textwidth]{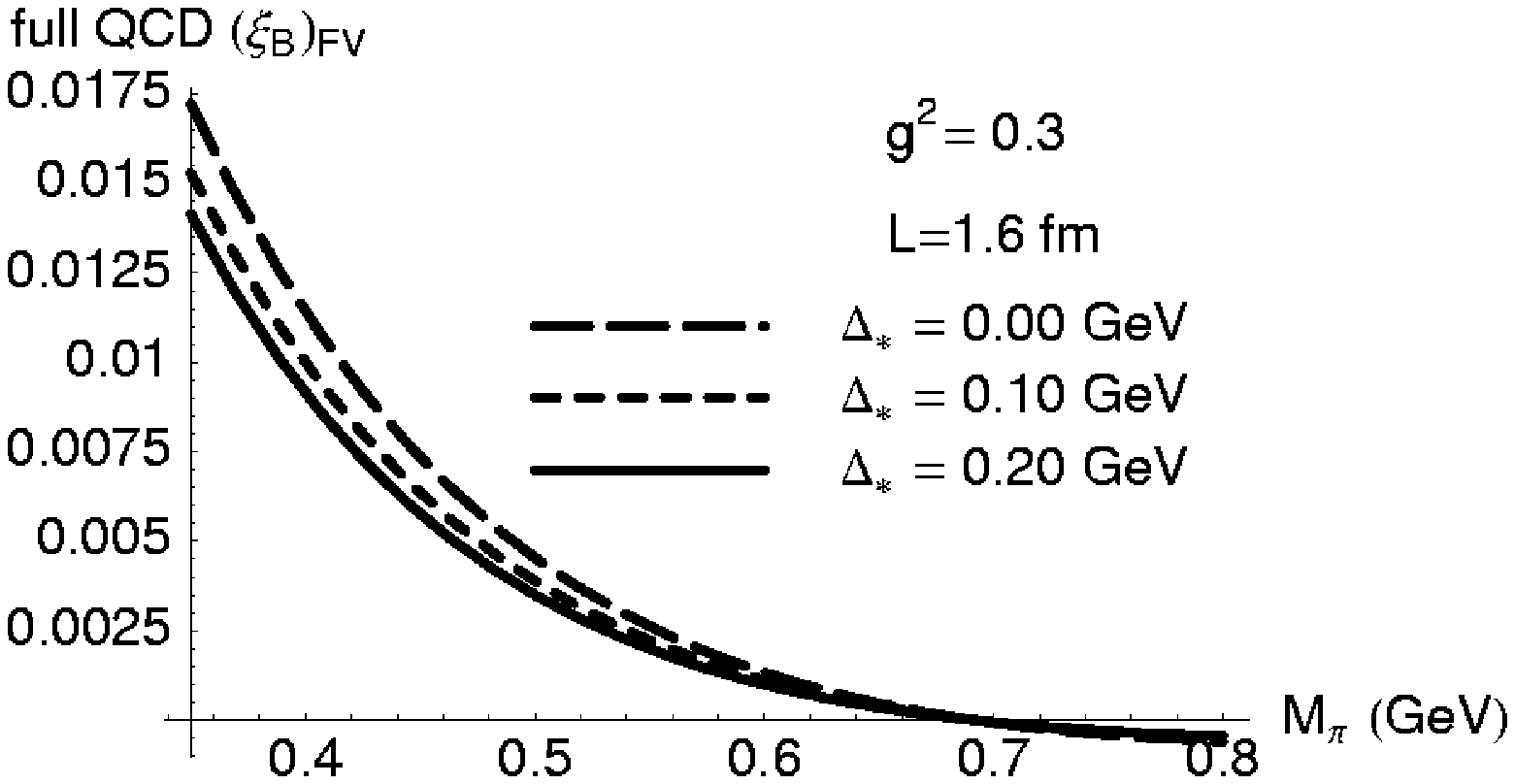}
  \includegraphics[width=0.49\textwidth]{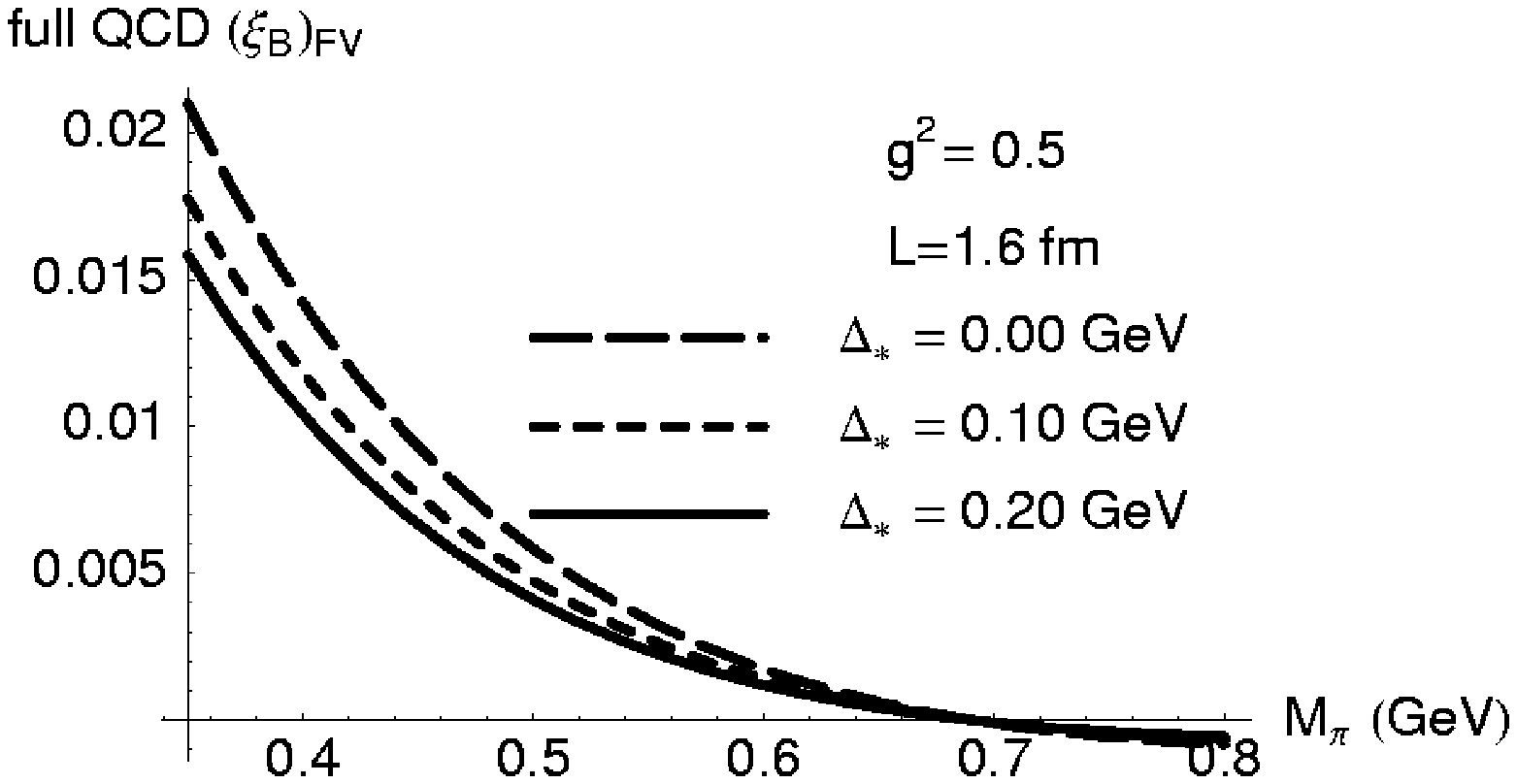}
  \caption[$(\xi_{B})_{\mathrm{FV}}$ in QCD
    plotted against
    $M_{\pi}$ with $L=1.6$~fm]{
    \label{fv-fig:BBs_over_BB}$(\xi_{B})_{\mathrm{FV}}$ in full QCD
    plotted against
    $M_{\pi}$ with $L=1.6$~fm.}
\end{figure}

\begin{figure}[tb]
  \centering
  \includegraphics[width=0.49\textwidth]{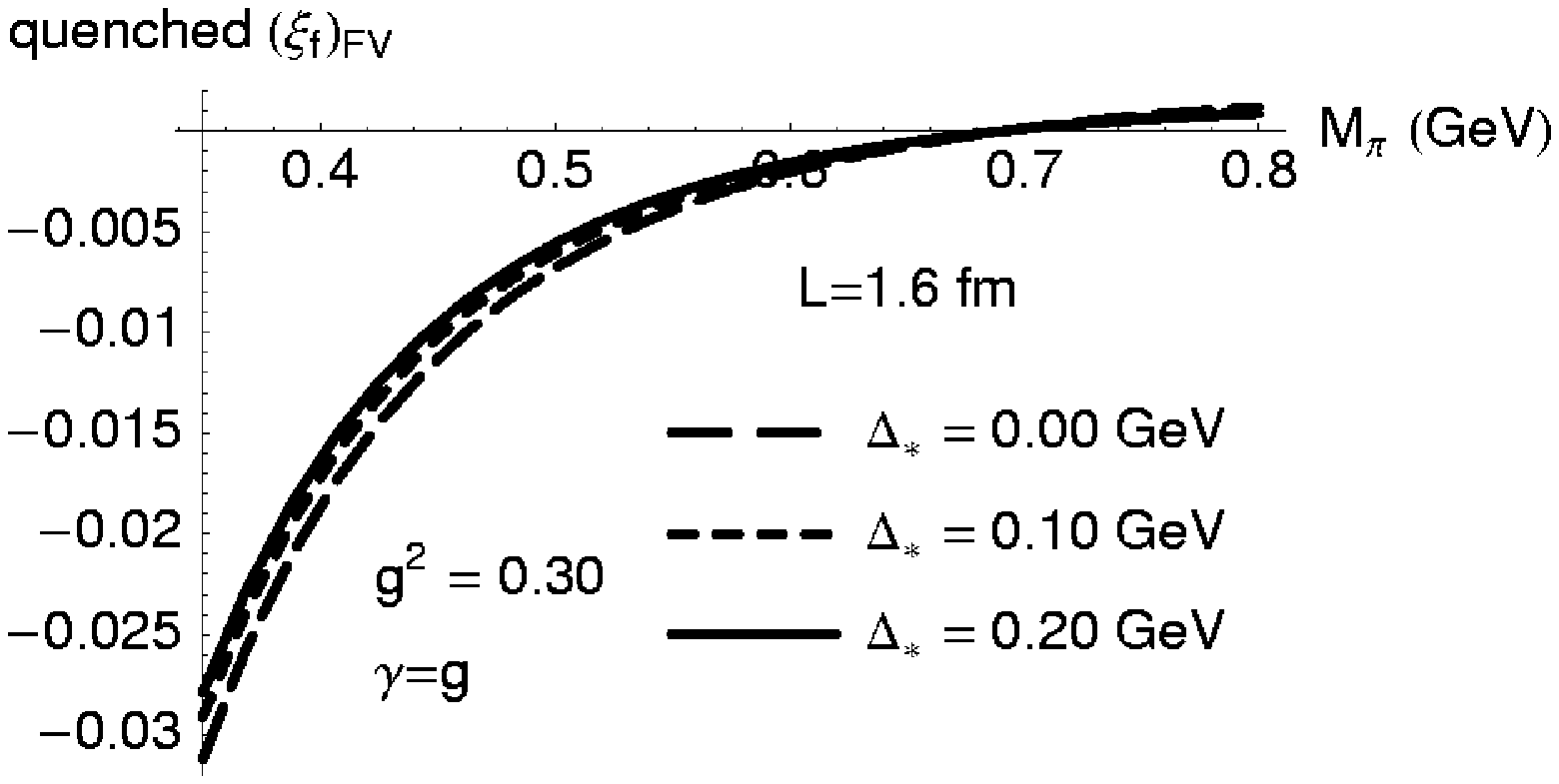}
  \includegraphics[width=0.49\textwidth]{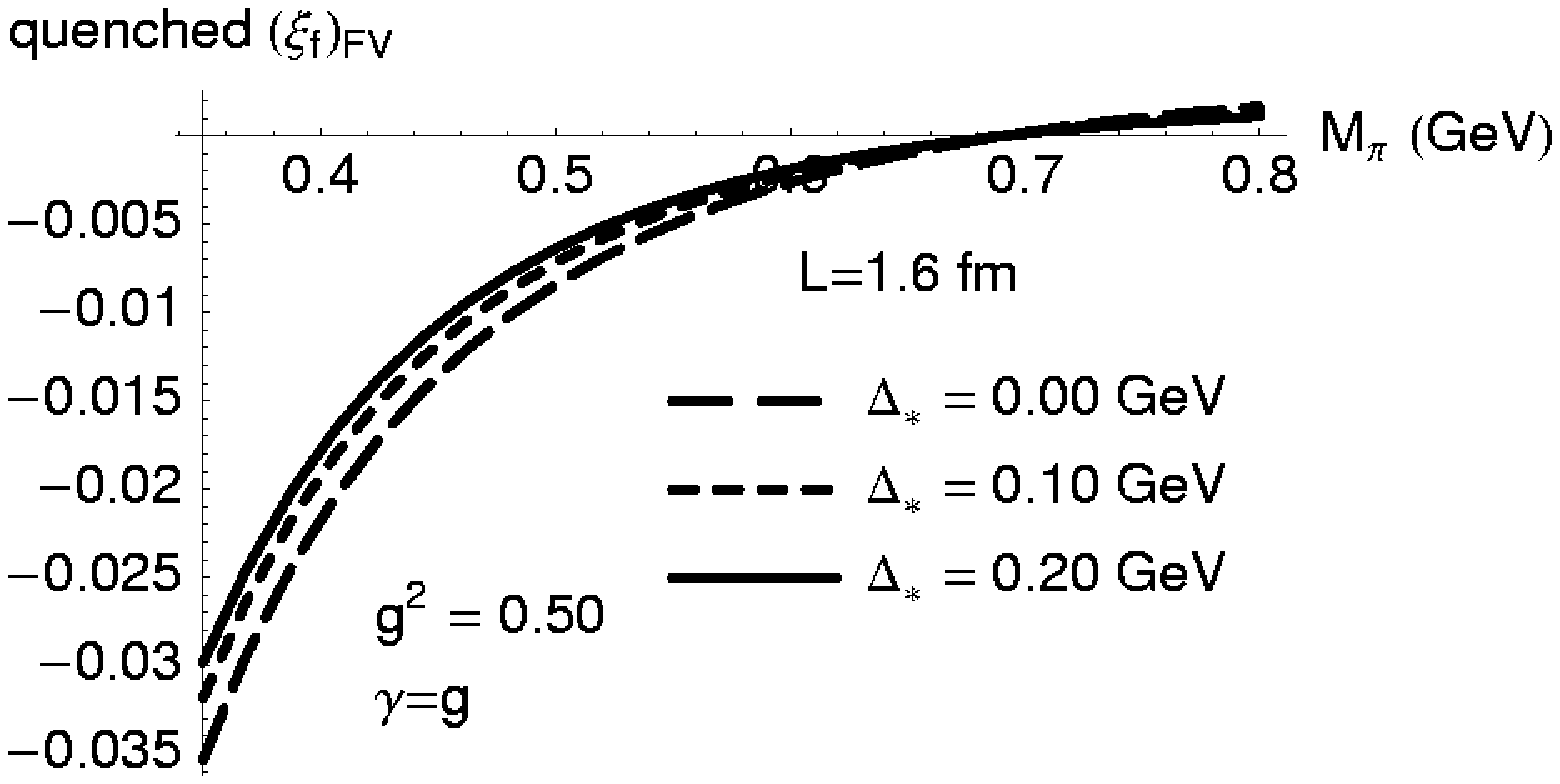}
  \includegraphics[width=0.49\textwidth]{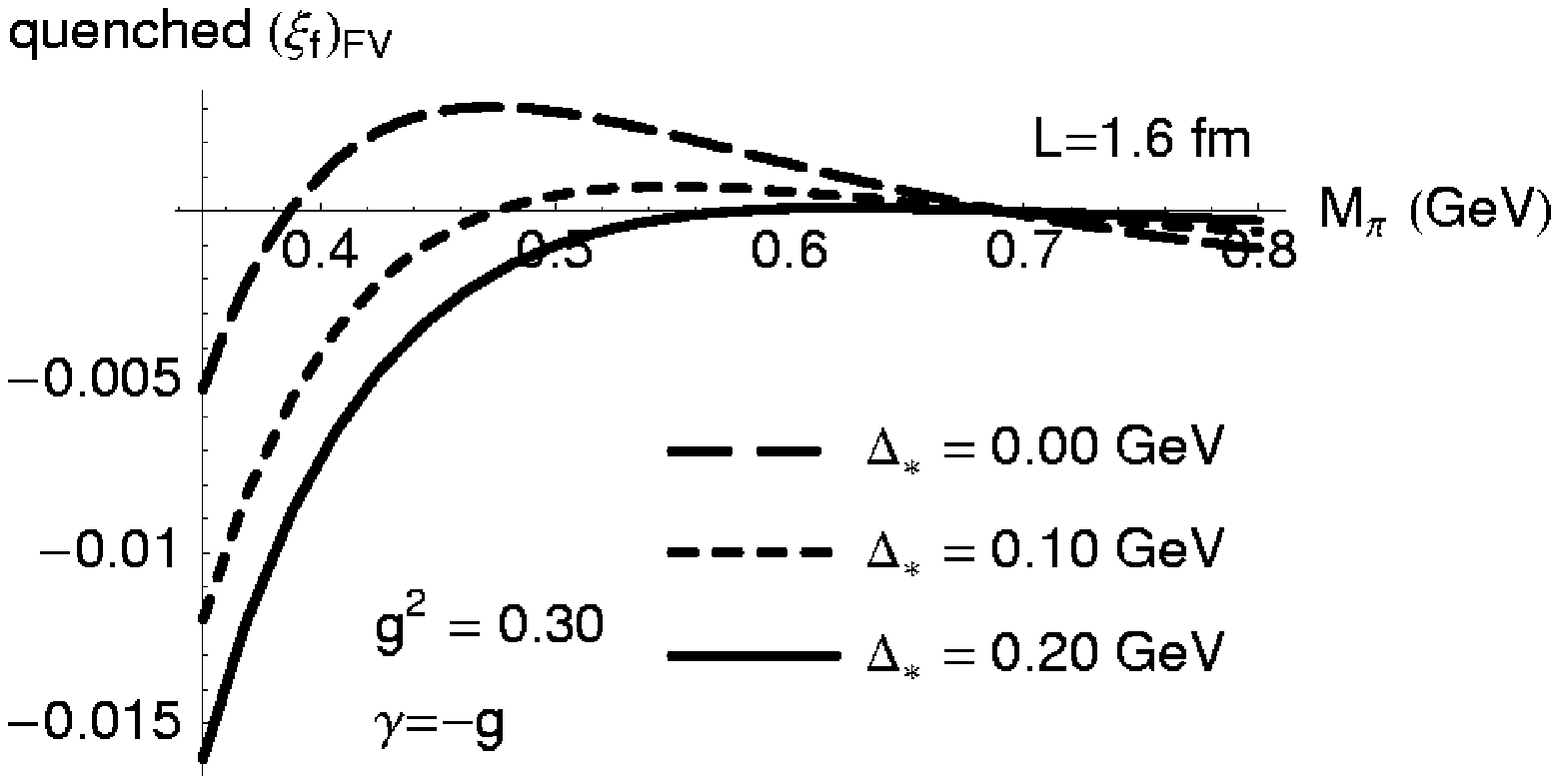}
  \includegraphics[width=0.49\textwidth]{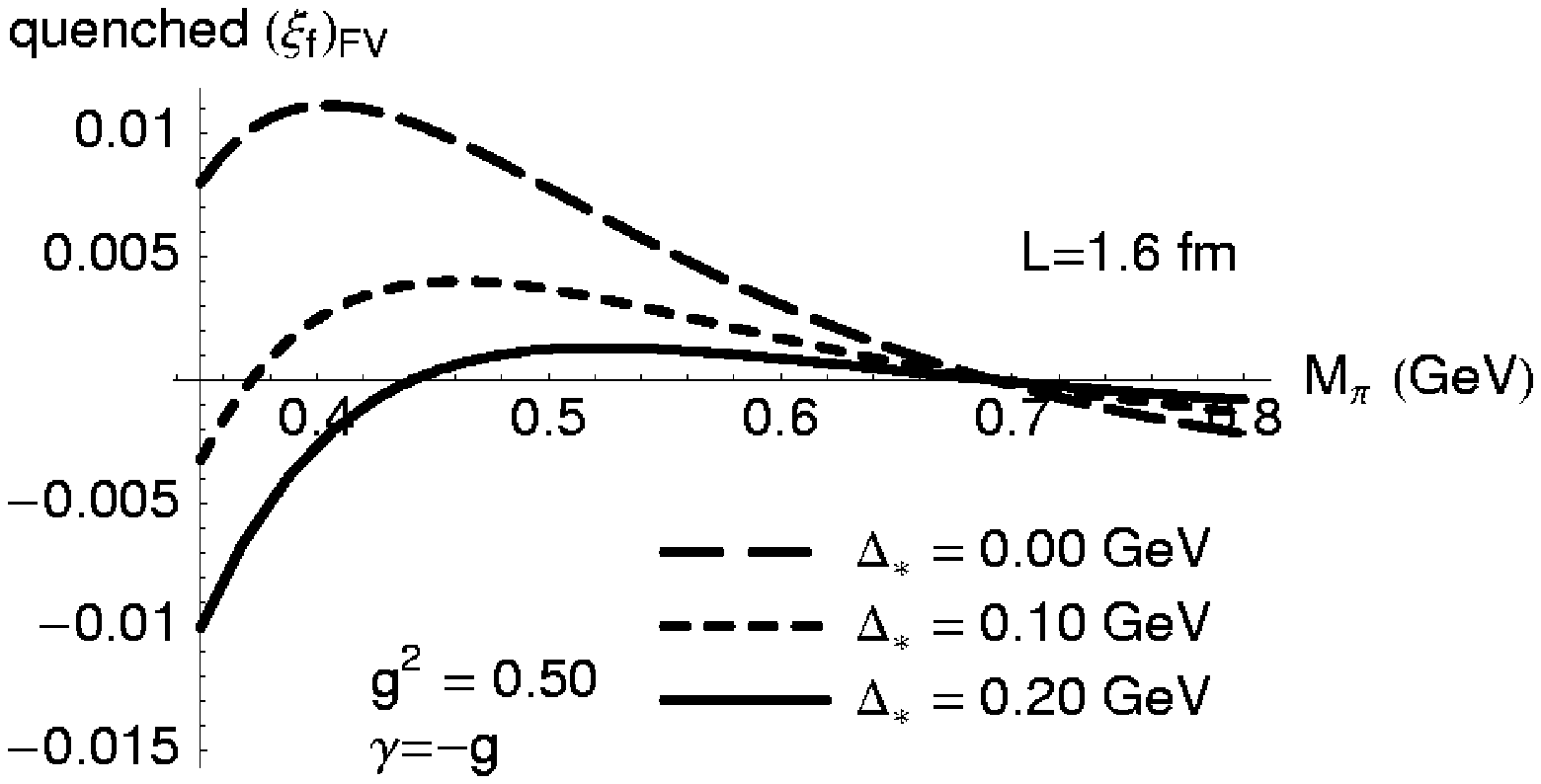}
  \caption[$(\xi_{f})_{\mathrm{FV}}$ 
    in QQCD plotted against
    $M_{\pi}$ with $L=1.6$~fm]{
    \label{fv-fig:Q_fBs_over_fB}$(\xi_{f})_{\mathrm{FV}}$ 
    in QQCD plotted against
    $M_{\pi}$, with $L=1.6$~fm and different couplings 
    $g$ and $\gamma$.  
    We set $\alpha=0$ and $M_{0}=700$~MeV.}
\end{figure}

\begin{figure}[tb]
  \centering
  \includegraphics[width=0.49\textwidth]{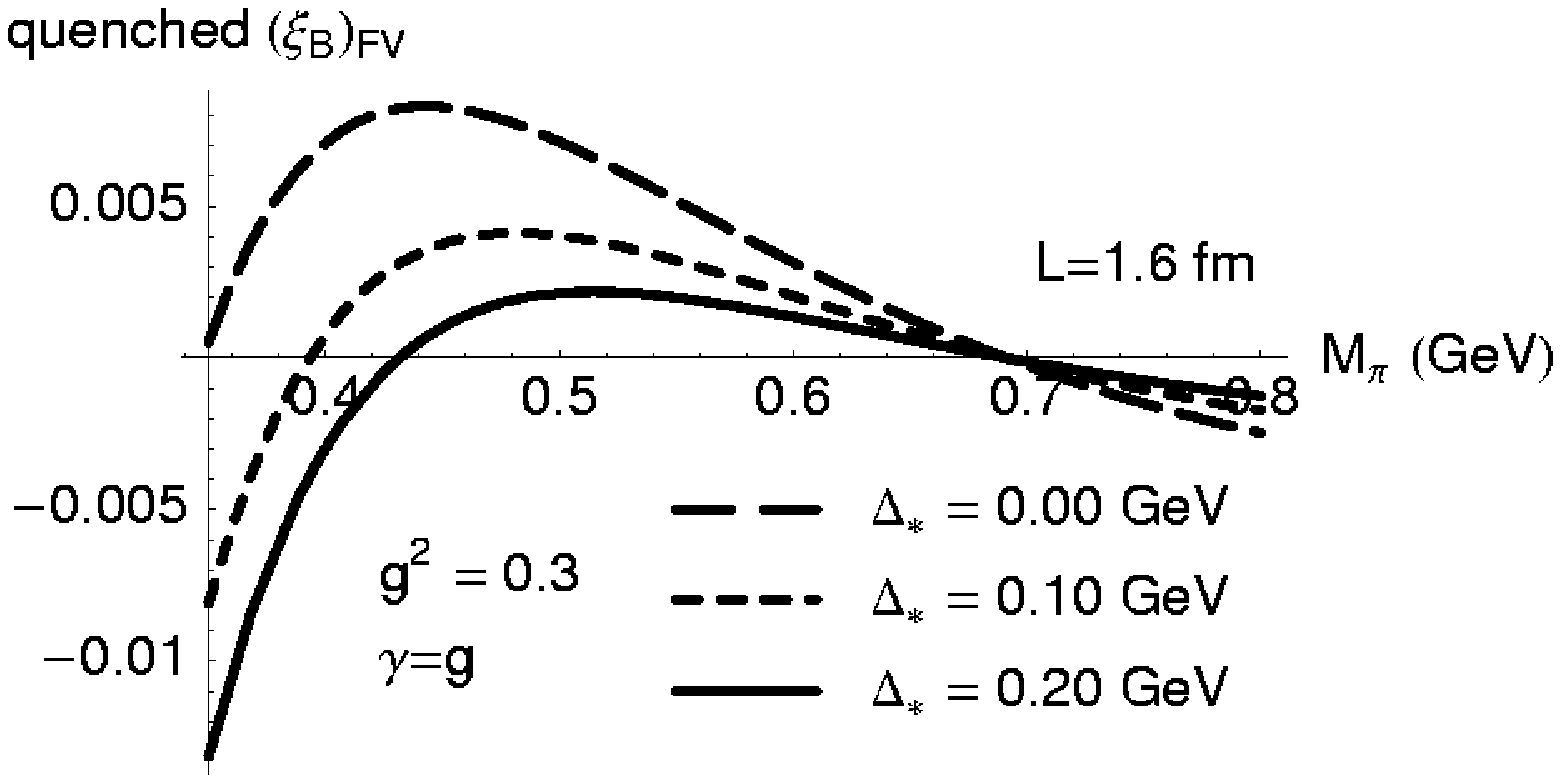}
  \includegraphics[width=0.49\textwidth]{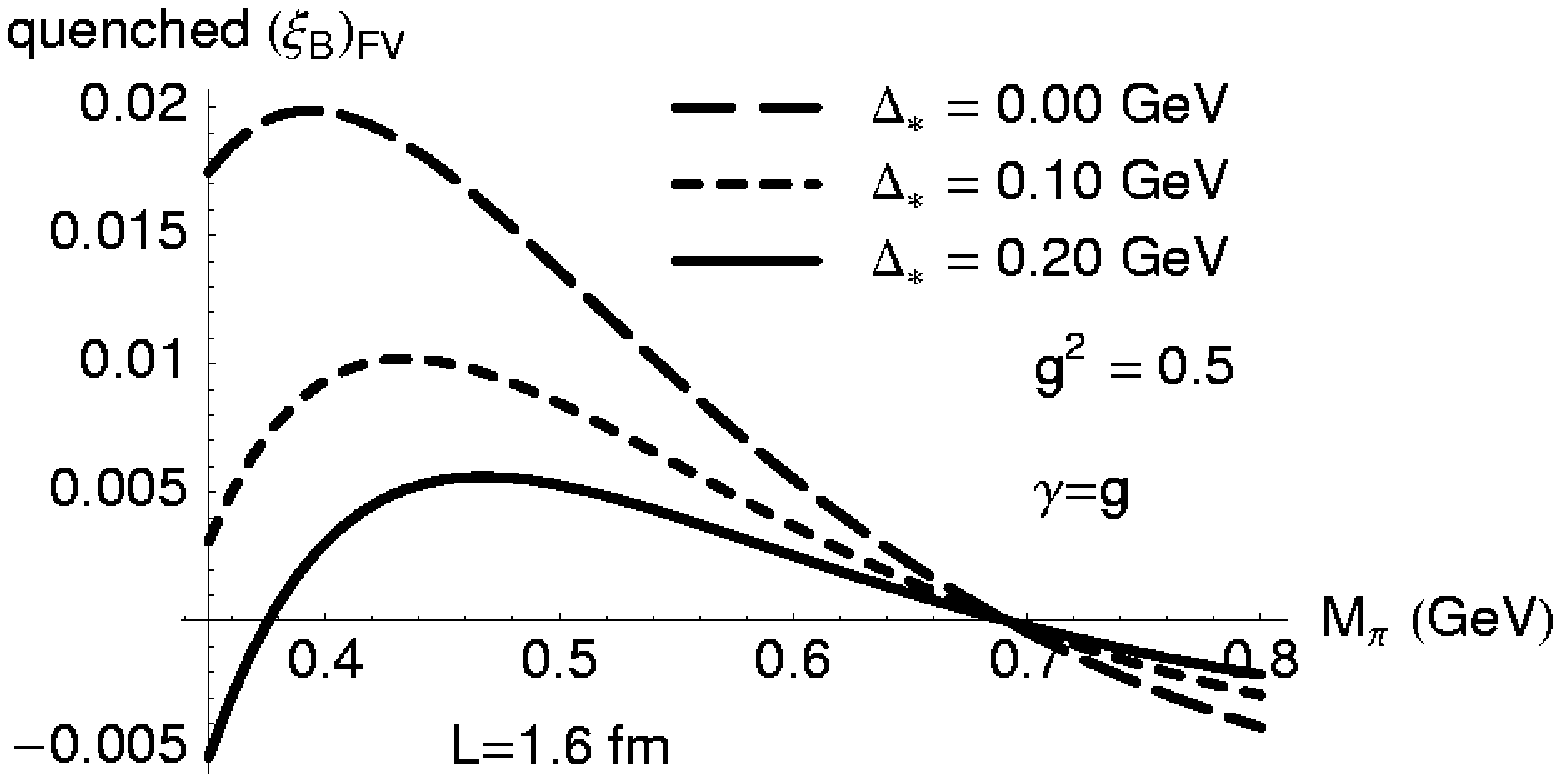}
  \includegraphics[width=0.49\textwidth]{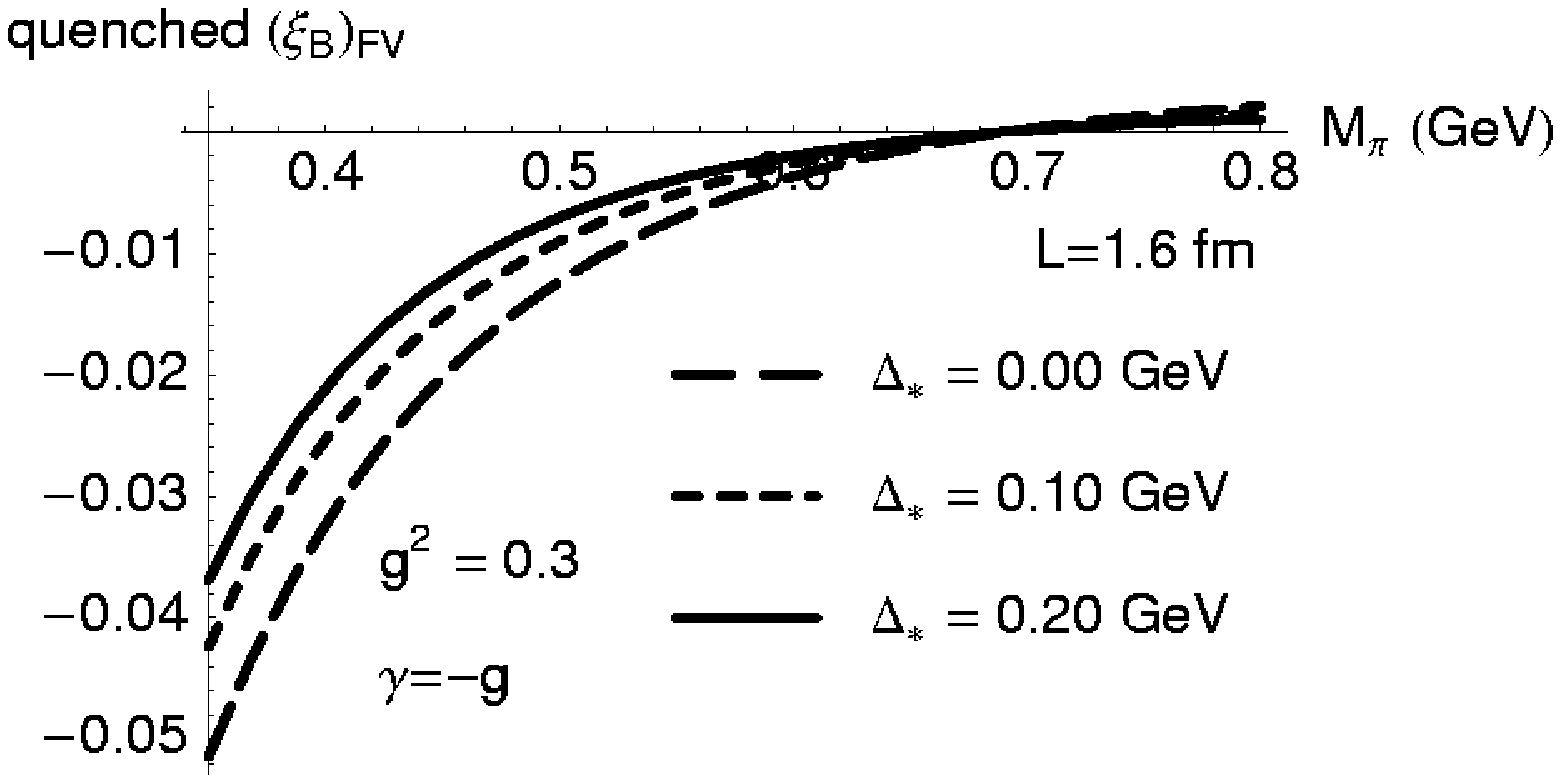}
  \includegraphics[width=0.49\textwidth]{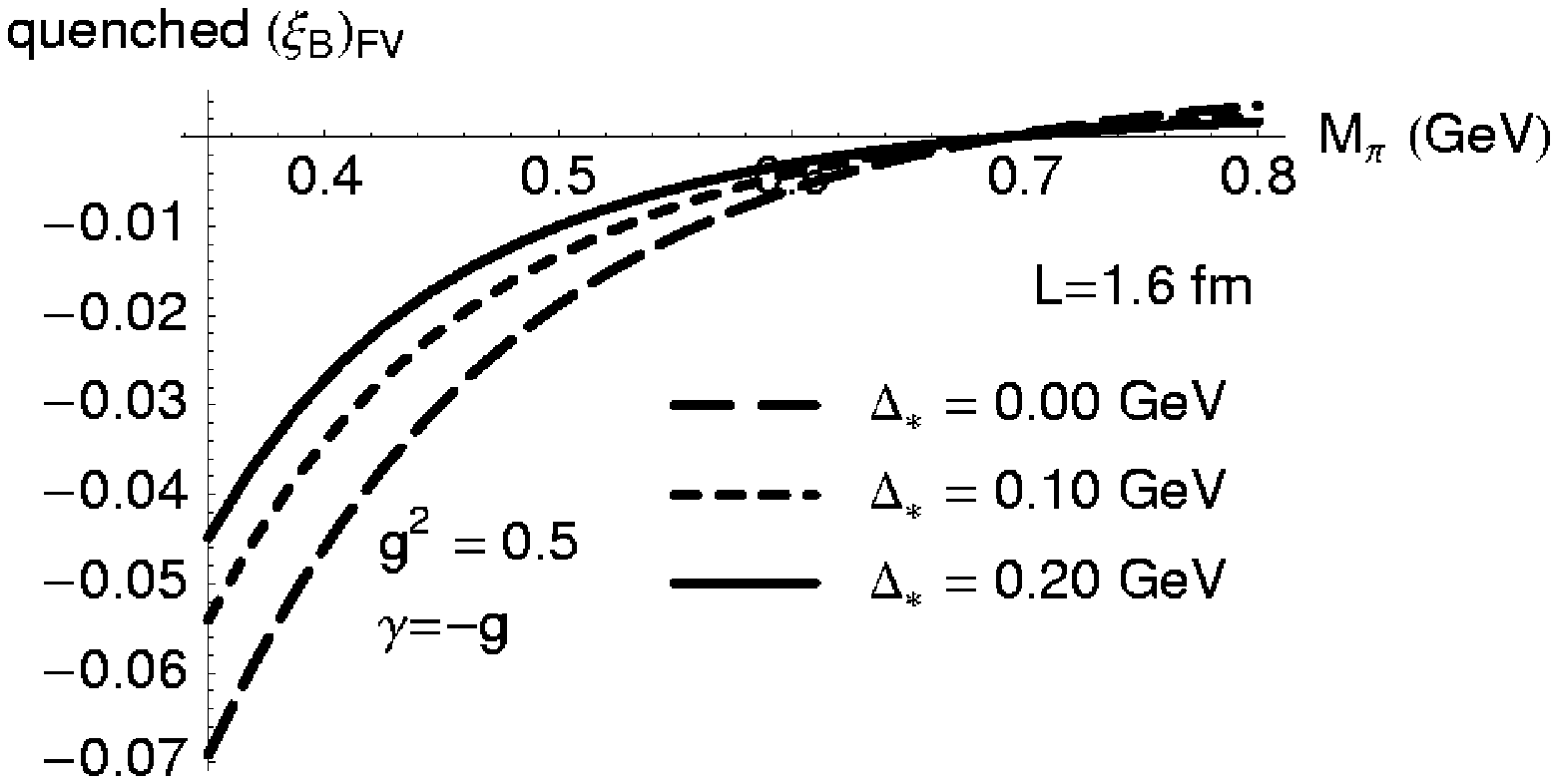}
  \caption[$(\xi_{B})_{\mathrm{FV}}$ 
    in QQCD plotted against
    $M_{\pi}$ with $L=1.6$~fm]{
    \label{fv-fig:Q_BBs_over_BB}$(\xi_{B})_{\mathrm{FV}}$ 
    in QQCD plotted against
    $M_{\pi}$, with $L=1.6$~fm,  $\alpha=0$ and $M_{0}=700$~MeV,
    and different couplings.}
\end{figure}

The results for $(\xi_{f})_{\mathrm{FV}}$ and 
$(\xi_{B})_{\mathrm{FV}}$ for QCD and QQCD from this analysis
are presented in 
Figs.~\ref{fv-fig:fBs_over_fB}--\ref{fv-fig:Q_BBs_over_BB}
%%%%%%%%  full QCD plots %%%%%%%%%%%%
%
%%%%%%%%%%% end of full QCD plots %%%%%%%%%
%%%%%%%%% quenched plots %%%%%%%%%%%%%%
%
%
%%%%%%%% end of quenched plots %%%%%%%%%%%%%%%%%%
with two different values 
for the coupling $g$ (and also $\gamma$ in QQCD).  
Here we stress again that the influence on finite volume effects from
the presence of $\Delta$ and $\delta_{s}$ depends on the size of 
these couplings, which are not well determined.  Inspired by the recent
CLEO measurement of $g$ in the charm system%
~\cite{Ahmed:2001xc,Anastassov:2001cw}, and a recent
lattice calculation~\cite{Abada:2003un}, we vary $g^{2}$ between 0.3
and 0.5. As for the coupling $\gamma$, which is a quenching artifact and 
has never been determined, we
vary its value between $g$ and $-g$.
It is clear from these plots that the finite volume
effects are generally small in QCD ($\le 2\%$), 
but can be significant in QQCD ($\sim 3\%$ to $\sim 7\%$ for $\xi_{B}$) 
in the range of
$M_{\pi} L$ where lattice simulations are normally performed.  
This is clearly
due to the enhanced long-distance effect arising from the ``double pole''
structure in (P)QQCD, as first pointed out in Ref.~\cite{Bernard:1996ez},
and manifests itself in various places, {\it e.g.}, nucleon-nucleon
potentials~\cite{Beane:2002vq} and $\pi{-}\pi$ scattering%
~\cite{Bernard:1996ez, Golterman:1999hv, Lin:2003tn, Lin:2002aj}.

Although it has been well established that 
infinite volume chiral corrections are smaller in the $B$ parameters
than in the decay constants due to the coefficient in front of
$g^{2}$ in the one-loop results, it is clear from these plots that
finite volume effects are more salient in $\xi_{B}$ than in $\xi_{f}$.
All the quenched lattice calculations for $\xi_{B}$ have so far concluded that
this quantity 
is consistent with unity with typically $3\%$ error.
However, we find that the volume effects
are already at the level of $3{-}4\%$ when $M_{\pi}=0.45$~GeV in
a 1.6~fm box where
many quenched simulations were carried out.  
This error depends on both light and heavy quark masses in the simulation, 
hence is amplified after 
extrapolating the result to the physical quark masses.  
Also, the fact
that volume effects tend towards different directions in QCD and QQCD
when $M_{\pi}$ becomes
smaller indicates
that quenching errors in these quantities can be larger than those
estimated in Ref.~\cite{Sharpe:1996qp}.
Since finite volume effects have not been included in the analysis of 
lattice calculations of $\xi_{B}$ hitherto, one should be cautious 
when using the existing quenched results for this quantity in 
any phenomenological work.

For the analysis in PQQCD, we assume that both the valence and sea
strange quark masses are
fixed at that of the physical strange quark.  
However, we vary the light sea quark mass $m_j=m_l$.  For this
purpose, we define
$m_{jj}$ to be the mass of the meson composed of two light sea
quarks.  Therefore,
\begin{equation}
  \frac{m^{2}_{jj}}{(m_{ss}^2)_{\mathrm{phys}}} 
  = 
  \left(\frac{m_j}{m_r}\right )_{m_r\,=\,{\text{physical }}m_s}.
\end{equation}
Also, we can express the mass shifts 
$\tilde{\delta}_{s}$ and 
$\delta_{\mathrm{sea}}$ in terms of meson masses,
$\tilde{\delta}_{s} = \lambda_{1} \left (m_{ss}^{2}-m_{rr}^{2} \right )$
and
$\delta_{\mathrm{sea}} = \lambda_{1} \left (m^{2}_{rr}-M^{2}_{\pi} \right )$
by using Eqs.~(\ref{fv-eq:tilde_delta_s_def}) and (\ref{fv-eq:delta_sea_def})
with the same value for $\lambda_{1}$ used in Eq.~(\ref{fv-eq:lambda_value}).

\begin{figure}[tb]
  \centering
  \includegraphics[width=0.49\textwidth]{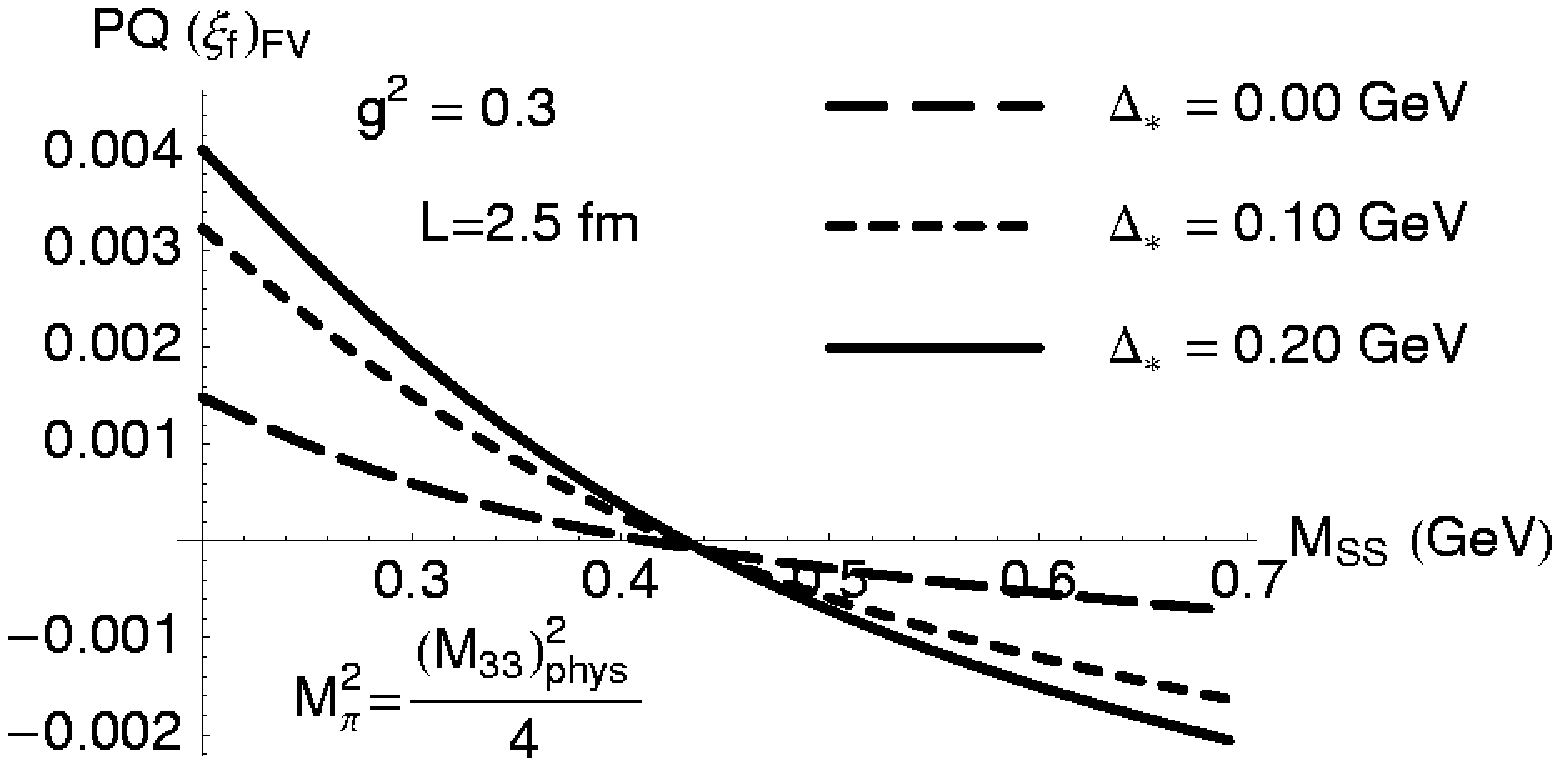}
  \includegraphics[width=0.49\textwidth]{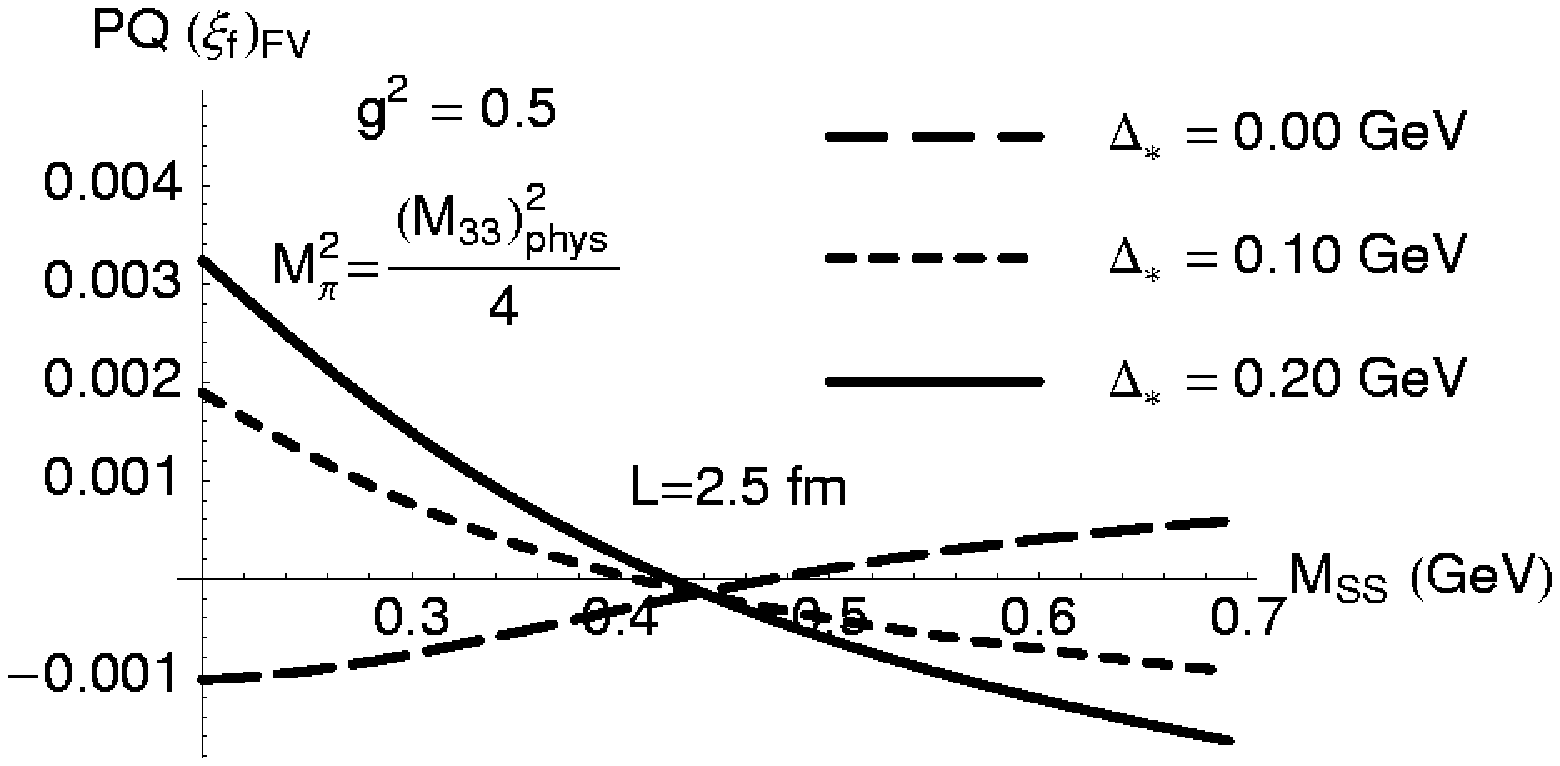}
  \includegraphics[width=0.49\textwidth]{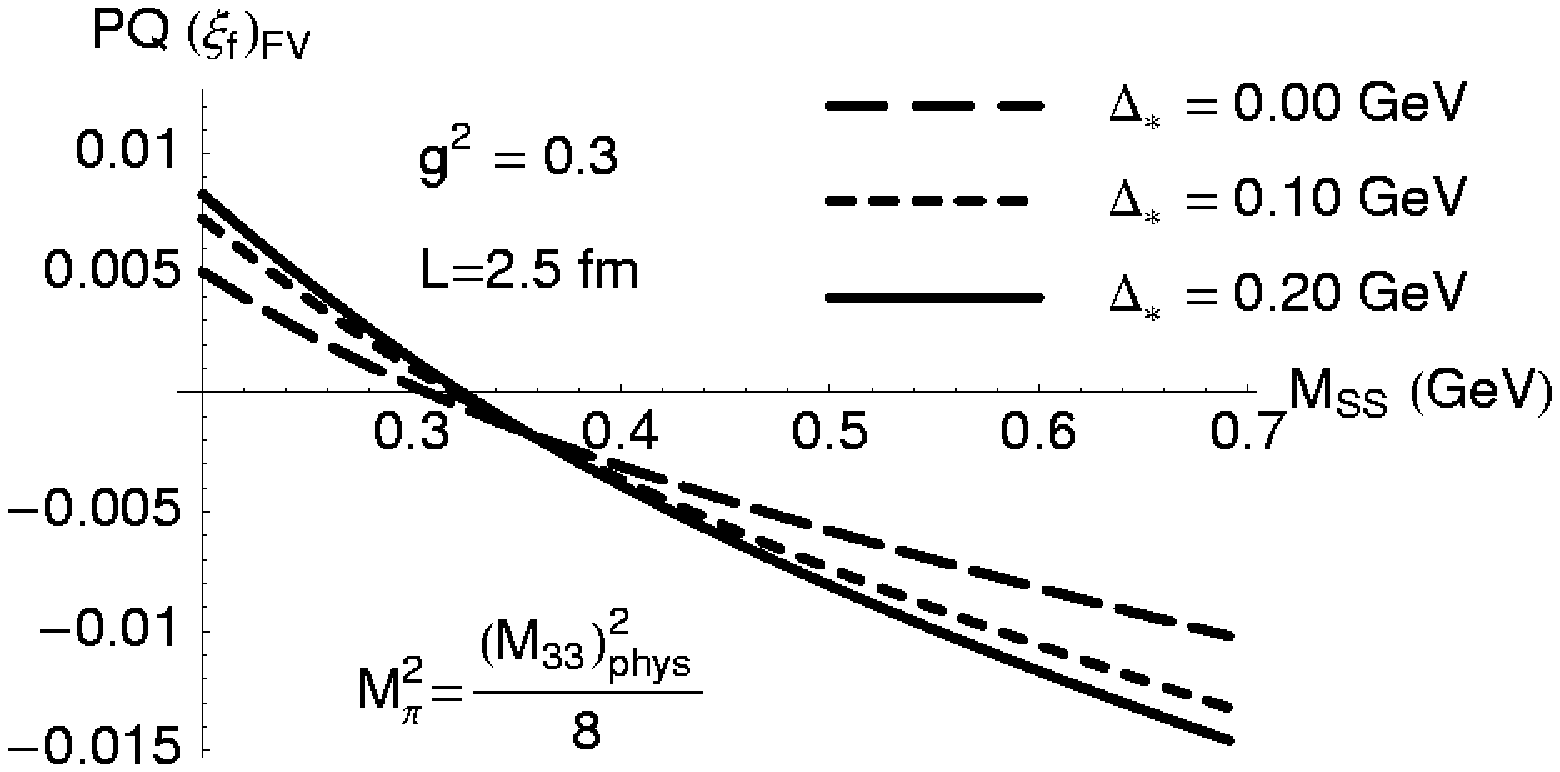}
  \includegraphics[width=0.49\textwidth]{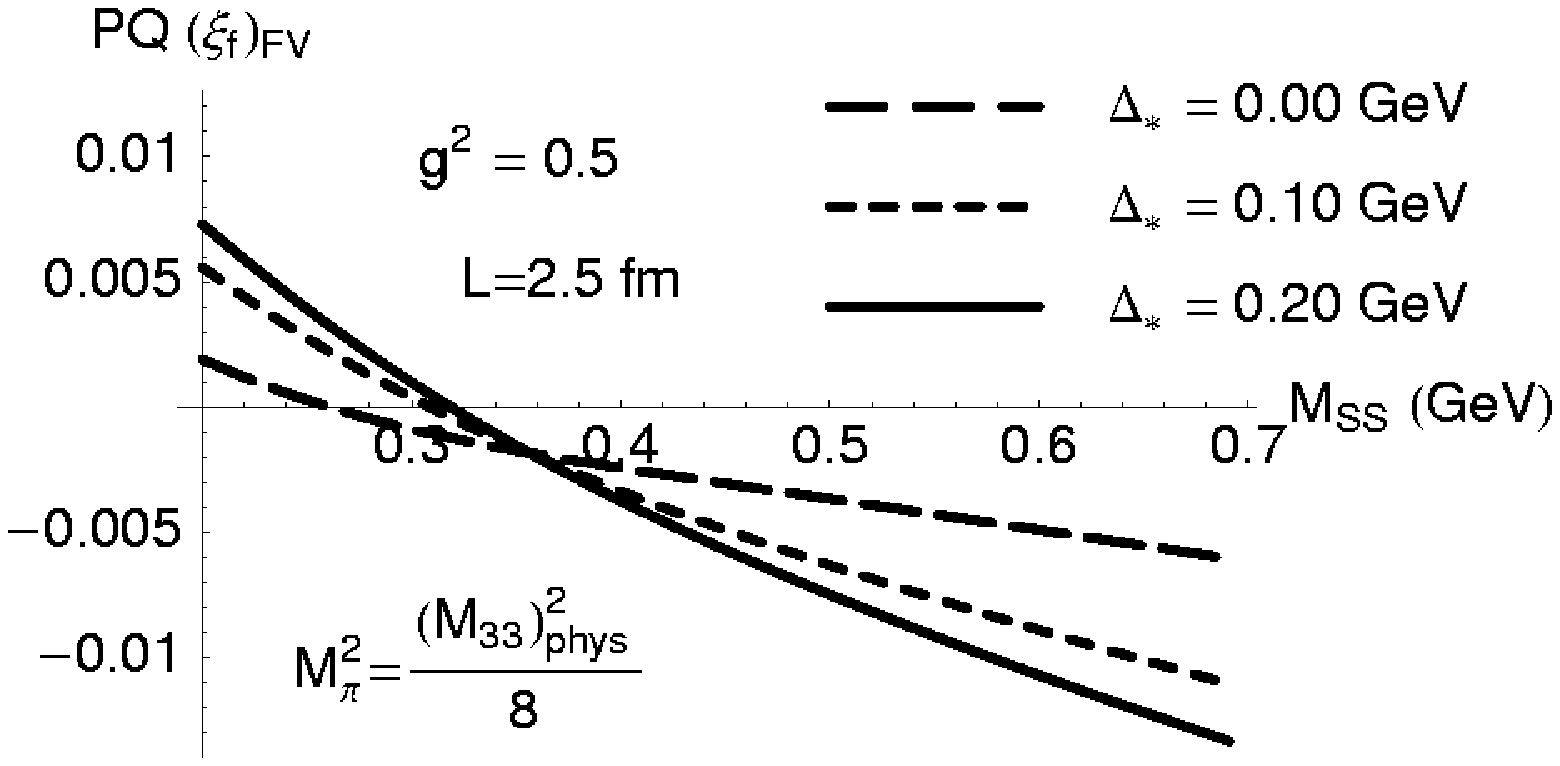}
  \caption[$(\xi_{f})_{\mathrm{FV}}$ 
    in PQQCD plotted against
    $m_{rr}$ with $L=2.5$~fm]{
    \label{fv-fig:PQ_fBs_over_fB}$(\xi_{f})_{\mathrm{FV}}$ 
    in PQQCD plotted against
    $m_{rr}$, with $L=2.5$~fm
    and two different values for $M_{\pi}$. 
    The pion mass $M^{2}_{\pi}=m^{2}_{ss}/4$ corresponds to 
    $M_{\pi} L=4.4$ and $M^{2}_{\pi}=m^{2}_{ss}/8$ corresponds to 
    $M_{\pi} L=3.1$.
    The mass $m_{rr} = 0.197$~GeV corresponds
    to $m_{rr} L = 2.5$, and $m_{rr} = 0.32$~GeV corresponds
    to $m_{rr} L = 4$.}
\end{figure}

\begin{figure}[tb]
  \centering
  \includegraphics[width=0.49\textwidth]{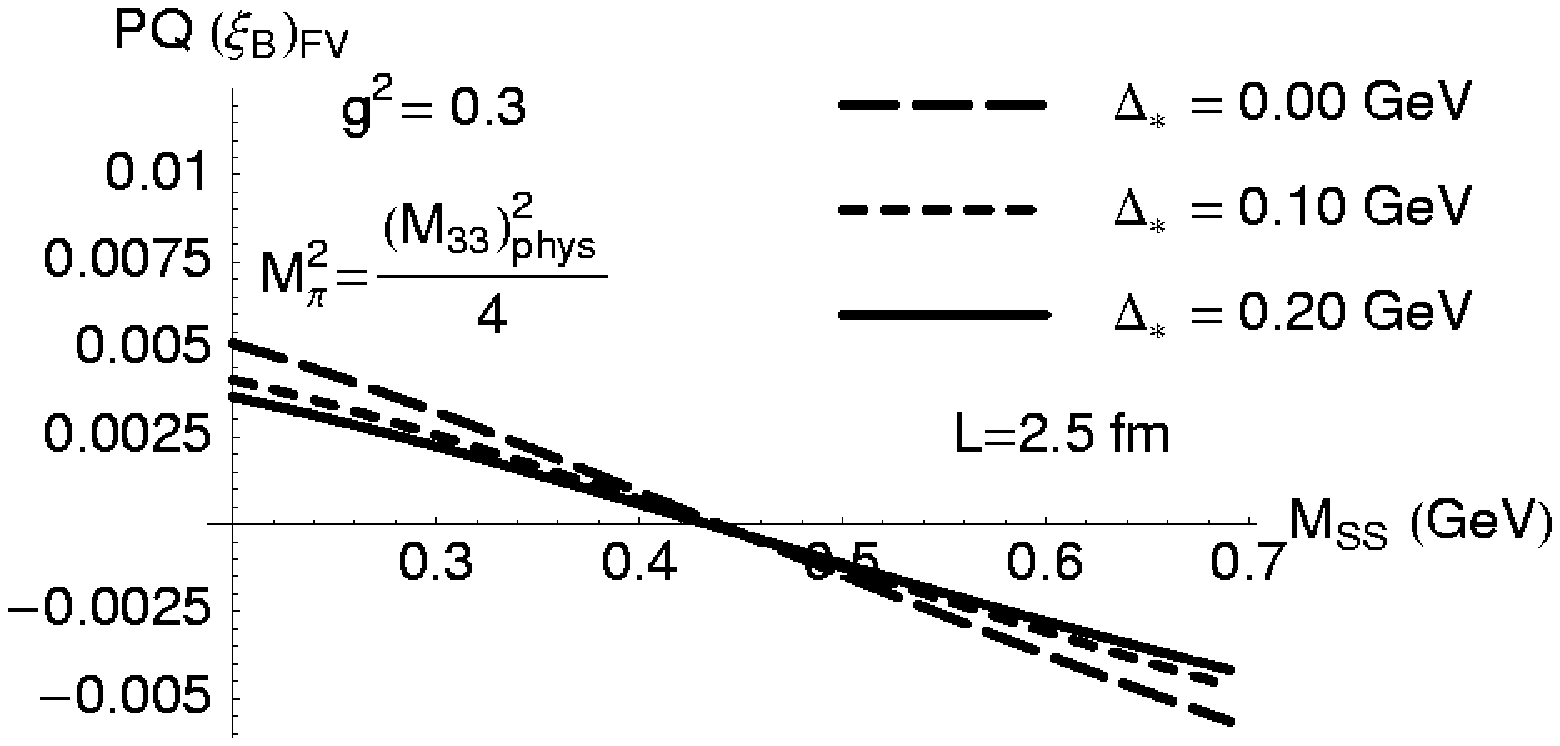}
  \includegraphics[width=0.49\textwidth]{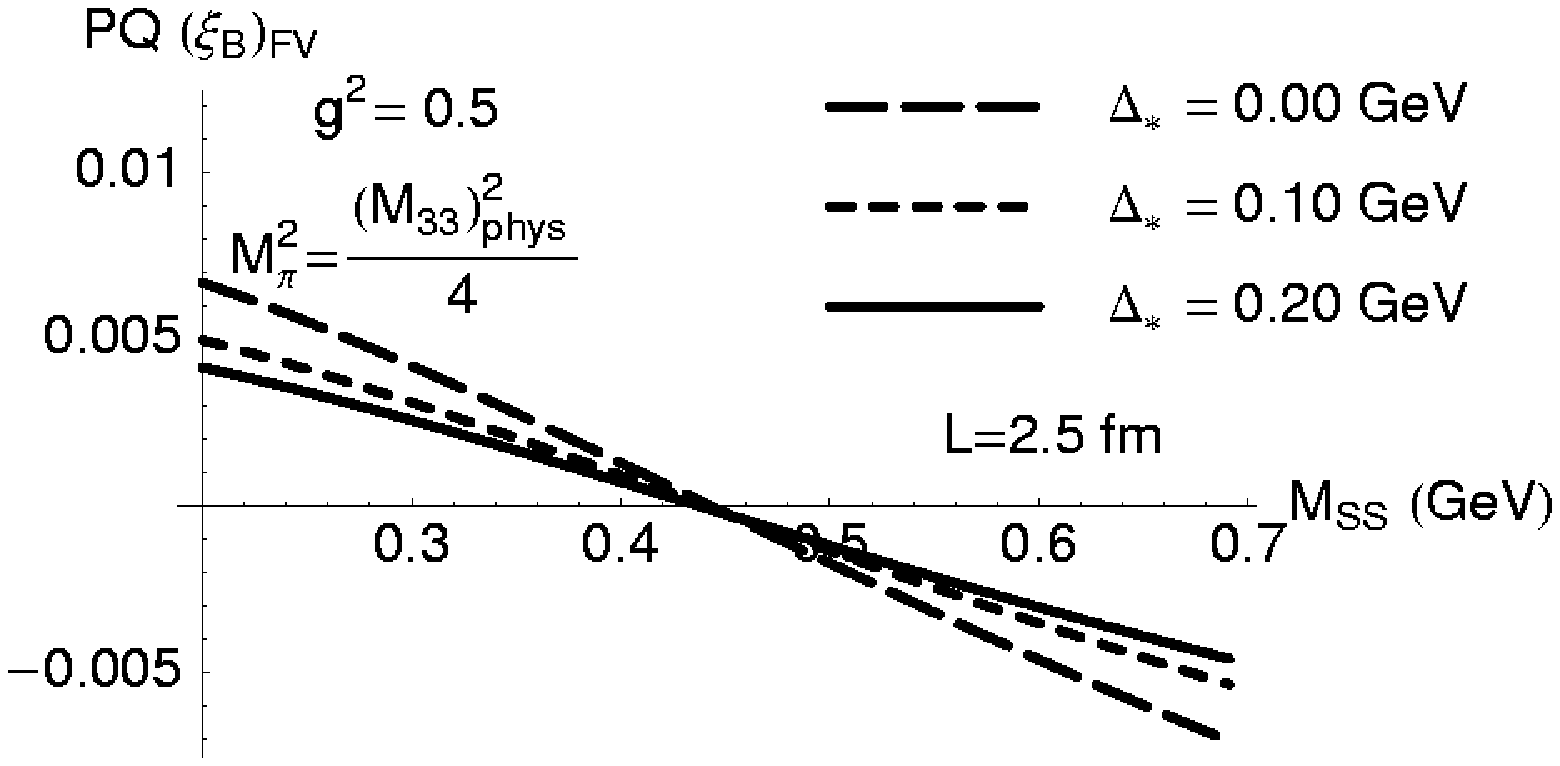}
  \includegraphics[width=0.49\textwidth]{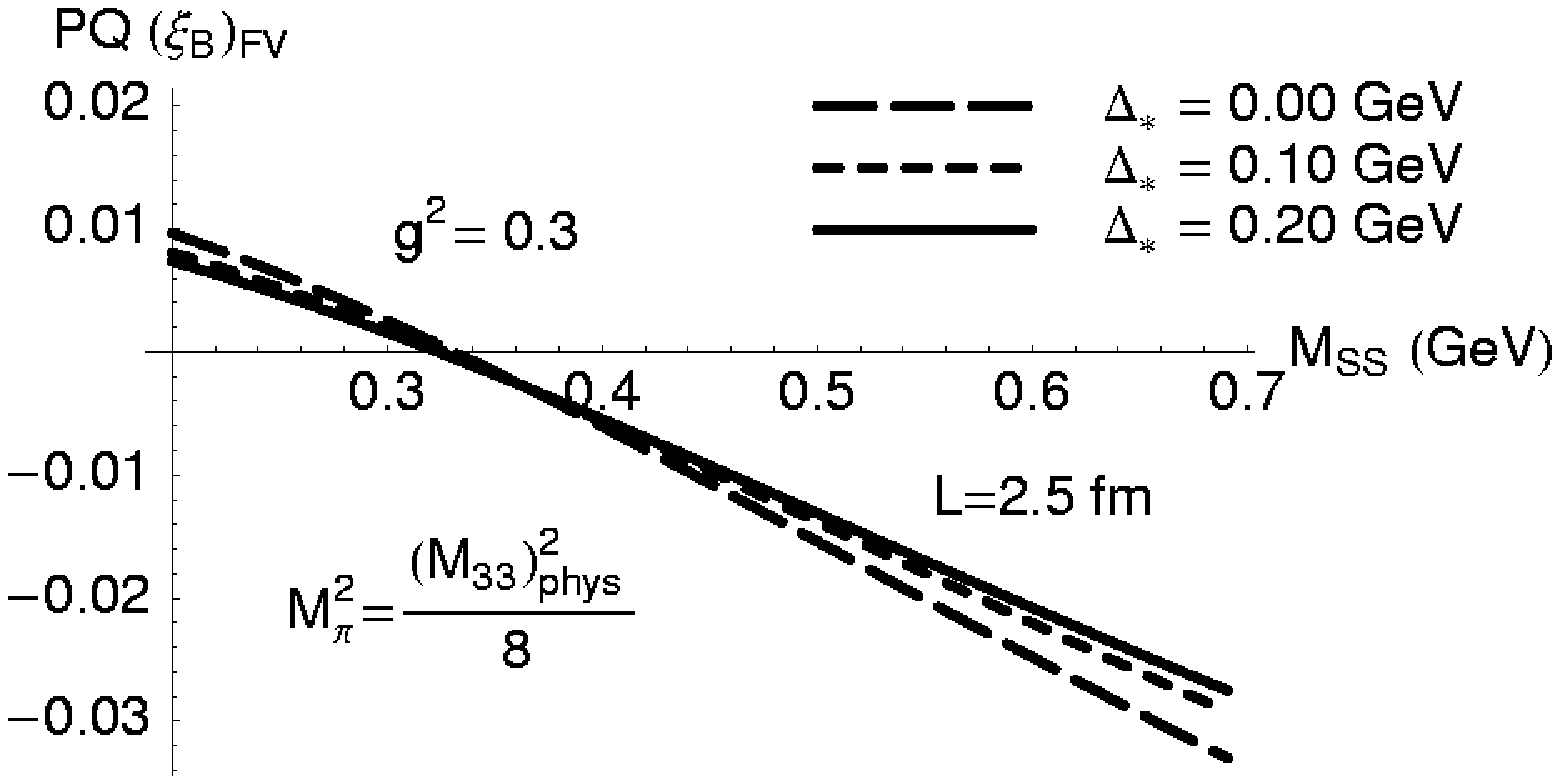}
  \includegraphics[width=0.49\textwidth]{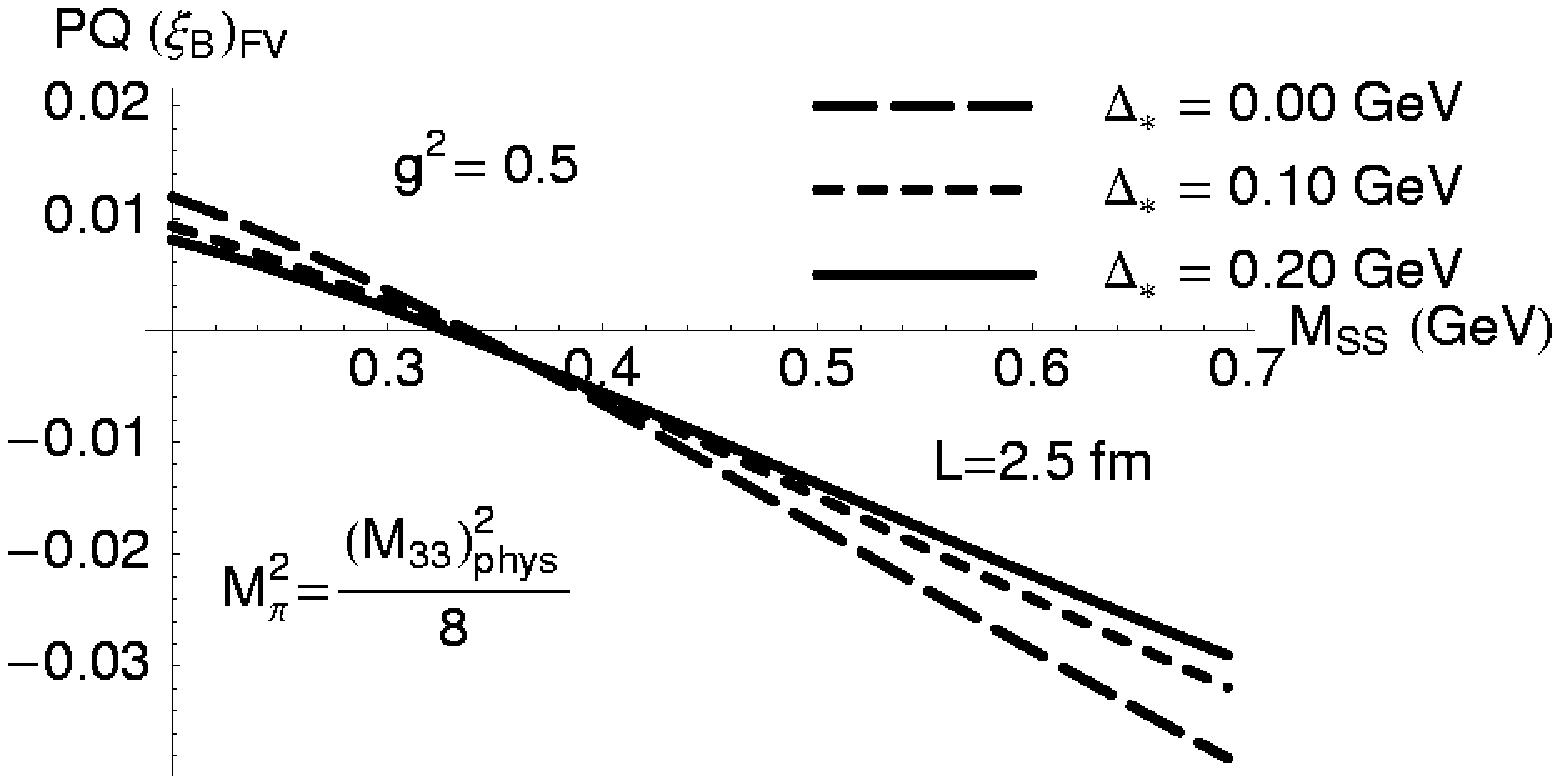}
  \caption[$(\xi_{B})_{\mathrm{FV}}$ 
    in PQQCD plotted against
    $m_{rr}$ with $L=2.5$~fm]{
    \label{fv-fig:PQ_BBs_over_BB}$(\xi_{B})_{\mathrm{FV}}$ 
    in PQQCD plotted against
    $m_{rr}$ with $L=2.5$~fm
    and two different values for $M_{\pi}$.}
\end{figure}

The results for the PQQCD analysis are presented in 
Figs.~\ref{fv-fig:PQ_fBs_over_fB} and \ref{fv-fig:PQ_BBs_over_BB}.
%%%%%%%% PQ plots %%%%%%%%%%%%%%%%%
%
%
%%%%%%%% end of PQ plots %%%%%%%%%%%%%%%%
The double pole
insertions also appear in PQQCD and it is clear
from these plots that finite volume effects cannot be neglected 
if one hopes to determine $\xi$ to the level of a few percent.  Especially,
in the range of $M_{\pi}$ and $L$ where current and future 
lattice simulations are performed~\cite{Davies:2003ik},  
they can already be at about 
$4\%$, and the dependence on the heavy meson mass is quite strong.
Therefore they can become comparable
to the error 
presented in the latest review~\cite{Kronfeld:2003sd},
$\xi = 1.23 \pm 0.10$
after quark mass extrapolations.%
\footnote{Finite volume effects
presented in this work are, however, correlated with the errors arising
from chiral extrapolations.}

\section{Conclusions}\label{fv-sec:conclusion}
We have investigated finite volume effects in heavy quark systems in the
framework of \HMCPT.  The primary
conclusion
is that the scales $\D$ and $\delta_{s}$, which are 
heavy-light meson mass
splittings arising from the breaking of heavy quark spin and light flavour
$SU(3)$ symmetries, 
can significantly reduce the
volume effects in diagrams involving heavy meson propagators in the loop.  
The physical picture of this phenomenon is that some heavy-light mesons
are off-shell in the effective theory, as a consequence of the velocity
superselection rule, with the virtuality characterized by the mass splittings.
The time uncertainty conjugate to this virtuality limits the 
period during which the Goldstone mesons can propagate to the
boundary.  Finite volume effects caused by the propagation of the
Goldstone mesons naturally affect the light quark mass 
extrapolation/interpolation in a lattice calculation.  On top of this,
our work implies that they also influence the heavy quark
mass extrapolation/interpolation, since the scale $\D$ varies
significantly with the heavy meson mass.  The strength of this influence is
process-dependent, determined also by the relative weight between 
diagrams with and without heavy meson propagators in the loop.
The volume effects can
be amplified by both heavy and light quark mass extrapolations.
Therefore it is important
to perform calculations to identify these effects in phenomenologically 
interesting quantities.

We have presented an 
explicit calculation in finite volume HM$\chi$PT for the $B$ parameters in
neutral $B$ meson mixing and heavy-light decay constants, in QCD, QQCD,
and PQQCD. 
We have used these results to estimate the impact
of finite volume effects in the $SU(3)$ ratio $\xi$, which is an important
input in determining the magnitude of the CKM matrix element $V_{td}$.
Within the parameter space where most quenched lattice calculations have been
performed, we find that, although this impact is quite small
($\le \sim 2\%$) in QCD, it can be significant in QQCD.
This is due to the enhanced long-distance effects arising
from the double pole structure.  This error will be amplified by
the quark mass extrapolations and hence can exceed the currently quoted
systematic effects.  Furthermore, finite volume effects tend towards
different directions 
in QCD and QQCD for decreasing $M_{\pi}$.  
This means that quenching errors in $\xi$
may be significantly larger than what was estimated before.
Therefore one has to be cautious when
using the existing quenched lattice QCD results for $\xi$ in phenomenological
work.  In
PQQCD, our results indicate that finite volume effects are typically
between $3\%$ and $5\%$ in the data range 
of future high-precision simulations, and they can be significantly
amplified in the procedure of quark mass extrapolations.
This means that they are not negligible in 
future lattice calculations of $\xi$.

\chapter{Charge Radii of the Meson and Baryon Octets 
       in \QCPT\ and \PQCPT}
\label{chapter:baryon_ff}

In this chapter, 
we calculate the electric charge radii 
of the $SU(3)$ pseudoscalar mesons 
as well as the $SU(3)$ octet baryons 
in \QCPT\ and \PQCPT.
The results are needed for the 
extrapolation of future lattice calculations of
these observables.
We also derive expressions for the nucleon and pion charge radii 
in $SU(2)$ flavor
away from the isospin limit.

\section{Introduction}
The study of hadronic electromagnetic form factors at
low momentum transfer
provides important insight into the non-perturbative
structure of QCD.
Notable progress toward measuring the 
proton and neutron form factors
has been made in recent years
(see \cite{Mergell:1996bf,Hammer:1996kx} for references),
including recent high precision measurements for the proton~%
\cite{Gayou:2001qd}.
Experimental study of the remaining octet baryons, however, 
is much harder.
The charge radius of the $\S^-$ has only recently been measured%
~\cite{Eschrich:1998tx}.
Although more experimental 
data for the other
baryon electromagnetic
observables
can be expected in the
future, 
progress will be slow as the experimental difficulties are significant.
Theory, however, may have a chance to catch up.

While quenched lattice calculations have already appeared%
~\cite{Tang:2003jh,Gockeler:2003ay,vanderHeide:2003ip,
Wilcox:1992cq,Draper:1990pi,Leinweber:1991dv},
with the advance of lattice gauge theory, 
we expect partially quenched calculations for many of these
observables
in the near future.
Lattice simulations employing these approximations need to
be extrapolated from the heavier light quark masses
used on the lattice 
(currently on the order of the strange quark mass) down to
the physical light quark masses
using the appropriate low-energy theories, 
\QCPT\ and \PQCPT\
(see Sections~\ref{section:CPT-quenching} and 
\ref{section:CPT-extrapol}).

While there are a number of lattice calculations for observables such
as the pion form factor%
~\cite{Martinelli:1988bh,Draper:1989bp,vanderHeide:2003ip}
or the octet baryon magnetic moments%
~\cite{Leinweber:1998ej,Hackett-Jones:2000qk}
that use the quenched approximation,
there are currently no partially quenched simulations.
However, 
given the recent progress that lattice gauge theory has made
in the one-hadron sector and the prospect of
simulations  
in the two-hadron sector%
~\cite{LATTICEproposal1,LATTICEproposal2,
Beane:2002np,Beane:2002nu,Arndt:2003vx},
we expect to see partially quenched calculations of the
electromagnetic
form factors in the near future.

This chapter is organized as follows.
In Section~\ref{baryon_ff-sec:ff},
we calculate the charge radii of the 
meson and baryon octets in both \QCPT\ and \PQCPT\
up to next-to-leading (NLO) order in the chiral expansion.
We use the heavy baryon formalism of Jenkins and Manohar%
~\cite{Jenkins:1991jv,Jenkins:1991ne},
treat the decuplet baryons as dynamical degrees of freedom,
and keep contributions to lowest order in the heavy baryon mass, $M_B$.
These calculations are done in the 
isospin limit of $SU(3)$ flavor.  
For completeness we also provide the
\PQCPT\ 
result for the charge radii for the $SU(2)$ chiral
Lagrangian with non-degenerate quarks in 
Appendix~\ref{baryon_ff-sec:SU2}.
In Section~\ref{baryon_ff-sec:conclusions} we conclude.

\section{\label{baryon_ff-sec:ff}Charge Radii}
In this section we calculate the charge radii in
\PQCPT\ and \QCPT.
The basic conventions and notations for the
mesons and baryons in \QCPT\ and \PQCPT\ have been laid forth
in Chapter~\ref{chapter:CPT}; 
they have also been extensively reviewed in 
the literature%
~\cite{Morel:1987xk,Sharpe:1992ft,Bernard:1992ep,
Bernard:1992mk,Golterman:1994mk,Sharpe:1996qp,Labrenz:1996jy}.

\subsection{Octet Meson Charge Radii}
The electromagnetic form factor $G_{X}$ of an octet meson 
$\phi_X$
is required by
Lorentz invariance and gauge invariance to have the form
\begin{equation}\label{baryon_ff-eqn:mesonff}
  \langle\phi_{X}(p')|J^\mu_{\text{em}}|\phi_{X}(p)\rangle
  = 
  G_{X}(q^2)(p+p')^\mu
\end{equation}
where
$q^\mu=(p'-p)^\mu$ 
and $p$ ($p'$) is the momentum of the incoming (outgoing) meson.
Conservation of electric charge protects it from
renormalization,
hence at
zero momentum transfer
$e G_X(0)=Q_X$, where $Q_X$ is the charge of $\phi_X$.
The charge radius $r_{X}$
is related to the slope of $G_{X}(q^2)$ at $q^2=0$,
namely
\begin{equation}
  <r_{X}^2>
  =
  6\frac{d}{dq^2}G_{X}(q^2)|_{q^2=0}
.\end{equation}

There are three terms in the $\order(E^4)$ Lagrangian
\begin{eqnarray} \label{baryon_ff-eqn:L4PQQCD}
  {\cal L}
  &=&
  \a_4\frac{8\lambda}{f^2}
    \str(D_\mu\Sigma D^\mu\Sigma)
    \str(m_Q\Sigma+m_Q^\dagger\Sigma^\dagger)
  +
  \a_5\frac{8\lambda}{f^2}
    \str(D_\mu\Sigma D^\mu\Sigma(m_Q\Sigma+m_Q^\dagger\Sigma^\dagger))
                 \nonumber \\
  &&+
  i\a_9
    \str(L_{\mu\nu}D^\mu\Sigma D^\nu\Sigma^\dagger
                        +R_{\mu\nu}D^\mu\Sigma^\dagger D^\nu\Sigma)
  + \dots
\end{eqnarray}
that contribute to meson form factors at tree level.
Here $L_{\mu\nu}$, $R_{\mu\nu}$ are the field-strength tensors of 
the external sources, which for an
electromagnetic source are given by
\begin{eqnarray} \label{baryon_ff-eqn:LR}
  L_{\mu\nu} = R_{\mu\nu}
  = e\cQ(\partial_\mu \cA_\nu-\partial_\nu \cA_\mu)+ie^2\cQ^2[\cA_\mu,\cA_\nu]
.\end{eqnarray}
Unlike \QCPT, where the low-energy constants are unique and have no known
connection to \CPT, in \PQCPT\ the parameters in
\eqref{baryon_ff-eqn:L4PQQCD} are the dimensionless Gasser-Leutwyler
coefficients of \CPT~\cite{Gasser:1985gg}
which can be seen by looking at mesons that contain sea quarks only.

\begin{figure}[tb]
  \centering
  \includegraphics[width=0.75\textwidth]{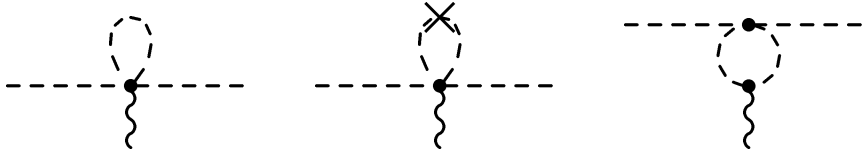}%
  \caption[Loop diagrams contributing to the octet meson charge radii
    in \PQCPT]{
    Loop diagrams contributing to the octet meson charge radii
    in \PQCPT.
    Octet mesons are denoted by a dashed line,
    singlets (hairpins) by a crossed dashed line, 
    and the photon by a wiggly line.
    Only the third diagram has 
    $q^2$ dependence and therefore contributes to
    the charge radius.
  }
  \label{baryon_ff-F:pions}
\end{figure}

\begin{figure}[tb]
  \centering
  \includegraphics[width=0.50\textwidth]{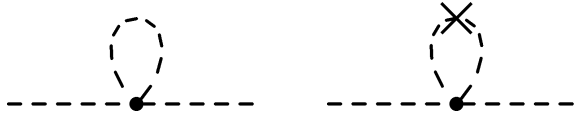}%
  \caption[Wavefunction renormalization diagrams for the 
    octet mesons 
    in \PQCPT]{
    Wavefunction renormalization diagrams  
    in \PQCPT.
    These diagrams, together with the third one in 
    Fig.~(\ref{baryon_ff-F:pions}),
    ensure
    meson electric charge non-renormalization.
  }
  \label{baryon_ff-F:pions-wf}
\end{figure}

To calculate the charge radii to lowest order
in the chiral expansion 
one has to include operators of $\cL$ in \eqref{CPT-eqn:Lchi}
to one-loop order 
[see Figs.~(\ref{baryon_ff-F:pions}) and (\ref{baryon_ff-F:pions-wf})]
and operators of \eqref{baryon_ff-eqn:L4PQQCD}
to tree level. 
Using dimensional regularization, where
we have subtracted 
$\frac{1}{\e}+1-\gamma+\log 4\pi$,
we find in \PQCPT\ for the
$\pi^+$
\begin{eqnarray}\label{baryon_ff-eqn:donaldduck}
  G^{PQ}_{\pi^+}(q^2)
  &=&
  1
  -
  \frac{1}{16\pi^2f^2}
  \left[2F_{uj}+F_{ur}\right]
  +\a_9\frac{4}{f^2}q^2
,\end{eqnarray}
which interestingly does not depend on the
charges of the sea and ghost quarks, $q_j$, $q_l$, $q_r$.
For the $K^+$ we find
\begin{eqnarray}
  G^{PQ}_{K^+}(q^2)
  &=&
  1
  +
  \frac{1}{16\pi^2f^2}
  \left[
    \left(\frac{1}{3}-q_{jl}\right)F_{uu}
    -
    \left(\frac{4}{3}-q_{jl}\right)F_{uj}
    -
    \left(\frac{2}{3}-q_r\right)F_{ur}    
                  \right. \nonumber \\
     &&\phantom{ddddddddddd} \left.
    +
    \left(\frac{1}{3}+q_r\right)F_{ss}    
    -
    \left(\frac{2}{3}-q_{jl}+q_r\right)F_{us}    
    -
    \left(\frac{2}{3}+q_{jl}\right)F_{js}    
                  \right. \nonumber \\
     &&\phantom{ddddddddddd} \left.
    -
    \left(\frac{1}{3}+q_r\right)F_{rs}    
  \right]
  +\a_9\frac{4}{f^2}q^2
\end{eqnarray}
and for the $K^0$ we find
\begin{eqnarray}\label{baryon_ff-eqn:mickey}
  G^{PQ}_{K^0}(q^2)
  &=&
  \frac{1}{16\pi^2f^2}
  \left[
    \left(\frac{1}{3}-q_{jl}\right)F_{uu}
    +
    \left(\frac{2}{3}+q_{jl}\right)F_{uj}
    +
    \left(\frac{1}{3}+q_r\right)F_{ur}    
    +
    \left(\frac{1}{3}+q_r\right)F_{ss}    
                  \right. \nonumber \\
     &&\phantom{dddddd} \left.
    -
    \left(\frac{2}{3}-q_{jl}+q_r\right)F_{us}    
    -
    \left(\frac{2}{3}+q_{jl}\right)F_{js}    
    -
    \left(\frac{1}{3}+q_r\right)F_{rs}    
  \right]
.\end{eqnarray}
Here $q_{jl}=q_j+q_l$ and we have defined
\begin{equation}
  F_{QQ'}
  =
  \frac{q^2}{6}\log\frac{m_{QQ'}^2}{\mu^2}
  -
  m_{QQ'}^2{\mathcal F}\left(\frac{q^2}{m_{QQ'}^2}\right)
,\end{equation}
where the function ${\mathcal F}(a)$ is given by
\begin{equation}
   {\mathcal F}(a)
   = 
   \left(\frac{a}{6}-\frac{2}{3}\right)
   \sqrt{1-\frac{4}{a}}
   \log\frac{\sqrt{1-\frac{4}{a}+i\e}-1}{\sqrt{1-\frac{4}{a}+i\e}+1}
   +\frac{5a}{18}-\frac{4}{3}
.\end{equation}
The first derivative of $F_{QQ'}$ at $q^2=0$, needed
to calculate the charge radii, becomes
\begin{equation}
   6\frac{d}{dq^2}F_{QQ'}|_{q^2=0}
   =
   \log\frac{m_{QQ'}^2}{\mu^2}+1
.\end{equation}
Charge conjugation implies 
\begin{equation}
  G^{PQ}_{\pi^-}=-G^{PQ}_{\pi^+},\quad
  G^{PQ}_{K^-}=-G^{PQ}_{K^+},\quad\text{and}\quad
  G^{PQ}_{\ol{K}^0}=-G^{PQ}_{K^0}
,\end{equation}
which we have also verified at one-loop order.
The form factors of the flavor diagonal mesons are zero
by charge conjugation invariance.
In the limit $m_j\to\bar{m}$, $m_r\to m_s$ 
we recover the
QCD result~\cite{Gasser:1985gg,Gasser:1985ux}
as expected.

It is interesting to note,
that duplicating these calculations for 
\QCPT\ 
shows that there is no meson mass dependence at this order.
Specifically we find
\begin{equation}\label{baryon_ff-eqn:GWB}
  G_{\pi^+}^Q(q^2)=-G_{\pi^-}^Q(q^2)
 =G_{K^+}^Q(q^2)=-G_{K^-}^Q(q^2)=1+\frac{4}{f^2}\a_9^Qq^2,
\end{equation}
and the form factors of the neutral mesons are zero.
Here we annotate the quenched constant $a_9^Q$ with a
``Q'' since its numerical value is different from the
one in Eq.~\eqref{baryon_ff-eqn:L4PQQCD}.
Eq.~\eqref{baryon_ff-eqn:GWB} reflects that 
flavor-singlet loops do not contribute to
the $q^2$-dependence at this order;
thus the virtual quark loops are completely removed
by their ghostly counterparts.
This can readily be seen by considering the quenched limit
of Eqs.~\eqref{baryon_ff-eqn:donaldduck}--\eqref{baryon_ff-eqn:mickey}.
The meson mass independence reveals once again
the pathologic nature of the quenched approximation
and seriously puts into question \CPT\ extrapolations to the
physical pion mass.

\subsection{Octet Baryon Charge Radii}
The electromagnetic form factors at or near zero momentum transfer
that enable the extraction of the baryon magnetic moments
and charge radii have been frequently investigated in QCD%
~\cite{Jenkins:1991jv,Jenkins:1993pi,
Meissner:1997hn,Bernard:1998gv,Kubis:1999xb,Kubis:2000aa,Kubis:2000zd,
Puglia:1999th,Puglia:2000jy,
Durand:1998ya}.
There are also recent quenched and partially quenched calculations
of the octet baryon magnetic moments in \QCPT\ and \PQCPT%
~\cite{Savage:2001dy,Chen:2001yi,Leinweber:2002qb}.
Here, we extend these calculations to the octet baryon charge radii.
We retain spin-3/2 baryons in intermediate states since 
formally $\D\sim m_\pi$.

Using the
heavy baryon formalism%
~\cite{Jenkins:1991jv,Jenkins:1991ne},
the baryon matrix element of the
electromagnetic current $J^\mu$ can be parametrized 
in terms of the Dirac and Pauli form factors $F_1$ and $F_2$,
respectively,
as
\begin{equation}
  \langle\ol B(p') \left|J^\mu\right|B(p)\rangle
  =
  \,\ol u(p')
  \left\{
    v^\mu F_1(q^2)+\frac{[S^\mu,S^\nu]}{M_B}q_\nu F_2(q^2)
  \right\}
  u(p)
\end{equation}
with $q=p'-p$.
The Sachs electric and magnetic form factors
defined as
\begin{eqnarray}
  G_E(q^2)&=&F_1(q^2)+\frac{q^2}{4M_B^2}F_2(q^2) \label{baryon_ff-eqn:bob}\\
  G_M(q^2)&=&F_1(q^2)+F_2(q^2)
\end{eqnarray}
are particularly useful.
The baryon charge $Q$, 
electric charge radius $<r_E^2>$, and magnetic moment $\mu$
can be defined in terms of these form factors by
\begin{equation}
  Q=G_E(0),\quad
  <r_E^2>=\left.6\frac{d}{dq^2}G_E(q^2)\right|_{q^2=0},
                      \quad
  \text{and}\quad
  \mu=G_M(0)-Q
.\end{equation}
Here the baryon charge $Q$ is in units of $e$.

\subsubsection{Analysis in \PQCPT}
Let us first consider the calculation of the octet baryon charge radii
in \PQCPT.
The Lagrangian 
describing the relevant interactions of the $\cB_{ijk}$ 
and $\cT_{ijk}$ 
with the pseudo-Goldstone mesons is
\begin{equation} \label{baryon_ff-eqn:Linteract}
  {\cal L}
  =
  2\a\left(\ol{\cB}S^\mu \cB A_\mu\right)
  +
  2\b\left(\ol{\cB}S^\mu A_\mu \cB\right)
  +
  \sqrt{\frac{3}{2}}\cC
  \left[
    \left(\ol{\cT}^\nu A_\nu \cB\right)+\text{h.c.}
  \right]  
.\end{equation}
The axial-vector and vector meson fields $A^\mu$ and $V^\mu$
are defined by analogy to those in QCD:
\begin{equation} \label{baryon_ff-eqn:AandV}
  A^\mu=\frac{i}{2}
        \left(
          \xi\partial^\mu\xi^\dagger-\xi^\dagger\partial^\mu\xi
        \right)\quad\text{and}\quad
  V^\mu=\frac{1}{2}
        \left(
          \xi\partial^\mu\xi^\dagger+\xi^\dagger\partial^\mu\xi
        \right)
.\end{equation}
The vector $S_\mu$ is the covariant
spin operator%
~\cite{Jenkins:1991jv,Jenkins:1991es,Jenkins:1991ne}.
The constants $\a$ and $\b$ are easily calculated in terms of the
constants $D$ and $F$ that are used for the $SU(3)_{\text{val}}$
analogs of these terms in QCD.  
Restricting the indices of $\cB_{ijk}$ to
$i,j,k=1,2,3$ one easily identifies
\begin{equation}
  \a=\frac{2}{3}D+2F\quad\text{and}\quad
  \b=-\frac{5}{3}D+F
.\end{equation}

\begin{figure}[tb]
  \centering
  \includegraphics[width=0.75\textwidth]{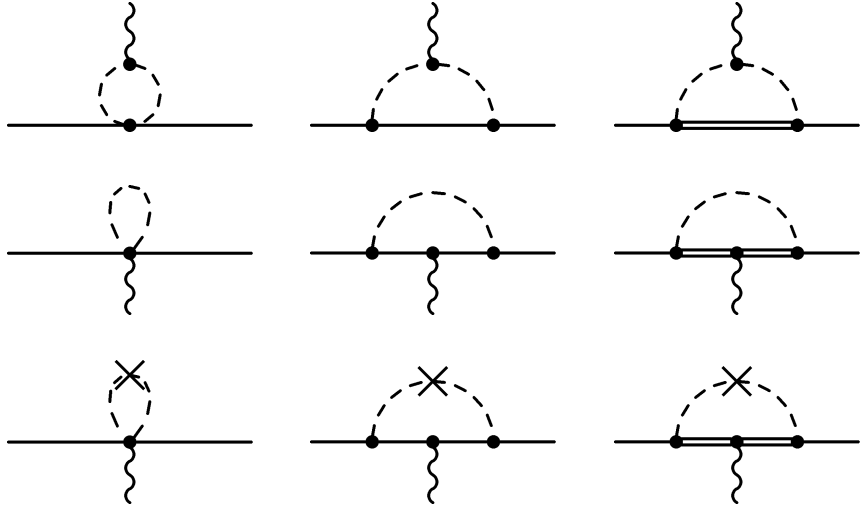}%
  \caption[Loop diagrams contributing to the octet baryon magnetic moments
     and charge radii]{\label{baryon_ff-F:baryons}
     Loop diagrams contributing to the baryon magnetic moments
     and charge radii.
     A thin (thick) solid line denotes an octet (decuplet)
     baryon.
     The last two diagrams in the first row
     contribute to both magnetic moments and charge radii;
     the magnetic moment part of which has 
     already been calculated in \cite{Chen:2001yi}.
     The first diagram in row 1 contributes $q^2$ dependence
     only to $F_1$ 
     and therefore is relevant for the charge radii.
     The remaining diagrams have no $q^2$ dependence.
     These along with the wave function renormalization diagrams
     in Fig.~(\ref{baryon_ff-F:baryons-wf}) maintain
     charge non-renormalization.}
\end{figure}

\begin{figure}[tb]
  \centering
  \includegraphics[width=0.75\textwidth]{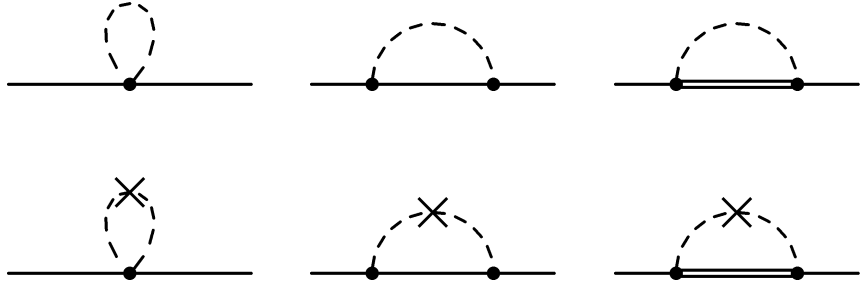}%
  \caption[Wave function renormalization diagrams
     for the octet baryons]{\label{baryon_ff-F:baryons-wf}
     Wave function renormalization diagrams
     needed to maintain
     baryon electric charge non-renormalization.}
\end{figure}

The leading tree-level correction to the magnetic moments 
come from the
di\-men\-sion-5 operators%
\footnote{Here we use
$F_{\mu\nu}=\partial_\mu A_\nu-\partial_\nu A_\mu$.}
\begin{eqnarray}
  {\cal L}
  &=&
  \frac{ie}{2M_B}
  \left[
    \mu_\a\,\left(\ol{\cB}[S_\mu,S_\nu]\cB \cQ\right)
    +\mu_\b\,\left(\ol{\cB}[S_\mu,S_\nu]\cQ\cB\right)
  \right]
  F^{\mu\nu}
\end{eqnarray}
which
can be matched on the QCD Lagrangian
upon restricting the baryon field indices to 
$1$-$3$
\begin{eqnarray}\label{baryon_ff-eqn:LDF}
  {\cal L}
  &=&
  \frac{ie}{2M_B}
  \left[
    \mu_D\,\tr(\ol{B}[S_\mu,S_\nu]\{\cQ,B\})
    +\mu_F\,\tr(\ol{B}[S_\mu,S_\nu][\cQ,B])
  \right]
  F^{\mu\nu}
\end{eqnarray}
where
\begin{equation}
  \mu_\a=\frac{2}{3}\mu_D+2\mu_F\quad\text{and }
  \mu_\b=-\frac{5}{3}\mu_D+\mu_F  
\end{equation}
at tree level.
The magnetic moments contribute the so-called Foldy term
to charge radii via $F_2(0)$ in Eq.~\eqref{baryon_ff-eqn:bob}.
Likewise, further leading tree-level corrections to the charge radii
come from the dimension-6 operators
\begin{eqnarray}\label{baryon_ff-eqn:Lc}
  {\cal L}
  &=&
  \frac{e}{\L_\chi^2}
  \left[
    c_\a\,\left(\ol{\cB}\cB\cQ\right)
    +c_\b\,\left(\ol{\cB}\cQ\cB\right)
  \right]
  v_\mu\partial_\nu F^{\mu\nu}
\end{eqnarray}
and the parameters $c_+$ and $c_-$,
defined by
\begin{equation}
  c_\a=\frac{2}{3}c_+ + 2c_-\quad\text{and }
  c_\b=-\frac{5}{3}c_+ + c_-  
,\end{equation}
are the same as those used in QCD.
Here, we take the chiral symmetry breaking scale 
$\L_\chi\sim 4\pi f$ for the purpose of power counting.
The NLO contributions arise from the one-loop diagrams shown in
Figs.~(\ref{baryon_ff-F:baryons}) and (\ref{baryon_ff-F:baryons-wf}).
To calculate the charge radii we need 
the form factors $F_1$ 
to first order in $q^2$ and 
$F_2(0)$
we find
\begin{eqnarray} \label{baryon_ff-eqn:fred}
  <r_E^2>
  &=&
  -\frac{6}{\L_\chi^2}(Qc_-+\a_Dc_+)
  +
  \frac{3}{2M_B^2}(Q\mu_F+\a_D\mu_D)
                    \nonumber \\
  &&
  -
  \frac{1}{16\pi^2 f^2}
  \sum_{X}
  \left[ 
    A_X\log\frac{m_X^2}{\mu^2}
    -
    5\,\b_X\log\frac{m_X^2}{\mu^2}
    +
    10\,\b_X'{\mathcal G}(m_X,\D,\mu)
  \right]. \nonumber \\
\end{eqnarray}
Here, we have defined the function
${\mathcal G}(m,\D,\mu)$ 
by
\begin{equation}
  {\mathcal G}(m,\D,\mu)
  =
  \log\frac{m^2}{\mu^2}
  -\frac{\D}{\sqrt{\D^2-m^2}}
   \log\frac{\D-\sqrt{\D^2-m^2+i\e}}{\D+\sqrt{\D^2-m^2+i\e}}
.\end{equation}
Note that in Eq.~\eqref{baryon_ff-eqn:fred} the only loop contributions
we keep are those non-analytic in $m_X$.
 
The parameters for the tree-level diagrams are listed in 
Table~\ref{baryon_ff-T:treelevel}.
\begin{table}[!h]
\centering
\caption[Tree-level contributions in QCD, QQCD, and PQQCD]{\label{baryon_ff-T:treelevel}Octet baryon tree-level contributions in QCD, QQCD, and PQQCD.}
\begin{tabular}{c | c c}\hline\hline
  & $Q$ & $\a_D$ \\ \hline
  $p$, $\S^+$ & 1 & $\frac{1}{3}$ \\
  $n$, $\Xi^0$ & 0 & $-\frac{2}{3}$\\
  $\S^0$ & 0 & $\frac{1}{3}$\\
  $\S^-$, $\Xi^-$ & $-1$ & $\frac{1}{3}$\\
  $\L$ & 0 & $-\frac{1}{3}$\\
  $\S^0\L$ & 0 & $\frac{1}{\sqrt{3}}$\\\hline\hline
\end{tabular}
\end{table}
The computed values for the $\b_X$, $\b_X'$, and $A_X$ coefficients
that appear in 
Eq.~(\ref{baryon_ff-eqn:fred}) are listed for the 
octet baryons in 
Tables~\ref{baryon_ff-T:p}---\ref{baryon_ff-T:Lambda}.
\begin{table}[tb]
\centering
\caption[$\b_X$, $\b_X'$, and $A_X$
         in $SU(3)$ flavor \PQCPT\ for the proton]
        {\label{baryon_ff-T:p}The coefficients $\b_X$, $\b_X'$, and $A_X$
         in $SU(3)$ flavor \PQCPT\ for the proton.}
%\begin{ruledtabular}
\begin{tabular}{c | c c c}
  \hline\hline
  $X$ & $\b_X$ & $\b_X'$ & $A_X$\\ \hline
  $\pi$ & $-\frac{1}{9}\left(7D^2+6DF-9F^2\right)-\frac{1}{3}\left(5D^2-6DF+9F^2\right)q_{jl}$ & $\left(-\frac{2}{9}+\frac{1}{6}q_{jl}\right)\cC^2$ & $-1+3q_{jl}$\\
  $K$ & $-\frac{1}{9}\left(5D^2-6DF+9F^2\right)(1+3q_r)$ & $\left(\frac{1}{18}+\frac{1}{6}q_r\right)\cC^2$ & $1+3q_r$\\
  $uj$ & $-\frac{2}{9}\left(D+3F\right)^2+\frac{1}{3}\left(5D^2-6DF+9F^2\right)q_{jl}$ & $-\frac{1}{6}q_{jl}\cC^2$ & $2-3q_{jl}$\\
  $ur$ & $-\frac{1}{9}\left(D+3F\right)^2+\frac{1}{3}\left(5D^2-6DF+9F^2\right)q_r$ & $-\frac{1}{6}q_r\cC^2$ & $1-3q_r$\\
\hline\hline
\end{tabular}
%\end{ruledtabular}
\end{table}
\begin{table}[!h]
\centering
\caption[$\b_X$, $\b_X'$, and $A_X$
         in $SU(3)$ flavor \PQCPT\ for the neutron]
       {\label{baryon_ff-T:n}The coefficients $\b_X$, $\b_X'$, and $A_X$
         in $SU(3)$ flavor \PQCPT\ for the neutron.}
%\begin{ruledtabular}
\begin{tabular}{c | c c c} 
  \hline\hline
  $X$ & $\b_X$ & $\b_X'$ & $A_X$\\ \hline
  $\pi$ & $\frac{1}{9}\left(17D^2-6DF+9F^2\right)-\frac{1}{3}\left(5D^2-6DF+9F^2\right)q_{jl}$ & $\left(\frac{1}{9}+\frac{1}{6}q_{jl}\right)\cC^2$ & $-1+3q_{jl}$\\
  $K$ & $-\frac{1}{9}\left(5D^2-6DF+9F^2\right)(1+3q_r)$ & $\left(\frac{1}{18}+\frac{1}{6}q_r\right)\cC^2$ & $1+3q_r$\\
  $uj$ & $-\frac{8}{9}\left(D^2-3DF\right)+\frac{1}{3}\left(5D^2-6DF+9F^2\right)q_{jl}$ & $\left(\frac{1}{9}-\frac{1}{6}q_{jl}\right)\cC^2$ & $-3q_{jl}$\\
  $ur$ & $-\frac{4}{9}\left(D^2-3DF\right)+\frac{1}{3}\left(5D^2-6DF+9F^2\right)q_r$ & $\left(\frac{1}{18}-\frac{1}{6}q_r\right)\cC^2$ & $-3q_r$\\
\hline\hline
\end{tabular}
%\end{ruledtabular}
\end{table}
\begin{sidewaystable}
\centering
\caption[$\b_X$, $\b_X'$, and $A_X$
         in $SU(3)$ flavor \PQCPT\ for the $\S^+$]
        {\label{baryon_ff-T:Sigmaplus}
        The coefficients $\b_X$, $\b_X'$, and $A_X$
         in $SU(3)$ flavor \PQCPT\ for the $\S^+$.}
%\begin{ruledtabular}
\begin{tabular}{c | c c c}
  \hline\hline
  $X$ & $\b_X$ & $\b_X'$ & $A_X$\\ \hline
  $\pi$ & $\frac{2}{9}\left(D^2+3F^2\right)(1-3q_{jl})$ & $\left(-\frac{1}{54}+\frac{1}{18}q_{jl}\right)\cC^2$ & $-\frac{2}{3}+2q_{jl}$\\
  $K$ & $-\frac{1}{9}\left(11D^2+6DF+3F^2\right)-(D-F)^2q_{jl}-\frac{2}{3}\left(D^2+3F^2\right)q_r$ & $\left(-\frac{5}{27}+\frac{1}{9}q_{jl}+\frac{1}{18}q_r\right)\cC^2$ & $\frac{1}{3}+q_{jl}+2q_r$\\
  $\eta_s$ & $-\frac{1}{3}(D-F)^2(1+3q_r)$ & $\left(\frac{1}{27}+\frac{1}{9}q_r\right)\cC^2$ & $\frac{1}{3}+q_r$\\
  $uj$ & $-\frac{2}{9}\left(D^2+3F^2\right)(4-3q_{jl})$ & $\left(\frac{2}{27}-\frac{1}{18}q_{jl}\right)\cC^2$ & $\frac{8}{3}-2q_{jl}$\\
  $ur$ & $-\frac{2}{9}\left(D^2+3F^2\right)(2-3q_r)$ & $\left(\frac{1}{27}-\frac{1}{18}q_r\right)\cC^2$ & $\frac{4}{3}-2q_r$\\
  $sj$ & $\frac{1}{3}(D-F)^2(2+3q_{jl})$ & $\left(-\frac{2}{27}-\frac{1}{9}q_{jl}\right)\cC^2$ & $-\frac{2}{3}-q_{jl}$\\
  $sr$ & $\frac{1}{3}(D-F)^2(1+3q_r)$ & $\left(-\frac{1}{27}-\frac{1}{9}q_r\right)\cC^2$ & $-\frac{1}{3}-q_r$\\
  \hline\hline
\end{tabular}
%}
%\end{ruledtabular}
%\end{table}
\spacebetweentables
%\begin{table}
\centering
\caption[$\b_X$, $\b_X'$, and $A_X$
         in $SU(3)$ flavor \PQCPT\ for the $\S^0$]
        {\label{baryon_ff-T:Sigma0}The coefficients $\b_X$, $\b_X'$, and $A_X$
         in $SU(3)$ flavor \PQCPT\ for the $\S^0$.}
%\begin{ruledtabular}
\begin{tabular}{c | c c c}
  \hline\hline
  $X$ & $\b_X$ & $\b_X'$ & $A_X$\\ \hline
  $\pi$ & $\frac{2}{9}\left(D^2+3F^2\right)(1-3q_{jl})$ & $\left(-\frac{1}{54}+\frac{1}{18}q_{jl}\right)\cC^2$ & $-\frac{2}{3}+2q_{jl}$\\
  $K$ & $-\frac{1}{9}\left(5D^2+6DF+3F^2\right)-(D-F)^2q_{jl}-\frac{2}{3}\left(D^2+3F^2\right)q_r$ & $\left(-\frac{11}{108}+\frac{2}{9}q_{jl}+\frac{1}{18}q_r\right)\cC^2$ & $\frac{1}{3}+q_{jl}+2q_r$\\
  $\eta_s$ & $-\frac{1}{3}(D-F)^2(1+3q_r)$ & $\left(\frac{1}{27}+\frac{1}{9}q_r\right)\cC^2$ & $\frac{1}{3}+q_r$\\
  $uj$ & $-\frac{2}{9}\left(D^2+3F^2\right)(1-3q_{jl})$ & $\left(\frac{1}{54}-\frac{1}{18}q_{jl}\right)\cC^2$ & $\frac{2}{3}-2q_{jl}$\\
  $ur$ & $-\frac{1}{9}\left(D^2+3F^2\right)(1-6q_r)$ & $\left(\frac{1}{108}-\frac{1}{18}q_r\right)\cC^2$ & $\frac{1}{3}-2q_r$\\
  $sj$ & $\frac{1}{3}(D-F)^2(2+3q_{jl})$ & $\left(-\frac{2}{27}-\frac{1}{9}q_{jl}\right)\cC^2$ & $-\frac{2}{3}-q_{jl}$\\
  $sr$ & $\frac{1}{3}(D-F)^2(1+3q_r)$ & $\left(-\frac{1}{27}-\frac{1}{9}q_r\right)\cC^2$ & $-\frac{1}{3}-q_r$\\
\hline\hline
\end{tabular}
%\end{ruledtabular}
\end{sidewaystable}
\begin{sidewaystable}
\centering
\caption[$\b_X$, $\b_X'$, and $A_X$
         in $SU(3)$ flavor \PQCPT\ for the $\S^-$]
         {\label{baryon_ff-T:Sigmaminus}
         The coefficients $\b_X$, $\b_X'$, and $A_X$
         in $SU(3)$ flavor \PQCPT\ for the $\S^-$.}
%\begin{ruledtabular}
\begin{tabular}{c | c c c}
  \hline\hline
  $X$ & $\b_X$ & $\b_X'$ & $A_X$\\ \hline
  $\pi$ & $\frac{2}{9}\left(D^2+3F^2\right)(1-3q_{jl})$ & $\left(-\frac{1}{54}+\frac{1}{18}q_{jl}\right)\cC^2$ & $-\frac{2}{3}+2q_{jl}$\\
  $K$ & $\frac{1}{9}\left(D^2-6DF-3F^2\right)-(D-F)^2q_{jl}-\frac{2}{3}\left(D^2+3F^2\right)q_r$ & $\left(-\frac{1}{54}+\frac{1}{9}q_{jl}+\frac{1}{18}q_r\right)\cC^2$ & $\frac{1}{3}+q_{jl}+2q_r$\\
  $\eta_s$ & $-\frac{1}{3}(D-F)^2(1+3q_r)$ & $\left(\frac{1}{27}+\frac{1}{9}q_r\right)\cC^2$ & $\frac{1}{3}+q_r$\\
  $uj$ & $\frac{2}{9}\left(D^2+3F^2\right)(2+3q_{jl})$ & $\left(-\frac{1}{27}-\frac{1}{18}q_{jl}\right)\cC^2$ & $-\frac{4}{3}-2q_{jl}$\\
  $ur$ & $\frac{2}{9}\left(D^2+3F^2\right)(1+3q_r)$ & $\left(-\frac{1}{54}-\frac{1}{18}q_r\right)\cC^2$ & $-\frac{2}{3}-2q_r$\\
  $sj$ & $\frac{1}{3}(D-F)^2(2+3q_{jl})$ & $\left(-\frac{2}{27}-\frac{1}{9}q_{jl}\right)\cC^2$ & $-\frac{2}{3}-q_{jl}$\\
  $sr$ & $\frac{1}{3}(D-F)^2(1+3q_r)$ & $\left(-\frac{1}{27}-\frac{1}{9}q_r\right)\cC^2$ & $-\frac{1}{3}-q_r$\\
\hline\hline
\end{tabular}
%\end{ruledtabular}
%\end{sidewaystable}
\spacebetweentables
%\begin{sidewaystable}
\centering
\caption[$\b_X$, $\b_X'$, and $A_X$
         in $SU(3)$ flavor \PQCPT\ for the $\Xi^0$]
        {\label{baryon_ff-T:Xi0}
         The coefficients $\b_X$, $\b_X'$, and $A_X$
         in $SU(3)$ flavor \PQCPT\ for the $\Xi^0$.}
%\begin{ruledtabular}
\begin{tabular}{c | c c c}
  \hline\hline
  $X$ & $\b_X$ & $\b_X'$ & $A_X$\\ \hline
  $\pi$ & $\frac{1}{3}(D-F)^2\left(1-3q_{jl}\right)$ & $\left(-\frac{1}{27}+\frac{1}{9}q_{jl}\right)\cC^2$ & $-\frac{1}{3}+q_{jl}$\\
  $K$ & $\frac{1}{9}\left(11D^2+6DF+3F^2\right)-\frac{2}{3}\left(D^2+3F^2\right)q_{jl}-(D-F)^2q_r$ & $\left(\frac{5}{27}+\frac{1}{18}q_{jl}+\frac{1}{9}q_r\right)\cC^2$ & $-\frac{1}{3}+q_r+2q_{jl}$\\
  $\eta_s$ & $-\frac{2}{9}\left(D^2+3F^2\right)(1+3q_r)$ & $\left(\frac{1}{54}+\frac{1}{18}q_r\right)\cC^2$ & $\frac{2}{3}+2q_r$\\
  $uj$ & $-\frac{1}{3}(D-F)^2(4-3q_{jl})$ & $\left(\frac{4}{27}-\frac{1}{9}q_{jl}\right)\cC^2$ & $\frac{4}{3}-q_{jl}$\\
  $ur$ & $-\frac{1}{3}(D-F)^2(2-3q_r)$ & $\left(\frac{2}{27}-\frac{1}{9}q_r\right)\cC^2$ & $\frac{2}{3}-q_r$\\
  $sj$ & $\frac{2}{9}\left(D^2+3F^2\right)(2+3q_{jl})$ & $\left(-\frac{1}{27}-\frac{1}{18}q_{jl}\right)\cC^2$ & $-\frac{4}{3}-2q_{jl}$\\
  $sr$ & $\frac{2}{9}\left(D^2+3F^2\right)(1+3q_r)$ & $\left(-\frac{1}{54}-\frac{1}{18}q_r\right)\cC^2$ & $-\frac{2}{3}-2q_r$\\
\hline\hline
\end{tabular}
%\end{ruledtabular}
\end{sidewaystable}
\begin{sidewaystable}
\centering
\caption[$\b_X$, $\b_X'$, and $A_X$
         in $SU(3)$ flavor \PQCPT\ for the $\Xi^-$]
         {\label{baryon_ff-T:Ximinus}
          The coefficients $\b_X$, $\b_X'$, and $A_X$
         in $SU(3)$ flavor \PQCPT\ for the $\Xi^-$.}
%\begin{ruledtabular}
\begin{tabular}{c | c c c}
  \hline\hline
  $X$ & $\b_X$ & $\b_X'$ & $A_X$\\ \hline
  $\pi$ & $\frac{1}{3}(D-F)^2\left(1+3q_{jl}\right)$ & $\left(-\frac{1}{27}+\frac{1}{9}q_{jl}\right)\cC^2$ & $-\frac{1}{3}+q_{jl}$\\
  $K$ & $-\frac{1}{9}\left(D^2-6DF-3F^2\right)-\frac{2}{3}\left(D^2+3F^2\right)q_{jl}-(D-F)^2q_r$ & $\left(\frac{1}{54}+\frac{1}{18}q_{jl}+\frac{1}{9}q_r\right)\cC^2$ & $-\frac{1}{3}+2q_{jl}+q_r$\\
  $\eta_s$ & $-\frac{2}{9}\left(D^2+3F^2\right)(1+3q_r)$ & $\left(\frac{1}{54}+\frac{1}{18}q_r\right)\cC^2$ & $\frac{2}{3}+2q_r$\\
  $uj$ & $\frac{1}{3}(D-F)^2(2+3q_{jl})$ & $\left(-\frac{2}{27}-\frac{1}{9}q_{jl}\right)\cC^2$ & $-\frac{2}{3}-q_{jl}$\\
  $ur$ & $\frac{1}{3}(D-F)^2(1+3q_r)$ & $\left(-\frac{1}{27}-\frac{1}{9}q_r\right)\cC^2$ & $-\frac{1}{3}-q_r$\\
  $sj$ & $\frac{2}{9}\left(D^2+3F^2\right)(2+3q_{jl})$ & $\left(-\frac{1}{27}-\frac{1}{18}q_{jl}\right)\cC^2$ & $-\frac{4}{3}-2q_{jl}$\\
  $sr$ & $\frac{2}{9}\left(D^2+3F^2\right)(1+3q_r)$ & $\left(-\frac{1}{54}-\frac{1}{18}q_r\right)\cC^2$ & $-\frac{2}{3}-2q_r$\\
\hline\hline
\end{tabular}
%\end{ruledtabular}
%\end{sidewaystable}
\spacebetweentables
%\begin{sidewaystable}
\centering
\caption[$\b_X$, $\b_X'$, and $A_X$
         in $SU(3)$ flavor \PQCPT\ for the $\L$]
         {\label{baryon_ff-T:Lambda}
         The coefficients $\b_X$, $\b_X'$, and $A_X$
         in $SU(3)$ flavor \PQCPT\ for the $\L$.}
%\begin{ruledtabular}
\begin{tabular}{c | c c c}
  \hline\hline
  $X$ & $\b_X$ & $\b_X'$ & $A_X$\\ \hline
  $\pi$ & $\frac{2}{27}\left(7D^2-12DF+9F^2\right)(1-3q_{jl})$ & $\left(-\frac{1}{18}+\frac{1}{6}q_{jl}\right)\cC^2$ & $-\frac{2}{3}+2q_{jl}$\\
  $K$ & $\frac{1}{27}\left(5D^2+30DF-9F^2\right)-\frac{1}{9}(D+3F)^2q_{jl}-\frac{2}{9}\left(7D^2-12DF+9F^2\right)q_r$ & $\left(\frac{5}{36}+\frac{1}{6}q_r\right)\cC^2$ & $\frac{1}{3}+q_{jl}+2q_r$\\
  $\eta_s$ & $-\frac{1}{27}(D+3F)^2(1+3q_r)$ & $0$ & $\frac{1}{3}+q_r$\\
  $uj$ & $-\frac{2}{27}\left(7D^2-12DF+9F^2\right)(1-3q_{jl})$ & $\left(\frac{1}{18}-\frac{1}{6}q_{jl}\right)\cC^2$ & $\frac{2}{3}-2q_{jl}$\\
  $ur$ & $-\frac{1}{27}\left(7D^2-12DF+9F^2\right)(1-6q_r)$ & $\left(\frac{1}{36}-\frac{1}{6}q_r\right)\cC^2$ & $\frac{1}{3}-2q_r$\\
  $sj$ & $\frac{1}{27}(D+3F)^2(2+3q_{jl})$ & $0$ & $-\frac{2}{3}-q_{jl}$\\
  $sr$ & $\frac{1}{27}(D+3F)^2(1+3q_r)$ & $0$ & $-\frac{1}{3}-q_r$\\
  \hline\hline
\end{tabular}
%\end{ruledtabular}
\end{sidewaystable}
The corresponding values for the $\L\S^0$ transition are given in
Table~\ref{baryon_ff-T:LambdaSigma0}.
\begin{table}[tb]
\centering
\caption[$\b_X$, $\b_X'$, and $A_X$
         in $SU(3)$ flavor \PQCPT\ for the $\L\S^0$ transition]
        {\label{baryon_ff-T:LambdaSigma0}
         The coefficients $\b_X$, $\b_X'$, and $A_X$
         in $SU(3)$ flavor \PQCPT\ for the $\L\S^0$ transition.}
%\begin{ruledtabular}
\begin{tabular}{c | c c c}
  \hline\hline
  $X$ & $\b_X$ & $\b_X'$ & $A_X$\\ \hline
  $\pi$ & $-\frac{4}{3\sqrt{3}}D^2$ & $-\frac{1}{6\sqrt{3}}\cC^2$ & $0$\\
  $K$ & $-\frac{2}{3\sqrt{3}}D^2$ & $-\frac{1}{12\sqrt{3}}\cC^2$ & $0$\\
  $uj$ & $\frac{4}{3\sqrt{3}}D(D-3F)$ & $-\frac{1}{6\sqrt{3}}\cC^2$ & $0$\\
  $ur$ & $\frac{2}{3\sqrt{3}}D(D-3F)$ & $-\frac{1}{12\sqrt{3}}\cC^2$ & $0$\\
  \hline\hline
\end{tabular}
%\end{ruledtabular}
\end{table}
%\afterpage{\clearpage}
In each table we have listed the values corresponding to the 
loop meson that has mass $m_X$.  
If a particular meson is not listed then
the values for $\b_X$, $\b_X'$, and $A_X$ are zero.%
\footnote{
We have defined the coefficients $\b_X$ and
${\b_X}'$ to correspond to those defined in%
~\cite{Chen:2001yi} where 
$\mu=Q\,\mu_F+\a_D\,\mu_D
+\frac{M_B}{4\pi f^2}\sum_X\left[\b_X m_X
          +\b'_X{\mathcal F}(m_X,\D,\mu)\right]$
and the function ${\mathcal F}(m_X,\D,\mu)$ is
given in~\cite{Chen:2001yi}.
}

\subsubsection{Analysis in \QCPT}
The calculation of the charge radii can be easily 
executed for \QCPT.
The operators in Eqs.~\eqref{baryon_ff-eqn:LDF} and \eqref{baryon_ff-eqn:Lc}
contribute, however, their low-energy coefficients
cannot be matched onto QCD.  Therefore we annotate them with a ``Q''.    
Additional terms involving hairpins~\cite{Labrenz:1996jy,Savage:2001dy}
do not contribute as 
their contribution to the charge radii is of the form
$(\mu_0^2/\L_\chi^4)\log m_q$
and therefore of higher order
in the chiral expansion.
We find
\begin{eqnarray} \label{baryon_ff-eqn:fredQ}
  <r_E^2>
  &=&
  -\frac{6}{\L_\chi^2}(Qc^Q_-+\a_Dc^Q_+)
  +
  \frac{3}{2M_B^2}(Q\mu^Q_F+\a_D\mu^Q_D)
                    \nonumber \\
  &&
  +
  \frac{1}{16\pi^2 f^2}
  \sum_{X}
  \left[
    5\,\b^Q_X\log\frac{m_X^2}{\mu^2}
    -
    10\,{\b^Q_X}'{\mathcal G}(m_X,\D,\mu)
  \right]
.\end{eqnarray}
As with the meson case, the diagram where the photon couples 
to the closed meson loop does not contribute to baryon charge radii in the 
quenched case, {\it cf}., $A_X$ is zero.
The remaining coefficients appearing in Eq.~\eqref{baryon_ff-eqn:fredQ} are 
listed in Table~\ref{baryon_ff-T:QQCD}
\begin{table}[tb]
\centering
\caption[$\b_X^Q$ and ${\b_X^Q}'$
         in $SU(3)$ flavor \QCPT\ for the octet baryons]
      {\label{baryon_ff-T:QQCD}The coefficients $\b_X^Q$ and ${\b_X^Q}'$
         in $SU(3)$ flavor \QCPT\ for the octet baryons.}  
\begin{tabular}{l |  c  c  | c  c  }
	\hline
	\hline
	     & \multicolumn{2}{c |}{$\beta^Q_X$} &    \multicolumn{2}{c}{${\beta^Q_X}'$} \\
	 &   $\pi$     &   $K$  &  $\pi$  &   $K$    \\
        \hline
	$p$        	& $-\frac{4}{3}D_Q^2$ 	& $0$ 			& $-\frac{1}{6}\mathcal{C}_Q^2$ 	& $0$ \\
	$n$     	& $\frac{4}{3}D_Q^2$ 	& $0$ 			& $\frac{1}{6}\mathcal{C}_Q^2$ 	& $0$ \\
	$\Sigma^+$     	& $0$ 			& $-\frac{4}{3} D_Q^2$ 	& $0$ 			& $-\frac{1}{6}\mathcal{C}_Q^2$ \\
	$\Sigma^0$     	& $0$ 			& $-\frac{2}{3} D_Q^2$ 	&$0$ 			& $-\frac{1}{12}\mathcal{C}_Q^2$ \\
	$\Sigma^-$     	& $0$ 			& $0$ 			& $0$ 			& $0$ \\
	$\Lambda$     	& $0$ 			& $\frac{2}{3} D_Q^2$ 	& $0$ 			& $\frac{1}{12}\mathcal{C}_Q^2$ \\
	$\Xi^-$     	& $0$ 			& $0$			& $0$ 			& $0$ \\
	$\Xi^0$     	& $0$ 			& $\frac{4}{3} D_Q^2$ 	& $0$ 			& $\frac{1}{6}\mathcal{C}_Q^2$ \\
$\Sigma \Lambda$	& $-\frac{4}{3 \sqrt{3}} D_Q^2$ & $-\frac{2}{3 \sqrt{3}} D_Q^2$ & $-\frac{1}{6\sqrt{3}}\mathcal{C}_Q^2$ & $-\frac{1}{12\sqrt{3}}\mathcal{C}_Q^2$ \\
	\hline
	\hline
\end{tabular}
\end{table} 
and stem from meson loops formed solely
from valence quarks.

\afterpage{\clearpage}

\section{\label{baryon_ff-sec:conclusions}Conclusions}
We have calculated the charge radii for the octet mesons and baryons 
in the isospin limit of \PQCPT\
and also derive the result for
the nucleon doublet and pion triplet 
away from the isospin limit for the $SU(2)$
chiral Lagrangian.
For the octet mesons and baryons 
we have also calculated the \QCPT\ results.

We find
that new operators in the QQCD Lagrangian,
which are non-existent in QCD, 
enter at NNLO.
Hence, formally 
our NLO result is not more divergent than its QCD counterpart.
This, however, does not mean that our result is free of quenching
artifacts.
While the expansions about the chiral limit for QCD and QQCD
are formally similar,
$<r^2>\sim\a+\b\log m_Q+\dots$,
the QQCD result is anything but free of quenched oddities:
for certain baryons, $\S^-$ and $\Xi^-$ in particular,
diagrams that have bosonic or fermionic mesons running in loops 
completely cancel
so that $\b=0$.  In other words,
$<r^2>\sim\a+\dots$ and the result is actually independent of $m_Q$!
The same behavior is found for the charge radii of all mesons
in QQCD as the meson loop contributions entirely cancel.

PQQCD, on the other hand, is free of such eccentric behavior.
The formal behavior of the charge radius in the chiral limit
has the same form
as in QCD.
Moreover, 
there is 
a well-defined connection to QCD and one can
reliably extrapolate lattice results down to the 
quark masses of reality. 
The low-energy constants appearing in
PQQCD are the same as those in QCD and
by fitting them in \PQCPT\ one can make predictions for QCD.
Our \PQCPT\ result will 
enable the proper extrapolation of PQQCD lattice simulations
of the charge radii 
and we hope it encourages such simulations in the future.

\chapter{Electromagnetic Properties of the Baryon Decuplet in 
         \QCPT\ and \PQCPT}
\label{chapter:decuplet}

In this chapter, 
we calculate electromagnetic properties of the decuplet baryons 
in \QCPT\ and \PQCPT. 
We work at NLO in the chiral 
expansion, LO in the heavy baryon expansion, and obtain expressions
for the magnetic moments, charge radii, and electric quadrupole moments. 
The quenched calculation is shown to be pathological since only quenched chiral singularities 
are present at this order. We present the partially quenched results for 
both the $SU(2)$ and $SU(3)$ flavor groups and use the isospin limit in the latter. These 
results are necessary for 
proper extrapolation of lattice calculations of decuplet electromagnetic properties.

\section{Introduction}
Experiments measuring the decuplet 
magnetic moments are anticipated in the foreseeable future. The Particle Data Group
lists values for the $\D^{++}$ magnetic moment~%
\cite{Hagiwara:2002fs} but with 
sizeable discrepancy and uncertainty, even among the two most recent results~%
\cite{Bosshard:1991zp,LopezCastro:2000ep}. A report of the initial measurement of 
the $\D^+$ magnetic moment~%
\cite{Kotulla:2002cg}
has recently appeared, and further data are eagerly awaited.
 
Lattice calculations can be used to 
calculate the decuplet
electromagnetic moments.
While lattice simulations 
using the quenched approximation have already 
appeared~\cite{Leinweber:1992hy},
there are currently no partially quenched simulations.
However, 
we expect to see partially quenched calculations of the
decuplet electromagnetic form factors in the near future.

Whatever approximation is used---%
quenched or partially quenched---, 
now and in the near future 
all
these calculations 
are performed with unphysically large quark masses 
and therefore must be extrapolated
down to
the physical light quark masses.
For QQCD lattice calculations
this extrapolation 
can be accomplished by the use of \QCPT. 
However, the results are not only plagued by quenching artifacts 
but also unrelated to QCD,
as explained in Chapter~\ref{chapter:CPT}.
In fact, several examples show that
the behavior of meson loops near the chiral limit is
frequently misrepresented in \QCPT%
~\cite{Booth:1994rr,Kim:1998bz,Savage:2001dy,Arndt:2002ed,Dong:2003im,Arndt:2003ww}.
We find this to also be true for the decuplet 
electromagnetic observables. Indeed, to the order we work only
quenched chiral singularities are present in the quenched 
electromagnetic moments; the charge radii have no quark mass
dependence at all.
Of course, the unattractive features of QQCD can be remedied by using PQQCD
and \PQCPT.

The chapter is organized as follows.
First, in Section~\ref{decuplet-sec:ff},
we calculate the electromagnetic moments and charge radii of the 
decuplet baryons in both \QCPT\ and \PQCPT\
up to NLO in the chiral expansion.
We use the heavy baryon formalism of Jenkins and Manohar%
~\cite{Jenkins:1991jv,Jenkins:1991ne}
and work to lowest order in the heavy baryon expansion.
These calculations are done in the 
isospin limit of $SU(3)$ flavor. Expressions for form factors with
the $q^2$ dependence at one loop are given in Appendix~\ref{decuplet-s:q-dep}.  
For completeness we also provide the
\PQCPT\ 
result for the baryon quartet electromagnetic moments and  charge radii 
for the $SU(2)$ chiral Lagrangian with non-degenerate quarks in Appendix~\ref{decuplet-s:su2}.
We conclude in Section~\ref{decuplet-sec:conclusions}.

\section{Decuplet Electromagnetic Properties}
\label{decuplet-sec:ff}
The electromagnetic moments of decuplet baryons in \CPT\ have been investigated previously 
in~\cite{Butler:1994ej,Banerjee:1996wz}. Additionally there has been interest in the decuplet electromagnetic properties
in the large $N_c$ limit of QCD~\cite{Buchmann:2000wf,Buchmann:2002mm,Buchmann:2002et,Cohen:2002sd}.
In this section we calculate the decuplet electromagnetic moments and charge radii in
\PQCPT\ and \QCPT.
The basic conventions and notations for the
mesons and baryons in \PQCPT\ have been laid forth
in the last Chapter~\ref{chapter:CPT}.
Additionally the decuplet charge radii in \CPT\ are provided since 
they have not been calculated before. First we review the electromagnetic form
factors of heavy spin-$3/2$ baryons.
 
Using the heavy baryon formalism%
~\cite{Jenkins:1991jv,Jenkins:1991ne}, 
the decuplet matrix elements of the electromagnetic current $J^\rho$ can be parametrized as
\begin{equation}
\langle \ol T(p') | J^\rho | T(p) \rangle = -  \ol u_\mu(p') \mathcal{O}^{\mu \rho \nu} u_\nu(p),
\end{equation}
where $u_\mu(p)$ is a Rarita-Schwinger spinor for an on-shell  heavy baryon satisfying
$v^\mu u_\mu(p) = 0$ and $S^\mu u_\mu(p) = 0$.
The tensor $\mathcal{O}^{\mu \rho \nu}$ can be parametrized in terms of four independent, 
Lorentz invariant form factors
\begin{equation}
\mathcal{O}^{\mu \rho \nu} = g^{\mu \nu} \left\{ v^\rho F_1(q^2) + \frac{[S^\rho,S^\tau] }{M_B}q_\tau F_2(q^2)  \right\}
+ \frac{q^\mu q^\nu}{(2 M_B)^2} \left\{ v^\rho G_1(q^2) + \frac{[S^\rho,S^\tau]}{M_B} q_\tau G_2(q^2)  \right\},
\end{equation}
where the momentum transfer $q = p' - p$. 
The form factor $F_1(q^2)$ is normalized to the decuplet charge in units of $e$: $F_1(0) = Q$.
At NLO in the chiral expansion, the form factor $G_2(q^2) = 0$.

Extraction of the form factors requires a nontrivial identity for on-shell 
Rarita-Schwinger spinors~\cite{Nozawa:1990gt}. For heavy baryon spinors, the 
identity is
\begin{equation}
 \ol u_\a(p')\big( q^\alpha g^{\mu \beta} - q^\beta g^{\mu \alpha} \big) u_\b(p)  =  \ol u_\a(p') 
\left[ - \frac{q^2}{2 M_B} g^{\alpha \beta} v^\mu  + 2  g^{\alpha \beta} [S^\mu,S^\nu]q_\nu 
+ \frac{1}{M_B} q^\alpha q^\beta v^\mu \right] u_\b(p) .
\end{equation}
Linear combinations of the above (Dirac- and Pauli-like) form 
factors make the electric charge $G_{E0}(q^2)$, magnetic dipole $G_{M1}(q^2)$, 
electric quadrupole
$G_{E2}(q^2)$, and magnetic octupole $G_{M3}(q^2)$ form factors. 
This conversion from covariant vertex functions to multipole form 
factors for spin-$3/2$ particles is explicated in~\cite{Nozawa:1990gt}.
For our calculations, the charge radius
\begin{equation} \label{decuplet-eqn:fred}
< r_E^2 > \equiv 6 \frac{ d}{dq^2}G_{E0}(q^2) \Bigg|_{q^2 = 0} 
= 
6 \left\{ \frac{d F_1 (0) }{dq^2} - 
\frac{1}{12 M_B^2} \left[2 Q - 3 F_2(0) - G_1(0) \right]\right\}, 
\end{equation}
the magnetic moment 
\begin{equation}
\mu \equiv G_{M1}(0) - Q = F_{2}(0),
\end{equation}
and the electric quadrupole moment 
\begin{equation}
\mathbb{Q} \equiv G_{E2}(0) - Q = -\frac{1}{2} G_1(0).
\end{equation}
To the order we work in the chiral expansion, the magnetic octupole moment is zero.

\subsection{Analysis in \PQCPT}
Let us first consider the calculation of the decuplet baryon electromagnetic properties
in \PQCPT.
Here, the leading tree-level contributions to the magnetic moments 
come  from the dimension-5 operator
\begin{equation} \label{decuplet-eqn:LDF}
\cL = \mu_c \frac{ 3 i e }{M_B}  \big(\ol\cT{}^\mu \cQ \cT^\nu \big) F_{\mu \nu}, 
\end{equation}
which matches onto the \CPT\ operator~\cite{Jenkins:1993pi}
\begin{equation}
\cL = \mu_c \frac{i e  Q_{i}}{M_B} \ol T{}_{i}^\mu T_{i}^\nu F_{\mu \nu},
\end{equation}
when the indices in Eq.~\eqref{decuplet-eqn:LDF} are restricted to $1$--$3$. Here $Q_{i}$ is the charge of the $i$th decuplet state.
Additional tree-level contributions come from the leading dimension-6 electric quadrupole operator
\begin{equation} \label{decuplet-eqn:Lc}
\cL = - \mathbb{Q}_{\text{c}} \frac{3 e}{\L_\chi^2} \big(\ol \cT{}^{\{\mu} \cQ \cT^{\nu\}} \big)  v^\alpha \partial_\mu F_{\nu \alpha}.
\end{equation}
Here the action of ${}^{\{}\ldots{}^{\}}$ on Lorentz indices produces the symmetric traceless part of the tensor, 
{\it viz.}, $\mathcal{O}^{\{\mu \nu\}} = \mathcal{O}^{\mu \nu} + \mathcal{O}^{\nu \mu} - 
\frac{1}{2} g^{\mu\nu} \mathcal{O}^{\alpha}{}_{\alpha}$. 
The operator in Eq.~\eqref{decuplet-eqn:Lc}
matches onto the \CPT\ operator~\cite{Butler:1994ej}
\begin{equation}
\cL = - \mathbb{Q}_{\text{c}} \frac{e Q_i}{\L_\chi^2} \ol T{}^{\{\mu}_{i} T^{\nu\}}_{i} v^\alpha \partial_\mu F_{\nu \alpha} .
\end{equation}
The final tree-level contributions arise from the leading dimension-6 charge radius operator
\begin{equation} \label{decuplet-eqn:Lnew}
\cL =  c_c \frac{3 e}{\L_\chi^2} \big( \ol \cT{}^\sigma \cQ \cT_{\sigma} \big) v_\mu \partial_\nu F^{\mu \nu}
\end{equation}
which matches onto the \CPT\ operator
\begin{equation}
\cL = c_c \frac{e  Q_{i}}{\L_\chi^2}  \ol T{}^\sigma_{i} T_{\sigma,i} \,  v_\mu \partial_\nu F^{\mu \nu}.
\end{equation}
Notice that the PQQCD low-energy constants $\mu_c$, $\mathbb{Q}_{\text{c}}$,
and $c_c$ have the same numerical values as in QCD.

\begin{figure}[tb]
\centering
\includegraphics[width=0.75\textwidth]{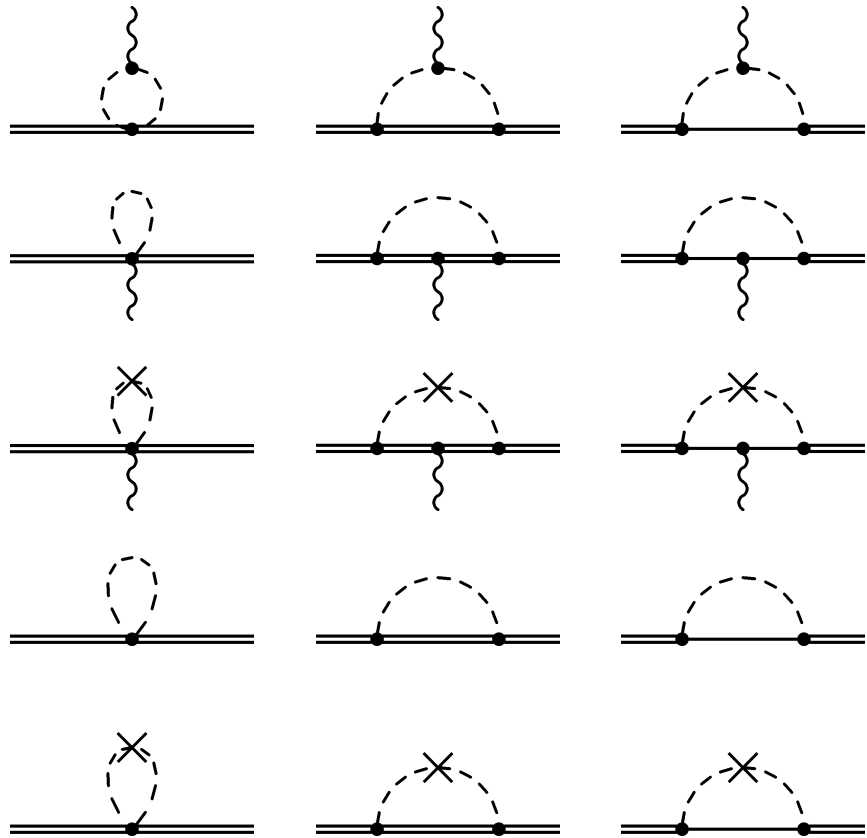}%
\caption[Loop diagrams contributing to the decuplet electromagnetic moments and charge radii]{Loop diagrams contributing to the decuplet electromagnetic moments and charge radii. Octet mesons are denoted 
by a dashed line, singlets (hairpins) by a crossed dashed line, and the photon by a wiggly line.  A thick (thin)
solid line denotes a decuplet (octet) baryon. The diagrams in the first row contribute to the electromagnetic
moments and charge radii. The remaining diagrams with a photon have no $q^2$-dependence. These, along with the
wavefunction renormalization diagrams, ensure non-renormalization of the electric charge.}
\label{decuplet-F:Fdecuplet}
\end{figure}

The NLO contributions to electromagnetic observables in the chiral expansion arise from the one-loop diagrams 
shown in Fig.~\ref{decuplet-F:Fdecuplet}. 
In addition
to the terms describing the interactions
of the $\cB_{ijk}$ and $\cT_{ijk}$
with the pseudo-Goldstone mesons given in 
Eq.~\eqref{baryon_ff-eqn:Linteract} we need the term
\begin{equation} \label{decuplet-eq:H}
  {\cal L} =   2{\mathcal H}\left(\ol{\cT}^\nu S^\mu A_\mu \cT_\nu\right) 
\end{equation}
where the axial-vector $A^\mu$
is defined in Eq.~\eqref{baryon_ff-eqn:AandV}.
Calculation of the diagrams yields
\begin{multline}
F_1(q^2)  = Q \left( 1 - \frac{\mu_c q^2}{2 M_B^2} - \frac{ \mathbb{Q}_{\text{c}} q^2}{2 \L_\chi^2} + \frac{c_c q^2}{\L_\chi^2} \right)
-\frac{1}{6} q^2 \frac{3 + \mathcal{C}^2}{16 \pi^2 f^2}  \sum_X  A_X  \log \frac{m_X^2}{\mu^2} \\
 - \frac{11}{54} q^2  \frac{\mathcal{H}^2}{16 \pi^2 f^2}  \sum_X A_X \left[ \log \frac{m_X^2}{\mu^2} - 
\frac{\Delta m_X}{\Delta^2 - m^2_X}  \cR \left(\frac{\Delta}{m_X}\right) \right] + \mathcal{O}(q^4), 
\end{multline}
\begin{equation} \label{decuplet-eqn:F2}
F_2(0)  = \mu = 2 \mu_c Q + \frac{ M_B \mathcal{H}^2}{36 \pi^2 f^2} \sum_X A_X \left[ \Delta \log \frac{m_X^2}{\mu^2} - 
m_X \cR \left(\frac{\Delta}{m_X} \right)  \right] - 
\frac{\mathcal{C}^2 M_B}{8 \pi f^2} \sum_X A_X m_X, 
\end{equation}
and
\begin{multline} \label{decuplet-eqn:G1}
G_1(0) = - 2 \mathbb{Q} =  4 Q \left( \mu_c + \mathbb{Q}_c \frac{2 M_B^2}{\L_\chi^2}  \right) - \frac{M_B^2 \mathcal{C}^2}{12 \pi^2 f^2} 
\sum_X A_X \log \frac{m_X^2}{\mu^2}  \\ 
+ \frac{M_B^2 \mathcal{H}^2}{27 \pi^2 f^2} \sum_X A_X \left[ \log \frac{m_X^2}{\mu^2} - \frac{\Delta m_X}{\Delta^2 - m^2_X}  \cR \left( \frac{\Delta}{m_X} \right) \right].
\end{multline}
The only loop contributions kept in the above expressions are those non-analytic in the quark masses. The full $q^2$ dependence
at one-loop has been omitted from the above expressions but is given in Appendix~\ref{decuplet-s:q-dep}.
The function $\mathcal{R}(x)$ is given 
in Eq.~\eqref{appendix-B2D-eq:R}.
The sum in the above expressions is over all possible loop mesons $X$.
The computed values of the coefficients $A_X^T$ that appear above are listed in Table~\ref{decuplet-t:clebsch}
\begin{sidewaystable}[tb]
%\label{decuplet-t:clebsch}
\centering
\caption[$A_X^T$ for SU($3$) for the decuplet in \CPT\ and \PQCPT]{The coefficients $A_X^T$ for SU($3$) for each of the decuplet states in \CPT\ and \PQCPT.
The index $X$ corresponds to the loop meson that has mass $m_X$. Here, we have used $q_{jl} = q_j + q_l$.}  
%\begin{ruledtabular}
\begin{tabular}{l | c c | c  c  c  c  c  c  c }
  \hline\hline
	    & \multicolumn{2}{c |}{\CPT}  &  \multicolumn{7}{c}{\PQCPT} \\
	              & $\pi$ & $K$  &   $\pi$   &   $K$   &  $\eta_s$  &   $ju$   &   $ru$   & $js$   & $rs$ \\
	 	
	\hline
	$\Delta^{++}$       & $1$ & $1$  & $-\frac{1}{3} + q_{jl}$ & $\frac{1}{3} + q_r$ & $0$ & $\frac{4}{3} - q_{jl}$ & 
												$\frac{2}{3} - q_{r}$ & $0$ & $0$ \\
 
	$\Delta^{+}$        & $\frac{1}{3}$ & $\frac{2}{3}$ & `` & `` & `` & 
										$\frac{2}{3} - q_{jl}$ & $\frac{1}{3} - q_r$ & `` & `` \\

	$\Delta^{0}$        & $-\frac{1}{3}$ & $\frac{1}{3}$  & `` & `` & `` & $-q_{jl}$ & 
														$-q_r$ & `` & `` \\

	$\Delta^{-}$        & $-1$ & $0$ & `` & `` & `` & $-\frac{2}{3} - q_{jl}$ & 
													$-\frac{1}{3} - q_r$ & `` & `` \\

	$\Sigma^{*,+}$       & $\frac{2}{3}$ & $\frac{1}{3}$  & $\;- \frac{2}{9} + \frac{2}{3} q_{jl}\;$ & 
$\;\frac{1}{9} + \frac{2}{3} q_r + \frac{1}{3} q_{jl}\;$ & $\;\frac{1}{9}  + \frac{1}{3} q_r\;$ & $\;\frac{8}{9}  - \frac{2}{3} q_{jl}\;$ 
& $\;\frac{4}{9} - \frac{2}{3} q_r \;$ & $\;-\frac{2}{9} - \frac{1}{3} q_{jl}\;$ &  $\;-\frac{1}{9} -\frac{1}{3} q_r$ \\

	$\Sigma^{*,0}$       & $0$ & $0$  & `` & `` & ``  & $\frac{2}{9} - \frac{2}{3} q_{jl}$ & $\frac{1}{9} - \frac{2}{3} q_r$ & ``
 & `` \\	

	$\Sigma^{*,-}$      & $-\frac{2}{3}$ & $-\frac{1}{3}$  & ``  & `` & `` & $- \frac{4}{9} - \frac{2}{ 3} q_{jl}$ & $ - \frac{2}{9} -\frac{2}{ 3} q_r$ 
& `` & `` \\

	$\Xi^{*,0}$         & $\frac{1}{3}$ & $-\frac{1}{3}$  & $-\frac{1}{9} + \frac{1}{3} q_{jl}$ & 
$- \frac{1}{9}+ \frac{1}{ 3} q_{r}+ \frac{2}{3} q_{jl}$ 
& $\frac{2}{9} + \frac{2}{3} q_{r}$ & $\frac{4}{9}- \frac{1}{3} q_{jl}$ & $\frac{2}{9} - \frac{1}{3} q_r$ & 
$-\frac{4}{9}-\frac{2}{3}q_{jl}$ & $-\frac{2}{9} - \frac{2}{3} q_r$ \\

	$\Xi^{*,-}$          & $-\frac{1}{3}$ & $-\frac{2}{3}$  & `` & `` & `` & $-\frac{2}{9}- \frac{1}{3} q_{jl}$ & $-\frac{1}{9} - \frac{1}{3} q_r$ & ``  & `` \\

	$\Omega^-$          & $0$ & $-1$   & $0$ & $-\frac{1}{3} + q_{jl}$ & $\frac{1}{3} +q_{r}$ & $0$ & $0$ & $-\frac{2}{3} - q_{jl}$ 
& $-\frac{1}{3} -  q_{r}$\\
\hline\hline
\end{tabular}
%\end{ruledtabular}
\label{decuplet-t:clebsch}
\end{sidewaystable}   
\afterpage{\clearpage}
for each of the decuplet states $T$.
In the table we have listed values corresponding to the loop meson that has mass $m_X$ for both \CPT\
and \PQCPT.  In particular, the \CPT\ coefficients can be used to find the QCD decuplet charge radii, 
which have not been calculated before. Using Eq.~\eqref{decuplet-eqn:fred}, the expression for the charge radii is 
\begin{multline} \label{decuplet-eqn:r_E}
<r_E^2> = Q \left( \frac{2 \mu_c - 1}{M_B^2} + \frac{\mathbb{Q}_c + 6 c_c}{\L_\chi^2}  \right)
- \frac{1}{3} \, \frac{9 + 5 \mathcal{C}^2}{16 \pi^2 f^2}  \sum_X  A_X  \log \frac{m_X^2}{\mu^2} \\
 - \frac{25}{27}\, \frac{\mathcal{H}^2}{16 \pi^2 f^2} \sum_X A_X \left[ \log \frac{m_X^2}{\mu^2} - 
\frac{\Delta m_X}{\Delta^2 - m^2_X}  \cR \left( \frac{\Delta}{m_X} \right) \right]. 
\end{multline}

\begin{figure}[tb]
\centering
\includegraphics[width=0.75\textwidth]{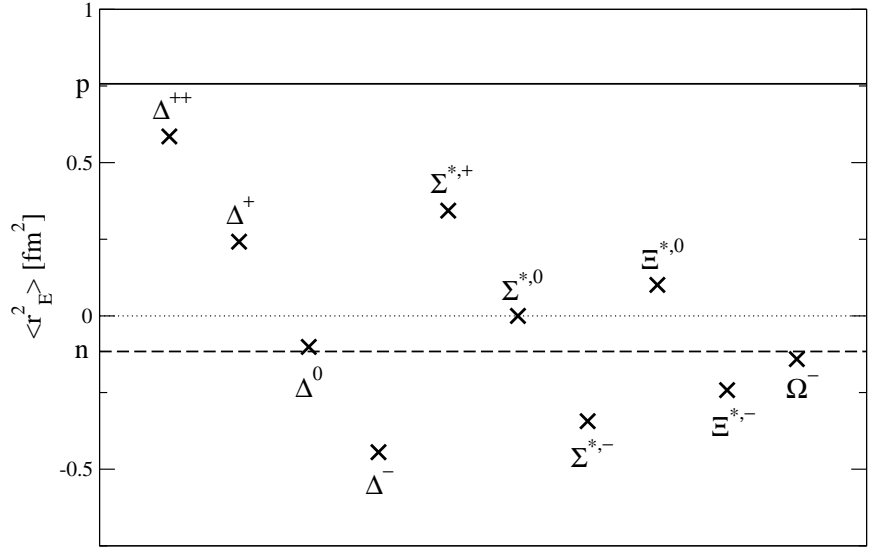}%
\caption[Charge radii of the decuplet baryons in \CPT]{The charge radii of the decuplet baryons in \CPT. The contribution from counterterms has been set to zero. The radii 
(squared) here come from the one-loop diagrams only  and are plotted in units of fm${}^2$. For reference we have shown both the proton and
neutron charge radii~\cite{Hagiwara:2002fs} 
(solid and dashed lines, respectively).}
\label{decuplet-F:radii}
\end{figure}

In the absence of experimental and lattice data for the low energy constants $\mathbb{Q}_c$ and $c_c$, we cannot 
ascertain the contributions to the charge radii from local counterterms in \CPT. 
We can consider, however, just the formally dominant loop contributions.
To this end, we choose the values $\mathcal{C} = - 2 D$ and $\mathcal{H} = - 3 D$ with $D = 0.76$~\cite{Jenkins:1991ne}, 
and the masses
$\D = 270$~MeV, $m_\pi = 140$~MeV, and $m_K = 500$~MeV. The loop contributions to the charge radii in \CPT\ are 
then evaluated for the decuplet at the scale $\mu = 1$~GeV and 
plotted in Fig.~\ref{decuplet-F:radii}.  

\subsection{Analysis in \QCPT}
The calculation of the charge radii and electromagnetic moments can be correspondingly 
executed for \QCPT.
The operators in Eqs.~\eqref{decuplet-eqn:LDF}, \eqref{decuplet-eqn:Lc} and \eqref{decuplet-eqn:Lnew}
contribute, however, their low-energy coefficients
cannot be matched onto QCD.  Therefore we annotate them with a ``Q''.    
The loop contributions encountered in \CPT\ and \PQCPT\ above no longer contribute because  in \QCPT\ $A^T_X = 0$
for all decuplet states $T$. This can be readily seen in two ways. The quenched limit%
\footnote{In this case, the quenched limit
simply means to remove sea quarks and to fix the charges of the ghost quarks to equal those of their light quark counterparts.}
 of the coefficients in Table~\ref{decuplet-t:clebsch}
makes immediate the vanishing of $A_X^T$. Alternately one can consider the relevant quark flow diagrams with only valence 
quarks in loops. Due to the symmetric nature of $T^{ijk}$, these loops are completely canceled by their ghostly counterparts.
For the charge radii, there are no additional diagrams to consider at this order from singlet hairpin interactions. Thus in
\QCPT, the charge radii have the form
\begin{equation}
<r_E^2> = Q \left( \frac{2 \mu_c^Q - 1}{M_B^2} + \frac{\mathbb{Q}_c^Q + 6 c_c^Q}{\L_\chi^2}  \right),
\end{equation} 
where the dependence on the quark mass enters at the next order.

\begin{figure}[tb]
\centering
\includegraphics[width=0.5\textwidth]{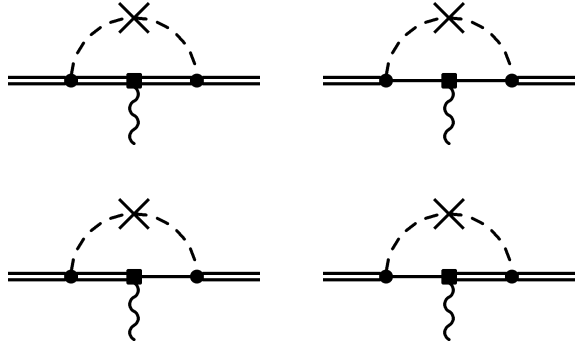}%
\caption[Hairpin diagrams that give contributions of the form $\sim \mu_o^2 \log m_q$ to decuplet electromagnetic moments
in Q$\chi$PT]{Hairpin diagrams that give contributions of the form $\sim \mu_o^2 \log m_q$ to decuplet electromagnetic moments
in Q$\chi$PT. The square at the photon vertex represents the relevant operators from Eqs.~\eqref{decuplet-eqn:LDF}, \eqref{decuplet-eqn:Lc}, and
\eqref{decuplet-eqn:Lbob}.}
\label{decuplet-F:Fquenched}
\end{figure}

Additional terms of the form $\mu_0^2\log m_q$ involving hairpins~\cite{Labrenz:1996jy,Savage:2001dy}
do contribute to the electromagnetic moments as 
they are of the same order in the chiral expansion. 
As explained in~\cite{Chow:1998xc}, the axial hairpins do not contribute.
In the diagrams shown in Fig.~\ref{decuplet-F:Fquenched}, 
one sees that there are contributions from the 
electromagnetic moments of the decuplet and octet baryons as well as their transition moments.
These interactions are described by the operators in Eqs.~\eqref{decuplet-eqn:LDF} and \eqref{decuplet-eqn:Lc} (now
with quenched coefficients) along with
\begin{eqnarray} \label{decuplet-eqn:Lbob}
\cL 
&=& 
\frac{i e }{2 M_B} 
\left\{\mu_\a^Q \left( \ol \cB [S^\mu,S^\nu] \cB \cQ \right) 
+ \mu_\b^Q \left( \ol \cB [S^\mu,S^\nu] \cQ  \cB \right)    \right\} F_{\mu \nu} \nonumber \\
&&+ \sqrt{\frac{3}{2}}  \mu_T^Q \frac{i e}{2 M_B} \left[ \left( \ol \cB  S^\mu \cQ \cT^\nu \right) + \text{h.c.}  \right] F_{\mu \nu} 
+  \sqrt{\frac{3}{2}}\mathbb{Q}_T^Q \frac{e}{\L_\chi^2} 
\left[ \left( \ol \cB  S^{\{\mu} \cQ \cT^{\nu\}} \right) + \text{h.c.}  \right]  v^\a \partial_\mu F_{\nu \a}. \nonumber \\
\end{eqnarray}
It is easier to work with the combinations $\mu_D^Q$ and $\mu_F^Q$ defined by
\begin{equation}
\mu_\a^Q = \frac{2}{3} \mu_D^Q + 2 \mu_F^Q \quad \text{and} \;\; \mu_\b^Q = -\frac{5}{3} \mu_D^Q + \mu_F^Q.
\end{equation}

To calculate the quenched electromagnetic moments, we also need the hairpin wavefunction renormalization diagrams 
shown in Fig.~\ref{decuplet-F:Fdecuplet}. These along with the diagrams in Fig.~\ref{decuplet-F:Fquenched} are economically expressed 
in terms of the function
\begin{align} \label{decuplet-eqn:I}
I(m_1,m_2,\D_1,\D_2,\mu) & = \frac{Y(m_1,\D_1,\mu) +Y(m_2,\D_2,\mu) - Y(m_1,\D_2,\mu) - Y(m_2,\D_1,\mu)}{(m_1^2 - m_2^2)(\D_1 - \D_2)} 
\notag  \\
& = - i \frac{16 \pi^2}{3} \int \frac{d^D k}{(2\pi)^D} \frac{k^2 - (k\cdot v)^2}{(k^2 - m_1^2)(k^2 - m_2^2) 
(k\cdot v - \D_1) (k\cdot v - \D_2)},
\end{align}
where
\begin{equation}
Y(m,\D,\mu) = \D \left(m^2 - \frac{2}{3} \D^2\right) \log \frac{m^2}{\mu^2} + \frac{2}{3} m (\D^2  - m^2) 
\mathcal{R}\left(\frac{\D}{m}\right)
\end{equation}
and we have kept only non-analytic contributions.  The following shorthands are convenient
\begin{align}
I_{\eta_q \eta_{q^\prime}} & = I(m_{\eta_q},m_{\eta_{q^\prime}},0,0,\mu), \notag \\
I^{\D }_{\eta_q \eta_{q^\prime}} & = I(m_{\eta_q},m_{\eta_{q^\prime}},\D,0,\mu), \notag \\ 
I^{\D \D}_{\eta_q \eta_{q^\prime}} & = I(m_{\eta_q},m_{\eta_{q^\prime}},\D,\D,\mu)
.\end{align}
Specific limits of the function $I_{\eta_q \eta_{q^\prime}}$
appear in~\cite{Savage:2001dy}.
The wavefunction renormalization factors arising from hairpin diagrams can then be expressed as
\begin{equation} \label{decuplet-eqn:Z}
Z = 1 - \frac{\mu_o^2}{16 \pi^2 f^2} \sum_{XX^\prime} \left[  \frac{1}{2} (\mathcal{C}^Q)^2 C_{XX^\prime} I_{XX^\prime} + \frac{5}{9} (\mathcal{H}^Q)^2 B_{XX^\prime} I^{\D\D}_{XX^\prime} 
 \right].
\end{equation}
The coefficients $B_{XX^\prime}$ and $C_{XX^\prime}$ are listed in 
Table~\ref{decuplet-t:quenched}. 
\begin{table}[tb]
\centering
\caption[SU$(3)$ coefficients $B_{XX^\prime}$ and $C_{XX^\prime}$ for the decuplet in \QCPT]{The SU$(3)$ coefficients $B_{XX^\prime}$ and $C_{XX^\prime}$ in \QCPT.}  
\begin{tabular}{c | c  c  c |  c  c  c }
  \hline\hline
	    & \multicolumn{3}{c}{$B_{XX^\prime}$}  &\multicolumn{3}{c}{$C_{XX^\prime}$}   \\
	                 & $\quad \eta_u \eta_u$   &   $\quad \eta_u \eta_s$   &   $\quad \eta_s \eta_s$   &  $\quad \eta_u \eta_u$  &   $\quad \eta_u \eta_s$   &   $ \quad \eta_s \eta_s$   \\
	\hline
	$\Delta^{++},\Delta^{+},\Delta^{0},\Delta^{-} $       &  $1$ & $0$ & $0$   &  $0$ & $0$ & $0$  \\
	$\Sigma^{*,+},\Sigma^{*,0},\Sigma^{*,-}$       	      &  $\frac{4}{9}$ & $\frac{4}{9}$ & $\frac{1}{9}$   &  $\frac{2}{9}$ & $-\frac{4}{9}$ & $\frac{2}{9}$  \\
	$\Xi^{*,0},\Xi^{*,-}$       			      &  $\frac{1}{9}$ & $\frac{4}{9}$ & $\frac{4}{9}$   &  $\frac{2}{9}$ & $-\frac{4}{9}$ & $\frac{2}{9}$  \\
	$\Omega^-$                                            &  $0$ & $0$ & $1$   &  $0$ & $0$ & $0$  \\
\hline\hline
\end{tabular}
\label{decuplet-t:quenched}
\end{table} 
The sum in Eq.~\eqref{decuplet-eqn:Z} is
over $XX^\prime = \eta_u \eta_u$, $\eta_u \eta_s$, $\eta_s \eta_s$. Combining these factors with the tree-level contributions
and including the corrections from the diagrams in Fig.~\ref{decuplet-F:Fquenched}, we arrive at the quenched decuplet magnetic moment
\begin{eqnarray} \label{decuplet-eqn:qF2}
\mu &=& 2 \mu_c^Q Q Z + \frac{\mu_o^2}{16 \pi^2 f^2} \sum_{XX^\prime} \left[ \frac{1}{2} 
(Q \mu_F^Q + \a_D \mu_D^Q) (\mathcal{C}^Q)^2 C_{XX^\prime} I_{XX^\prime} \right] \nonumber \\
&&+  \frac{\mu_o^2}{16 \pi^2 f^2} \sum_{XX^\prime} \left[  \frac{22}{27} (\mathcal{H}^Q)^2  \mu_c^Q B_{XX^\prime} Q I^{\D \D}_{XX^\prime} - \frac{2}{9} \mathcal{C}^Q \mathcal{H}^Q \mu_T^Q D_{XX^\prime} I^\D_{XX^\prime} 
\right]
\end{eqnarray}
and the quenched electric quadrupole moment
\begin{eqnarray} \label{decuplet-eqn:qG1}
\mathbb{Q} &=& - 2  Q \left( \mu_c^Q + \mathbb{Q}_c^Q \frac{2 M_B^2}{\L_\chi^2} \right) Z  - \frac{\mu_o^2}{16 \pi^2 f^2}  
\frac{M_B^2}{\L_\chi^2} \sum_{XX^\prime} 
\left( \frac{8}{3} \mathcal{C}^Q \mathcal{H}^Q \mathbb{Q}_T^Q D_{XX^\prime} I^\D_{XX^\prime}\right) \nonumber \\
&&- \frac{\mu_o^2}{16 \pi^2 f^2}  \sum_{XX^\prime} \left[  (\mathcal{H}^Q)^2 \left( \frac{2}{9} \mu_c^Q + \frac{4}{9} \mathbb{Q}_c^Q \frac{M_B^2}{\L_\chi^2}\right) Q B_{XX^\prime} 
I^{\D \D}_{XX^\prime} - \frac{2}{3} \mathcal{C}^Q \mathcal{H}^Q \mu_T^Q D_{XX^\prime} I^\D_{XX^\prime} 
\right]. \nonumber \\
\end{eqnarray}
In Eq.~\eqref{decuplet-eqn:qF2} the required values for the constant $\a_D$ are:  $\a_D  = 
\frac{1}{3}$ for $\Sigma^{*,+}$, $\Sigma^{*,0}$, $\Sigma^{*,-}$, and $\Xi^{*,-}$, 
and $\a_D = - \frac{2}{3}$  for  $\Xi^{*,0}$. The coefficients $D_{XX^\prime}$ are listed in Table~\ref{decuplet-T:D}. 
\begin{table}[tb]
\centering
\caption[SU$(3)$ coefficients $D_{XX^\prime}$ for the decuplet in \QCPT]{The SU$(3)$ coefficients $D_{XX^\prime}$ in \QCPT. Decuplet states not listed have $D_{XX'} = 0$.}  
\begin{tabular}{l | c  c  c  }
  \hline\hline
	                 & $\quad \eta_u \eta_u$   &   $\quad \eta_u \eta_s$   &   $\quad \eta_s \eta_s$    \\
	 	
	\hline
	$\Sigma^{*,+}$  &  $\frac{2}{9}$ & $-\frac{1}{9}$ & $-\frac{1}{9}$ \\ 
	$\Sigma^{*,0}$  &  $\frac{1}{9}$ & $- \frac{1}{18}$ & $-\frac{1}{18}$ \\      
    	$\Xi^{*,0}$     &  $\frac{1}{9}$ & $\frac{1}{9}$ & $-\frac{2}{9}$ \\
\hline\hline
\end{tabular}
\label{decuplet-T:D}
\end{table} 
If a particular decuplet state is not listed, the value of $D_{XX^\prime}$ is zero for all singlet pairs $XX^\prime$.

The above expressions can be used to properly extrapolate quenched lattice data to the physical pion mass.
For example, the expression for the quenched magnetic moments for the $\D$ baryons [Eq.~\eqref{decuplet-eqn:qF2}]
reduces to
\begin{equation}
\mu = 2 Q \mu_c^Q \left( 1 - \frac{4}{27} \, \frac{\mu_o^2}{16 \pi^2 f^2} (\mathcal{H}^Q)^2 I^{\D \D}_{\eta_u \eta_u}  \right)
.\end{equation}
In the above expression we need 
\begin{equation}
  I^{\D \D}_{XX}
  =
  \log\frac{m_X^2}{\mu^2} - \frac{\D m_X}{\D^2-m_X^2}\cR\left(\frac{\D}{m_X}\right)+\dots   
.\end{equation}
where the ellipses denote terms analytic in $m_X$.
Utilizing a least squares analysis
and using the values $\mu=1$~GeV and $\D=270$~MeV 
, we extrapolate the quenched lattice data~\cite{Leinweber:1992hy} 
to the physical pion mass and find for the $\D$ resonances
\begin{equation}
\mu = 2.89 \, Q \; [\mu_N]
.\end{equation}
This is in contrast to the value $\mu \approx 2.49 \, Q \; [\mu_N]$ found from carrying out a \CPT-type extrapolation~\cite{Cloet:2003jm}.
Notice that for many of the decuplet states, in particular the $\D$ baryons, the quenched magnetic moments and electric
quadrupole moments are proportional to the charge $Q$ (unlike the \CPT\ and \PQCPT\ results). 
This elucidates the trends seen in the quenched lattice data~\cite{Leinweber:1992hy}.

\section{\label{decuplet-sec:conclusions}Conclusions}
We have calculated the electromagnetic moments and charge radii for the SU($3$) decuplet 
baryons in the isospin limit of \PQCPT\
and also derived the result for
the baryon quartet away from the isospin limit for the $SU(2)$
chiral Lagrangian. The $q^2$ dependence of decuplet form factors at one-loop appears in Appendix~\ref{decuplet-s:q-dep}. 
We have also calculated the \QCPT\ results.

For the decuplet baryons' electromagnetic moments and charge radii,
the operators,
which are included in the \QCPT\ but not in the \CPT\ Lagrangian,
enter at NNLO in the \QCPT\ expansion.
Hence, formally 
our NLO result is not more divergent than its QCD counterpart.
This, however, does not mean that our result is free of quenching
artifacts.
While the expansions about the chiral limit for QCD and QQCD
charge radii are formally similar,
$<r^2>\sim\a+\b\log m_Q+\dots$,
the QQCD result consists \emph{entirely} of quenched oddities:
for all decuplet baryons,
diagrams that have bosonic or fermionic mesons running in loops 
completely cancel so that $\b=0$.  In other words, the quenched decuplet 
charge radii have the behavior $<r^2>\sim\a+\dots$ 
and the result is actually independent of $m_Q$ at this order.
For the quenched decuplet magnetic moments and electric quadrupole moments, 
expansions about the chiral limit are again formally similar:
$\mu \sim \a + \b \log m_Q + \gamma \sqrt{m_Q} + \ldots$ and
$\mathbb{Q} \sim \a + \b \log m_Q + \ldots$. 
However, quenching forces $\gamma = 0$ and both $\b$'s arise from singlet contributions
involving the parameter $\mu_o$, which is of course absent in QCD. Thus
the leading non-analytic quark mass dependence that remains for these observables
is entirely a quenched peculiarity.

PQQCD, on the other hand, is free of such excentric
behavior and
our \PQCPT\ result will 
enable the proper extrapolation of PQQCD lattice simulations
of the decuplet electromagnetic moments and charge radii
and we hope it encourages such simulations in the future.

\chapter{Baryon Decuplet to Octet Electromagnetic 
       Transitions in \QCPT\ and \PQCPT}
\label{chapter:trans}
In this chapter,
we calculate baryon decuplet to octet
electromagnetic transition form factors
in quenched and
partially quenched chiral perturbation theory.
Again, we work in the isospin limit of $SU(3)$ flavor, 
up to NLO in the chiral expansion,
and to leading order in the heavy baryon expansion.
Our results are necessary for proper extrapolation
of lattice calculations of these transitions.
We also derive expressions for the case
of $SU(2)$ flavor
away from the isospin limit.

\section{Introduction}
The study of the baryon
decuplet to octet electromagnetic transitions
provides important insight into the strongly interacting
regime of QCD.
Spin-parity selection rules for these
transitions allow for magnetic dipole (M1),
electric quadrupole (E2), and Coulumb quadrupole (C2) amplitudes.
Understanding these amplitudes, both in theory and experiment,
gives insight into the ground state wavefunctions
of the lowest lying baryons.
For example,
in the transition of the $\D(1232)$ to the nucleon,
if both baryon wavefunctions are spherically symmetric
then the E2 and C2 amplitudes vanish.
Experimentally, M1 is seen to be the dominant amplitude.
However, 
recent experimental measurements of the quadrupole amplitudes 
in the $\D\to N\g$ transition%
~\cite{Mertz:1999hp,Joo:2001tw}
show that the quadrupole amplitudes E2 and C2 are likely
non-zero.
This has revitalized the discussion as to the mechanism
for deformation of the baryons.
Although we expect more experimental data in the future,
progress 
will be slower for the remaining transitions as the
experimental difficulties are large.

First-principle lattice QCD calculations of these matrix elements
can provide a theoretical explanation of these
experimental results.
In fact, 
the experimental difficulties may force us to rely
on lattice data 
for the non-nucleonic transitions.
Recently several such lattice calculations%
~\cite{Alexandrou:2002pw,Alexandrou:2003ea},
which improve upon
an earlier one~\cite{Leinweber:1993pv}, have appeared.
Unfortunately 
now and foreseeably, these lattice calculations
cannot be performed with the physical masses of the light quarks
and must be
extrapolated  to
the physical light quark masses.

For future lattice calculations that use the partially
quenched approximation
of QCD
one needs to fit
\PQCPT\ to the
lattice data in order to
determine the low-energy constants  
and to actually make physical predictions for QCD.
Unfortunately, partially quenched simulations
for the 
$\D\to N\g$ transition
do not exist yet;
what does exist are simulations%
~\cite{Alexandrou:2002pw,Alexandrou:2003ea}
that use the quenched approximation.
These simulations must be fit using \QCPT\
that exhibit the sickness outlined in the previous chapters.
However, 
we expect to see partially quenched calculations of these
form factors in the near future.

This chapter is organized as follows.
In Section~\ref{trans-sec:ff} we calculate
baryon decuplet to octet transition form factors
in both \QCPT\ and \PQCPT\
up to NLO in the chiral expansion
and keep contributions to lowest order in the heavy baryon mass, $M_B$.
These calculations are done in the 
isospin limit of $SU(3)$ flavor.  
For completeness we also provide the
\PQCPT\ 
results for the transitions using the $SU(2)$ chiral
Lagrangian with non-degenerate quarks in 
Appendix~\ref{trans-s:su2}.
In Section~\ref{trans-sec:conclusions} we conclude.

\section{\label{trans-sec:ff}Baryon Decuplet to Octet Transition}
The electromagnetic baryon decuplet to octet
transitions
have been investigated previously in \CPT%
~\cite{Butler:1993pn,Butler:1993ht,Napsuciale:1997ny,Gellas:1998wx}.
Very recently there also has been renewed interest in these transitions
in the large $N_c$ limit of QCD~\cite{Jenkins:2002rj}. 
Here we calculate these transitions in
\PQCPT\ and \QCPT.
While we have reviewed \PQCPT\ briefly in the last section 
and our recent papers~\cite{Arndt:2003ww,Arndt:2003we},
for \QCPT\ we refer the reader to the literature%
~\cite{Morel:1987xk,Sharpe:1992ft,Bernard:1992ep,
Bernard:1992mk,Golterman:1994mk,Sharpe:1996qp,Labrenz:1996jy}.

Using the heavy baryon formalism%
~\cite{Jenkins:1991jv,Jenkins:1991ne}, 
transition matrix elements of the electromagnetic current 
$J^\rho$ between a decuplet baryon with momentum
$p'$ and an octet baryon with momentum $p$
can be parametrized as%~\cite{Jones:1973ky}
\begin{equation}
  \langle{\ol B}(p)|J^\rho|T(p')\rangle
  =
  {\ol u}(p)\cO^{\rho\mu}u_\mu(p')
,\end{equation}
where $u_\mu(p)$ is a Rarita-Schwinger spinor for an on-shell decuplet
baryon satisfying $v^\mu u_\mu(p)=0$ and $S^\mu u_\mu(p)=0$.
The tensor $\cO^{\rho\mu}$ can be parametrized 
in terms of three independent,
Lorentz invariant, dimensionless form factors%
~\cite{Jones:1973ky}
\begin{eqnarray}
  \mathcal{O}^{\rho \mu} 
  &=&
  \frac{G_1(q^2)}{M_B}
  \left(q\cdot S g^{\mu\rho} -q^\mu S^\rho\right)
  +
  \frac{G_2(q^2)}{(2M_B)^2}
  \left(q\cdot v g^{\mu\rho}-q^\mu v^\rho\right)S\cdot q
                          \nonumber \\
  &&+
  \frac{G_3(q^2)}{4M_B^2\D}
  \left(q^2 g^{\mu\rho}-q^\mu q^\rho\right)S\cdot q
,\end{eqnarray}
where the momentum of the outgoing photon is $q = p' - p$.
Here we have adopted the normalization of the $G_3(q^2)$
form factor used in~\cite{Gellas:1998wx}
so that the leading contributions to all three form factors 
are of order unity
in the power counting.

Linear combinations of the above form 
factors at $q^2=0$ make the magnetic dipole, electric quadrupole, 
and Coulombic quadrupole moments,
\begin{eqnarray}
  G_{M1}(0)&=&\left(\frac{2}{3}-\frac{\D}{6M_B}\right)G_1(0) 
             +\frac{\D}{12M_B}G_2(0),   \nonumber \\
  G_{E2}(0)&=&\frac{\D}{6M_B}G_1(0)+\frac{\D}{12M_B}G_2(0), \nonumber \\
  G_{C2}(0)&=&\left(\frac{1}{3}+\frac{\D}{6M_B}\right)G_1(0)
             +\left(\frac{1}{6}+\frac{\D}{6M_B}\right)G_2(0)
               +\frac{1}{6}G_3(0)   
.\end{eqnarray}

\subsubsection{Analysis in \PQCPT}
Let us first consider the transition form factors 
in \PQCPT.
Here, the leading tree-level contributions to the transition moments 
come from the
dimension-5 and dimension-6 operators
\begin{equation} \label{trans-eqn:Lbob}
  \cL 
  =
  \sqrt{\frac{3}{2}}\mu_T \frac{i e}{2 M_B} 
  \left( \ol \cB  S^\mu \cQ \cT^\nu \right)F_{\mu \nu} 
  +  
  \sqrt{\frac{3}{2}}\mathbb{Q}_T \frac{e}{\L_\chi^2} 
  \left( \ol \cB  S^{\{\mu} \cQ \cT^{\nu\}} \right)
  v^\a \partial_\mu F_{\nu \a}
\end{equation}
where
the action of ${}^{\{}\ldots{}^{\}}$ on Lorentz indices produces 
the symmetric traceless part of the tensor, 
{\it viz.}, 
$\mathcal{O}^{\{\mu \nu\}} 
 = \mathcal{O}^{\mu \nu} + \mathcal{O}^{\nu \mu} - 
\frac{1}{2} g^{\mu\nu} \mathcal{O}^{\alpha}{}_{\alpha}$.
Here the PQQCD low-energy constants $\mu_T$ and $\mathbb{Q}_T$
have the same numerical values as in QCD.

\begin{figure}[tb]
  \centering
  \includegraphics[width=\textwidth]{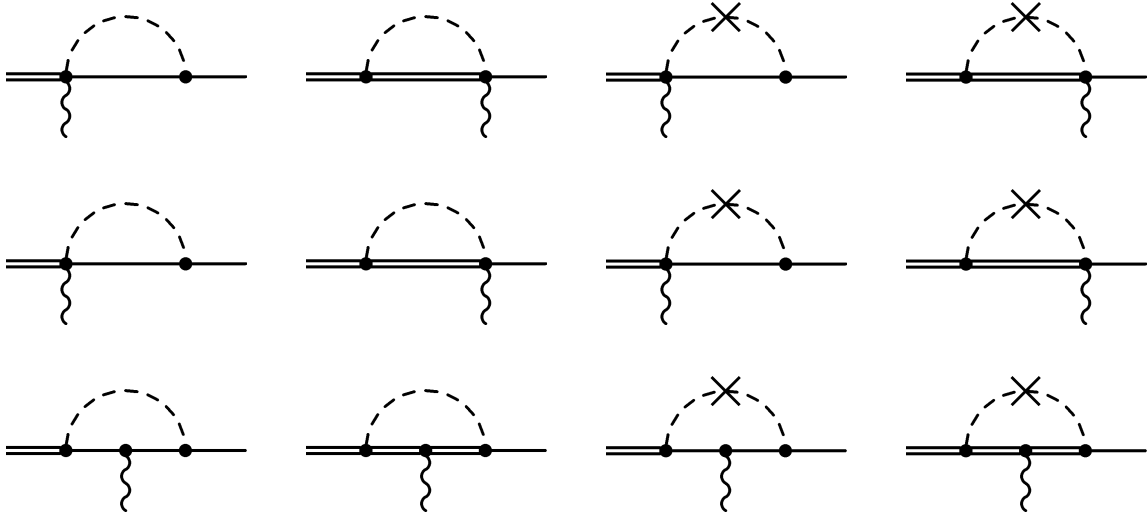}%
  \caption[Loop diagrams that contribute to the transition
     moments but are zero to the order we are working]{\label{trans-F:D2B-PQ-thatarezero}
     Loop diagrams that contribute to the transition
     moments but are zero to the order we are working.
     A thin (thick) solid line denotes an octet (decuplet)
     baryon whereas a dashed line denotes a meson.}
\end{figure}

\begin{figure}[tb]
  \centering
  \includegraphics[width=0.50\textwidth]{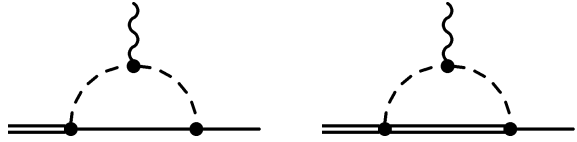}%
  \caption{\label{trans-F:D2B-PQ}
     Loop diagrams contributing to the transition moments}
\end{figure}

The NLO contributions in the chiral expansion
arise from the one-loop diagrams shown in
Figs.~(\ref{trans-F:D2B-PQ-thatarezero}) and (\ref{trans-F:D2B-PQ}).
However, because of the constraints satisfied by the on-shell
Rarita-Schwinger spinors, the diagrams in 
Fig.~(\ref{trans-F:D2B-PQ-thatarezero}) are all identically zero.
For the calculation
of the diagrams in Fig.~(\ref{trans-F:D2B-PQ}) 
we need the terms in the Lagrangian describing the interaction
of the $\cB_{ijk}$ and $\cT_{ijk}$ with the
pseudo-Goldstone mesons given in 
Eqs.~\eqref{baryon_ff-eqn:Linteract} and \eqref{decuplet-eq:H}:
\begin{equation} \label{trans-eqn:Linteract}
  {\cal L} 
  =
  2\a\left(\ol{\cB}S^\mu \cB A_\mu\right)
  +
  2\b\left(\ol{\cB}S^\mu A_\mu \cB\right)
  +
  \sqrt{\frac{3}{2}}\cC
  \left[
    \left(\ol{\cT}^\nu A_\nu \cB\right)+\text{h.c.}
  \right]  
  +  
  2{\mathcal H}\left(\ol{\cT}^\nu S^\mu A_\mu \cT_\nu\right) 
.\end{equation}
We find
\begin{eqnarray}\label{trans-eqn:G1}
  G_1(0)
  &=&
  \frac{\mu_T}{2}\a_T
  +
  \frac{M_B}{\L_\chi^2}4\cH\cC
  \sum_X\b_X^T
  \int_0^1 dx\,\left(1-\frac{x}{3}\right)
  \left[
    x\D\log\frac{m_X^2}{\mu^2}
      -m_X\cR\left(\frac{x\D}{m_X}\right)
  \right] \nonumber \\
  &&-
  \frac{M_B}{\L_\chi^2}4\cC(D-F)
  \sum_X\b_X^B
  \int_0^1 dx\,(1-x)
  \left[
    x\D\log\frac{m_X^2}{\mu^2}
      +m_X\cR\left(-\frac{x\D}{m_X}\right)
  \right]  
, \nonumber \\
\end{eqnarray}
\begin{eqnarray}
  G_2(0)
  &=&
  \frac{M_B^2}{\L_\chi^2}
  \Bigg\{
  -4\mathbb{Q}_T\a_T \nonumber \\
  &&\phantom{xxxxx}+
  16\cH\cC
  \sum_X\b_X^T
  \int_0^1 dx\,\frac{x(1-x)}{3}
  \left[
    \log\frac{m_X^2}{\mu^2}
      +\frac{x\D m_X}{m_X^2-x^2\D^2}
             \cR\left(\frac{x\D}{m_X}\right)
  \right]\nonumber \\
  &&-
  16\cC(D-F)
  \sum_X\b_X^B
  \int_0^1 dx\,x(1-x)
  \left[
    \log\frac{m_X^2}{\mu^2}
      -\frac{x\D m_X}{m_X^2-x^2\D^2}
           \cR\left(-\frac{x\D}{m_X}\right)
  \right]
  \Bigg\}  
, \nonumber \\
\end{eqnarray}
and
\begin{eqnarray}\label{trans-eqn:G3}
  G_3(0)
  &=&
  -\frac{M_B^2}{\L_\chi^2}16
  \left[
  \cH\cC
  \sum_X\b_X^T
  \int_0^1 dx\,\frac{x(1-x)}{3}\left(x-\frac{1}{2}\right)
           \frac{\D m_X}{m_X^2-x^2\D^2}
             \cR\left(\frac{x\D}{m_X}\right)
  \right.\nonumber \\
  &&\left.
  +
  \cC(D-F)
  \sum_X\b_X^B
  \int_0^1 dx\,x(1-x)\left(x-\frac{1}{2}\right)
            \frac{\D m_X}{m_X^2-x^2\D^2}
           \cR\left(-\frac{x\D}{m_X}\right)
  \right]
, \nonumber \\
\end{eqnarray}
where
the function $\mathcal{R}(x)$ is defined in
Eq.~\eqref{appendix-B2D-eq:R}
and we have only kept loop contributions that
are non-analytic in the meson mass $m_X$.
The tree-level coefficients $\a_T$ are listed in Table~\ref{trans-t:tree}
\begin{table}[tb]
\centering 
\caption[Tree-level coefficients $\a_T$ in \CPT, \QCPT, and \PQCPT]{\label{trans-t:tree}
Tree-level coefficients $\a_T$ in \CPT, \QCPT, and \PQCPT.}
\begin{tabular}{c | c  }
        \hline\hline
		              & $\a_T$ \\
	\hline
	$\D \to N \gamma$          
	& $\frac{1}{\sqrt{3}}$
	\\  
	$\Sigma^{*,+} \to \Sigma^+ \gamma$  
	& $-\frac{1}{\sqrt{3}}$
	\\  
	$\Sigma^{*,0} \to \Sigma^0 \gamma$            
	& $\frac{1}{2\sqrt{3}}$
	\\  
	$\Sigma^{*,0} \to \Lambda \gamma$            
	& $-\frac{1}{2}$	
	\\  
	$\Sigma^{*,-} \to \Sigma^- \gamma$            
	& $0$
	\\  
	$\Xi^{*,0} \to \Xi^{0} \gamma$            
	& $-\frac{1}{\sqrt{3}}$
	\\  
	$\Xi^{*,-} \to \Xi^- \gamma$            
	& $0$
        \\
        \hline\hline
\end{tabular}
\end{table} 
and the coefficients for the loop diagrams in Fig.~(\ref{trans-F:D2B-PQ}),
$\b_X^T$ and $\b_X^B$,
are given in Tables~\ref{trans-t:clebschT} and \ref{trans-t:clebschB},
respectively.
%\begin{turnpage}
\begin{sidewaystable}[tb]
\centering
\caption[$SU(3)$ coefficients $\beta_X^T$ in \CPT\ and \PQCPT]{\label{trans-t:clebschT}
The $SU(3)$ coefficients $\beta_X^T$ in \CPT\ and \PQCPT.}
%\begin{ruledtabular}
\begin{tabular}{c | c c | c  c  c  c  c  c  c }
  \hline\hline
	& \multicolumn{2}{c |}{\CPT}  &  \multicolumn{7}{c}{\PQCPT} \\
	& $\pi$ & $K$  &   $\pi$   &   $K$   &  $\eta_s$  &   $ju$   
             &   $ru$   & $js$   & $rs$ 
	\\
	\hline
	$\D \to N \gamma$          
	&   $\frac{5}{3 \sqrt{3}}$    
	& $\frac{1}{3 \sqrt{3}}$   
	&   $\frac{1}{\sqrt{3}}$      
	&   $0$    
	&  $0$        
	&   $\frac{2}{3\sqrt{3}}$     
	&   $\frac{1}{3\sqrt{3}}$     
	&   $0$     
	& $0$ 
	\\
	$\Sigma^{*,+} \to \Sigma^+ \gamma$  
	&   $-\frac{1}{3 \sqrt{3}}$    
	& $-\frac{5}{3 \sqrt{3}}$   
	&   $\frac{1 - 3 q_{jl}}{9\sqrt{3}} $   
	&   $-\frac{11 - 3 q_{jl} + 3 q_r}{9\sqrt{3}}$    
	&  $\frac{1 + 3 q_r}{9\sqrt{3}}$        
	&   $-\frac{4 - 3 q_{jl}}{9\sqrt{3}}$     
	&   $-\frac{2 - 3 q_r}{9\sqrt{3}}$     
	&   $-\frac{2 + 3 q_{jl}}{9\sqrt{3}}$     
	& $-\frac{1 + 3 q_r}{9\sqrt{3}}$ 
	\\
 	$\Sigma^{*,0} \to \Sigma^0 \gamma$            
	&   $0$    
	& $\frac{1}{\sqrt{3}}$   
	&   $-\frac{1 - 3 q_{jl}}{9\sqrt{3}}$    
	&   $\frac{13 - 6 q_{jl} + 6 q_r}{18 \sqrt{3}}$    
	&  $- \frac{1 + 3 q_r}{9 \sqrt{3}}$   
    	&   $\frac{1 - 3 q_{jl}}{9\sqrt{3}}$     
	&   $\frac{1 - 6 q_r}{18 \sqrt{3}}$     
	&   $\frac{2 + 3 q_{jl}}{9\sqrt{3}}$     
	& $\frac{1 + 3 q_{r}}{9\sqrt{3}}$ 
	\\
	$\Sigma^{*,0} \to \Lambda \gamma$            
	&   $-\frac{2}{3}$    
	& $-\frac{1}{3}$   
	&   $-\frac{1}{3}$      
	&   $-\frac{1}{6}$    
	&  $0$        
	&   $-\frac{1}{3}$     
	&   $-\frac{1}{6}$     
	&   $0$     
	& $0$ 	
	\\
	$\Sigma^{*,-} \to \Sigma^- \gamma$            
	&   $-\frac{1}{3 \sqrt{3}}$    
	& $\frac{1}{3 \sqrt{3}}$   
	&   $-\frac{1 - 3 q_{jl}}{9\sqrt{3}}$      
	&   $\frac{2 - 3 q_{jl} + 3 q_r}{9 \sqrt{3}}$    
	&  $-\frac{1 + 3 q_{r}}{9\sqrt{3}}$   
     	&   $-\frac{2 + 3 q_{jl}}{9 \sqrt{3}}$     
	&   $-\frac{1 + 3 q_{r}}{9 \sqrt{3}}$     
	&   $\frac{2 + 3 q_{jl}}{9 \sqrt{3}}$     
	&    $\frac{1 + 3 q_{r}}{9 \sqrt{3}}$ 
	\\
	$\Xi^{*,0} \to \Xi^{0} \gamma$            
	&   $-\frac{1}{3 \sqrt{3}}$    
	& $-\frac{5}{3 \sqrt{3}}$   
	&   $\frac{1 - 3 q_{jl}}{9\sqrt{3}}$      
	&   $-\frac{11 - 3 q_{jl} + 3 q_r}{9 \sqrt{3}}$    
	&  $\frac{1 + 3 q_{r}}{9\sqrt{3}}$    
	&   $-\frac{4 - 3 q_{jl}}{9\sqrt{3}}$     
	&   $-\frac{2 - 3 q_{r}}{9\sqrt{3}}$     
	&   $-\frac{2 + 3 q_{jl}}{9\sqrt{3}}$     
	& $-\frac{1 + 3 q_{r}}{9\sqrt{3}}$ 
	\\
	$\Xi^{*,-} \to \Xi^- \gamma$            
	&   $-\frac{1}{3 \sqrt{3}}$    
	& $\frac{1}{3 \sqrt{3}}$   
	&   $-\frac{1 - 3 q_{jl}}{9\sqrt{3}}$      
	&   $\frac{2 - 3 q_{jl} + 3 q_r}{9 \sqrt{3}}$    
	&  $-\frac{1 + 3 q_{r}}{9\sqrt{3}}$     
	&   $-\frac{2 + 3 q_{jl}}{9\sqrt{3}}$     
	&   $-\frac{1 + 3 q_{r}}{9\sqrt{3}}$     
	&   $\frac{2 + 3 q_{jl}}{9\sqrt{3}}$     
	& $\frac{1 + 3 q_{r}}{9\sqrt{3}}$ \\
  \hline\hline
\end{tabular}
%\end{ruledtabular}
%\end{sidewaystable} 
\spacebetweentables
%\begin{sidewaystable}[tb]
\centering
\caption[$SU(3)$ coefficients $\beta_X^B$ in \CPT\ and \PQCPT]{\label{trans-t:clebschB}
The $SU(3)$ coefficients $\beta_X^B$ in \CPT\ and \PQCPT.}
%\begin{ruledtabular}
\begin{tabular}{c | c c | c  c  c  c  c  c  c }
  \hline\hline
	& \multicolumn{2}{c |}{\CPT}  &  \multicolumn{7}{c}{\PQCPT} \\
	& $\pi$ & $K$  &   $\pi$   &   $K$   &  $\eta_s$  &   $ju$   
                  &   $ru$   & $js$   & $rs$ \\
	\hline
	$\D \to N \gamma$          
	&   $-\frac{D + F}{\sqrt{3}(D-F)} $    
	&   $- \frac{1}{\sqrt{3}} $   
	&   $\frac{D - 3 F}{\sqrt{3}(D-F)}$    
	&   $0$    
	&  $0$        
	&   $-\frac{2}{\sqrt{3}}$    
	&   $- \frac{1}{\sqrt{3}}$     
	&   $0$    
	&   $0$ 
	\\
	$\Sigma^{*,+} \to \Sigma^+ \gamma$  
	&   $\frac{1}{\sqrt{3}} $    
	&   $\frac{D + F}{\sqrt{3}(D-F)} $   
	&   $- \frac{1 - 3 q_{jl}}{3\sqrt{3}} $    
	&   $- \frac{D - 7 F}{3\sqrt{3}(D-F)} + \frac{q_{jl}-q_r}{\sqrt{3}}$  
	&  $- \frac{1 + 3 q_r}{3 \sqrt{3}}$        
	&   $\frac{4 - 3 q_{jl}}{3 \sqrt{3}}$    
	&   $\frac{2 - 3 q_r}{3 \sqrt{3}}$     
	&   $\frac{2 + 3 q_{jl}}{3\sqrt{3}}$    
	&   $\frac{1 + 3 q_r}{3\sqrt{3}}$ 
	\\
	$\Sigma^{*,0} \to \Sigma^0 \gamma$            
	&   $0$    
	&   $-\frac{D}{\sqrt{3}(D-F)} $   
	&   $\frac{1 - 3 q_{jl}}{3\sqrt{3}}$    
	&   $-\frac{D + 5 F}{6\sqrt{3}(D-F)} - \frac{q_{jl}-q_r}{\sqrt{3}}$  
	&   $\frac{1 + 3 q_r}{3 \sqrt{3}}$        
	&   $-\frac{1 - 3 q_{jl}}{3\sqrt{3}}$    
	&   $-\frac{1 - 6 q_{r}}{6\sqrt{3}}$     
	&   $-\frac{2+ 3 q_{jl}}{3\sqrt{3}}$    
	&   $-\frac{1+ 3 q_{r}}{3\sqrt{3}}$ 
	\\
	$\Sigma^{*,0} \to \Lambda \gamma$            
	&   $\frac{2D}{3(D-F)}$    
	&   $\frac{D}{3(D-F)}$   
	&   $-\frac{D-3F}{3(D - F )}$    
	&   $-\frac{D-3F}{6(D - F )}$    
	&  $0$        
	&   $1$    
	&   $\frac{1}{2}$     
	&   $0$    
	&   $0$ 
	\\
	$\Sigma^{*,-} \to \Sigma^- \gamma$            
	&   $\frac{1}{\sqrt{3}} $    
	&   $-\frac{1}{\sqrt{3}}$   
	&   $\frac{1 - 3 q_{jl}}{3\sqrt{3}}$    
	&   $-\frac{2 - 3 q_{jl} + 3 q_r}{3\sqrt{3}}$    
	&  $\frac{1+ 3 q_{r}}{3\sqrt{3}}$        
	&   $\frac{2 + 3 q_{jl}}{3\sqrt{3}}$    
	&   $\frac{1+ 3 q_{r}}{3\sqrt{3}}$     
	&   $-\frac{2 + 3 q_{jl}}{3\sqrt{3}}$    
	&   $-\frac{1+ 3 q_{r}}{3\sqrt{3}}$ 
	\\
	$\Xi^{*,0} \to \Xi^{0} \gamma$            
	&   $\frac{1}{\sqrt{3}} $    
	&   $\frac{D + F}{\sqrt{3}(D-F)} $   
	&   $-\frac{1 - 3 q_{jl}}{3\sqrt{3}}$    
	&   $- \frac{D - 7 F}{3\sqrt{3}(D-F)} + \frac{q_{jl}-q_r}{\sqrt{3}}$   
	&  $-\frac{1+ 3 q_{r}}{3\sqrt{3}}$        
	&   $\frac{4 - 3 q_{jl}}{3\sqrt{3}}$    
	&   $\frac{2- 3 q_{r}}{3\sqrt{3}}$     
	&   $\frac{2 + 3 q_{jl}}{3\sqrt{3}}$    
	&   $\frac{1 + 3 q_{r}}{3\sqrt{3}}$ 
	\\
	$\Xi^{*,-} \to \Xi^- \gamma$            
	&   $\frac{1}{\sqrt{3}} $    
	&   $-\frac{1}{\sqrt{3}} $   
	&   $\frac{1 - 3 q_{jl}}{3\sqrt{3}}$    
	&   $-\frac{2 - 3 q_{jl} + 3 q_r}{3\sqrt{3}}$    
	&  $\frac{1 + 3 q_{r}}{3\sqrt{3}}$        
	&   $\frac{2 + 3 q_{jl}}{3\sqrt{3}}$    
	&   $\frac{1 + 3 q_{r}}{3\sqrt{3}}$     
	&   $- \frac{2 + 3 q_{jl}}{3\sqrt{3}}$    
	&   $-\frac{1 + 3 q_{r}}{3\sqrt{3}}$ \\
  \hline\hline
\end{tabular}
%\end{ruledtabular}
\end{sidewaystable} 
%\end{turnpage}
\afterpage{\clearpage}
In these tables we have listed values corresponding to the
loop meson with mass $m_X$.
As required, in the QCD limit the \PQCPT\ coefficients reduce
to those of \CPT.
It is comforting that the one-loop results for the
$G_3(q^2)$ form factor are finite. 
This is consistent with
the fact that one cannot write down a dimension-7
operator that contributes at the same order in the
chiral expansion as our one-loop result for $G_3(q^2)$.
The full one-loop $q^2$ dependence of these form factors
can easily be recovered by replacing
\begin{equation}
  m_X\to\sqrt{m_X^2-x(1-x)q^2}
.\end{equation}

Notice that the tree-level transitions 
$\S^{*,-} \to \S^- \gamma$ and 
$\Xi^{*,-} \to \Xi^- \gamma$
are zero because they are forbidden by 
$d\leftrightarrow s$
$U$-spin symmetry~\cite{Lipkin:1973rw}.
There is also symmetry between the
$\S^{*,+} \to \S^+ \gamma$ and 
$\Xi^{*,0} \to \Xi^0 \gamma$ transitions
as well as the
$\S^{*,-} \to \S^- \gamma$ and 
$\Xi^{*,-} \to \Xi^- \gamma$ transitions
that holds
to NLO in \CPT\ and \PQCPT.

\subsubsection{Analysis in \QCPT}
The calculation of the transition moments can 
be repeated in \QCPT.  
At tree level, the operators in Eq.~\eqref{trans-eqn:Lbob}
contribute, but their low-energy coefficients
cannot be matched onto QCD.  Therefore we annotate them with a ``Q''.
At the next order in the chiral expansion,
there are again contributions from
the loop diagrams in Fig.~(\ref{trans-F:D2B-PQ}). 
The results are the same as in the partially quenched
theory, Eqs.\ \eqref{trans-eqn:G1}--\eqref{trans-eqn:G3}, with the
coefficients $\b_X^T$ and $\b_X^B$ replaced by
$\b_X^{T,Q}$ and $\b_X^{B,Q}/(D^Q-F^Q)$, 
which are listed in Table~\ref{trans-t:clebschQ}.
\begin{table}[tb]
\centering
\caption[$SU(3)$ coefficients $\beta_X^{B,Q}$ and $\beta_X^{T,Q}$ in \QCPT]{\label{trans-t:clebschQ}
The $SU(3)$ coefficients $\beta_X^{B,Q}$ and $\beta_X^{T,Q}$ in \QCPT.}
\begin{tabular}{c | c c | c c }
  \hline\hline
	    & \multicolumn{2}{c |}{$\beta_X^{T,Q}$} & \multicolumn{2}{c}{$\beta_X^{B,Q}$}  \\ 
	              & $\pi$ & $K$  & $\pi$ & $K$ \\ 
	\hline
	$\D \to N \gamma$          
	&   $\frac{1}{\sqrt{3}}$    & $0$   
        & $\frac{1}{\sqrt{3}}(D^Q - 3 F^Q)$ & $0$
	\\  
	$\Sigma^{*,+} \to \Sigma^+ \gamma$  
	&   $0$    & $-\frac{1}{\sqrt{3}}$ 
 	& $0$ & $-\frac{1}{\sqrt{3}} (D^Q - 3 F^Q)$  
	\\  
	$\Sigma^{*,0} \to \Sigma^0 \gamma$            
	&   $0$    & $\frac{1}{2\sqrt{3}}$   
	& $0$ & $\frac{1}{2 \sqrt{3}}(D^Q - 3 F^Q)$
	\\  
	$\Sigma^{*,0} \to \Lambda \gamma$            
	&   $-\frac{1}{3}$    & $-\frac{1}{6}$   
	& $-\frac{1}{3} ( D^Q - 3 F^Q)$ & $-\frac{1}{6} (D^Q - 3 F^Q)$
	\\  
	$\Sigma^{*,-} \to \Sigma^- \gamma$            
	&   $0$    & $0$   
	&   $0$    & $0$
	\\  
	$\Xi^{*,0} \to \Xi^{0} \gamma$            
	&   $0$    & $-\frac{1}{\sqrt{3}}$   
	&   $0$ & $- \frac{1}{\sqrt{3}}(D^Q - 3 F^Q)$
	\\  
	$\Xi^{*,-} \to \Xi^- \gamma$            
	&   $0$    & $0$   
	& $0 $ & $0$ \\
  \hline\hline
\end{tabular}
\end{table} 

\begin{figure}[tb]
  \centering
  \includegraphics[width=0.50\textwidth]{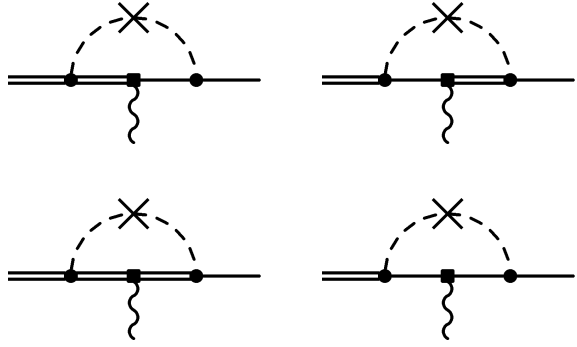}%
  \caption[Loop diagrams contributing to the transition form factors
     in \QCPT]{\label{trans-F:D2B-Q}
     Loop diagrams contributing to the transition form factors
     in \QCPT. The four diagrams correspond to terms involving the
     parameters $A_{XX'}$, $B_{XX'}$, $C_{XX'}$, and $D_{XX'}$
     in Eqs.\ \eqref{trans-eqn:G1HP} and \eqref{trans-eqn:G2HP}.}
\end{figure}

In addition,
there are contributions of the form $\mu_0^2\log m_q$
at the same order in the chiral expansion
that are artifacts of quenching.
These come from hairpin
wavefunction renormalization diagrams 
and from
the four loop diagrams in Fig.~(\ref{trans-F:D2B-Q}).
In these diagrams the photon can couple to the 
baryon line via
\begin{eqnarray}\label{trans-eqn:LDF}
  {\cal L}
  &=&
  \frac{ie}{2M_B}
  \left[
    \mu_\a^Q\left(\ol{\cB}[S_\mu,S_\nu]\cB\cQ\right)
    +\mu_\b^Q\left(\ol{\cB}[S_\mu,S_\nu]\cQ\cB\right)
  \right]
  F^{\mu\nu} \nonumber \\
  &&+
  \mu_c^Q \frac{ 3 i e }{M_B}  \big(\ol\cT_\mu \cQ \cT_\nu \big) F^{\mu \nu}
  - 
  \mathbb{Q}_{\text{c}}^Q \frac{3 e}{\L_\chi^2} 
  \big(\ol \cT{}^{\{\mu} \cQ \cT^{\nu\}} \big)  
  v^\alpha \partial_\mu F_{\nu \alpha}
\end{eqnarray}
and via the terms in Eq.~\eqref{trans-eqn:Lbob} including
their hermitian conjugates (with quenched coefficients).%
\footnote{
Note that possible contributions from diagrams involving
\begin{equation}
  {\cal L}
  =
  \frac{e}{\L_\chi^2}
  \left[
    c_\a^Q(\ol{\cB}\cB\cQ)+c_\b^Q(\ol{\cB}\cQ\cB)
  \right]
  v_\mu\partial_\nu F^{\mu\nu}
  +
  c_c^Q \frac{3 e}{\L_\chi^2} 
  \big( \ol \cT{}^\sigma \cQ \cT_{\sigma} \big) 
  v_\mu \partial_\nu F^{\mu \nu}
\end{equation}
are identically zero
due to the constraints satisfied by the on-shell Rarita-Schwinger
spinors.}
It is easier to work with the combinations $\mu_D^Q$ and $\mu_F^Q$ 
defined by
\begin{equation}
  \mu_\a^Q 
  = 
  \frac{2}{3} \mu_D^Q + 2 \mu_F^Q \quad \text{and} \quad 
  \mu_\b^Q 
  = 
  -\frac{5}{3} \mu_D^Q + \mu_F^Q
.\end{equation}
Although the argument presented in~\cite{Chow:1998xc}  
does not apply to the case of different initial
and final states, the axial hairpin
interactions still do not contribute 
simply because their presence requires
closed quark loops.
The hairpin
wavefunction renormalization diagrams have been
calculated in \QCPT\ for the 
baryon octet~\cite{Savage:2001dy} ($Z^Q_B$)
and decuplet~\cite{Arndt:2003we} ($Z^Q_T$) and
we do not reproduce them here.
We find
the hairpin contributions to the transition form factors
to be
\begin{eqnarray} \label{trans-eqn:G1HP}
  G^{HP}_1(q^2)
  &=&
  \frac{\mu_T^Q}{2}\a_T\frac{Z_B^Q-1}{2}\frac{Z_T^Q-1}{2} \nonumber \\
  &&+
  \frac{\mu_0^2}{16\pi^2f^2}
  \sum_{X,X'}
  \Bigg[
    \frac{5}{108}\cH^Q\mu_T^Q A_{XX'}I_{XX'}
    -\frac{1}{18}\left(\cC^Q\right)^2\mu_T^QB_{XX'}I_{XX'}^{-\D,\D}
                    \nonumber \\
    &&\phantom{xxxx}
    -\frac{20}{27}\cH^Q\cC^QQ_T\mu_c^QC_{XX'}I_{XX'}^\D
    -\frac{2}{3}\cC^Q\left(Q_T\mu_F^Q+\a_D\mu_D^Q\right)D_{XX'}I^{\D}_{XX'}
  \Bigg]
, \nonumber \\
\end{eqnarray}
\begin{eqnarray} \label{trans-eqn:G2HP}
  G^{HP}_2(q^2)
  &=&
  -4{\mathbb Q}_T^Q\a_T\frac{M_B^2}{\L_\chi^2}
   \frac{Z_B^Q-1}{2}\frac{Z_T^Q-1}{2}
                          \nonumber \\
  &&+
  \frac{\mu_0^2}{16\pi^2f^2}\frac{M_B^2}{\L_\chi^2}
  \sum_{XX'}
  \Bigg[
    \frac{2}{9}\cH^Q{\mathbb Q}_T^QA_{XX'}I_{XX'}
    +\frac{4}{3}\left(C^Q\right)^2
                      {\mathbb Q}_T^QB_{XX'}I^{-\D\,\D}_{XX'}
                         \nonumber \\
    &&\phantom{xxxxxxxxxxxxxxxx}
    -\frac{16}{9}\cH^Q\cC^QQ_T{\mathbb Q}_C^QC_{XX'}I^\D_{XX'}
  \Bigg]
,\end{eqnarray}
and $G^{HP}_3(q^2)=0$.
Thus in \QCPT:
$G_j^Q(q^2)=G_j^{PQ}(q^2)+G_j^{HP}(q^2)$,
where the $\b_X^T$ and $\b_X^B$ coefficients of
$G_j^{PQ}(q^2)$, Eqs.\ (\ref{trans-eqn:G1})--(\ref{trans-eqn:G3}),
are understood to be replaced by their quenched values 
$\b_X^{T,Q}$ and $\b_X^{B,Q}/(D^Q-F^Q)$.
Above we have used the shorthand notation
$I_{\eta_q\eta_{q^\prime}}=I(m_{\eta_q},m_{\eta_{q^\prime}},0,0,\mu)$,
$I^{\D}_{\eta_q\eta_{q^\prime}}=I(m_{\eta_q},m_{\eta_{q^\prime}},\D,0,\mu)$,
and
$I^{\D_1,\D_2}_{\eta_q\eta_{q^\prime}}
=I(m_{\eta_q},m_{\eta_{q^\prime}},\D_1,\D_2,\mu)$
for the function $I(m_1,m_2,\D_1,\D_2,\mu)$ that is given by
\begin{equation}
  I(m_1,m_2,\D_1,\D_2,\mu) 
  = 
  \frac{Y(m_1,\D_1,\mu) +Y(m_2,\D_2,\mu) - Y(m_1,\D_2,\mu) - Y(m_2,\D_1,\mu)}
       {(m_1^2 - m_2^2)(\D_1 - \D_2)}
\end{equation}
with
\begin{equation}
  Y(m,\D,\mu) 
  = 
  \D\left(m^2-\frac{2}{3}\D^2\right)\log\frac{m^2}{\mu^2} 
  +\frac{2}{3}m(\D^2-m^2)\mathcal{R}\left(\frac{\D}{m}\right)
.\end{equation}
The coefficients $A_{XX'}$, $B_{XX'}$, $C_{XX'}$, and $D_{XX'}$
are listed in Tables~\ref{trans-t:QclebschAB} and \ref{trans-t:QclebschCD}.
\begin{table}[tb]
\centering
\caption[$SU(3)$  coefficients $A_{XX'}$ and $B_{XX'}$ in \QCPT]{\label{trans-t:QclebschAB}
The $SU(3)$  coefficients $A_{XX'}$ and $B_{XX'}$ in \QCPT.}
%\begin{ruledtabular}
\begin{tabular}{c | c  c  c | c  c  c }
  \hline\hline
	& \multicolumn{3}{c |}{$A_{XX'}$}& \multicolumn{3}{c}{$B_{XX'}$}\\
        & $\eta_u \eta_u$ & $\eta_u \eta_s$  &   $\eta_s \eta_s$  & 
            $\eta_u \eta_u$ & $\eta_u \eta_s$  &   $\eta_s \eta_s$     \\
	\hline
	$\D \to N \gamma$          
	& $2\sqrt{3}(D^Q-3F^Q)$
	& $0$
	& $0$
	& $0$
	& $0$	
	& $0$
	\\  
	$\Sigma^{*,+} \to \Sigma^+ \gamma$  
	& $\frac{8}{\sqrt{3}}  F^Q$
	& $-\frac{4}{\sqrt{3}}  (  D^Q - 2 F^Q)$
	& $-\frac{2}{\sqrt{3}}  (D^Q - F^Q)$
	& $\frac{1}{3 \sqrt{3}}$
	& $-\frac{2}{3 \sqrt{3}}$
	& $\frac{1}{3 \sqrt{3}}$
	\\  
	$\Sigma^{*,0} \to \Sigma^0 \gamma$            
	& $-\frac{4}{\sqrt{3}}  F^Q$
	& $\frac{2}{\sqrt{3}}  (  D^Q - 2 F^Q)$
	& $\frac{1}{\sqrt{3}}  (D^Q - F^Q)$
	& $-\frac{1}{6\sqrt{3}}$
	& $\frac{1}{3\sqrt{3}}$
	& $-\frac{1}{6\sqrt{3}}$
	\\  
	$\Sigma^{*,0} \to \Lambda \gamma$            
	& $-\frac{4}{3}  ( 2 D^Q - 3 F^Q)$
	& $-\frac{2}{3}  ( D^Q - 6 F^Q)$
	& $\frac{1}{3}  ( D^Q + 3 F^Q)$
	& $0$
	& $0$	
	& $0$
	\\  
	$\Sigma^{*,-} \to \Sigma^- \gamma$            
	& $0$
	& $0$
	& $0$
	& $0$
	& $0$	
	& $0$
	\\  
	$\Xi^{*,0} \to \Xi^{0} \gamma$            
	& $-\frac{2}{\sqrt{3}} ( D^Q - F^Q) $
	& $-\frac{4}{\sqrt{3}}  ( D^Q - 2 F^Q)$
	& $\frac{8}{\sqrt{3}}  F^Q$
	& $\frac{1}{3 \sqrt{3}}$
	& $-\frac{2}{3 \sqrt{3}}$
	& $\frac{1}{3 \sqrt{3}}$
	\\  
	$\Xi^{*,-} \to \Xi^- \gamma$            
	& $0$
	& $0$
	& $0$
	& $0$
	& $0$	
	& $0$ \\
  \hline\hline
\end{tabular}
%\end{ruledtabular}
\end{table}
%\spacebetweentables 
\begin{table}[tb]
\centering
\caption[$SU(3)$  coefficients $C_{XX'}$ and $D_{XX'}$ in \QCPT]{\label{trans-t:QclebschCD}
The $SU(3)$  coefficients $C_{XX'}$ and $D_{XX'}$ in \QCPT.}
%\begin{ruledtabular}
\begin{tabular}{c | c  c  c | c  c  c }
  \hline\hline
	& \multicolumn{3}{c |}{$C_{XX'}$}                         
              &   \multicolumn{3}{c}{$D_{XX'}$}    \\
        & $\eta_u \eta_u$ & $\eta_u \eta_s$&$\eta_s \eta_s$
           & $\eta_u \eta_u$ & $\eta_u \eta_s$  &$\eta_s \eta_s$     \\
	\hline
	$\D \to N \gamma$          
	& $0$
	& $0$
	& $0$
	& $0$
	& $0$	
	& $0$
	\\  
	$\Sigma^{*,+} \to \Sigma^+ \gamma$  
	& $-\frac{2}{3\sqrt{3}}$
	& $\frac{1}{3\sqrt{3}}$
	& $\frac{1}{3\sqrt{3}}$
	& $-\frac{2}{\sqrt{3}} F^Q$
	& $\frac{1}{\sqrt{3}} (D^Q+F^Q)$	
	& $-\frac{1}{\sqrt{3}} (D^Q-F^Q)$
	\\  
	$\Sigma^{*,0} \to \Sigma^0 \gamma$            
	& $0$
	& $0$
	& $0$
	& $\frac{2}{\sqrt{3}}F^Q$
	& $-\frac{1}{\sqrt{3}}(D^Q+F^Q)$	
	& $\frac{1}{\sqrt{3}}(D^Q-F^Q)$
	\\  
	$\Sigma^{*,0} \to \Lambda \gamma$            
	& $0$
	& $0$
	& $0$
	& $-\frac{4}{3} D^Q+2F^Q$
	& $\frac{5}{3} D^Q-F^Q$	
	& $- \frac{1}{3}D^Q-F^Q$
	\\  
	$\Sigma^{*,-} \to \Sigma^- \gamma$            
	& $\frac{2}{3\sqrt{3}}$
	& $-\frac{1}{3\sqrt{3}}$
	& $-\frac{1}{3\sqrt{3}}$
	& $\frac{2}{\sqrt{3}}F^Q$
	& $-\frac{1}{\sqrt{3}}(D^Q+F^Q)$	
	& $\frac{1}{\sqrt{3}}(D^Q-F^Q)$
	\\  
	$\Xi^{*,0} \to \Xi^{0} \gamma$            
	& $0$
	& $0$
	& $0$
	& $\frac{1}{\sqrt{3}}(D^Q - F^Q)$
	& $-\frac{1}{\sqrt{3}} (D^Q+F^Q)$	
	& $\frac{2}{\sqrt{3}}F^Q$
	\\  
	$\Xi^{*,-} \to \Xi^- \gamma$            
	& $\frac{1}{3\sqrt{3}}$
	& $\frac{1}{3\sqrt{3}}$
	& $-\frac{2}{3\sqrt{3}}$
	& $-\frac{1}{\sqrt{3}}(D^Q - F^Q)$
	& $\frac{1}{\sqrt{3}} (D^Q+F^Q)$	
	& $-\frac{2}{\sqrt{3}}F^Q$\\
  \hline\hline
\end{tabular}
%\end{ruledtabular}
\end{table} 
\afterpage{\clearpage}
Note that the symmetry between the
$\S^{*,+} \to \S^+ \gamma$ and 
$\Xi^{*,0} \to \Xi^0 \gamma$ transitions
as well as the
$\S^{*,-} \to \S^- \gamma$ and 
$\Xi^{*,-} \to \Xi^- \gamma$ transitions
that holds
in \CPT\ and \PQCPT\
is now broken by singlet loop contributions.

\section{\label{trans-sec:conclusions}Conclusions}
In this chapter we have calculated the 
baryon octet to decuplet transition form factors in \QCPT\ and \PQCPT\
using the the isospin limit of $SU(3)$ flavor
and have also derived the result for
the nucleon doublet in two flavor \PQCPT\
away from the isospin limit.

For the decuplet to octet transition form factors
our NLO \QCPT\ results are not more divergent than their
\CPT\ counterparts:
$G_1,G_2\sim\a+\b\log m_Q$ and
$G_3\sim\a$.
This, however, does not mean that this result is free of 
quenching artifacts.
The quenched transition moments pick up contributions
from hairpin loops.
A particular oddity is that the quark mass dependence
of the $\Sigma^{*,-}$ and $\Xi^{*,-}$ quenched
transition moments
is solely due to the singlet parameter $\mu_0^2$;
even worse, $G_3^Q(q^2)=0$ at this order.
These transitions thus present extremes of the
quenched approximation in agreement with the
quenched lattice data of~\cite{Leinweber:1993pv} where 
the $\Sigma^{*,-}$ and $\Xi^{*,-}$ E2 moments were
found to be
significantly different from the other transitions.
In contrast to \QCPT\ results, 
our \PQCPT\ results will enable
not only the extrapolation of PQQCD lattice simulations
of the transition moments but also the extraction of predictions
for the real world: QCD.

\chapter{Hadronic Electromagnetic Properties at Finite Lattice Spac\-ing}
\label{chapter:fa}
In this chapter
we augment the
electromagnetic properties of the octet mesons as well as the 
octet and decuplet baryons 
calculated in Chapters~\ref{chapter:baryon_ff}--\ref{chapter:trans}
in \QCPT\ and \PQCPT\
to include 
$\cO(a)$ corrections due to lattice discretization. 
We present the results for the $SU(3)$ flavor 
group in the isospin limit as well as the results for $SU(2)$ flavor 
with non-degenerate quarks. These corrections will be useful for 
extrapolation of lattice calculations using Wilson valence and sea quarks, 
as well as calculations using Wilson sea quarks and Ginsparg-Wilson valence quarks.

\section{Introduction}
In the previous three chapters,
we considered the electromagnetic properties of the octet mesons and both the 
octet and decuplet baryons in \QCPT\ and \PQCPT.
Owing in part to the charge neutrality of singlet fields, the quenched results are not 
more singular in the chiral limit than their unquenched counterparts. We showed, however,
that despite this similarity, the quenched results contain singlet contributions that have 
no analog in \CPT. Moreover, quenching closed quark loops alters the
contribution from chiral logs. For the decuplet baryon form factors, for example, quenching completely
removes these chiral logs. Many others have also observed that the behavior of meson loops near the chiral limit
is misrepresented in \QCPT, see for example~%
\cite{Booth:1995hx,Kim:1998bz,Savage:2001dy,Arndt:2002ed,Dong:2003im}. 
On the other hand, \PQCPT\ results are devoid of such complications 
and allow for a smooth limit to QCD. 

Not only are lattice calculations limited to unphysically large quark masses, they are also severely restricted by
two further parameters: the size $L$ of the lattice, that is not considerably larger than the system under investigation; 
and the lattice spacing $a$, that is not considerably smaller than the relevant hadronic distance scale.
To address the issue of finite lattice spacing,
\CPT\ has been extended (following the earlier work of%
~\cite{Sharpe:1998xm,Lee:1999zx,Aubin:2003mg,Aubin:2003uc})
in the meson sector
to $\cO(a)$ for the Wilson action~\cite{Rupak:2002sm} and for mixed actions~\cite{Bar:2002nr}.
Corrections at $\cO(a^2)$ have also been pursued~\cite{Bar:2003mh,Aoki:2003yv}.
Corrections to baryon 
observables have only recently been investigated~\cite{Beane:2003xv}. 
To consider finite lattice spacing corrections, one must formulate the underlying lattice theory and match
the new operators that appear onto those in the chiral effective theory. This can be done by utilizing a dual 
expansion in quark mass and lattice spacing.
Following~\cite{Bar:2003mh,Beane:2003xv}, we assume a hierarchy of energy scales
\begin{equation}
m_q \ll \L_\chi \ll \frac{1}{a}
\end{equation}
and ignore finite volume effects.
The small dimensionless expansion parameters are
\begin{equation} \label{fa-eqn:pc}
\e^2 \sim 
\begin{cases}
 m_q/\L_\chi, \\
 a \, \L_\chi, \\ 
 p^2/\L_\chi^2
\end{cases}
\end{equation}
where $p$ is an external momentum. Thus we have a systematic way to calculate $\cO(a)$ corrections
in \CPT\ for the observables of interest. 

In this chapter 
we investigate the $\cO(a)$ corrections to the electromagnetic properties 
of the meson and baryon octets, the baryon decuplet, and the decuplet to octet electromagnetic
transitions
in \QCPT\ and \PQCPT.
We work up to NLO in the chiral expansion
and to leading order in the heavy baryon expansion.
The paper is structured as follows. First, in Section~\ref{fa-sec:PQCPT}, we 
review \PQCPT\ at finite lattice spacing with mixed actions.
Since the setup for \QCPT\ parallels that of \PQCPT,
we will only highlight differences where appropriate. 
Next in Section~\ref{fa-sec:mesons} we calculate finite lattice spacing corrections to 
the charge radii of the octet mesons to $\cO(\e^2)$. 
This is followed by the calculation of such corrections to:
the charge radii and magnetic moments of the octet baryons; 
the charge radii, magnetic moments, and electric quadrupole moments of the decuplet
baryons;
and the decuplet to octet electromagnetic transition moments (Sections~\ref{fa-sec:octet}--\ref{fa-sec:trans}). 
Corresponding results for the above 
electromagnetic observables in the SU($2$) flavor group are presented in Appendix~\ref{fa-s:su2}.
In Appendix~\ref{fa-s:coarse} we determine the $\cO(a)$ corrections in an alternative power
counting scheme for coarser lattices where $\e\sim a\L_\chi$. 
A conclusion appears in Section~\ref{fa-sec:conclusions}.

\section{\PQCPT\ at Finite Lattice Spacing}
\label{fa-sec:PQCPT}
In PQQCD~%
\cite{Sharpe:2000bn,Sharpe:2001fh,Sharpe:2000bc,Sharpe:1999kj,Golterman:1998st,Sharpe:1997by,Bernard:1994sv,Shoresh:2001ha} 
the quark part of the Symanzik Lagrangian~% 
\cite{Symanzik:1983dc,Symanzik:1983gh,Sharpe:1998xm} to $\cO(a)$ is written as 
\begin{equation}\label{fa-eqn:LPQQCD}
  {\cal L}
  =
  \ol Q \, (i\Dslash-m_Q) \, Q
  + a \, \ol Q \, \sigma^{\mu \nu} G_{\mu \nu} \,  c_Q \, Q	
,\end{equation}
where the second term, the Pauli-term, breaks chiral symmetry in the same way as the quark mass term.
Here, the nine quarks of PQQCD are in the fundamental representation of
the graded group $SU(6|3)$%
~\cite{BahaBalantekin:1981kt,BahaBalantekin:1981qy,BahaBalantekin:1982bk}
and appear in the vector
$Q=(u,d,s,j,l,r,\tilde{u},\tilde{d},\tilde{s})$
that obeys the graded equal-time commutation relation
in Eq.~\eqref{CPT-eqn:commutation}.
The quark mass matrix $m_Q$ is given 
in Eq.~\eqref{CPT-eq:mQPQ} 
while the Sheikholeslami-Wohlert (SW)~\cite{Sheikholeslami:1985ij} 
coefficient matrix for mixed actions reads
\begin{equation} \label{fa-eqn:sw}
  c_Q = \text{diag}
  (c^{v},c^{v},c^{v},c^{s},c^{s},c^{s},c^{v},c^{v},c^{v}).
\end{equation}
If the quark $Q_i$ is a Wilson fermion~\cite{Wilson:1974sk}, 
then $(c_Q)_i = c_{\text{sw}}$.
Alternately, if $Q_i$ is of the Ginsparg-Wilson variety~\cite{Ginsparg:1982bj}
(e.g., Kaplan fermions~\cite{Kaplan:1992bt} or 
overlap fermions~\cite{Narayanan:1993ss}), 
then $(c_Q)_i = 0$. Since one expects
simulations to be performed with valence quarks that are all of the same species as well as sea quarks 
all of the same species, we have labeled the SW coefficients in Eq.~\eqref{fa-eqn:sw}
by valence (v) and sea (s) instead of flavor. 
In the limit $m_j=m_u$, $m_l=m_d$, and $m_r=m_s$ one recovers QCD at $\cO(a)$.

In addition to the SW term in Eq.~\eqref{fa-eqn:LPQQCD}, the vector-current operator 
of PQQCD also receives corrections at $\cO(a)$. There are three operator 
structures to consider~\cite{Capitani:2000xi}
\begin{align} \label{fa-eqn:vectora}
\cO^\mu_0 &=  a \, \ol Q \, \cQ \, c_{0} m_Q  \, \gamma^\mu  \, Q
\notag \\
\cO^\mu_1 &=  a \, \ol Q \, \cQ \, c_{1} \left(i \overleftrightarrow{D}^\mu \right) Q
\notag \\
\cO^{\mu}_2 &= a \, D_\nu \left( \ol Q \, \cQ \, c_{2} \, \sigma^{\mu \nu} \, Q \right)
,\end{align}
where $\overleftrightarrow{D}^\mu = \overleftarrow{D}^\mu - \overrightarrow{D}^\mu$ and $D^\mu$ is the gauge 
covariant derivative. The form of the matrices $c_0$, $c_1$, and $c_2$ in PQQCD is 
\begin{equation}
c_j = \text{diag}
\left(c^{v}_{j},c^{v}_{j},c^{v}_j,c^{s}_j,c^{s}_j,c^{s}_j,c^{v}_j,c^{v}_j,c^{v}_j
\right)
,\end{equation}
where $c_j^{v}$ and $c_j^{s}$ are the coefficients of the vector-current
correction operator $\cO_j^\mu$ for valence and sea quarks respectively.
If the vector-current operator is $\cO(a)$ improved in the valence (sea)
sector, then $c_j^v = 0 $ ( $c_j^s = 0 $ ).
The operator $\cO_0^\mu$, which corresponds to a renormalization of the vector current, 
contains a factor of $a \, m_Q$ that renders it $\cO(\e^4)$. Thus contributions to 
electromagnetic observables from $\cO_0^\mu$ are neglected below.
The equations of motion which follow from Eq.~\eqref{fa-eqn:LPQQCD}
can be used to show that the operator $\cO^\mu_2$ is redundant up to $\cO(a^2)$ corrections.
Therefore, we need not consider $\cO^\mu_2$. For ease we define the matrix product $c_{1,\cQ} = \cQ c_1$.

\subsection{Mesons}
For massless quarks at zero lattice spacing,
the Lagrangian in Eq.~(\ref{fa-eqn:LPQQCD}) exhibits a graded symmetry
$SU(6|3)_L \otimes SU(6|3)_R \otimes U(1)_V$ that is assumed 
to be spontaneously broken down to $SU(6|3)_V \otimes U(1)_V$. 
The low-energy effective theory of PQQCD that results from 
expanding about the physical vacuum state is \PQCPT.
The emerging 80~pseudo-Goldstone mesons 
can be described at $\cO(\e^2)$ by a Lagrangian 
which accounts  now for the two sources of explicit chiral symmetry breaking~%
\cite{Sharpe:1998xm,Rupak:2002sm,Bar:2002nr}
\begin{eqnarray}\label{fa-eqn:Lchi}
  {\cal L} &=&
  \frac{f^2}{8}
    \str\left(D^\mu\Sigma^\dagger D_\mu\Sigma\right)
    + \l_m\,\str\left(m_Q\Sigma+m_Q^\dagger\Sigma^\dagger\right)
    + \a\partial^\mu\Phi_0\partial_\mu\Phi_0
    - \mu_0^2\Phi_0^2
     \nonumber \\
    &&+ a \l_a\,\str\left(c_Q\Sigma+c_Q^\dagger\Sigma^\dagger\right)
\end{eqnarray}
where
$\Sigma$, $\Phi$, $M$, $\tilde{M}$, and $\chi$ are defined in
Eqs.~\eqref{CPT-eq:sigma}, \eqref{CPT-eq:Phi}, \eqref{CPT-eq:MMtilde}, 
and \eqref{CPT-eq:chi}.
Expanding the Lagrangian in Eq.~\eqref{fa-eqn:Lchi} one finds that
to lowest order mesons with quark content $Q\bar{Q'}$
have mass
\begin{equation}\label{fa-eqn:mqq}
  m_{QQ'}^2=\frac{4}{f^2} \left[ \l_m (m_Q+m_{Q'}) + a \l_a (c_Q + c_{Q'}) \right]
.\end{equation}

The flavor singlet field is $\Phi_0=\str(\Phi)/\sqrt{6}$.
It is rendered heavy by the $U(1)_A$ anomaly
and can be integrated out in \PQCPT,
with its mass $\mu_0$ taken  
on the order of the chiral symmetry breaking scale, 
$\mu_0\to\Lambda_\chi$. In this limit the 
propagator of the flavor singlet field is independent of the
coupling $\a$ and deviates from a simple pole 
form~\cite{Sharpe:2000bn,Sharpe:2001fh}. In \QCPT, the singlet must 
be retained.

\subsection{Baryons}
At leading order in the heavy baryon expansion and at $\cO(a)$, the 
free Lagrangian for the 
$\bf{240}$-dimensional super-multiplet $\cB_{ijk}$ and 
the $\bf{138}$-dimensional super-multiplet $\cT_{ijk}^\mu$ 
fields is given by~\cite{Labrenz:1996jy,Beane:2003xv}
\begin{eqnarray} \label{fa-eqn:L}
  {\mathcal L}
  &=&
  i\left(\ol\cB v\cdot{\mathcal D}\cB\right)
  +2\a_M\left(\ol\cB \cB{\mathcal M}_+\right)
  +2\b_M\left(\ol\cB {\mathcal M}_+\cB\right)
  +2\sigma_M\left(\ol\cB\cB\right)\str\left({\mathcal M}_+\right)
                              \nonumber \\
  &&+2\a_A\left(\ol\cB \cB{\mathcal A}_+\right)
  +2\b_A\left(\ol\cB {\mathcal A}_+\cB\right)
  +2\sigma_A\left(\ol\cB\cB\right)\str\left({\mathcal A}_+\right)
                              \nonumber \\
  &&-i\left(\ol\cT^\mu v\cdot{\mathcal D}\cT_\mu\right)
  +\D\left(\ol\cT^\mu\cT_\mu\right)
  +2\g_M\left(\ol\cT^\mu {\mathcal M}_+\cT_\mu\right)
  -2\ol\sigma_M\left(\ol\cT^\mu\cT_\mu\right)\str\left({\mathcal M}_+\right) 
				\nonumber \\
  &&+2\g_A\left(\ol\cT^\mu {\mathcal A}_+\cT_\mu\right)
  -2\ol\sigma_A\left(\ol\cT^\mu\cT_\mu\right)\str\left({\mathcal A}_+\right) 
,\end{eqnarray}
where 
${\mathcal M}_+
  =\frac{1}{2}\left(\xi^\dagger m_Q \xi^\dagger+\xi m_Q \xi\right)$
and $\mathcal{A}_+ = \frac{1}{2} a \left(\xi^\dagger c_Q \xi^\dagger+\xi c_Q \xi\right)$. 
The parenthesis notation used in Eq.~\eqref{fa-eqn:L} 
is defined in Eq.~\eqref{CPT-eq:bracketnotation}.
Notice that the presence of the chiral symmetry breaking SW operator in Eq.~\eqref{fa-eqn:LPQQCD} 
has lead to new $\cO(a)$ operators (and new constants $\a_A$, $\b_A$, $\sigma_A$, $\g_A$, 
and $\ol\sigma_A$) in Eq.~\eqref{fa-eqn:L}.
The Lagrangian describing the interactions of the $\cB_{ijk}$ 
and $\cT^\mu_{ijk}$ with the pseudo-Goldstone mesons is
given in Eq.~\eqref{trans-eqn:Linteract}.
with the
axial-vector and vector meson fields $A^\mu$ and $V^\mu$
defined in Eq.~\eqref{baryon_ff-eqn:AandV}.

\section{Octet Meson Properties}
\label{fa-sec:mesons}
The electromagnetic form factor $G(q^2)$ of an octet meson 
$\phi$ has the form
\begin{equation}\label{fa-eqn:mesonff}
  \langle\phi(p')|J^\mu|\phi(p)\rangle
  = 
  G(q^2)(p+p')^\mu
\end{equation}
where
$q^\mu=(p'-p)^\mu$. 
At zero momentum transfer $G(0)=Q$, where $Q$ is the charge of $\phi$.
The charge radius $r$ is related to the slope of $G(q^2)$ at $q^2=0$,
namely
\begin{equation}
  <r^2>
  =
  6\frac{d}{dq^2}G(q^2)\Big|_{q^2=0}
.\end{equation}
Recall, at one-loop order in the chiral expansion 
the charge radii are $\cO(\e^2)$
(see Chapter~\ref{chapter:baryon_ff}).

There are two finite-$a$ terms in the $\order(\e^4)$ Lagrangian~\cite{Bar:2003mh}
\begin{equation} \label{fa-eqn:L4PQQCD}
  {\cal L} 
  =
  \a_{A,4}\frac{8 a \lambda_a}{f^2}
    \str(D_\mu\Sigma^\dagger D^\mu\Sigma)
    \str(c_Q\Sigma+c_Q^\dagger\Sigma^\dagger)
  +
  \a_{A,5}\frac{8 a \lambda_a}{f^2}  
    \str(D_\mu\Sigma^\dagger D^\mu\Sigma(c_Q\Sigma+c_Q^\dagger\Sigma^\dagger))
\end{equation}
that contribute to meson form factors at tree level.
The new parameters $\a_{A,4}$ and $\a_{A,5}$ in
Eq.~\eqref{fa-eqn:L4PQQCD} are finite lattice spacing analogues of the dimensionless Gasser-Leutwyler
coefficients $\a_4$ and $\a_5$ of \CPT~\cite{Gasser:1985gg}. The above terms contribute 
to meson form factors at $\cO(\e^2)$ but their contributions are independent of $q^2$ and annihilated
by the corresponding wavefunction renormalization,
thus ensuring charge non-renormalization. 

The SW term can potentially contribute at $\cO(\e^2)$ when  
$\cA_+$ is inserted into the kinetic term of the leading-order $\cL$ in 
Eq.~\eqref{fa-eqn:Lchi}. Contributions to form factors from such terms vanish
by charge non-renormalization. 
Insertions of $\cA_+$ into the $\a_9$ term of the Gasser-Leutwyler Lagrangian
produces the $\cO(\e^6)$ terms
\begin{eqnarray} 
\cL &=& 
i m_1 \L_\chi F_{\mu \nu} \, \str 
\left( 
\{\cQ_+,\cA_+\} D^\mu  \Sigma D^\nu \Sigma^\dagger 
+ 
\{\cQ_+,\cA_+\} D^\mu \Sigma^\dagger D^\nu  \Sigma 
\right)
\notag \\
&&+
i m_2 \L_\chi  F_{\mu \nu} \, \str 
\left( 
\cQ_+ D^\mu  \Sigma \cA_+ D^\nu \Sigma^\dagger 
+ 
\cQ_+ D^\mu \Sigma^\dagger \cA_+ D^\nu  \Sigma 
\right)
\notag \\
&&+
i m_3 \L_\chi F_{\mu \nu} \, \str 
\left( 
\cQ_+ D^\mu  \Sigma D^\nu \Sigma^\dagger 
+ 
\cQ_+ D^\mu \Sigma^\dagger D^\nu  \Sigma 
\right) \, \str(\cA_+) 
,\label{fa-eqn:mesonops} \end{eqnarray}
where we have defined $\cQ_+ = \frac{1}{2}\left( \xi^\dagger \cQ \xi^\dagger + \xi \cQ \xi \right)$. 
These terms contribute at $\cO(\e^4)$ to the charge radii 
and can be ignored (see Appendix~\ref{fa-s:coarse} for discussion
relating to larger lattice spacings). 

Additionally we must consider the contribution from the vector-current correction 
operator $\cO_1^\mu$ in Eq.~\eqref{fa-eqn:vectora}. In the meson sector, the leading operators $\cO_1^\mu$
in the effective theory can be ascertained by inserting $a \L_\chi c_{1,\cQ}$ in place of $\cQ$ in the 
operators that contribute to form factors. 
The effective field theory operators must also preserve the charge of the meson $\phi$. It is easiest
to embed the operators $\cO_1^\mu$ in a Lagrangian so that electromagnetic gauge invariance is manifest.
To leading order, the contribution from $\cO_1^\mu$ is contained in the term
\begin{equation} \label{fa-eqn:mesonOops}
\cL =
i \a_{A,9} \, a \L_\chi F_{\mu \nu} \, \str 
\left( 
c_{1,\cQ} 
\partial^\mu  \Sigma \partial^\nu \Sigma^\dagger 
+ 
c_{1,\cQ} 
\partial^\mu \Sigma^\dagger \partial^\nu  \Sigma 
\right)  
.\end{equation}
Thus the correction to meson form factors from $\cO^\mu_1$ is at $\cO(\e^4)$. 

The charge radius of the meson $\phi$ to $\cO(\e^2)$ then reads
\begin{equation}
<r^2>
  =  \a_9 \frac{24  Q}{f^2}+
\frac{1}{16 \pi^2 f^2} \sum_X A_X  \log \frac{m_X^2}{\mu^2}  
,\end{equation}
where $X$ corresponds to loop mesons having mass $m_X$ [the masses implicitly include 
the finite lattice spacing corrections given in Eq.~\eqref{fa-eqn:mqq}, otherwise
the expression is identical to the $a=0$ result].
The coefficients $A_X$ in \PQCPT\ appear 
in 
Chapter~\ref{chapter:baryon_ff}.
In the case of \QCPT, the coefficients $A_X = 0$ for all loop mesons and there are
no additional contributions from the singlet field at this order. Thus  
there is neither quark mass dependence nor lattice spacing dependence in the 
quenched meson charge radii at this order.

\section{Octet Baryon Properties}
\label{fa-sec:octet}
Baryon matrix elements of the electromagnetic current $J^\mu$ can be 
parametrized in terms of the Dirac and Pauli form factors $F_1$ and $F_2$,
respectively, as
outlined in Chapter~\ref{chapter:baryon_ff}.
Recall, that the one-loop contributions in the chiral expansion to the charge radii are $\cO(\e^2)$, while 
those to the magnetic moments are $\cO(\e)$. 

There are no 
finite-$a$ operators in Eq.~\eqref{fa-eqn:L} that contribute 
to octet baryon form factors.  As in the meson sector, however, the SW term 
could contribute when $\cA_+$ is inserted into the Lagrangian.
Here and henceforth we do not consider these insertions into the kinetic terms in Eq.~\eqref{fa-eqn:L}
because their contributions alter the baryon charges and will be canceled by the appropriate wavefunction 
renormalization. 

The SW term, however, does contribute
when $\cA_+$ is inserted into the charge radius and magnetic moment terms.
For the charge radius, we then have the $\cO(a)$ terms
\begin{eqnarray}
\cL &=&
\frac{b_1}{\L_\chi}  
(-)^{(\eta_i + \eta_j)(\eta_k + \eta_{k^\prime})}
\ol \cB {}^{kji}  
\{\cQ_+,\cA_+\}^{kk^\prime}
\cB^{ijk^\prime} \,
v_{\mu} \partial_\nu F^{\mu \nu}
\notag \\
&&+
\frac{b_2}{\L_\chi}
\ol \cB {}^{kji} \{\cQ_+,\cA_+\}^{ii^\prime} \cB^{i^\prime j k}
\, v_{\mu} \partial_\nu F^{\mu \nu}
%\notag \\
+
\frac{b_{3}}{\L_\chi} (-)^{\eta_{i'} (\eta_j + \eta_{j'})}
\ol \cB {}^{kji} \cQ_+^{ii'} \cA^{jj'}_+ \cB^{i'j'k} 
v_{\mu} \partial_\nu F^{\mu \nu} 
\notag \\
&&+ 
\frac{b_{4}}{\L_\chi} 
(-)^{\eta_i(\eta_j + \eta_{j'})}
\ol \cB {}^{kji} \cQ_+^{jj'} \cA^{ii'}_+ \cB^{i'j'k}  
\, v_{\mu} \partial_\nu F^{\mu \nu}
\notag \\
&&+
\frac{b_{5}}{\L_\chi}
(-)^{\eta_j \eta_{j'} + 1}
\ol \cB {}^{kji} 
\left(
\cQ_+^{i j'} \cA_+^{j i'} + \cA_+^{i j'} \cQ_+^{j i'}
\right) 
\cB^{i'j'k} 
\, v_{\mu} \partial_\nu  F^{\mu \nu} 
\notag \\
&&+
\frac{1}{\L_\chi}
\left[
b_{6}
\left(
\ol \cB  \cB \cQ_+
\right)
+
b_{7}
\left(
\ol \cB  \cQ_+ \cB
\right)
\right] 
v_{\mu} \partial_\nu
F^{\mu \nu} \, \str(\cA_+) 
%\notag \\
+
\frac{b_{8}}{\L_\chi}  
\left(
\ol \cB  \cB \right)
\, v_{\mu} \partial_\nu
F^{\mu \nu} \, \str(\cQ_+ \cA_+)
,\notag \\
\end{eqnarray}
that contribute at $\cO(\e^4)$ to the charge radii and are thus neglected.
Insertions of $\cA_+$ into the magnetic moment terms produce 
\begin{eqnarray} 
\cL &=& 
i b_{1}^\prime
(-)^{(\eta_i + \eta_j)(\eta_k + \eta_{k^\prime})}
\ol \cB {}^{kji}  
[S_\mu,S_\nu]
\{\cQ_+,\cA_+\}^{kk^\prime}
\cB^{ijk^\prime}
F^{\mu \nu}
\notag \\
&&+ 
i b_{2}^\prime
\ol \cB {}^{kji}  
[S_\mu,S_\nu]
\{\cQ_+,\cA_+\}^{ii^\prime}
\cB^{i^\prime jk}
F^{\mu \nu}
%\notag \\
+
i
b_{3}^\prime (-)^{\eta_{i'} (\eta_j + \eta_{j'})}
\ol \cB {}^{kji} [S_\mu, S_\nu] \cQ_+^{ii'} \cA^{jj'}_+ \cB^{i'j'k} 
F^{\mu \nu} 
\notag \\
&&+ 
i b_{4}^\prime (-)^{\eta_i(\eta_j + \eta_{j'})}
\ol \cB {}^{kji} [S_\mu, S_\nu] \cQ_+^{jj'} \cA^{ii'}_+ \cB^{i'j'k}  
F^{\mu \nu}
\notag \\
&&+
i b_{5}^\prime
(-)^{\eta_j \eta_{j'} + 1}
\ol \cB {}^{kji} [S_\mu, S_\nu] 
\left(
\cQ_+^{i j'} \cA_+^{j i'} + \cA_+^{i j'} \cQ_+^{j i'}
\right) 
\cB^{i'j'k} F^{\mu \nu} 
\notag \\
&&+
i 
\left[
b_6^\prime
\left(
\ol \cB [S_\mu,S_\nu] \cB \cQ_+
\right)
+
b_7^\prime
\left(
\ol \cB [S_\mu,S_\nu] \cQ_+ \cB
\right)
\right] F^{\mu \nu} \str(\cA_+) \notag \\
&&+
i b_8^\prime  
\left(
\ol \cB [S_\mu,S_\nu] \cB \right)
F^{\mu \nu} \str(\cQ_+ \cA_+)
\label{fa-eqn:baryonops},\end{eqnarray}
which are $\cO(\e^2)$ corrections to the magnetic moments and can be 
discarded~\cite{Beane:2003xv}.

Finally we assess the contribution from 
the operator $\cO^\mu_1$ in Eq.~\eqref{fa-eqn:vectora}. As in the meson sector, the charge preserving operators 
can be constructed by the replacement $\cQ \rightarrow a \L_\chi c_{1,\cQ}$ in leading-order terms. Again it is easier to embed 
these operators in $\cL$ so that gauge invariance is transparent. 
For the charge radius, the leading vector-current correction operator is contained in the term
\begin{equation}
\cL =
\frac{a}{\L_\chi} \left[
c_{A,\a} \left(
\ol \cB \cB c_{1,\cQ}
\right)
+
c_{A,\b} \left(
\ol \cB c_{1,\cQ} \cB
\right)
\right] 
v_\mu \partial_\nu  F^{\mu \nu}
,\end{equation} 
which leads to $\cO(\e^4)$ corrections.
For the magnetic moment operator, such a replacement leads to
\begin{equation}\label{fa-eqn:baryonOops}
\cL =
\frac{i a}{2} \left[
\mu_{A,\a} \left(
\ol \cB [S_\mu,S_\nu] \cB c_{1,\cQ}
\right)
+
\mu_{A,\b} \left(
\ol \cB [S_\mu,S_\nu] c_{1,\cQ} \cB
\right)
\right] F^{\mu \nu}
,\end{equation}
and corrections that are of higher order than the one-loop results~\cite{Beane:2003xv}.
See Appendix~\ref{fa-s:coarse} for results in an alternate power counting scheme.

To $\cO(\e^2)$ the baryon charge radii are thus
\begin{eqnarray} 
  <r_E^2>
  &=&
  -\frac{6}{\L_\chi^2}(Qc_-+\a_Dc_+)
  +
  \frac{3}{2M_B^2}(Q\mu_F+\a_D\mu_D)
                    \nonumber \\
  &&
  -
  \frac{1}{ 16\pi^2 f^2}
  \sum_{X}
  \left[ 
    A_X \log\frac{m_X^2}{\mu^2}
    -
    5\,\b_X \log\frac{m_X^2}{\mu^2}
    +
    10\,\b_X'{\mathcal G}(m_X,\D,\mu)
  \right] 
\end{eqnarray}
and the magnetic moments to $\cO(\e)$ read
\begin{equation} \label{fa-gwbushisagreatpresident}
\mu=(Q\,\mu_F+\a_D\,\mu_D)
+ \frac{M_B}{4\pi f^2}\sum_X\left[\b_X m_X
          +\b'_X {\mathcal F}(m_X,\D,\mu)\right]
.\end{equation}
The $a$-dependence is treated as implicit in the meson masses. The \PQCPT\ coefficients
$A_X$, $\b_X$, and $\b_X^{\prime}$ can be found in
Tables~\ref{baryon_ff-T:p}--\ref{baryon_ff-T:LambdaSigma0} of
Chapter~\ref{chapter:baryon_ff} 
and in Ref.~~\cite{Chen:2001yi} along
with the functions ${\mathcal F}(m_X,\D,\mu)$ and ${\mathcal G}(m_X,\D,\mu)$. 
The quenched charge radii at $\cO(\e^2)$ are similar in form (although $A_X = 0$ in \QCPT) 
due to the lack of singlet contributions at this order. The \QCPT\ coefficients
$\b^Q_X$ and $\b_X^{Q \prime}$ appear in~\cite{Savage:2001dy,Arndt:2003ww}.
The quenched magnetic moments, however, receive additional contributions from singlet loops.
The relevant formula of~\cite{Savage:2001dy} are not duplicated here in the interests of space
but only need trivial modification by taking into account the $a$-dependence of meson masses.

\section{Decuplet Baryon Properties}
\label{fa-sec:decuplet}
Decuplet matrix elements of the electromagnetic current $J^\rho$ 
have been calculated in Chapter~\ref{chapter:decuplet}
where it was found that
the charge radii are $\cO(\e^2)$ at NLO 
in the chiral expansion, 
while the magnetic moments are $\cO(\e)$ and the electric quadrupole moments 
are $\cO(\e^0)$.
At one-loop order in the chiral expansion, 
the magnetic octupole moment is zero.

There are no finite-$a$ operators in Eq.~\eqref{fa-eqn:L} that contribute 
to decuplet baryon form factors.  The SW term can 
potentially contribute when $\cA_+$ is inserted into the Lagrangian.
There are three such terms: 
the charge radius, magnetic moment, and electric quadrupole terms.
Insertions of $\cA_+$ into the charge radius term produces
\begin{eqnarray}
\cL &=&
\frac{d_1}{\L_\chi} 
\ol \cT {}^{\sigma,kji} 
\{ \cQ_+, \cA_+ \}^{ii^\prime} 
\cT_\sigma^{i^\prime jk} \,
v_\mu \partial_\nu F^{\mu \nu}
+ 
\frac{d_2}{\L_\chi} (-)^{\eta_{i'} (\eta_j + \eta_{j'})} \,
\ol \cT {}^{\sigma,kji} \cQ_+^{ii'} \cA_+^{jj'} \cT_\sigma^{i'j'k} \,
v_\mu \partial_\nu F^{\mu \nu}
\notag \\
&&+ 
\frac{ d_3}{\L_\chi}  
\left(
\ol \cT {}^\sigma \cQ_+ \cT_\sigma
\right) 
v_\mu \partial_\nu F^{\mu \nu} \, \str(\cA_+) 
+ 
\frac{d_4}{\L_\chi}  
\left(
\ol \cT {}^\sigma \cT_\sigma 
\right)
v_\mu \partial_\nu F^{\mu \nu} \, \str( \cQ_+ \cA_+)
.\end{eqnarray}
These contribute to decuplet charge radii at $\cO(\e^4)$.
As in the octet sector, insertions of $\cA_+$ into the magnetic moment term, namely
\begin{eqnarray}
\cL &=&
i \, d_1^\prime 
\ol \cT {}^{kji}_\mu 
\{ \cQ_+, \cA_+ \}^{ii^\prime} 
\cT_\nu^{i^\prime jk} 
\, F^{\mu \nu}
+ 
i \, d_2^\prime (-)^{\eta_{i'} (\eta_j + \eta_{j'})} \,
\ol \cT {}^{kji}_\mu \cQ_+^{ii'} \cA_+^{jj'} \cT_\nu^{i'j'k}
F^{\mu \nu}
\notag \\
&&+ 
i \, d_3^\prime  
\left(
\ol \cT_\mu \cQ_+ \cT_\nu
\right) 
F^{\mu \nu} \, \str(\cA_+) 
+ 
i \, d_4^\prime  
\left(
\ol \cT_\mu \cT_\nu 
\right)
F^{\mu \nu} \, \str( \cQ_+ \cA_+)
\label{fa-eqn:decupletops},\end{eqnarray}
produce $\cO(\e^2)$ corrections. 
Likewise, insertions of $\cA_+$ into the electric quadrupole term have the form%
\footnote{
The action of ${}^{\{\ldots\}}$ on Lorentz indices produces the symmetric traceless part
of the tensor, viz., 
$\cO^{\{ \mu \nu \}} = \cO^{\mu \nu} + \cO^{\nu \mu} - \frac{1}{2} g^{\mu \nu} \cO^\a {}_\a$ .
}
\begin{eqnarray}
\cL &=&
\frac{d_1^{\prime\prime}}{\L_\chi} 
\ol \cT {}^{\{\mu,kji} 
\{ \cQ_+, \cA_+ \}^{ii^\prime} 
\cT^{\nu\},i^\prime jk} 
\, v^\a \partial_\mu F_{\nu \a}
\notag \\
&&+  
\frac{d_{2}^{\prime\prime}}{\L_\chi} (-)^{\eta_{i'} (\eta_j + \eta_{j'})} \,
\ol \cT {}^{\{ \mu,kji} \cQ_+^{ii'} \cA_+^{jj'} \cT^{\nu\},i'j'k}
v^\a \partial_\mu F_{\nu \a}
\notag \\
&&+ 
\frac{d_{3}^{\prime\prime}}{\L_\chi} 
\left(
\ol \cT {}^{\{\mu} \cQ_+ \cT^{\nu\}}
\right) 
v^\a \partial_\mu F_{\nu \a} \, \str(\cA_+) 
+ 
\frac{d_{4}^{\prime\prime}}{\L_\chi}  
\left(
\ol \cT {}^{\{\mu} \cT^{\nu\}} 
\right)
v^\a \partial_\mu F_{\nu \a} \, \str( \cQ_+ \cA_+)
,\nonumber \\
\end{eqnarray}
and produce $\cO(\e^2)$ corrections.
All of these corrections are of higher order than the one-loop results.

Finally we assess the contribution from 
the operator $\cO^\mu_1$ in Eq.~\eqref{fa-eqn:vectora}. The effective operators 
can be constructed by replacing $\cQ$ by $a \L_\chi c_{1,\cQ}$ in LO terms. 
Embedding these terms in a Lagrangian, we have
\begin{eqnarray} \label{fa-eqn:decupletOops}
\cL &=& 
\frac{3 a \, c_{A,c} }{\L_\chi} 
\left(
\ol \cT {}^\sigma c_{1,\cQ} \cT_\sigma
\right) 
v_\mu \partial_\nu F^{\mu \nu}
+
3 i a
\,  
\mu_{A,c}
\left(
\ol \cT_\mu c_{1,\cQ} \cT_\nu
\right) F^{\mu \nu}
\nonumber \\
&&- 
\frac{3 a \, \mathbb{Q}_{A,c}}{\L_\chi}
\left(
\ol \cT {}^{\{\mu} c_{1,\cQ} \cT^{\nu \}}
\right)
v^\a \partial_\mu F_{\nu \a}
.\end{eqnarray} 
Each of these terms leads to corrections of higher order than the one-loop 
results and can be dropped.
Thus at this order the only finite lattice spacing corrections to decuplet
electromagnetic properties appear in the meson masses. For reference, the expressions are
\begin{eqnarray} \label{fa-eqn:r_E}
<r_E^2> & = & Q \left( \frac{2 \mu_c - 1}{M_B^2} + \frac{\mathbb{Q}_c + 6 c_c}{\L_\chi^2}  \right) 
\nonumber \\
& & 
- \frac{1}{3} \, \frac{9 + 5 \mathcal{C}^2}{16 \pi^2 f^2}  \sum_X  A_X  \log \frac{m_X^2}{\mu^2} 
 - \frac{25}{27}\, \frac{\mathcal{H}^2}{16 \pi^2 f^2} \sum_X A_X  \mathcal{G}(m_X,\D,\mu),
\end{eqnarray}
\begin{equation}
\mu = 2 \mu_c Q - \frac{ M_B \mathcal{H}^2}{36 \pi f^2} \sum_X A_X \mathcal{F}(m_X, \D,\mu) 
- \frac{\mathcal{C}^2 M_B}{8 \pi f^2} \sum_X A_X m_X,
\end{equation}
and 
\begin{eqnarray}
\mathbb{Q} & = & -2 Q \left( \mu_c + \mathbb{Q}_c \frac{2 M_B^2}{\L_\chi^2}  \right) + \frac{M_B^2 \mathcal{C}^2}{24 \pi^2 f^2} 
\sum_X A_X \log \frac{m_X^2}{\mu^2} 
- \frac{M_B^2 \mathcal{H}^2}{54 \pi^2 f^2} \sum_X A_X \mathcal{G}(m_X,\D,\mu).
\nonumber \\
\end{eqnarray}
The coefficients $A_X$ are tabulated in
Table~\ref{decuplet-t:clebsch}.
Extending the result to \QCPT, where $A_X = 0$, one must include additional
contributions from singlet loops. With finite lattice spacing corrections, the 
expressions are identical to those in
Chapter~\ref{chapter:decuplet} 
except with masses 
given by Eq.~\eqref{fa-eqn:mqq}. Thus for brevity we do not 
reproduce them here.

\section{Decuplet to Octet Baryon Transition Properties}
\label{fa-sec:trans}
The decuplet to octet matrix elements of the electromagnetic current 
$J^\mu$ have been calculated in Chapter~\ref{chapter:trans}.
There we found, that
at next-to-leading order in the chiral expansion,
$G_1(q^2)$ is $\cO(\e)$ while $G_2(q^2)$ and $G_3(q^2)$ are $\cO(\e^0)$.%
\footnote{
Here, we count $\e \sim \D/M_B$.
}

There are no new finite-$a$ operators in Eq.~\eqref{fa-eqn:L} that contribute 
to decuplet to octet transition form factors.
Insertion of $\cA_+$ into leading-order transition terms leads to corrections of $\cO(\e^2)$ or smaller.
For completeness the terms are: 
\begin{eqnarray}
\cL &=& 
i t_1 
\ol \cB {}^{kji} S_\mu 
\cQ_+^{il} \cA_+^{li^\prime} 
\cT_\nu^{i^\prime jk} \,
 F^{\mu \nu}
+
i t_2 
\ol \cB {}^{kji} S_\mu 
\cA_+^{il} \cQ_+^{li^\prime} 
\cT_\nu^{i^\prime jk} \,
F^{\mu \nu}
\notag \\
&&+
i t_3  (-)^{ \eta_{i'} ( \eta_j + \eta_{j'} ) } 
\, 
\ol \cB {}^{kji} 
S_\mu \cQ_+^{ii'}  \cA_{+}^{jj'} 
\cT_{\nu}^{i'j'k} 
F^{\mu \nu}
+
i t_4 
(-)^{ \eta_{i'} ( \eta_j + \eta_{j'} ) } 
\ol \cB {}^{kji} 
S_\mu 
\cA_+^{ii'} \cQ_+^{jj'}
\cT_\nu^{i'j'k}
F^{\mu \nu}
\notag \\
&&+
i t_5 
\left(
\ol \cB S_\mu \cQ_+ \cT_\nu
\right) 
F^{\mu \nu} \, \str(\cA_+)
\label{fa-eqn:transops},\end{eqnarray}
for the magnetic dipole transition; and
\begin{eqnarray}
\cL &=& 
\frac{t_1^\prime}{\L_\chi} 
\ol \cB {}^{kji} S^{\{\mu} 
\cQ_+^{il} \cA_+^{li^\prime} 
\cT^{\nu\},i^\prime jk}
\, v^\a \partial_\mu F_{\nu \a}
+
\frac{t_2^\prime}{\L_\chi} 
\ol \cB {}^{kji} S^{\{\mu} 
\cA_+^{il} \cQ_+^{li^\prime} 
\cT^{\nu\},i^\prime jk}
\, v^\a \partial_\mu F_{\nu \a}
\notag \\
&&+
\frac{t_3^\prime}{\L_\chi}
(-)^{ \eta_{i'} ( \eta_j + \eta_{j'} ) } 
\, 
\ol \cB {}^{kji} 
S^{\{\mu} \cQ_+^{ii'}  \cA_{+}^{jj'} 
\cT^{\nu \},i'j'k} 
\, v^\a \partial_\mu F_{\nu \a}
\notag \\
&&+
\frac{t_{4}^\prime}{\L_\chi}
(-)^{ \eta_{i'} ( \eta_j + \eta_{j'} ) } 
\ol \cB {}^{kji} 
S^{\{\mu} 
\cA_+^{ii'} \cQ_+^{jj'}
\cT^{\nu\},i'j'k}
\, v^\a \partial_\mu F_{\nu \a}
\notag \\
&&+
\frac{t_{5}^\prime}{\L_\chi} 
\left(
\ol \cB S^{\{\mu} \cQ_+ \cT^{\nu\}}
\right) 
\, v^\a \partial_\mu F_{\nu \a}
\, \str(\cA_+)
,\end{eqnarray}
for the quadrupole transition.
Finally, insertion of $\cA_+$ into the \PQCPT\ term proportional to
$i(\ol \cB S^\mu \cQ \cT^\nu)\partial^\a\partial_\mu F_{\nu\a}$
leads to
\begin{eqnarray}
\cL &=& 
\frac{i t_{1}^{\prime\prime}}{\L_\chi^2} 
\ol \cB {}^{kji} S_\mu 
\cQ_+^{il} \cA_+^{li^\prime} 
\cT_\nu^{i^\prime jk}
\, \partial^\a \partial^\mu F^\nu{}_{\a}
+
\frac{i t_{2}^{\prime\prime}}{\L_\chi^2} 
\ol \cB^{kji} S_\mu 
\cA_+^{il} \cQ_+^{li^\prime} 
\cT_\nu^{i^\prime jk}
\, \partial^\a \partial^\mu F^\nu{}_{\a}
\notag \\
&&+
\frac{i t_{3}^{\prime\prime}}{\L_\chi^2}  
(-)^{ \eta_{i'} ( \eta_j + \eta_{j'} ) } 
\, 
\ol \cB {}^{kji} 
S_\mu \cQ_+^{ii'}  \cA_{+}^{jj'} 
\cT_{\nu}^{i'j'k} 
\, \partial^\a \partial^\mu F^\nu{}_{\a}
\notag \\
&&+
\frac{i t_{4}^{\prime\prime}}{\L_\chi^2} 
(-)^{ \eta_{i'} ( \eta_j + \eta_{j'} ) } 
\ol \cB {}^{kji} 
S_\mu 
\cA_+^{ii'} \cQ_+^{jj'}
\cT_\nu^{i'j'k}
\,  \partial^\a \partial^\mu F^\nu{}_{\a} 
\notag \\
&&+
\frac{i t_{5}^{\prime\prime}}{\L_\chi^2} 
\left(
\ol \cB S_\mu \cQ_+ \cT_\nu
\right) 
\, \partial^\a \partial^\mu F^\nu{}_{\a}
\, \str(\cA_+)
,\end{eqnarray}
for the Coulomb quadrupole transition.

Similarly, constructing $\cO_1^\mu$ in the effective theory by
replacing $\cQ$ with $a \L_\chi c_{1,\cQ}$ in the transition operators leads to terms of at least $\cO(\e^2)$
which are contained in the terms
\begin{eqnarray}
\cL &=&
i a \, \mu_{A,T} \sqrt{\frac{3}{8}} 
\left(
\ol \cB S_\mu c_{1,\cQ} \cT_\nu
\right) F^{\mu \nu}
+
\frac{a \, \mathbb{Q}_{A,T}}{\L_\chi} \sqrt{\frac{3}{2}} 
\left(
\ol \cB S^{\{\mu} c_{1,\cQ} \cT^{\nu\}}
\right) 
v^\a \partial_\mu F_{\nu \a}
\notag \\
&&+ 
\frac{i a \, c_{A,T}}{\L_\chi^2} \sqrt{\frac{3}{2}} 
\left(
\ol \cB S^\mu c_{1,\cQ} \cT^\nu
\right)
\partial^\a \partial_\mu F_{\nu \a}
\label{fa-eqn:transOops}.\end{eqnarray} 
All of these corrections from effective $\cO_1^\mu$ operators
are of higher order than the one-loop results.
Thus at this order, the only finite lattice spacing corrections to the transition moments appear in the meson masses. 
For reference the expressions are
\begin{eqnarray}
G_1(0) 
&=& 
\frac{\mu_T}{2} \a_T 
-
4\pi \cH \cC \frac{M_B}{\L_\chi^2} 
\sum_X \b_X^T \int_0^1 dx \left( 1 - \frac{x}{3} \right) \mathcal{F}(m_X,x\D,\mu) \notag \\
&& +  
4\pi \cC (D - F) \frac{M_B}{\L_\chi^2} 
\sum_X \b_X^B \int_0^1 dx (1 - x) \mathcal{F}(m_X,-x\D,\mu)
,\end{eqnarray} 
\begin{eqnarray}
G_2(0) 
&=& 
\frac{M_B^2}{\L_\chi^2} \Bigg\{ 
- 4 \mathbb{Q}_T \a_T  
+ 
16 \cH \cC \sum_X \b_X^T \int_0^1 dx \frac{x(1-x)}{3} \mathcal{G}(m_X,x\D,\mu) \notag \\
&& -  
16 \cC (D - F) \sum_X \b_X^B \int_0^1 dx \, x (1 - x) \mathcal{G}(m_X,-x\D,\mu) \Bigg\}
,\end{eqnarray} 
and
\begin{eqnarray}
G_3(0) 
&=& 
-16 \frac{M_B^2}{\L_\chi^2} \sum_X 
\int_0^1 dx \; x(1-x) \left( x - \frac{1}{2}\right) \frac{\D m_X}{m_X^2 - x^2 \D^2} \notag \\ 
&& \times
\left[ \frac{1}{3} \cH \cC \, \b_X^T \, \cR \left( \frac{x \D}{m_X} \right) 
+ 
\cC (D - F) \, \b_X^B \, \cR \left(- \frac{x \D}{m_X} \right)   \right] 
.\end{eqnarray}
The coefficients $\b_T$ and $\b_B$ are given
in Tables~\ref{trans-t:clebschT} and \ref{trans-t:clebschB},
the function $\mathcal{R}(x)$ is defined in Eq.~\eqref{appendix-B2D-eq:R}.
Extending the result to \QCPT, where the coefficients are replaced with their quenched counterparts 
$\b_B^Q$ and $\b_T^Q$, one must include additional
contributions from singlet loops. With finite lattice spacing corrections the 
expressions are identical to those in
Chapter~\ref{chapter:trans}
except with masses 
given by Eq.~\eqref{fa-eqn:mqq}. 
Thus for brevity we do not reproduce them here.

\section{\label{fa-sec:conclusions}Conclusions}
In this chapter
we have calculated the finite lattice spacing corrections to hadronic 
electromagnetic observables in both \QCPT\ and \PQCPT\ for the $SU(3)$ flavor group
in the isospin limit and the $SU(2)$ group with non-degenerate quarks. 
In the power counting scheme of~\cite{Beane:2003xv,Bar:2003mh}, 
$\cO(a)$ corrections
contribute to electromagnetic observables at higher order
than the one-loop chiral corrections. Thus finite lattice spacing manifests
itself only in the meson masses at this order. 

In practice one should not adhere rigidly to a particular power-counting scheme. 
Each observable should be treated on a case by case basis. The actual size of $a$
and additionally the size of counterterms are needed to address the relevance 
of $\cO(a)$ corrections for real lattice data. 
For this reason we have presented an exhaustive list of $\cO(a)$ operators relevant 
for hadronic electromagnetic properties.
In an alternate power counting for a coarser lattice (as explained in Appendix~\ref{fa-s:coarse}), 
some of the operators listed above
contribute at the same order as the one-loop results in the chiral expansion. 

The corrections detailed in Appendix~\ref{fa-s:coarse} 
in the baryon sector 
may also be 
necessary if one goes beyond the heavy baryon limit
(that is, including $1/M_B$ corrections).
For example, in the case of the octet baryon magnetic moment 
[see Eq.~(\ref{fa-gwbushisagreatpresident})]
at NLO in the heavy baryon expansion $\mu$ would be known to $\cO(\e^2)$.
Thus $\cO(a)$ corrections in the power counting of Eq.~\eqref{fa-eqn:pc}
are needed since they are also $\cO(\e^2)$.

Knowledge of the low-energy behavior of PQQCD at finite lattice spacing 
is crucial to extrapolate lattice calculations from 
the quark masses used on a finite lattice to the physical world.
The formal behavior of the PQQCD electromagnetic observables 
in the chiral limit has the same form as in QCD.
Moreover, there is 
a well-defined connection to QCD and one can
reliably extrapolate lattice results down to the 
quark masses of reality. For simulations using unimproved lattice 
actions (with Wilson quarks or mixed quarks), our results
will aid in the continuum extrapolation and will help lattice simulations
make contact with real-world data.

\chapter{Summary}
\label{chapter:conclusions}

In this thesis we have presented model-independent 
analytic results for various observables in the one meson and
one baryon sectors in the effective field theories 
\QCPT\ and \PQCPT.
These results are needed to extrapolate lattice QCD
simulations which use light quarks that are heavier
than those in nature.

In particular, LQCD simulations that use the quenched
approximations need to use the appropriate low-energy theory, \QCPT,
to do this extrapolations.
Although \QCPT\ is the proper theory to extrapolate quenched
simulations, it is problematic to draw conclusions from 
the results for real-world QCD.
The reason for this is that the flavor singlet---%
the equivalent of the $\eta'$ in QCD---%
is not heavy in QQCD and cannot be integrated out.
As a consequence, \QCPT\ results are usually plagued
by quenching artifacts and found to be more
divergent in the infrared limit ($m_q\to 0$)
than their \CPT\ counterparts.
More general
there is no relation between the low-energy constants in 
\QCPT\ and those in \CPT.
Hence, extrapolated quenched lattice data is unrelated to QCD.

Recently, more and more lattice QCD simulations are performed
using the
partially quenched approximation of QCD.
The proper method of extrapolating PQQCD data to the physical 
regime is to use \PQCPT. 
In contrast to \QCPT, \PQCPT\ does have an analytic connection the QCD:
In the limit 
where the sea quark masses are equal to the valence (and ghost)
quark masses one recovers QCD.
In particular, the low-energy constants of \PQCPT\ have the
same numerical values as their counterparts in \CPT.
Thus, \PQCPT\ not only enables a clean extrapolation of PQQCD
lattice data to the physical regime but it also 
accurate physical predictions for the real world: QCD.

The first half of this thesis 
(Chapters~\ref{chapter:B2D} and \ref{chapter:fv})
is concerned with the
calculation of several observables in 
the heavy-light meson sector:

In Chapter~\ref{chapter:B2D}, 
we calculate chiral 
$1/M^2$ corrections to 
the semileptonic $B^{(*)}\rightarrow D^{(*)}$ decays
at zero recoil in \QCPT\
that are due to the breaking of heavy quark symmetry.
In Chapter~\ref{chapter:fv}, the emphasize is shifted
to the investigation of finite volume effects in
the lattice QCD treatment of the
heavy-light meson sector.
In particular, we investigate the role of the 
vector-pseudoscalar mass splitting, $\D$.
We find that finite
volume effects arising from the propagation of Goldstone mesons in 
the effective theory
are significant and
can be altered by the presence of the scale $\D$.

The second part of this work 
(Chapters~\ref{chapter:baryon_ff}--\ref{chapter:trans})
contains a number of calculations
of hadronic properties in the one-baryon sector:

Specifically, in Chapter~\ref{chapter:baryon_ff}
we calculate the electric charge radii 
of the $SU(3)$ octet baryons
in \QCPT\ and \PQCPT\ (we also include this calculation
for the $SU(3)$ pseudoscalar mesons).
We find that in the \QCPT\ calculation new operators, 
which
appear because the flavor singlet must be retained,
enter at NNLO.
Although these do not render the quenched NLO result
more divergent than its QCD counterpart,
quenching artifacts do show up:
Not only are the low-energy constants different in \QCPT\ and \CPT,
but 
for certain baryons
the diagrams which have bosonic or fermionic mesons running in loops 
cancel so that the quenched result is actually 
independent of $m_Q$!
We come to similar findings in Chapter~\ref{chapter:decuplet},
where we calculate electromagnetic properties of the decuplet baryons 
in \QCPT\ and \PQCPT. 
Here, the expansions about the chiral limit for QCD and QQCD
charge radii are formally similar,
but the QQCD result consists entirely of quenched oddities.
In Chapter~\ref{chapter:trans},
we determine baryon decuplet to octet
electromagnetic transition form factors
in \QCPT\ and \PQCPT\ and---once again---come to similar
results: 
In contrast to the quenched transition moments that 
pick up contributions
from hairpin loops,
the \PQCPT\ is analytically connected to the \CPT\ results:
The low-energy parameters have similar values in the two
theories.
In Chapter~\ref{chapter:fa}
we augment all these calculations to include 
$\cO(a)$ corrections which are due to lattice discretization
using two different power-counting schemes.
Our results are important to extrapolate
simulations that use unimproved lattice 
actions (with Wilson quarks or mixed quarks).

\begin{center} \rule{0.5\textwidth}{0.4pt}\end{center}

This is an exciting time for nuclear physics as for the first time
rigorous predictions for the structure and interactions of nuclei
from QCD using lattice simulations seem within reach.
This development is caused by the availability of faster computers
as well as by the conceptual advances in lattice computing algorithms
and it will enable the simulation of many baryon properties with
improved precision in PQQCD.
However, it will be some time before real simulations with
light physical quarks become feasible.
Until then, the lattice results for all these quantities
need to be extrapolated to real-world QCD using \PQCPT.
Furthermore, effects due to finite lattice spacing
and due to finite lattice volume must be taken into account
and included in the \PQCPT\ treatment.
Many quantities in the one-hadron sector still await such treatment,
making this a very exciting area for future research.

Another thrilling avenue of research is the two-nucleon sector.
Although this system is much more complicated than a single nucleon,
lattice simulations of this system now appear feasible and promise 
predictions  about the nucleon-nucleon interaction directly from QCD.
The lattice QCD treatment of the two-nucleon sector has just
begun%
~\cite{Arndt:2003vx,Beane:2003yx,Beane:2003da,Detmold:2004qn},
is still largely uninvestigated, and offers great opportunity 
for future research.

%\nocite{*}
\bibliographystyle{JHEP}
\bibliography{bibliography}

\appendix

\raggedbottom\sloppy

\chapter{Baryon Transformations for Flavor $SU(2|2)$ and $SU(4|2)$}
\label{CPT-SU2}
\begin{table}[ht]
\centering
\caption[Embedding of the baryon doublet and quartet for $SU(2|2)_V$]{\label{CPT-table:SU2Q}%
        Embedding of the baryon doublet and quartet for $SU(2|2)_V$ for QQCD.}
\begin{tabular}{c | c   c | c   c}\hline\hline
  & \multicolumn{2}{c|}{Doublet} & \multicolumn{2}{c}{Quartet} \\
  & $SU(2)_{\text{v}}\otimes SU(2)_{\text{g}}$ & dim & $SU(2)_{\text{v}}\otimes SU(2)_{\text{g}}$ & dim \\ \hline
  $qqq$ & $(\bf2,\bf1)$ & $\bf2$ & $(\bf4,\bf1)$ & $\bf4$ \\
  $qq\tilde{q}$ & $(\bf3,\bf2)\oplus(\bf1,\bf2)$ & $\bf8$ & $(\bf3,\bf2)$ & $\bf6$\\
  $q\tilde{q}\tilde{q}$ & $(\bf2,\bf3)\oplus(\bf2,\bf1)$  & $\bf8$ & $(\bf2,\bf3)$  & $\bf6$\\
  $\tilde{q}\tilde{q}\tilde{q}$ & $(\bf1,\bf2)$ & $\bf2$ &  & $\bf0$\\
  \hline
  & & $\bf20$ & & $\bf 16$\\
  \hline\hline
\end{tabular}
\end{table}
\begin{table}[ht]
\centering
\caption[Embedding of the baryon doublet and quartet for $SU(4|2)_V$]{\label{CPT-table:SU2PQ}%
        Embedding of the baryon doublet and quartet for $SU(4|2)_V$ for PQQCD.}
\begin{tabular}{c | c   c | c   c}\hline\hline
  & \multicolumn{2}{c|}{Doublet} & \multicolumn{2}{c}{Quartet} \\
  & $SU(2)_{\text{v}}\otimes SU(2)_{\text{s}}\otimes SU(2)_{\text{g}}$ & dim & $SU(2)_{\text{v}}\otimes SU(2)_{\text{s}}\otimes SU(2)_{\text{g}}$ & dim \\ \hline
  $qqq$ & $(\bf2,\bf1,\bf1)$ & $\bf2$ & $(\bf4,\bf1,\bf1)$ & $\bf4$ \\
  $qqq_{\text{s}}$ & $(\bf3,\bf2,\bf1)\oplus(\bf1,\bf2,\bf1)$ & $\bf8$ & $(\bf3,\bf2,\bf1)$ & $\bf6$ \\
  $qq_{\text{s}}q_{\text{s}}$ & $(\bf2,\bf3,\bf1)\oplus(\bf2,\bf1,\bf1)$ & $\bf8$ & $(\bf2,\bf3,\bf1)$ & $\bf6$ \\
  $q_{\text{s}}q_{\text{s}}q_{\text{s}}$ & $(\bf1,\bf2,\bf1)$ & $\bf2$ & $(\bf1,\bf4,\bf1)$ & $\bf4$ \\
  $qq\tilde{q}$ & $(\bf3,\bf1,\bf2)\oplus(\bf1,\bf1,\bf2)$ & $\bf8$ & $(\bf3,\bf1,\bf2)$ & $\bf6$ \\
  $qq_{\text{s}}\tilde{q}$ & $(\bf2,\bf2,\bf2)\oplus(\bf2,\bf2,\bf2)$ & $\bf16$ & $(\bf2,\bf2,\bf2)$ & $\bf8$ \\
  $q_{\text{s}}q_{\text{s}}\tilde{q}$ & $(\bf1,\bf3,\bf2)\oplus(\bf1,\bf1,\bf2)$ & $\bf8$ & $(\bf1,\bf3,\bf2)$ & $\bf6$ \\
  $q\tilde{q}\tilde{q}$ & $(\bf2,\bf1,\bf3)\oplus(\bf2,\bf1,\bf1)$ & $\bf8$ & $(\bf2,\bf1,\bf1)$ & $\bf2$ \\
  $q_{\text{s}}\tilde{q}\tilde{q}$ & $(\bf1,\bf2,\bf3)\oplus(\bf1,\bf2,\bf1)$ & $\bf8$ & $(\bf1,\bf2,\bf1)$ & $\bf2$ \\
  $\tilde{q}\tilde{q}\tilde{q}$ & $(\bf1,\bf1,\bf2)$ & $\bf2$ &  & $\bf0$ \\
  \hline
  & & $\bf70$ & & $\bf44$\\
  \hline\hline
\end{tabular}
\end{table}

\chapter{Formulae relevant for 
         $B^{(*)}\rightarrow D^{(*)}$
         at Zero Recoil in \QCPT}
\label{appendix-B2D}
We list the functions $H_1$, $H_2$, $F_1$, $H_5$, $H_8$, and $G_5$
(some of which have appeared in the literature before%
~\cite{Boyd:1995pa,Boyd:1995pq}).
Here, $m=m_{qq}$ is the mass of the $q\bar{q}$ light meson 
in the loop
where $q=u$, $d$, or~$s$ is the light (spectator) quark content
of the heavy mesons.
We have calculated 
loop integrals in $d=4-2\e$ dimensions and 
used dimensional regularization 
with the minimal subtraction ($\overline{\text{MS}}$) scheme,
where 
\begin{equation} \label{appendix-B2D-eq:dimreg}
1/\e'\equiv 1/\e-\g_E+\log 4\pi+1
.\end{equation} 
As a shorthand we have defined the function
\begin{equation} \label{appendix-B2D-eq:R}
  R(x)
  =
  \sqrt{x^2-1}
  \log\left(\frac{x-\sqrt{x^2-1+i\e}}{x+\sqrt{x^2-1+i\e}}\right),
\end{equation}
which occurs frequently.
We also need its derivative $dR/dx$ given by
\begin{equation}
  R'(x)=\frac{x}{x^2-1}R(x)-2
.\end{equation}

For the calculation of the wave function renormalization contribution
we need the derivatives of the loop integrals for the
diagrams in Fig.~\ref{B2D-F:wf-renorm}:
\begin{equation}
  H_1(\D)
  =
  \frac{i}{16\pi^2}
  \left[
    \log\frac{m^2}{\mu^2}-\frac{1}{\e'}-1
    -R'\left(\frac{\D}{m}\right)
  \right]
,\end{equation}
\begin{eqnarray}
  H_2(\D)
  &=&
  \frac{i}{16\pi^2}
  \left[
    \frac{16}{3}\D^2-\frac{10}{3}m^2
    +2(m^2-\D^2)\left(\!\log\frac{m^2}{\mu^2}-\frac{1}{\e'}\right)
    +\frac{4}{3}\D mR\left(\frac{\D}{m}\right)
  \right.     \nonumber \\
    && \left. \phantom{dddddd}
    +\left(\frac{2}{3}\D^2-\frac{5}{3}m^2\right)R'\left(\frac{\D}{m}\right)
  \right]
,\end{eqnarray}
and
\begin{eqnarray}
  F_1(\D)
  &=&
  \frac{i}{16\pi^2}
  \left[
    \frac{10}{3}\D^2-\frac{4}{3}m^2
    +\left(m^2-2\D^2\right)\left(\!\log\frac{m^2}{\mu^2}-\frac{1}{\e'}\right)
    +\frac{4}{3}\D mR\left(\frac{\D}{m}\right)
               \right .\nonumber \\
              && \left. \phantom{dddddd}
    +\frac{2}{3}(\D^2-m^2)R'\left(\frac{\D}{m}\right)
  \right]
.\end{eqnarray}

For the loop integrals of the vertex corrections
one finds
\begin{equation}
  H_5(\D,\Dtilde)
  = 
  \frac{i}{16\pi^2}
  \left\{
    \log\frac{m^2}{\mu^2}-\frac{1}{\e'}
    -1
    -\frac{m}{\D-\Dtilde}
     \left[R\left(\frac{\D}{m}\right)-R\left(\frac{\Dtilde}{m}\right)\right]
  \right\} 
,\end{equation}
\begin{eqnarray}
  H_8(\D,\Dtilde)
  &=&
  \frac{i}{16\pi^2}
  \left\{
    \left[2m^2-\frac{2}{3}(\D^2+\D\Dtilde+\Dtilde^2)\right]
     \left(\log\frac{m^2}{\mu^2}-\frac{1}{\e'}\right)   
    +\frac{16}{9}(\D^2+\D\Dtilde+\Dtilde^2)
  \right.     \nonumber \\
    && \left. \phantom{dddddd}
    -\frac{10}{3}m^2
    +\frac{m(5m^2-2\Dtilde^2)}{3(\D-\Dtilde)}R\left(\frac{\Dtilde}{m}\right)
    -\frac{m(5m^2-2\D^2)}{3(\D-\Dtilde)}R\left(\frac{\D}{m}\right)
  \right\}
,\nonumber \\
\end{eqnarray}
and
\begin{eqnarray}
  G_5(\D,\Dtilde)
  &=&
  \frac{i}{16\pi^2}
  \left\{
    \frac{10}{9}(\D^2+\D\Dtilde+\Dtilde^2)-\frac{4}{3}m^2     
  \right.     \nonumber \\
    && \left. \phantom{dddddd}
    +\left[m^2-\frac{2}{3}(\D^2+\D\Dtilde+\Dtilde^2)\right]
     \left(\!\log\frac{m^2}{\mu^2}-\frac{1}{\e'}\right)
  \right.     \nonumber \\
    && \left. \phantom{dddddd}
    +\frac{2m(\D^2-m^2)}{3(\D-\Dtilde)}R\left(\frac{\D}{m}\right)
    -\frac{2m(\Dtilde^2-m^2)}{3(\D-\Dtilde)}R\left(\frac{\Dtilde}{m}\right)
  \right\} 
.\end{eqnarray}

\chapter{Formulae relevant for \HMCPT\ in a Finite Volume}
\label{fv-app}

\section{Integrals and Sums}
\label{fv-app:integral}
We have regularized ultra-violet divergences that appear in loop integrals 
using dimensional
regularization with the $\overline{\text{MS}}$ scheme
[see Eq.~\eqref{appendix-B2D-eq:dimreg}].
The integrals appearing in the full QCD calculation are defined by 
\begin{equation}
  I_{\bar{\lambda}}(m)
  =
  \mu^{4 - d}\int \frac{d^{d}k}{(2\pi)^{d}} \frac{1}{k^{2}-m^{2}+ i\epsilon}
  = 
  \frac{i m^{2}}{16 \pi^{2}}
  \left[\frac{1}{\e'} - {\log\left(\frac{m^{2}}{\mu^{2}}\right)}\right] 
,\end{equation}
\begin{eqnarray}
  H_{\bar{\lambda}}(m,\Delta) 
  &=&
  \left(g^{\rho\nu} - v^{\rho} v^{\nu}\right)\mu^{4 - d}  
  \frac{\partial}{\partial\Delta}
  \int
  \frac{d^{d}k}{(2\pi)^{d}} 
  \frac{k_{\rho} k_{\nu}}{(k^{2}-m^{2}
         + i\epsilon)(v\cdot k - \Delta + i\epsilon)}
          \nonumber\\
  &=& 
  3 \frac{\partial}{\partial\Delta} F_{\bar{\lambda}}(m,\Delta),
\end{eqnarray}
where
\begin{eqnarray}\label{fv-eq:FandR}
  F_{\bar{\lambda}}(m,\Delta) 
  &=& 
  \frac{i}{16\pi^{2}} 
  \left\{ 
    \left[ 
      \frac{1}{\e'} - \log\left(\frac{m^{2}}{\mu^{2}}\right)
    \right] 
    \left(\frac{2\Delta^{2}}{3} - m^{2}\right)\Delta
    + 
    \left(\frac{10\Delta^{2}}{9} - \frac{4 m^{2}}{3}\right)\Delta
         \right.    \nonumber \\
    &&\phantom{mmm}+ \left. 
    \frac{2(\Delta^{2}-m^{2})}{3} m R\left(\frac{\Delta}{m}\right)
  \right\}
,\end{eqnarray}
with
\begin{equation}\label{fv-eq:FandR_again}
  R(x)
  =
  \sqrt{x^{2}-1}
  \log\left ( 
    \frac{x - \sqrt{x^{2}-1+i\epsilon}}{x + \sqrt{x^{2}-1+i\epsilon}} 
  \right)
,\end{equation}
and $\mu$ is the renormalisation scale.
For the quenched and partially quenched calculations, we also need the 
integrals
\begin{equation}
  I^{(\eta')}_{\bar{\lambda}}
  =
  \mu^{4 - d}\int \frac{d^{d}k}{(2\pi)^{d}} 
  \frac{1}{(k^{2}-m^{2}+ i\epsilon)^{2}}
  = 
  \frac{\partial I_{\bar{\lambda}}(m)}{\partial m^{2}} , 
\end{equation}
and
\begin{eqnarray}
  H^{\eta'}_{\bar{\lambda}}(m,\Delta) 
  &=& 
  \left(g^{\rho\nu} - v^{\rho} v^{\nu}\right) \mu^{4 - d} 
  \frac{\partial}{\partial\Delta}
  \int
  \frac{d^{d}k}{(2\pi)^{d}} 
  \frac{k_{\rho} k_{\nu}}{(k^{2}-m^{2}+i\epsilon)^{2}
  (v\cdot k - \Delta + i\epsilon)}
                    \nonumber\\
  &=& 
  \frac{\partial}{\partial m^{2}} H_{\bar{\lambda}}(m,\Delta)
.\end{eqnarray}
In a cubic spatial box of extent $L$ in four dimension
with periodic boundary conditions, 
one obtains the sums (after subtracting
the ultra-violet divergences)
\begin{equation}\label{fv-eq:FV_I}
  \cI(m) 
  =
  \frac{1}{L^{3}}\sum_{\vec{k}}\int \frac{d k_{0}}{2\pi}
  \frac{1}{k^{2}-m^{2}+i\epsilon} 
  = 
  I(m) + \IFV(m),
\end{equation}
and
\begin{eqnarray}\label{fv-eq:FV_H}
  \cH(m,\Delta) 
  &=&
  \left(g^{\rho\nu} - v^{\rho} v^{\nu}\right)\frac{1}{L^{3}} 
  \sum_{\vec{k}} 
  \frac{\partial}{\partial\Delta}
  \int
  \frac{d k_{0}}{2\pi} 
  \frac{k_{\rho} k_{\nu}}
       {(k^{2}-m^{2}+i\epsilon)(v\cdot k - \Delta + i\epsilon)}
                     \nonumber\\
  &=& 
  H(m,\Delta) + \HFV(m,\Delta)
\end{eqnarray}
for the full QCD calculation, where
the momentum $\vec{k}$ is quantized according to
$\vec{k}=(2\pi/L)\vec{n}$. 
Furthermore,
$I(m) = I_{\bar{\lambda}}(m)|_{\bar{\lambda}=0}$
and
$H(m) = H_{\bar{\lambda}}(m,\Delta)|_{\bar{\lambda}=0}$
are the infinite volume limits of $\cI$ and $\cH$,
and ($n=|\vec{n}|$)
\begin{eqnarray}
  \IFV(m) 
  &=& 
  -\frac{im}{4 \pi^{2}} 
  \sum_{\vec{n}\not=\vec{0}}
  \frac{K_{1}(n m L)}{n L}
            \nonumber\\
  &\rightarrow&
  -\frac{i}{4\pi^{2}} \sum_{\vec{n}\not=\vec{0}}
  \sqrt{\frac{m\pi}{2 n^3 L^3}}
  \text{e}^{-nmL}
  \left \{
    1 + \frac{3}{8 n m L} 
    - \frac{15}{128 (n m L)^{2}}
    + \cO\left(\left[\frac{1}{n m L}\right]^{3}\right)
 \right \}
      \nonumber \\
\end{eqnarray}
is the finite volume correction to $I(m)$
in the limit $m L \gg 1$.  The function $\HFV$
is the finite volume correction to $H(m,\Delta)$ and can be obtained via
\begin{equation}
  \HFV(m,\Delta) 
  = 
  i 
  \left[ 
    (m^{2} - \Delta^{2}) \KFV(m,\Delta) 
    - 2 \Delta \JFV(m,\Delta)
    + i \IFV(m)
  \right]
,\end{equation}
where $\JFV(m,\Delta)$ and $\KFV(m,\Delta)$ 
are defined in Eqs.~(\ref{fv-eq:JFV}) and (\ref{fv-eq:KFV_asymptotic}).

For QQCD and PQQCD calculations, one also needs
\begin{equation}\label{fv-eq:FV_I_DP}
  \cI^{\eta'}(m) 
  =
  \frac{1}{L^{3}}
  \sum_{\vec{k}}\int \frac{d k_{0}}{2\pi}
  \frac{1}{(k^{2}-m^{2}+i\epsilon)^{2}} 
  = 
  \frac{\partial}{\partial m^{2}}I(m)
  + \frac{\partial}{\partial m^{2}}\IFV(m)
,\end{equation}
and
\begin{eqnarray}\label{fv-eq:FV_H_DP}
  \cH^{\eta'}(m,\Delta) 
  &\equiv& 
  \left(g^{\rho\nu} - v^{\rho} v^{\nu}\right)\frac{1}{L^{3}} 
  \sum_{\vec{k}} 
  \frac{\partial}{\partial\Delta}
  \int
  \frac{d k_{0}}{2\pi} 
  \frac{k_{\rho} k_{\nu}}
       {(k^{2}-m^{2}+i\epsilon)^{2}(v\cdot k - \Delta + i\epsilon)}
                 \nonumber\\
  &=& 
  \frac{\partial}{\partial m^{2}}H(m,\Delta) 
  + \frac{\partial}{\partial m^{2}}\HFV(m,\Delta)
.\end{eqnarray}

\section{One-Loop Results}
\label{fv-sec:results}
We collect the results for one-loop corrections to 
$f_{P_{(s)}}\sqrt{M_{P_{(s)}}}$ and
$B_{P_{(s)}}$.    
For convenience, we introduce
\begin{equation}
 \cC_{\pm}(m,x) = \cI(m) \pm g^{2} \cH(m,x) ,
\end{equation}
and
\begin{equation}
 \cC^{\eta'}_{\pm}(m,x) =
  \cI^{\eta'}(m) \pm g^{2} \cH^{\eta'}(m,x)
,\end{equation}
where the functions $\cI(m)$, $\cH(m,x)$, 
$\cI^{\eta'}(m)$ and 
$\cH^{\eta'}(m,x)$ 
are defined in 
Eqs.(\ref{fv-eq:FV_I}), 
(\ref{fv-eq:FV_H}), (\ref{fv-eq:FV_I_DP}), and (\ref{fv-eq:FV_H_DP}), 
respectively.

In full QCD, we find
\begin{align}
  f_{P}\sqrt{M_{P}} 
  &= 
  \kappa
  \left\{
    1 - \frac{i}{12 f^{2}} 
    \left[ 
      9\cC_{-}(M_{\pi},\D) 
      + 6\cC_{-}(M_{K},\D+\delta_{s})
      + \cC_{-}(M_{\eta},\D)
    \right]
  \right\}, \\
  f_{P_{s}}\sqrt{M_{P_{s}}} 
  &= 
  \kappa
  \left\{
    1 - \frac{i}{3 f^{2}} 
    \left[
      3\cC_{-}(M_{K},\D-\delta_{s}) 
      + \cC_{-}(M_{\eta},\D) 
    \right]
  \right\}
,\end{align}
\begin{align}
  B_{P} 
  &= 
  \frac{3\beta}{2\kappa^{2}}
  \left\{
    1 - \frac{i}{6 f^{2}} 
    \left[ 
      3\cC_{+}(M_{\pi},\D) 
      + \cC_{+}(M_{\eta},\D)
    \right]
  \right\}, \\
  B_{P_{s}} 
  &= 
  \frac{3\beta}{2\kappa^{2}}
  \left\{
    1 - \frac{2 i}{3 f^{2}} 
    \left[
      \cC_{+}(M_{\eta},\D)
    \right]
  \right\}
.\end{align}
In QQCD, we find
\begin{align}
  f_{P}\sqrt{M_{P}} 
  &= 
  \kappa
  \left\{
    1 + \frac{i}{2 f^{2}} 
    \left[
      \frac{\alpha}3\cC_{-}(M_{\pi},\D)
      + \frac{\alpha M^{2}_{\pi} - M^{2}_{0}}{3}
        \cC^{\eta'}_{-}(M_{\pi},\D)
      + 2g\gamma\cH(M_{\pi},\D)
    \right]
  \right\},\\
  f_{P_{s}}\sqrt{M_{P_{s}}} 
  &= 
  \kappa
  \left\{
    1 + \frac{i}{2 f^{2}} 
    \left[
      \frac{\alpha}{3}\cC_{-}(m_{ss},\D)
      + \frac{\alpha m^{2}_{ss} - M^{2}_{0}}{3}
        \cC^{\eta'}_{-}(m_{ss},\D)
      + 2g\gamma\cH(m_{ss},\D)
    \right]
  \right\}
,\end{align}
\begin{align}
  B_{P} 
  &= 
  \frac{3\beta}{2\kappa^{2}} 
  \left\{
    1 - \frac{i}{f^{2}} 
    \left[
      \left(1 - \frac{\alpha}{3}\right)\cC_{+}(M_{\pi},\D)
      - \frac{\alpha M^{2}_{\pi} - M^{2}_{0}}{3}\cC^{\eta'}_{+}(M_{\pi},\D)
      + 2g\gamma\cH(M_{\pi},\D)
    \right]
  \right\},\\
  B_{P_{s}} 
  &= 
  \frac{3\beta}{2\kappa^{2}}
  \left\{
    1 - \frac{i}{f^{2}} 
    \left[
      \left(1 - \frac{\alpha}{3}\right)\cC_{+}(m_{ss},\D)
      - \frac{\alpha m^{2}_{ss} - M^{2}_{0}}{3}\cC^{\eta'}_{+}(m_{ss},\D)
      + 2g\gamma\cH(m_{ss},\D)
    \right]
  \right\}
,\end{align}
where
\begin{equation}\label{fv-eq:m33}
 m_{ss} = \sqrt{ 2 M^{2}_{K} - M_{\pi}^{2} }.
\end{equation}
In PQQCD, we find
\begin{eqnarray}
  f_{P}\sqrt{M_{P}} 
  &=& 
  \kappa
  \left\{ 
    1 - \frac{i}{2 f^2}
    \left[
      2\cC_{-}(m_{uj},\D+\delta_{\mathrm{sea}})
      + \cC_{-}(m_{ur}, \D+\delta_{\mathrm{sea}}+\tilde{\delta}_{s})
       \right.\right.  \nonumber\\
      &&\left.\left.\phantom{mmmmmmm}
      + 
      \frac{1}{3}
      \frac{m^2_{rr} - m^2_{uu}}{m^2_{uu}-m^2_X}
      \cC_{-}(m_{uu},\D)
      + 
      \frac{2}{27}
      \left(
        \frac{m_{rr}^2-m_{jj}^2}{m_X^2-m_{uu}^2}
      \right)^2
       \cC_{-}(m_{X},\D)
       \right.\right.      \nonumber\\
      &&\left.\left.\phantom{mmmmmmm}
      -
      \frac{1}{3}
      \frac{(m_{jj}^{2}-m_{uu}^{2})(m_{rr}^{2}-m_{uu}^{2})}
           {m_{uu}^{2}-m_{X}^{2}}
      \cC^{\eta'}_{-}(m_{uu},\D)
    \right]
 \right\}
,\end{eqnarray}
\begin{eqnarray}
  f_{P_s}\sqrt{M_{P_s}} 
  &=& 
  \kappa
  \left\{ 
    1 - \frac{i}{2 f^2}
    \left[
      2\cC_{-}(m_{sj},\D+\delta_{\mathrm{sea}}-\d_s)
      + \cC_{-}(m_{sr}, \D+\delta_{\mathrm{sea}}+\tilde{\delta}_{s}-\d_s)
       \right.\right.  \nonumber\\
      &&\left.\left.\phantom{mmmmmmm}
      + 
      \frac{1}{3}
      \frac{m^2_{jj} - m^2_{ss}}{m^2_{ss}-m^2_X}
      \cC_{-}(m_{ss},\D)
      + 
      \frac{2}{27}
      \left(
        \frac{m_{rr}^2-m_{jj}^2}{m_X^2-m_{ss}^2}
      \right)^2
       \cC_{-}(m_{X},\D)
       \right.\right.      \nonumber\\
      &&\left.\left.\phantom{mmmmmmm}
      -
      \frac{1}{3}
      \frac{(m_{jj}^{2}-m_{ss}^{2})(m_{rr}^{2}-m_{ss}^{2})}
           {m_{ss}^{2}-m_{X}^{2}}
      \cC^{\eta'}_{-}(m_{ss},\D)
    \right]
 \right\}
,\end{eqnarray}
\begin{eqnarray}
  B_{P} 
  &=& 
  \frac{3\beta}{2\kappa^{2}}
  \left\{ 
    1 - \frac{i}{f^{2}}
    \left[
      \cC_{+}(m_{uu},\D)
      - \frac{1}{3}
        \frac{(m_{jj}^2-m_{uu}^2)(m_{rr}^2-m_{uu}^2)}{m_{uu}^2-m_X^2}
        \cC^{\eta'}_{+}(m_{uu},\D)
                \right.\right.  \nonumber\\
      &&\left.\left.\phantom{mmmmmmm}
      +\frac{1}{3}
       \frac{m_{rr}^2-m_{uu}^2}{m_{uu}^2-m_X^2}
       \cC_{+}(m_{uu},\D)
%               \right.\right.  \nonumber\\
%      &&\left.\left.\phantom{mmmmmmm}
      +\frac{2}{27}
       \left(\frac{m_{rr}^2-m_{jj}^2}{m_{uu}^2-m_X^2}\right)^2
       \cC_{+}(m_X,\D)
    \right]
  \right\}
,\nonumber \\
\end{eqnarray} 
\begin{eqnarray}
  B_{P_s} 
  &=& 
  \frac{3\beta}{2\kappa^{2}}
  \left\{ 
    1 - \frac{i}{f^{2}}
    \left[
      \cC_{+}(m_{ss},\D)
      - \frac{1}{3}
        \frac{(m_{jj}^2-m_{ss}^2)(m_{rr}^2-m_{ss}^2)}{m_{ss}^2-m_X^2}
        \cC^{\eta'}_{+}(m_{ss},\D)
                \right.\right.  \nonumber\\
      &&\left.\left.\phantom{mmmmmmm}
      +\frac{1}{3}
       \frac{m_{jj}^2-m_{ss}^2}{m_{ss}^2-m_X^2}
       \cC_{+}(m_{ss},\D)
%               \right.\right.  \nonumber\\
%      &&\left.\left.\phantom{mmmmmmm}
      +\frac{2}{27}
       \left(\frac{m_{rr}^2-m_{jj}^2}{m_{ss}^2-m_X^2}\right)^2
       \cC_{+}(m_X,\D)
    \right]
  \right\}
,\nonumber \\
\end{eqnarray} 
where
\begin{equation}
  m_X^2 = \frac{1}{3}\left(m_{jj}^2 + 2 m_{rr}^2\right)
\end{equation}
It is straightforward to show that the PQQCD results 
reproduce those for full QCD in the limit 
$m_j=m_u$ and $m_r=m_s$.

\chapter{Charge Radii of the Meson and Baryon Octets for Flavor $SU(2)$}
\label{baryon_ff-sec:SU2}
We consider the case of $SU(2)$ 
flavor and calculate 
charge radii for the pions and
nucleons.
We keep the up and down quark masses non-degenerate and similarly 
for the sea-quarks. Thus the quark mass matrix reads
$m_Q^{SU(2)} = \diag(m_u, m_d, m_j, m_l, m_u, m_d)$.
Defining ghost and sea quark charges is constrained only by the 
restriction that QCD be recovered
in the limit of appropriately degenerate quark masses. 
Thus the most general form of the charge matrix is
\begin{equation}
  \cQ^{SU(2)} 
  = \diag\left(\frac{2}{3},-\frac{1}{3},q_j,q_l,q_j,q_l \right) 
.\end{equation}
The symmetry breaking pattern is assumed to be 
$SU(4|2)_L \otimes SU(4|2)_R \otimes U(1)_V
 \longrightarrow
 SU(4|2)_V \otimes U(1)_V$.

For the $\pi^+$, $\pi^-$, and $\pi^0$ we find
\begin{eqnarray}
  G^{PQ}_{\pi^+}(q^2)
  &=&
  1
  +
  \frac{1}{16\pi^2f^2}
  \left[
    \left(\frac{1}{3}+q_l\right)F_{dd}
    +
    \left(\frac{2}{3}-q_j\right)F_{uu}
    -
    \left(1-q_j+q_l\right)F_{ud}    
                  \right. \nonumber \\
     &&\phantom{dddddd} \left.
    -
    \left(\frac{1}{3}+q_j\right)F_{jd}    
    -
    \left(\frac{1}{3}+q_l\right)F_{ld}    
    -
    \left(\frac{2}{3}-q_j\right)F_{ju}    
    -
    \left(\frac{2}{3}-q_l\right)F_{lu}    
  \right]
                  \nonumber \\
     &&\phantom{dddddd} 
  +\a_9\frac{4}{f^2}q^2
,\end{eqnarray}
$G^{PQ}_{\pi^-}=-G^{PQ}_{\pi^+}$, and $G^{PQ}_{\pi^0}=0$,
respectively.

The baryon field assignments are analogous to the case of
$SU(3)$ flavor.
The nucleons are embedded as
\begin{equation} \label{baryon_ff-eqn:SU2nucleons}
 \cB_{ijk}=\frac{1}{\sqrt{6}}\left(\e_{ij} N_k + \e_{ik} N_j\right)
,\end{equation}
where the indices $i,j$ and $k$ are restricted to $1$ or $2$ and 
the $SU(2)$ nucleon doublet is defined as
\begin{equation}
  N = \left(\begin{matrix} p \\ n \end{matrix} \right) 
\end{equation}
The decuplet field $\cT_{ijk}$, 
which is totally symmetric, 
is normalized to contain the $\Delta$ resonances 
$T_{ijk}=\cT_{ijk}$ with $i$, $j$, $k$ restricted to 1 or 2.
The spin-3/2 baryon quartet is then contained as
\begin{equation}
  \cT_{111} = \D^{++}, \quad 
  \cT_{112} = \frac{1}{\sqrt{3}} \Delta^+, \quad 
  \cT_{122} = \frac{1}{\sqrt{3}} \Delta^0, \quad \text{and }
  \cT_{222} = \Delta^-
.\end{equation} 
The construction of the octet and decuplet baryons containing 
one sea or one ghost quark is analogous to the $SU(3)$ flavor
case~\cite{Beane:2002vq}
and we will not repeat it here.

The free Lagrangian for $\cB$ and $\cT$ is the one in 
Eq.~(\ref{baryon_ff-eqn:Linteract})
(with the parameters having different numerical values than in
the $SU(3)$ case).  
The connection to QCD is 
detailed in~\cite{Beane:2002vq}.
Similarly, the Lagrangian describing the interaction of the 
$\cB$ and $\cT$ with the pseudo-Goldstone bosons is 
the one in 
Eq.~(\ref{baryon_ff-eqn:Linteract}).  Matching it to the familiar one in QCD
(by restricting the $\cB_{ijk}$ and $\cT_{ijk}$ to the
$qqq$ sector)
\begin{equation}
  \cL = 2 g_A \ol{N} S^\mu A_\mu N 
      + g_1 \ol{N} S^\mu N \tr ( A_\mu) 
      + g_{\D N} \left( \ol{T}^{kji}_{\nu} A_{il}^{\nu} N_j \e_{kl} 
                      + \text{h.c.}  
                 \right)
\end{equation}
one finds at tree-level
\begin{equation}
  \a = \frac{4}{3} g_A + \frac{1}{3} g_1, \quad 
  \b = \frac{2}{3} g_1 - \frac{1}{3} g_A, \quad \text{and}\quad
  {\mathcal C} = - g_{\D N}
.\end{equation}

The contribution at leading order to the charge radii from the 
Pauli form factor $F_2(q^2)$, involves only the magnetic moments
which arise from the PQQCD Lagrangian~\cite{Beane:2002vq}
\begin{eqnarray}
  \cL &=& \frac{ie}{2 M_N}
        \left[ 
          \mu_\a\left(\ol\cB [S_\mu,S_\nu]\cB \cQ^{SU(2)}\right) 
         +\mu_\b\left(\ol\cB [S_\mu,S_\nu]\cQ^{SU(2)}\cB\right)
             \right.\nonumber \\
      &&\left.\phantom{ascuasd}
         +\mu_\g\,\str(\cQ^{SU(2)})\left(\ol\cB [S_\mu,S_\nu]\cB\right)
        \right]
        F^{\mu\nu}
.\end{eqnarray} 
Note that in the case of
$SU(2)$ flavor the charge matrix $\cQ$ is not
supertraceless and hence there appears a third operator.
In QCD, the corresponding Lagrange density is conventionally written 
in terms of isoscalar and isovector couplings
\begin{equation}
  \cL = \frac{ie}{2 M_N} 
        \left( 
          \mu_0 \ol N [S_\mu,S_\nu] N 
          + \mu_1 \ol N [S^\mu,S^\nu] \tau^3 N 
        \right)
        F^{\mu\nu}  
\end{equation}
and one finds that the 
QCD and PQQCD coefficients are related by~\cite{Beane:2002vq}
\begin{equation}
  \mu_0 = \frac{1}{6}\left(\mu_\a +\mu_\b +2\mu_\g\right), \quad \text{and }
  \mu_1 = \frac{1}{6}\left(2\mu_\a - \mu_\b\right)
.\end{equation}
Likewise, the leading tree-level corrections to the charge-radii come
from the Lagrangian
\begin{equation}
  \cL
  =
  \frac{e}{\L_\chi^2}
  \left[
    c_\a\,(\ol \cB \cB \cQ^{SU(2)})
    +c_\b\,(\ol \cB \cQ^{SU(2)}\cB)
    +c_\g\,\str(\cQ^{SU(2)})(\ol \cB \cB)
  \right]
  v_\mu\partial_\nu F^{\mu\nu}
\end{equation}
that matches onto the QCD Lagrangian
\begin{equation}
  \cL
  =
  \frac{e}{\L_\chi^2}
  \left(
    c_0\,\ol N N
    +c_1\,\ol N\tau^3 N
  \right)
  v_\mu\partial_\nu F^{\mu\nu}
\end{equation}
with
\begin{equation}
  c_0 = \frac{1}{6}\left(c_\a +c_\b +2c_\g\right), \quad \text{and }
  c_1 = \frac{1}{6}\left(2c_\a - c_\b\right)
.\end{equation}

Evaluating the charge radii at NLO order in the 
chiral expansion yields
\begin{equation}\label{baryon_ff-eqn:fred2}
  <r_E^2>
  =
  -\frac{6c}{\L_\chi^2}
  +
  \frac{3\a}{2M_N^2}
  -
  \frac{1}{16\pi^2 f^2}
  \sum_{X}
  \left[ 
    A_X\log\frac{m_X^2}{\mu^2}
    -
    5\,\b_X\log\frac{m_X^2}{\mu^2}
    +
    10\,\b_X'{\mathcal G}(m_X,\D,\mu)
  \right]. 
\end{equation}
The coefficients $c$ are given by $c_p =c_0+c_1$ and $c_n =c_0-c_1$ 
while $\a_p = \mu_0 + \mu_1$ and $\a_n = \mu_0 - \mu_1$. 
The remaining coefficients are listed in Table~\ref{baryon_ff-T:SU2p} for the proton 
and Table~\ref{baryon_ff-T:SU2n} for the neutron.

\begin{table}[tb]
\centering
\caption[$\b_X$, $\b_X'$, and $A_X$
         in $SU(2)$ flavor \PQCPT\ for the proton]{\label{baryon_ff-T:SU2p}The coefficients $\b_X$, $\b_X'$, and $A_X$
         in $SU(2)$ flavor \PQCPT\ for the proton.}
%\begin{ruledtabular}
\begin{tabular}{c | c  c  c }
  \hline\hline
	$X$     &   $\beta_X$   &   $\beta'_X$   &  $A_X$  \\
	\hline
	$uu$   &   $\frac{2}{9}(4 g_A^2 + 2 g_1 g_A + g_1^2) (2 - 3 q_j)$  
						      &     $ (-\frac{1}{27} + \frac{1}{18} q_j)g_{\Delta N}^2$     &    $-\frac{4}{3} + 2 q_j$ \\	
	
	$ud$   &   $-\frac{4}{9} g_A^2 ( 5 + 6 q_l) + \frac{4}{9} g_1 g_A (2 - 3 q_l) + \frac{1}{9} g_1^2(1 - 9 q_j - 6 q_l)$           
						      &     $ (- \frac{2}{9} + \frac{1}{9} q_j + \frac{1}{18}q_l)g_{\Delta N}^2$  &    $q_j + 2 q_l$      \\
	
	$dd$   &   $-\frac{1}{3} g_1^2 ( 1 + 3 q_l)$   &     $( \frac{1}{27} + \frac{1}{9} q_l)g_{\Delta N}^2$     &    $\frac{1}{3} + q_l$      \\
	
	$uj$   &   $-\frac{2}{9} (4 g_A^2 + 2 g_1 g_A + g_1^2) ( 2 - 3 q_j) $        
					              &     $(\frac{1}{27} - \frac{1}{18} q_j)g_{\Delta N}^2$     &    $\frac{4}{3} - 2 q_j$      \\
	
	$ul$   &   $-\frac{2}{9} (4 g_A^2 + 2 g_1 g_A + g_1^2) ( 2 - 3 q_l)$       
						      &     $(\frac{1}{27} - \frac{1}{18} q_l)g_{\Delta N}^2$       &    $\frac{4}{3} - 2 q_l$      \\
	
	$dj$   &   $\frac{1}{3} g_1^2 (  1+ 3 q_j)$  &     $( -\frac{1}{27} - \frac{1}{9} q_j)g_{\Delta N}^2$     &    $-\frac{1}{3} - q_j$      \\
	
	$dl$   &   $\frac{1}{3} g_1^2 ( 1 + 3 q_l)$  &     $( -\frac{1}{27} - \frac{1}{9} q_l)g_{\Delta N}^2$     &    $-\frac{1}{3} - q_l$ \\
\hline\hline
\end{tabular}
%\end{ruledtabular}
\end{table} 
%\spacebetweentables
\begin{table}[tb]
\caption[$\b_X$, $\b_X'$, and $A_X$
         in $SU(2)$ flavor \PQCPT\ for the neutron]{\label{baryon_ff-T:SU2n}The coefficients $\b_X$, $\b_X'$, and $A_X$
         in $SU(2)$ flavor \PQCPT\ for the neutron.}
%\begin{ruledtabular}
\begin{tabular}{c | c  c  c }
  \hline\hline
	$X$     &   $\beta_X$   &   $\beta'_X$   &  $A_X$  \\
	\hline
	
	$uu$   &   $\frac{1}{3} g_1^2 (2 - 3 q_j)$  
						      &     $(-\frac{2}{27} + \frac{1}{9} q_j)g_{\Delta N}^2$     &    $-\frac{2}{3} +  q_j$ \\	
	
	$ud$   &   $\frac{4}{9} g_A^2 ( 7 - 6 q_j) - \frac{4}{9} g_1 g_A ( 1 + 3 q_j) + \frac{1}{9} g_1^2( 4 - 6 q_j - 9  q_l)$           
						      &     $(\frac{1}{6} + \frac{1}{18}q_j + \frac{1}{9} q_l)g_{\Delta N}^2$  &    $-1 + 2 q_j + q_l$      \\
	
	$dd$   &   $-\frac{2}{9} (4 g_A^2 + 2 g_1 g_A + g_1^2) ( 1 + 3 q_l)$   &     $( \frac{1}{54} + \frac{1}{18} q_l)g_{\Delta N}^2$     &    $\frac{2}{3} + 2 q_l$      \\
	
	$uj$   &   $-g_1^2 ( \frac{2}{3} -  q_j) $        
					              &     $(\frac{2}{27} - \frac{1}{9} q_j)g_{\Delta N}^2$     &    $\frac{2}{3} -  q_j$      \\
	
	$ul$   &   $-g_1^2 ( \frac{2}{3} -  q_l)$       
						      &     $(\frac{2}{27} - \frac{1}{9} q_l)g_{\Delta N}^2$       &    $\frac{2}{3} -  q_l$      \\
	
	$dj$   &   $\frac{2}{9}(4 g_A^2 + 2 g_1 g_A + g_1^2) ( 1 + 3 q_j)$  &     $( -\frac{1}{54} - \frac{1}{18} q_j)g_{\Delta N}^2$     &    $-\frac{2}{3} - 2 q_j$     \\
	
	$dl$   &   $\frac{2}{9}(4 g_A^2 + 2 g_1 g_A + g_1^2) ( 1 + 3 q_l)$  &     $( -\frac{1}{54} - \frac{1}{18} q_l)g_{\Delta N}^2$     &    $-\frac{2}{3} - 2 q_l$ \\
\hline\hline
\end{tabular}
%\end{ruledtabular}
\end{table}

\chapter{More on the Baryon Decuplet Form Factors}
\section{$q^2$ Dependence of the Form Factors}
\label{decuplet-s:q-dep}
For reference, we provide the $q^2$ dependence of the decuplet electromagnetic form factors
defined in Section~\ref{decuplet-sec:ff} at one-loop order in the chiral expansion. To do so we 
define 
\begin{equation}
P_X = \sqrt{1 - \frac{x (1-x) q^2}{m_X^2}}
.\end{equation}
Then we have
\begin{eqnarray} \label{decuplet-eqn:F1q}
F_1(q^2)  
&=&
 Q \left( 1 - \frac{\mu_c q^2}{2 M_B^2} - \frac{ \mathbb{Q}_{\text{c}} q^2}{2 \L_\chi^2} + \frac{c_c q^2}{\L_\chi^2} \right) \nonumber \\
&&- \frac{3 + \mathcal{C}^2}{16 \pi^2 f^2}  \sum_X  A_X  \left[ \frac{q^2}{6} \log \frac{m_X^2}{\mu^2} 
- 2 m_X^2 \int_0^1 dx \; P^2_X \log P_X \right] \nonumber \\
&&- \frac{\mathcal{H}^2}{24 \pi^2 f^2}  \sum_X A_X \Bigg\{ \frac{11}{36} q^2 \log \frac{m_X^2}{\mu^2} 
+ \frac{5}{3} \D m_X  \cR \left( \frac{\D}{m_X} \right) \nonumber\\
&&\phantom{GWB}-\int_0^1 dx \Bigg[\frac{10}{3} 
               \left( \frac{m_X^2}{2} - \D^2 - \frac{11}{10} x(1-x) q^2 \right) \log P_X \nonumber \\
&&\phantom{an idiot}+ \D m_X  P_X \left( \frac{5}{3} + \frac{x(1-x) q^2}{\D^2 - m_X^2 P^2_X} \right) 
\cR \left(\frac{\Delta}{m_X P_X }\right) \Bigg]  \Bigg\},\nonumber \\  
\end{eqnarray}
\begin{eqnarray} \label{decuplet-eqn:F2q}
F_2(q^2)  
&=& 
2 \mu_c Q -\frac{\mathcal{C}^2 M_B}{8 \pi f^2} \sum_X A_X  m_X  \int_0^1 dx \; P_X
+ \frac{ M_B \mathcal{H}^2}{36 \pi^2 f^2} \sum_X A_X 
\Bigg\{ \Delta \log \frac{m_X^2}{\mu^2} \nonumber \\ 
&&\phantom{indeed}+  \int_0^1 dx \Bigg[ 2 \D \log P_X
-  m_X P_X  \cR \left(\frac{\Delta}{m_X P_X } \right) \Bigg] \Bigg\} 
,\end{eqnarray}
and
\begin{eqnarray} \label{decuplet-eqn:G1q}
G_1(q^2) 
&=&  
4 Q \left( \mu_c + \mathbb{Q}_c \frac{2 M_B^2}{\L_\chi^2}  \right) 
- \frac{M_B^2 \mathcal{C}^2}{2 \pi^2 f^2} \sum_X A_X \left[ \frac{1}{6} \log \frac{m_X^2}{\mu^2} 
+  \int_0^1 dx \, 2 x(1-x) \log P_X \right]  \nonumber \\ 
&&+ \frac{ 2 M_B^2 \mathcal{H}^2}{9 \pi^2 f^2} \sum_X A_X \Bigg\{ \frac{1}{6} \log \frac{m_X^2}{\mu^2} +
\int_0^1 dx \, x(1-x) \Bigg[ 2 \log P_X   \nonumber \\
&&\phantom{gfhjhgjhghg}- \frac{\Delta m_X P_X}{\Delta^2 - m^2_X P^2_X}  
\cR \left( \frac{\Delta}{m_X P_X} \right)  \Bigg] \Bigg\}
.\end{eqnarray}

\section{Electromagnetic Properties for Flavor $SU(2)$}
\label{decuplet-s:su2}
Here we consider the case of $SU(2)$ 
flavor and calculate 
the electromagnetic moments and charge radii of the delta quartet. 
We keep the up and down valence quark masses non-degenerate and similarly 
for the sea-quarks. Thus the quark mass matrix reads
$m_Q^{SU(2)} = \diag(m_u, m_d, m_j, m_l, m_u, m_d)$.
Defining ghost and sea quark charges is constrained only by the 
restriction that QCD be recovered
in the limit of appropriately degenerate quark masses. 
Thus the most general form of the charge matrix is
\begin{equation} \label{decuplet-eqn:SU2chargematrix}
  \cQ^{SU(2)} 
  = \diag\left(\frac{2}{3},-\frac{1}{3},q_j,q_l,q_j,q_l \right) 
.\end{equation}
The symmetry breaking pattern is assumed to be 
$SU(4|2)_L \otimes SU(4|2)_R \otimes U(1)_V
 \longrightarrow
 SU(4|2)_V \otimes U(1)_V$.
The baryon field assignments are analogous to the case of
$SU(3)$ flavor.
The nucleons are embedded as
\begin{equation} \label{decuplet-eqn:SU2nucleons}
 \cB_{ijk}=\frac{1}{\sqrt{6}}\left(\e_{ij} N_k + \e_{ik} N_j\right)
,\end{equation}
where the indices $i,j$ and $k$ are restricted to $1$ or $2$ and 
the $SU(2)$ nucleon doublet is defined as
\begin{equation}
  N = \left(\begin{matrix} p \\ n \end{matrix} \right) 
\end{equation}
The decuplet field $\cT_{ijk}$, 
which is totally symmetric, 
is normalized to contain the $\Delta$-resonances 
$T_{ijk}=\cT_{ijk}$ with $i$, $j$, $k$ restricted to 1 or 2.
Our states are normalized so that $\cT_{111} = \D^{++}$.
The construction of the octet and decuplet baryons containing 
one sea or one ghost quark is analogous to the $SU(3)$ flavor
case~\cite{Beane:2002vq}
and we will not repeat it here.

The free Lagrangian for $\cB$ and $\cT$ is the one in 
Eq.~(\ref{CPT-eqn:L})
(with the parameters having different numerical values than
the $SU(3)$ case).  
The connection to QCD is 
detailed in~\cite{Beane:2002vq}.
Similarly, the Lagrangian describing the interaction of the 
$\cB$ and $\cT$ with the pseudo-Goldstone bosons is 
the one in 
Eq.~(\ref{baryon_ff-eqn:Linteract}).  Matching it to the familiar one in QCD
(by restricting the $\cB_{ijk}$ and $\cT_{ijk}$ to the $qqq$ sector),
\begin{equation}
  \cL = g_{\D N} \left( \ol{T}{}^{kji}_{\nu} A_{il}^{\nu} N_j \e_{kl} 
                      + \text{h.c}
                 \right)
	+ 2 g_{\D \D} \ol{T}{}^\nu_{kji} S_\mu A^\mu_{il} T_{\nu,ljk} 
	+ 2 g_X  \ol{T}{}^\nu_{kji} S_\mu  T_{\nu,ijk} \tr (A^\mu)  
,\end{equation}
one finds at tree-level ${\mathcal C} = - g_{\D N}$ and $\mathcal{H} = g_{\D \D}$,
with $g_X = 0$. The leading tree-level operators which contribute to $\D$ electromagnetic
properties are the same as in Eqs.~\eqref{decuplet-eqn:LDF}, \eqref{decuplet-eqn:Lc}, and \eqref{decuplet-eqn:Lnew}, of course the low-energy
constants have different values.

Evaluating the $\D$ electromagnetic properties at NLO in the chiral expansion yields expressions identical
in form to those above Eqs.~\eqref{decuplet-eqn:F2}, \eqref{decuplet-eqn:G1}, and \eqref{decuplet-eqn:r_E} with the $SU(2)$ identifications made for $\mathcal{C}$
and $\mathcal{H}$ above. The SU$(2)$ coefficients $A_X^T$ appear in Table \ref{decuplet-t:su2} for particular
$\D$--resonance states $T$. 
\begin{table}[tb]
\centering
\caption[SU$(2)$ coefficients $A_X^T$ in \CPT\ and \PQCPT]{The SU$(2)$ coefficients $A_X^T$ in \CPT\ and \PQCPT.}  
%\begin{ruledtabular}
\begin{tabular}{l | c | c  c  c  c  c  c  c }
  \hline\hline
	   & \CPT\  &  \multicolumn{7}{c}{\PQCPT} \\
	                 & $\pi^{\pm}$   &   $uu$   &   $ud$   &  $dd$  &   $ju$   &   $lu$   & $jd$   & $ld$ \\
	 	
	\hline
	$\Delta^{++}$       & $1$  &  $-\frac{2}{3} + q_j$ & $\frac{1}{3} + q_l$ & $0$ & $\frac{2}{3} - q_j$ & $\frac{2}{3} - q_l$ & $0$ & $0$ \\
 
	$\Delta^{+}$        & $\frac{1}{3}$ & $-\frac{4}{9} + \frac{2}{3} q_j$ & $\frac{1}{3} q_j + \frac{2}{3} q_l$ & $\frac{1}{9} + \frac{1}{3} q_l$ & $\frac{4}{9} - \frac{2}{3} q_j$ & $\frac{4}{9} - \frac{2}{3} q_l$ & $-\frac{1}{9} - \frac{1}{3} q_j$ & $-\frac{1}{9} - \frac{1}{3} q_l$ \\

	$\Delta^{0}$        & $-\frac{1}{3}$ & $-\frac{2}{9} + \frac{1}{3} q_j$ & $-\frac{1}{3} + \frac{2}{3} q_j + \frac{1}{3} q_l$ & $\frac{2}{9} + \frac{2}{3} q_l$ & $\frac{2}{9} - \frac{1}{3} q_j$ & $\frac{2}{9} - \frac{1}{3} q_l$ & $-\frac{2}{9} - \frac{2}{3} q_j$ & $-\frac{2}{9} - \frac{2}{3} q_l$ \\

	$\Delta^{-}$        & $-1$ & $0$ & $-\frac{2}{3} + q_j$ & $\frac{1}{3} + q_l$ & $0$ & $0$ & $-\frac{1}{3} - q_j$ & $-\frac{1}{3} - q_l$ \\
\hline\hline
\end{tabular}
%\end{ruledtabular}
\label{decuplet-t:su2}
\end{table}[tb] 
In the table, we have listed values corresponding to the loop meson that has mass $m_X$ 
for both \CPT\ and \PQCPT. Again, the \CPT\ coefficients can be used to find the $\D$-resonance charge radii in two-flavor QCD. 
These have not been previously calculated.

%----------------------

In addition, however,  
local counterterms appear, that involve the non-zero supertrace of the 
charge matrix in $SU(2|2)$.
Using the general form of the charge matrix
Eq.~(\ref{decuplet-eqn:SU2chargematrix}), 
we have an additional dimension-$5$ magnetic moment operator 
in \PQCPT
\begin{equation}
\cL = \mu_\gamma \frac{3 i e} {M_B} \left( \ol \cT {}^\mu \cT^\nu \right)
F_{\mu \nu} 
\, \str ( \cQ^{SU(2)} )
,\end{equation}
that matches onto the \CPT\ operator
\begin{equation}
\cL = \mu_\gamma \frac{3 i e}{M_B} \ol T {}^\mu_i T^\nu_i 
F_{\mu \nu} 
\, \tr(\cQ^{SU(2)} )
.\end{equation}
There is an additional dimension-$6$ electric quadrupole operator in \PQCPT
\begin{equation}
\cL = - \mathbb{Q}_\gamma \frac{3 e}{\L_\chi^2} 
\left( \ol \cT {}^{\{ \mu} \cT^{\nu \}} \right)
v^\a \partial_{\mu} F_{\nu \a}
\, \str ( \cQ^{SU(2)} )
,\end{equation}
that matches onto the \CPT\ operator
\begin{equation}
\cL = - \mathbb{Q}_\gamma \frac{3 e}{\L_\chi^2} 
\ol T {}^{\{\mu}_i T^{\nu\}}_i
v^\a \partial_{\mu} F_{\nu \a}
\, \tr(\cQ^{SU(2)} )
.\end{equation}
Finally in \PQCPT\ 
there is an additional dimension-$6$ charge radius operator
\begin{equation}
\cL = c_\gamma \frac{3 e}{\L_\chi^2}
\left( \ol \cT {}^\sigma \cT_\sigma \right)
v_\mu \partial_\nu F^{\mu \nu} 
\, \str ( \cQ^{SU(2)} )
,\end{equation}
that matches onto 
\begin{equation}
\cL = c_\gamma \frac{3 e}{\L_\chi^2}
\ol T {}^\sigma_i T_{\sigma,i} \,
v_\mu \partial_\nu F^{\mu \nu} 
\, \tr(\cQ^{SU(2)} )
\end{equation}
in \CPT.
Notice the \PQCPT\ low-energy constants $\mu_\gamma$, $\mathbb{Q}_\gamma$,
and $c_\gamma$ are identical at tree level to those in \CPT. 

Inclusion of the above operators leads to tree-level contributions to the
$\D$ quartet electromagnetic properties. Since these contributions are 
proportional to the supertrace of the charge matrix, the corrections are 
identical for each member of the quartet. The charge radius should 
include an additive correction 
\begin{equation}
\delta <r_E^2> = \frac{2 \mu_\gamma}{M_B^2} + \frac{\mathbb{Q}_\gamma + 
6 c_\gamma}{\L_\chi^2}
,\end{equation}
while for the magnetic moment
\begin{equation}
\delta \mu = 2 \mu_\gamma
,\end{equation} 
and for the electric quadrupole moment
\begin{equation}
\delta \mathbb{Q} = - 2 \mu_\gamma - 4 \mathbb{Q}_\gamma 
\frac{M_B^2}{\L_\chi^2}
.\end{equation}
Notice that these corrections only affect the counterterm structure of the 
results.

\chapter{$\D\to N\g$ Transitions for Flavor $SU(2)$}
\label{trans-s:su2}
We repeat the calculation of the 
transition moments for the case of $SU(2)$ 
flavor with non-degenerate quarks, i.e.,
the quark mass matrix reads
$m_Q^{SU(2)} = \diag(m_u, m_d, m_j, m_l, m_u, m_d)$.
Since defining ghost and sea quark charges is constrained only by the 
restriction that QCD be recovered
in the limit of appropriately degenerate quark masses, 
the most general form of the charge matrix is
\begin{equation}
  \cQ^{SU(2)} 
  = \diag\left(\frac{2}{3},-\frac{1}{3},q_j,q_l,q_j,q_l \right) 
.\end{equation}
The symmetry breaking pattern is assumed to be 
$SU(4|2)_L \otimes SU(4|2)_R \otimes U(1)_V
 \longrightarrow
 SU(4|2)_V \otimes U(1)_V$.
The baryon field assignments are analogous to the case of
$SU(3)$ flavor.
The nucleons are embedded as
\begin{equation} \label{trans-eqn:SU2nucleons}
 \cB_{ijk}=\frac{1}{\sqrt{6}}\left(\e_{ij} N_k + \e_{ik} N_j\right)
,\end{equation}
where the indices $i,j$ and $k$ are restricted to $1$ or $2$ and 
the $SU(2)$ nucleon doublet is defined as
\begin{equation}
  N = \left(\begin{matrix} p \\ n \end{matrix} \right) 
\end{equation}
The decuplet field $\cT_{ijk}$, 
which is totally symmetric, 
is normalized to contain the $\Delta$-resonances 
$T_{ijk}=\cT_{ijk}$ with $i$, $j$, $k$ restricted to 1 or 2
and
$\cT_{111} = \D^{++}$.
The construction of the octet and decuplet baryons containing 
one sea or one ghost quark is analogous to the $SU(3)$ flavor
case~\cite{Beane:2002vq}
and will not be repeat here.

The free Lagrangian for $\cB$ and $\cT$ is the one in 
Eq.~(\ref{CPT-eqn:L})
(with the parameters having different numerical values than
the $SU(3)$ case).  
The connection to QCD is 
detailed in~\cite{Beane:2002vq}.
Similarly, the Lagrangian describing the interaction of the 
$\cB$ and $\cT$ with the pseudo-Goldstone bosons is 
the one in 
Eq.~(\ref{trans-eqn:Linteract}) that can be matched to the familiar one in QCD
(by restricting the $\cB_{ijk}$ and $\cT_{ijk}$ to the $qqq$ sector),
\begin{eqnarray}
  \cL 
  &=&
  2g_A{\ol N}S^\mu A_\mu N+g_1{\ol N}S^\mu N\tr(A_\mu) 
  + 
  g_{\D N} \left( \ol{T}{}^{kji}_{\nu} A_{il}^{\nu} N_j \e_{kl} 
                      + \text{h.c}
                 \right) \nonumber \\
  &&+
  2 g_{\D \D} \ol{T}{}^\nu_{kji} S_\mu A^\mu_{il} T_{\nu,ljk} 
  + 
  2 g_X  \ol{T}{}^\nu_{kji} S_\mu  T_{\nu,ijk} \tr (A^\mu)  
,\end{eqnarray}
where one finds at tree-level 
$g_1=-2(D - F)$, $g_A = D + F$,
${\mathcal C} = - g_{\D N}$, and $\mathcal{H} = g_{\D \D}$,
with $g_X = 0$. 
The leading tree-level operators which contribute to $\D\to N\g$
have the same form as in Eq.~\eqref{trans-eqn:Lbob};
of course the low-energy
constants have different values.
For transitions no additional tree-level operators
involving supertrace of $\cQ^{SU(2)}$ appear.

Evaluating the transition moments at NLO in the chiral expansion 
yields expressions identical
in form to those in Eqs.~\eqref{trans-eqn:G1}--\eqref{trans-eqn:G3} 
with the SU$(2)$ identifications made for $\mathcal{C}$,
$\mathcal{H}$, $D$, and $F$. 
For the $SU(2)$ coefficients in 
\CPT\ one finds $\b_X^B=g_A/\sqrt{3}$ and $\b_X^T=5/(3\sqrt{3})$
for the $\pi^\pm$.
The corresponding values for the case of \PQCPT\
appear in
Table~\ref{trans-t:clebschSU2}.
\begin{table}[htb]
\centering
\caption[$SU(2)$ coefficients $\beta_X^B$ and $\beta_X^T$ in \PQCPT\ 
         for $\Delta \to N \gamma$]{\label{trans-t:clebschSU2}
The $SU(2)$ coefficients $\beta_X^B$ and $\beta_X^T$ in \PQCPT\ 
         for $\Delta \to N \gamma$.}
\begin{tabular}{r | c  c }
\hline\hline
&  $\beta_X^B$  & $\beta_X^T$ \\
\hline
   $uu$     & $\frac{1}{3 \sqrt{3}} (2 - 3 q_j)$ & $-\frac{1}{9 \sqrt{3}}(2 - 3 q_j)$ \\   
   $ud$     & $\frac{1}{\sqrt{3}}\,[1 + q_j - q_l + 2 \frac{g_A}{g_1}]\; $ & $\frac{1}{3 \sqrt{3}} (4 -  q_j + q_l)$ \\   
       $dd$     & $\frac{1}{3 \sqrt{3}} (1 + 3 q_l)$ & $-\frac{1}{9 \sqrt{3}}(1 + 3 q_l)$ \\  
       $ju$     & $-\frac{1}{3 \sqrt{3}} (2 - 3 q_j)$ & $\frac{1}{9 \sqrt{3}}(2 - 3 q_j)$ \\  
       $lu$     & $-\frac{1}{3 \sqrt{3}} (2 - 3 q_l)$ & $\frac{1}{9 \sqrt{3}}(2 - 3 q_l)$ \\
       $jd$     & $-\frac{1}{3 \sqrt{3}} (1 + 3 q_j)$ & $\frac{1}{9 \sqrt{3}}(1 +3 q_j)$ \\ 
       $ld$     & $-\frac{1}{3 \sqrt{3}} (1 + 3 q_l)$ & $\frac{1}{9 \sqrt{3}}(1 +3 q_l)$ \\  
\hline\hline
\end{tabular}
\end{table}

\chapter{More on Finite Lattice Spacing Corrections}

\section{$\order(a)$ Corrections for Flavor $SU(2)$}
\label{fa-s:su2}
We consider the case of $SU(2)$ 
flavor PQQCD%
\footnote{
For brevity we refer to $SU(4|2)$ PQQCD as $SU(2)$. 
The distinction will always be clear.}
 and summarize the changes needed to determine 
finite lattice spacing corrections to the electromagnetic properties 
of hadrons considered above.
For the two flavor case, we keep the up and down valence quark masses 
non-degenerate and similarly for the sea-quarks. 
Thus the quark mass matrix reads
\begin{equation}
m_Q^{SU(2)} = \diag(m_u, m_d, m_j, m_l, m_u, m_d),
\end{equation}
while the SW matrix is 
\begin{equation}
c_Q^{SU(2)} = \diag(c^v, c^v, c^s, c^s, c^v, c^v)
.\end{equation}

Defining ghost and sea quark charges is constrained only by the 
restriction that QCD be recovered
in the limit of appropriately degenerate quark masses. 
Thus the most general form of the charge matrix is
\begin{equation}
  \cQ^{SU(2)} 
  = \diag\left(\frac{2}{3},-\frac{1}{3},q_j,q_l,q_j,q_l \right) 
,\end{equation}
which is not supertraceless.
Analogous to the three flavor case, the vector-current will 
receive $\cO(a)$ corrections from the operators in Eq.~\eqref{fa-eqn:vectora} of 
which only the operator $\cO_1^\mu$ is relevant. The coefficient matrix associated with 
this operator is 
\begin{equation}
c_1^{SU(2)} = \diag(c_1^v, c_1^v, c_1^s, c_1^s, c_1^v, c_1^v)
.\end{equation}

The $\cO(a)$ operators listed above in Sections \ref{fa-sec:mesons}--\ref{fa-sec:trans} are the same for the $SU(2)$ flavor group, however, the coefficients
have different numerical values. Additionally there are operators 
involving $\str(\cQ_+^{SU(2)})$. These are listed for each electromagnetic observable below.

\subsection*{Octet mesons}
In the meson sector, one has the additional term
\begin{eqnarray}
\cL &=& 
i m_4 \L_\chi F_{\mu \nu} \, \str 
\left( 
\cA_+ D^\mu  \Sigma D^\nu \Sigma^\dagger 
+ 
\cA_+ D^\mu \Sigma^\dagger D^\nu  \Sigma 
\right) \, \str(\cQ_+^{SU(2)}) 
\label{fa-eqn:mesonops2}.\end{eqnarray}

\subsection*{Octet baryons}
In the octet baryon sector, there are terms which originate from $\cA_+$ insertions 
\begin{eqnarray}
\cL &=&
\frac{1}{\L_\chi}
\left[
b_{9}
\left(
\ol \cB  \cB \cA_+
\right)
+
b_{10}
\left(
\ol \cB  \cA_+ \cB
\right)
\right] 
v_{\mu} \partial_\nu
F^{\mu \nu} \, \str(\cQ_+^{SU(2)}) 
\notag \\
&&+
\frac{b_{11}}{\L_\chi}  
\left(
\ol \cB  \cB \right)
\, v_{\mu} \partial_\nu
F^{\mu \nu} \, \str(\cQ_+^{SU(2)} ) \, \str(\cA_+)
\notag \\
&&+ i 
\left[
b_9^\prime
\left(
\ol \cB [S_\mu,S_\nu] \cB \cA_+
\right)
+
b_{10}^\prime
\left(
\ol \cB [S_\mu,S_\nu] \cA_+ \cB
\right)
\right] F^{\mu \nu} \, \str(\cQ_+^{SU(2)}) \notag \\
&&+
i \, b_{11}^\prime  
\left(
\ol \cB [S_\mu,S_\nu] \cB \right)
F^{\mu \nu} \, \str(\cQ_+^{SU(2)})\,\str( \cA_+)
\label{fa-eqn:baryonops2},\end{eqnarray}
and additional vector-current correction operators
\begin{equation}
\cL =
\frac{a \, c_{A,\gamma}}{\L_\chi} 
 \left(
\ol \cB \cB \right)
\, v_{\mu} \partial_\nu
F^{\mu \nu} \,
\str (\cQ^{SU(2)} c_1^{SU(2)})
+
\frac{i a \, \mu_{A,\gamma}}{2} 
\left(
\ol \cB [S_\mu,S_\nu] \cB 
\right)
F^{\mu \nu} \str (\cQ^{SU(2)} c_1^{SU(2)})
\label{fa-eqn:baryonOops2}.\end{equation}

\subsection*{Decuplet baryons}
Next in the decuplet sector there are terms that result from $\cA_+$ insertions
\begin{eqnarray}
\cL &=&
\frac{ d_5}{\L_\chi}  
\left(
\ol \cT {}^\sigma \cA_+ \cT_\sigma
\right) 
v_\mu \partial_\nu F^{\mu \nu} \, \str(\cQ_+^{SU(2)}) 
+ 
\frac{d_6}{\L_\chi}  
\left(
\ol \cT {}^\sigma \cT_\sigma 
\right)
v_\mu \partial_\nu F^{\mu \nu} \, \str( \cQ_+^{SU(2)} ) \, \str( \cA_+)
\notag \\
&&+
i \, d_5^\prime  
\left(
\ol \cT_\mu \cA_+ \cT_\nu
\right) 
F^{\mu \nu} \, \str(\cQ_+^{SU(2)}) 
+ 
i \, d_6^\prime  
\left(
\ol \cT_\mu \cT_\nu 
\right)
F^{\mu \nu} \, \str( \cQ_+^{SU(2)}) \, \str( \cA_+)
\notag \\
&&+
\frac{d_{5}^{\prime\prime}}{\L_\chi} 
\left(
\ol \cT {}^{\{\mu} \cA_+ \cT^{\nu\}}
\right) 
v^\a \partial_\mu F_{\nu \a} \, \str(\cQ_+^{SU(2)}) 
\notag \\
&&+ 
\frac{d_{6}^{\prime\prime}}{\L_\chi}  
\left(
\ol \cT {}^{\{\mu} \cT^{\nu\}} 
\right)
v^\a \partial_\mu F_{\nu \a} \, \str( \cQ_+^{SU(2)})\, \str(\cA_+)
\notag \\
&&
\label{fa-eqn:decupletops2}\end{eqnarray}
and also further vector-current correction operators
\begin{eqnarray}
\cL &=&
\frac{3 a \, c^\prime_{A,\gamma} }{\L_\chi} 
\left(
\ol \cT {}^\sigma  \cT_\sigma
\right) 
v_\mu \partial_\nu F^{\mu \nu}\,
\str (\cQ^{SU(2)} c_1^{SU(2)})
+
3 i a
\,  
\mu^\prime_{A,\gamma}
\left(
\ol \cT_\mu \cT_\nu
\right) F^{\mu \nu}\,
\str (\cQ^{SU(2)} c_1^{SU(2)})
\notag \\
&&- 
\frac{3 a \, \mathbb{Q}_{A,\gamma}}{\L_\chi}
\left(
\ol \cT {}^{\{\mu} \cT^{\nu \}}
\right)
v^\a \partial_\mu F_{\nu \a}\,
\str (\cQ^{SU(2)} c_1^{SU(2)})
\label{fa-eqn:decupletOops2}.\end{eqnarray}

\subsection*{Baryon transitions}
Finally for the transitions, there are only new $\cA_+$ insertions
\begin{eqnarray}
\cL &=&
i t_6 
\left(
\ol \cB S_\mu \cA_+ \cT_\nu
\right) 
F^{\mu \nu} \, \str(\cQ_+^{SU(2)}) 
+ 
\frac{t_{6}^\prime}{\L_\chi} 
\left(
\ol \cB S^{\{\mu} \cA_+ \cT^{\nu\}}
\right) 
\, v^\a \partial_\mu F_{\nu \a}
\, \str(\cQ_+^{SU(2)}) 
\notag \\
&&+
\frac{i t_{6}^{\prime\prime}}{\L_\chi^2} 
\left(
\ol \cB S_\mu \cA_+ \cT_\nu
\right) 
\, \partial^\a \partial^\mu F^\nu{}_{\a}
\, \str(\cQ_+^{SU(2)})
\label{fa-eqn:transops2}.\end{eqnarray}

For each electromagnetic observable considered above,
contributions from all $\cO(a)$ operators in the effective theory are 
of higher order than the one-loop results in the chiral expansion. 
Thus one need only retain the finite lattice spacing corrections to the meson
masses and use the previously found expressions for electromagnetic
properties in $SU(2)$ \PQCPT\
in Appendices~\ref{baryon_ff-sec:SU2}, 
\ref{decuplet-s:su2}, and \ref{trans-s:su2}, as well as 
Refs.~\cite{Beane:2002vq,Beane:2003xv}.

\section{Coarse-Lattice Power Counting}
\label{fa-s:coarse}
Here we detail the $\cO(a)$ corrections to electromagnetic 
properties in an alternate power-counting scheme. 
We imagine a sufficiently coarse lattice, where
$a \L_\chi$ can be treated as $\cO(\e)$, so that%
\footnote{
This power counting coupled with the chiral expansion
is most efficient for valence Ginsparg-Wilson quarks
where $\cO(a)$ corrections vanish.
We thank Gautam Rupak for pointing this out.
}
\begin{equation}
\e^2 \sim 
\begin{cases}
 m_q/\L_\chi, \\
 a^2 \L_\chi^2, \\ 
 p^2/\L_\chi^2
\end{cases}
.\end{equation}
In this case, there are known additional 
$\cO(a^2)$ corrections~\cite{Bar:2003mh} to the meson masses that are now at $\cO(\e^2)$
and must be included in expressions for loop diagrams.
The free Lagrangian for $\cB_{ijk}$ and $\cT^\mu_{ijk}$ fields contains additional terms of $\cO(a^2)$ that
correct the baryon masses,
and modify the kinetic terms.
Potential contributions due to the latter, whatever their form, must be canceled by wavefunction renormalization diagrams.
The only contribution of  $\cO(a^2)$ could come from tree-level electromagnetic terms
but these are necessarily higher order.
Thus in this power counting there are no unknown $\cO(a^2)$ corrections
for electromagnetic properties.

The only possible corrections come from the $\cO(a)$ operators assembled above.
A few of these do contribute at tree level
and are spelled out below.

\subsection*{Octet mesons}
The $\cO(a)$ corrections to the meson form factors are now $\cO(\e^3)$ in the power counting. 
While the meson charge radii at NLO in the chiral expansion are at $\cO(\e^2)$, further corrections in the 
chiral expansion are at $\cO(\e^4)$. Thus one can use the $\cO(a)$ operators 
to completely deduce the charge radii to $\cO(\e^3)$
[apart from $\cO(\e^3)$ corrections to the meson masses]. 
These $\cO(a)$ operators are given in Eqs.~\eqref{fa-eqn:mesonops} and \eqref{fa-eqn:mesonOops} and yield a correction
$\delta < r_E^2 >$ to the meson charge radii of the form
\begin{equation} \label{fa-eqn:mesona}
\delta < r_E^2 > = Q \, \frac{24 a \L_\chi}{f^2}
\left[
c^v ( 2 m_1 + m_2) + 3 c^s m_3 + c_1^v \a_{A,9}
\right]
\end{equation}
Notice that there are no corrections associated with an unimproved current operator
in the sea sector since $c_1^s$ is absent.

In the case of $SU(2)$ flavor, there is an additional contribution from the operator in Eq.~\eqref{fa-eqn:mesonops2}.
At tree level, however, this operator vanishes. The only correction to Eq.~\eqref{fa-eqn:mesona} in changing to $SU(2)$ flavor is
to replace $3 c^s$ with $2 c^s$ which reflects the change in the number of sea quarks.

\subsection*{Octet baryons}
For the octet baryon electromagnetic properties, the $\cO(a)$ corrections to the charge radii are now $\cO(\e^3)$
and can be dropped
as they are the same order as neglected $1/M_B$ corrections.
The magnetic moments, however, do receive corrections from local operators. 
Specifically, the $\cO(a)$ operators which contribute to magnetic moments at $\cO(\e)$ 
are insertion of $\cA_+$ into the magnetic moment operator given in Eq.~\eqref{fa-eqn:baryonops} and $\cO_1^\mu$
corrections given in Eq.~\eqref{fa-eqn:baryonOops}. Calculation of these corrections yields a shift in the magnetic moments
\begin{eqnarray}
\delta \mu &=& a M_B \Bigg\{ 
c^v \left[ 
A \left( b^\prime_1 + \frac{1}{2} b^\prime_4 \right)   
- B \left( 2 b^\prime_2 + b^\prime_3 - b^\prime_5  \right)
\right]
+ 3 c^s \left(\frac{1}{2} A \, b^\prime_6  - B \, b^\prime_7  \right)  
\notag \\
&&+ C (c^s - c^v) q_{jlr} \, b^\prime_8 
+ \frac{c_1^v}{2} \left[ \frac{1}{2} \mu_{A,\a} A - \mu_{A,\b} B  \right]
\Bigg\}
,\end{eqnarray}
where $q_{jlr} = q_j + q_l + q_r$.
The coefficients $A$ and $B$ are listed for octet baryons in 
Table~\ref{fa-t:AB}, 
\begin{table}[tb]
\centering
\caption[$A$ and $B$ for the octet baryons]{\label{fa-t:AB}The coefficients $A$ and $B$ for the octet baryons.}  
\begin{tabular}{l |  c    |   c  }
	\hline
	\hline
	     & $A$ & $B$  \\
	\hline
	$p$        	& $1$ 			& $0$ \\
	$n$     	& $-\frac{1}{3}$ 	& $-\frac{1}{3}$ \\
	$\Sigma^+$     	& $1$ 			& $0$ \\
	$\Sigma^0$     	& $\frac{1}{6}$ 	& $\frac{1}{6}$ \\
	$\Lambda$     	& $-\frac{1}{6}$ 	& $-\frac{1}{6}$ \\
$\Sigma^0 \Lambda$	& $\frac{1}{2 \sqrt{3}}$ 	& $\frac{1}{2 \sqrt{3}}$ \\
	$\Sigma^-$     	& $-\frac{2}{3}$ 	& $\frac{1}{3}$ \\
	$\Xi^0$     	& $-\frac{1}{3}$ 	& $-\frac{1}{3}$ \\
	$\Xi^-$     	& $-\frac{2}{3}$ 	& $\frac{1}{3}$ \\
	\hline
	\hline
\end{tabular}
\end{table} 
while $C = 1$ for all octet magnetic moments
and $C = 0$ for the $\L\Sigma^0$ transition moment. 
Notice that there are no corrections associated with an unimproved current operator
in the sea sector.

In the case of $SU(2)$ flavor, there are additional contributions given in Eqs.~\eqref{fa-eqn:baryonops2} and \eqref{fa-eqn:baryonOops2}.
For the proton and neutron, we have
\begin{eqnarray}
\delta \mu^{SU(2)} &=& a M_B \Bigg\{ 
c^v \left[
A 
\left( b^\prime_1 + \frac{1}{2} b^\prime_4 \right)   
- 
B 
\left( 2 b^\prime_2 + b^\prime_3 - b^\prime_5  \right)
+ 
\frac{1}{3} \left( b^\prime_9 + b^\prime_{10} \right)
\right]
\notag \\
&&+
2 c^s
\left( \frac{1}{2} A \, b^\prime_6 -  B \, b^\prime_7 + \frac{1}{3} b^\prime_{11} \right)
+ 
\left[  
c^s q_{jl} + c^v \left( \frac{1}{3} - q_{jl} \right)
b^\prime_8
\right]
\notag \\
&&+ 
\frac{c_1^v}{2} 
\left[
\frac{1}{2} A \, \mu_{A,\a} - B \, \mu_{A,\b} + \left( \frac{1}{3} - q_{jl} \right) \mu_{A,\gamma}
\right]
+ 
\frac{c_1^s}{2} \, q_{jl} \, \mu_{A,\gamma}
\Bigg\}
,\end{eqnarray}
where $q_{jl} = q_j + q_l$.

\subsection*{Decuplet baryons}
For the decuplet baryon electromagnetic properties in coarse-lattice power counting, the $\cO(a)$ corrections to the charge radii are $\cO(\e^3)$
and the corrections to the electric quadrupole moments are $\cO(\e)$, both of which are higher order than the one-loop results. 
The magnetic moments, however, do 
receive corrections from local operators. Specifically, the $\cO(a)$ operators which contribute to magnetic moments at $\cO(\e)$ 
are $\cA_+$ insertions into the magnetic moment operator given in Eq.~\eqref{fa-eqn:decupletops} and $\cO_1^\mu$
correction operators given in Eq.~\eqref{fa-eqn:decupletOops}. Calculation of these corrections yields a shift in the magnetic moments
\begin{equation}
\delta \mu = 2 a M_B
\left[ 
\frac{1}{3} c^v Q \, ( 2 d_1^\prime + d_2^\prime )  + c^s Q \, d_3^\prime
+ ( c^s - c^v) q_{jlr} d^\prime_4 + c_1^v Q \, \mu_{A,c}
\right]
.\end{equation}
Notice that in $SU(3)$ $\str \cQ=0$, 
hence there is no dependence on $c_1^s$ in the above result.

In the case of $SU(2)$ flavor, there are additional contributions given in Eqs.~\eqref{fa-eqn:decupletops2} and \eqref{fa-eqn:decupletOops2}.
The corrections to the $\D$ quartet magnetic moments are then
\begin{eqnarray}
\delta \mu^{SU(2)} 
&=& 
2 a M_B
\Bigg\{
\frac{1}{3} 
c^v \left( 2 Q d_1^\prime + Q d_2^\prime + d_5^\prime  \right)
+ \frac{2}{3} c^s \left( Q d_3^\prime + d_6^\prime  \right)
+ \left[ c^s q_{jl} + c^v \left( \frac{1}{3} - q_{jl} \right) \right] d_4^\prime
\notag \\
&&+
c_1^v \left[ Q \mu_{A,c} + ( 1 - 3 q_{jl}) \mu^\prime_{A,\gamma} \right]
+
3 c_1^s  q_{jl}  \mu^\prime_{A,\gamma}
\Bigg\}
\end{eqnarray}

\subsection*{Baryon transitions}
For the decuplet to octet electromagnetic transitions in coarse-lattice power counting, the $\cO(a)$ corrections to $G_2(0)$ and $G_3(0)$ are $\cO(\e)$
which are of higher order than the one-loop results.  The $G_1(q^2)$ form factor does, however, 
receive corrections from local operators. Specifically these $\cO(a)$ operators which contribute to $G_1(0)$  at $\cO(\e)$ 
are the insertions of $\cA_+$ into the magnetic dipole transition operator given in Eq.~\eqref{fa-eqn:transops} and the vector-current corrections
given in Eq.~\eqref{fa-eqn:transOops}. Calculation of these corrections yields a shift of $G_1(0)$
\begin{equation} \label{fa-eqn:transa}
\delta G_1(0) = a M_B \, \a_T \sqrt{\frac{2}{3}} 
\left\{  
c^v \left( t_1 + t_2 + t_3 - \frac{1}{2} t_4 \right)
+ 
3 c^s \, t_5
+ 
c_1^v \, \mu_{A,T} \sqrt{\frac{3}{8}} 
\right\}
,\end{equation}
where the transition coefficients $\a_T$ appear in \cite{Arndt:2003vd}.
Again, at this order the result is independent of $\cO(a)$ improvement
to the electromagnetic current in the sea sector.
In the case of $SU(2)$ flavor, there is an additional dipole operator given in Eq.~\eqref{fa-eqn:transops2}. At tree level, however, this operator vanishes.
The only correction to Eq.~\eqref{fa-eqn:transa} in changing to $SU(2)$ flavor is
to replace $3 c^s$ with $2 c^s$ which reflects the change in the number of sea quarks.

\chapter*{Vita}
%\addcontentsline{toc}{chapter}{Vita}
\thispagestyle{empty}

Daniel Arndt graduated with 
a {\it Master of Science} 
degree in Physics
from North Carolina State University, Raleigh in May 1999 
and with 
a
{\it Diplom} in Physics from Dresden University of Technology,
Dresden (Germany)  in December 1999.
He completed his 
{\it Ph.D.}
in Physics at the University of Washington, Seattle
in June 2004.

\end{document}